\documentclass[aps,pra,a4paper,showpacs,twocolumn,longbibliography,english,10pt,superscriptaddress]{revtex4-2}

\usepackage[dvips]{graphicx}
\usepackage{amsmath,amssymb}
\usepackage{amsfonts}
\usepackage{hyperref}
\usepackage[table,dvipsnames]{xcolor}
\usepackage[retainorgcmds]{IEEEtrantools}

\usepackage[normalem]{ulem}  
\usepackage{placeins}  
\usepackage{soul}      
\usepackage{comment}

\begin{document}
\newcommand{\Eq}[1]{Eq.~\eqref{#1}}
\newcommand{\Eqs}[1]{Eqs.~\eqref{#1}}
\newcommand{\diff}{\text{d}}
\newcommand{\ui}{\text{i}}
\newcommand{\iexp}{\text{i}}
\newcommand{\im}{\text{Im}}
\newcommand{\re}{\text{Re}}

\newcommand{\sys}{\text{sys}}
\newcommand{\eff}{\text{eff}}
\newcommand{\nlin}{\text{nlin}}
%
\newcommand{\sLC}{LC}
\newcommand{\thresh}{\text{th}}
\newcommand{\In}{\text{in}}
\newcommand{\Out}{\text{out}}
\newcommand{\wref}{\omega_{\text{ref}}}

\newcommand{\flqu}{\frac{\Phi_0}{2\pi}}
\newcommand{\iflqu}{\frac{2\pi}{\Phi_0}}
\newcommand{\uGHz}{\,\text{GHz}}
\newcommand{\uMHz}{\,\text{MHz}}
\newcommand{\ums}{\,\text{ms}}
\newcommand{\uns}{\,\text{ns}}
\newcommand{\umicroA}{\,\mu \text{A}}
\newcommand{\umicrom}{\,\mu \text{m}}
\newcommand{\ufF}{\,\text{fF}}
\newcommand{\unH}{\,\text{nH}}
\newcommand{\todo}[1]{{[\color{purple} {\bf todo:} #1]}}
\newcommand{\zheng}{\color{green}}

\title{
Modeling of simple bandpass filters: bandwidth broadening of Josephson parametric devices due to non-Markovian coupling to dressed transmission-line modes}

\author{R.~Yang}
\email{ruiyang@mail.sim.ac.cn	
}
\affiliation{State Key Laboratory of Materials for Integrated Circuits, Shanghai Institute of Microsystem and Information Technology, Chinese Academy of Sciences, 865 Changning Road, Shanghai 200050 China }

\author{Z.~Shi}
\affiliation{Institute for Quantum Computing and Department of Electrical \& Computer Engineering, University of Waterloo, Waterloo, ON N2L 3G1, Canada}

\author{Z.R.~Lin}
\email{zrlin@mail.sim.ac.cn}
\affiliation{State Key Laboratory of Materials for Integrated Circuits, Shanghai Institute of Microsystem and Information Technology, Chinese Academy of Sciences, 865 Changning Road, Shanghai 200050 China }
\affiliation{Shanghai Key Laboratory of Superconductor Integrated Circuit Technology, Shanghai Institute of Microsystem and Information Technology, Chinese Academy of Sciences, Shanghai 200050, China}

\author{W.~Wustmann}
\email{waltraut.wustmann@otago.ac.nz
}
\affiliation{Department of Physics and The Dodd-Walls Centre for Photonic and Quantum Technologies, University of Otago, Dunedin 9016, New Zealand}

\date{\today}

\begin{abstract}
Josephson parametric devices are widely used in superconducting quantum computing research but suffer from an inherent gain-bandwidth trade-off. 
This limitation is partly overcome by coupling the device to its input/output transmission line via a bandpass filter, leading to wider bandwidth at undiminished gain.
Here we perform a non-perturbative circuit analysis in terms of dressed transmission-line modes for representative resonant coupling circuits,
going beyond the weak-coupling treatment. 
The strong frequency dependence of the resulting coupling coefficients implies that the 
Markov approximation commonly employed in cQED analysis is inadequate.  
By retaining the full frequency dependence of the coupling, we arrive at a non-Markovian form of the quantum Langevin equation with the frequency-dependent complex-valued self-energy of the coupling in place of a single damping parameter. 
We also consistently generalize the input-output relations and unitarity conditions. 
Using the exact self-energies of elementary filter networks -- a series- and parallel-$\sLC$ circuit and a simple representative bandpass filter consisting of their combination  --
we calculate the generalized parametric gain factors. 
Compared with their Markovian counterpart, 
these gain profiles are strongly modified. 
We find bandwidth broadening not only in the established parameter regime, where the self-energy of the coupling is in resonance with the device and its real part has unity slope,
but also within off-resonant parameter regimes where the real part of the self-energy is large. 
Our results offer insight for the bandwidth engineering of Josephson parametric devices using simple coupling networks. 
\end{abstract}

\maketitle 


\section{Introduction}

\subsection{Background}

Superconducting quantum architectures rely heavily on Josephson parametric devices (JPD) \cite{JPA_StateoftheArt, ClerkETAL_review2010, KrantzETAL2019_review}.
First and foremost, they are used to amplify the weak microwave signals
containing readout information about the state of a qubit
\cite{ClerkETAL_review2010, KrantzETAL2019_review, LehnertETAL2007, LehnertETAL2008, YamamotoETAL2008, WalterETAL2017}.
Other prominent applications are for directional coupling
\cite{SliwaETAL2015, ChapmanETAL2017b, LecocqETAL2017, LecocqETAL2021, AbdoETAL2021},
noise squeezing \cite{YurkeETAL1988, EichlerETAL2011, OBrienETAL2023},
frequency conversion \cite{BergealETAL2010b, LecocqETAL2017, NaamanAumentado2022},
and entanglement generation \cite{BraunsteinETAL2005, FlurinETAL2012, PetrovninETAL2022}.
JPDs may be implemented either in resonant form,
based on a microwave resonator that incorporates few Josephson junctions (JJs),
or as a non-resonant, traveling-wave device 
where the signal passes through an extended nonlinear waveguide.
Some applications specifically rely on resonant JPDs,
in particular when operated in or near their highly nonlinear parametric oscillation regime 
\cite{WusShu2013, WusShu2017, SvenssonETAL2017, SvenssonETAL2018}.
While parametric amplification is  unstable in this regime,
it allows for direct qubit state discrimination,
either via mapping the qubit state onto one of two large-amplitude oscillation states \cite{YamamotoETAL2014, RosenthalETAL2021},
or by exploiting the qubit-state dependent dispersive shift of the resonator frequency
across the parametric oscillation threshold \cite{JBA, JBA2, KrantzETAL2016_NatComm}.
The parametric oscillation regime is also exploited in bosonic encodings \cite{Kerrcatqubit, Kerrcatqubit2} and for quantum annealing \cite{PuriETAL2017_JPO}.

\textcolor{black}{In its simplest form, the parametric resonator is coupled, e.g.~via a capacitor, to the transmission line (TL) which carries the input and output signals.
In this simple design, the bandwidth of the gain scales inversely
with its maximum amplitude,} and this bandwidth limitation necessitates a careful
tuning of the pump frequency for a particular signal frequency.
In contrast, the main benefit of the Josephson traveling-wave parametric amplifiers (JTWPAs) \cite{MacklinETAL2015, WhiteETAL2015}
is their large bandwidth which allows to process signals from a wide frequency
range and thus can enable multiplexed qubit readout \cite{ChenETAL2012, JergerETAL2012, ChapmanETAL2017}.
On the other hand, JTWPAs consist of hundreds of JJs
and place very high demands on fabrication accuracy.
Therefore, resonant JPDs with their relatively simple design and comparably low fabrication demands
remain important tools in superconducting quantum research and development.

Moreover, schemes to overcome the bandwidth limitation of resonant JPDs
have been developed and are now routinely implemented \cite{RoyETAL2015, MutusETAL2014, GrebelETAL2021, DuanETAL2021, WhiteETAL2023}.
The core principle of this bandwidth engineering is to
couple the JPD and the TL by a filter
that acts as a frequency-dependent environment to the JPD.
\textcolor{black}{In practice, such filters can be implemented by
continuous microwave devices such as TL stubs~\cite{pozar2011microwave}; for example, Ref.~\cite{RoyETAL2015} employs a combination of a $\lambda/2$-cavity and a $\lambda/4$-cavity.}
Their basic functionality can be modeled by lumped-element
coupling circuits, e.g.~a series-$\sLC$ resonator in the simplest case.
Further bandwidth widening can be achieved via more complex,
multi-pole bandpass filters known from electrical engineering \cite{NaamanAumentado2022}.
A filter may consist of a ladder of alternating parallel-$\sLC$ and series-$\sLC$ resonators, or a train of capacitively-coupled, parallel-$\sLC$
resonators \cite{NaamanAumentado2022} (see~Fig.~\ref{fig:filters}).

The pronounced resonant behavior of such resonator-based couplings,
\textcolor{black}{compared with simple capacitive coupling},
calls into question some approximations frequently invoked in cQED analysis.
In particular, the quantum Langevin dynamics is no longer adequately characterized
by a single, frequency-independent parameter (a damping rate),
resulting from the Markov approximation \cite{GarZol_book} which assumes the coupling is local in time. 
When we go beyond the Markov approximation
and retain the frequency dependence of the coupling constants,
the damping rate is replaced by a complex-valued and frequency-dependent self-energy
which describes both damping and frequency shift \cite{BruusFlensberg_book}.
\textcolor{black}{While the Markov approximation is usually justified in the weak-coupling limit \cite{GarZol_book}},
the frequency-dependent quantum Langevin equation in principle applies
also in regimes of stronger coupling.

The derivation of a well-defined non-Markovian model with non-divergent coupling coefficients is not trivial.
As we show in this paper, the frequency-dependent quantum Langevin equation is straightforward to derive if the 
coupling is described by a generic bilinear interaction Hamiltonian between 
the resonator amplitude(s) and the TL mode amplitudes, weighted by coupling coefficients. 
However,
the analysis of a concrete coupling circuit starts from the Lagrangian description in terms of node fluxes and charges. Under the frequently invoked weak-coupling assumption that the TL modes are unaffected by the coupling, a naive approach to obtain the interaction Hamiltonian from the Lagrangian may lead to divergent coupling coefficients.
These problems are discussed in detail in Ref.~\onlinecite{ParraRodriguezETAL2018},
where the authors also review various cQED approaches
to this common coupling situation.
The authors' solution is to allow the TL modes to be dressed by the parameters of the coupling.
The dressing is embodied by the boundary condition,
which is placed on the TL modes by the coupling circuit
and which involves characteristic length scale(s) of the coupling.
These finite length(s) in turn give rise to a natural intrinsic
cut-off of the coupling coefficients, thus avoiding divergences.
In case of coupling circuits with a strong (resonant) frequency dependence,
the possible divergence of the coupling coefficients is particularly problematic.
We therefore adopt an approach similar to that developed in Ref.~\onlinecite{ParraRodriguezETAL2018}
and allow for the TL to be dressed by the parameters of the coupling.
Using the dressed TL modes, the Hamiltonian for the coupled system
with the bilinear interaction term is then found non-perturbatively,
without having to restrict to a weak-coupling limit.
Note that such dressed continuum modes also appear under the name `scattering basis' in other contexts involving quantum impurities~\cite{MatveevETAL1993, YueETAL1994, LalETAL2002, DasETAL2004, NazarovGlazman2003, PolyakovGornyi2003, ShiAffleck2016, ShiKomijani2017}.

\subsection{Our approach and results}

We perform the dressed-TL mode analysis for several coupling circuits,
starting with the series-$\sLC$ coupling that can act as a basic band-pass filter
and captures the mechanism of JPD-bandwidth broadening \cite{RoyETAL2015}.
Surprisingly, series-$\sLC$ coupling has not been \textcolor{black}{studied in Ref.~\onlinecite{ParraRodriguezETAL2018}, though it was discussed briefly in Ref.~\cite{parrarodriguez_PhD_thesis} within the context of nonreciprocal circuits~\cite{ParraRodriguez2019canonical,ParraRodriguez2022canonical,Egusquiza2022algebraic,ParraRodriguez2025exact}}.
For comparison, we also briefly discuss the case of a parallel-$\sLC$ coupling circuit,
which has already been analyzed in Ref.~\onlinecite{ParraRodriguezETAL2018}.
Under resonant coupling -- when coupling and system resonance are aligned --
we find it to act as a band-stop filter, with the coupling rate
dropping to zero at the system frequency.
For completeness, we also present the results for
the purely capacitive and purely inductive coupling as limiting
cases of the parallel-$\sLC$ coupling.
Since these couplings have no internal resonance,
the frequency dependence of their coupling coefficients is relatively weak.
As a more complex coupling situation we consider a simple ladder filter
consisting of a series-$\sLC$ resonator followed by a parallel-$\sLC$ resonator.
In networks like these, the terminal resonator of the coupling circuit,
i.e.~the one facing the TL (e.g.~the parallel-$\sLC$ or series-$\sLC$ resonator),
determines the boundary condition for the dressed TL modes.
The remaining resonators of the coupling circuit can be treated
together with the system of interest, such as a JPD resonator, as a hybridized system.
\textcolor{black}{
This is similar to the coupled-mode treatment of filters \cite{AumentadoETAL2015, NaamanAumentado2022, coupled_modes_Purcell_filter}
which approximates a filter as an array of coupled $\sLC$ resonators situated between the TL and the system, and performs the Markov approximation on the resonator directly coupled to the TL. However,
our approach does not invoke the Markov approximation and is thus
more broadly applicable, for instance in stronger coupling situations.}

\textcolor{black}{
After deriving the Hamiltonian of the dressed-TL modes with bilinear interaction term, we analyze the resulting self-energies of the coupling. 
For the chosen coupling circuits, we can exactly solve the spectral integrals defining the self-energy. 
We also derive suitable resonant approximations which have more accessible analytic forms. 
The self-energy of the coupling is an important quantity to characterize the system response to input signals from the TL. 
While in the Markov approximation, 
its real and imaginary parts are assumed to be zero and constant, respectively,
the derived exact self-energies are frequency-dependent and allow to 
go beyond the Markov approximation. 
For the chosen resonator-based couplings, the frequency-dependence of the self-energy can then lead to significant deviations in the system response as function of the input frequency from the Markov approximation.
In particular, this is the case
whenever it changes rapidly with frequency at the system frequency, or if its real part is comparable in magnitude to the imaginary part there.  
On the other hand, if neither of these criteria is fulfilled, the system response approximately agrees with the Markov limit. 
}

The dressed-mode analysis is independent of the particular
finite system coupled to the TL.
Our results are thus relevant for a wide range of cQED settings,
in particular in strong-coupling situations.
Here, we are concentrating on the implications for JPD performance.
For concreteness, we restrict to the case
of a (lumped-element) single-mode Josephson Parametric Amplifier (JPA)
that is flux-pumped under degenerate parametric resonance \cite{RosenthalETAL2021, LinETAL2013_JPA, YamamotoETAL2014, KrantzETAL2016_NatComm, WusShu2013}.
Generalizations to `current-pumping' \cite{VijayETAL2011, BergealETAL2010b}
and to non-degenerate pumping of a multi-mode device \cite{BergealETAL2010b, SimoenETAL2015, WusShu2017, BengtssonETAL2018} are in principle straightforward.
\textcolor{black}{The JPA is operated as an amplifier of input signals in the regime below the parametric instability threshold~\cite{WusShu2013}, which we find to be modified by the frequency-dependent self-energy relative to the Markov limit. 
}
Using the series-$\sLC$ coupling, our dressed-mode analysis
confirms the established `resonant slope-matching' coupling condition
as suitable for bandwidth broadening \cite{RoyETAL2015}.
This condition is fulfilled when the coupling resonance
aligns with that of the system and at the same time the real part of the
coupling self-energy has unity slope.
Additionally, however, we find that even far off-resonant coupling can
allow favorable bandwidth broadening, 
as demonstrated for the series-$\sLC$ coupling
if the real part of the
self-energy at the system resonance is \textcolor{black}{of comparable magnitude to} the imaginary part.
Similarly, for the coupling via the simple ladder filter,
large bandwidth broadening is achieved
at finite detuning between the two hybridized system resonances,
while the pump is also significantly detuned.
These new coupling regimes widen the parameter scope for
practical devices and are worth further investigation.

This article is organized as follows:
Section~\ref{sec:summary_seriesLC_andfilter}
summarizes the dressed-mode diagonalization of the coupled resonator-TL system
for the series-$\sLC$ coupling circuit as well as the simple ladder filter coupling circuit,
giving their resulting coupling coefficients.
Derivation details are deferred to Appendices \ref{app:seriesLC_TLmodes} and \ref{app:parallelLC_beforeseriesLC_TLmodes}.
A similar analysis for the
case of purely parallel-$\sLC$ coupling is given in App.~\ref{app:parallelLC_TLmodes}.
In Sec.~\ref{sec:frequencydependentcoupling} the quantum Langevin
equations for the general case of frequency-dependent coupling are derived,
as well as the consistent input-output relations and the unitarity conditions. We also compare with the corresponding results
under the Markov approximation, i.e. when assuming frequency-independent coupling.
At the end of the section, we calculate the frequency-dependent self-energy -- the key to modeling beyond the Markov approximation -- corresponding to the coupling circuits in Sec.~\ref{sec:summary_seriesLC_andfilter}. Details of this calculation are relegated to App.~\ref{app:selfenergy_variouscouplings}, which also discusses the self-energy of the purely parallel-$\sLC$ coupling.
In Sec.~\ref{sec:paramp_frequencydependentcoupling} we apply our results to the particular case of a degenerate JPA.
The resulting parametric amplification properties are presented in Sec.~\ref{sec:results_bandwidthbroadening},
in comparison with their Markovian counterparts.

%

\section{Dressed-mode formalism for representative coupling circuits}
\label{sec:summary_seriesLC_andfilter}

\begin{figure}\centering
\includegraphics[width=0.95\columnwidth]{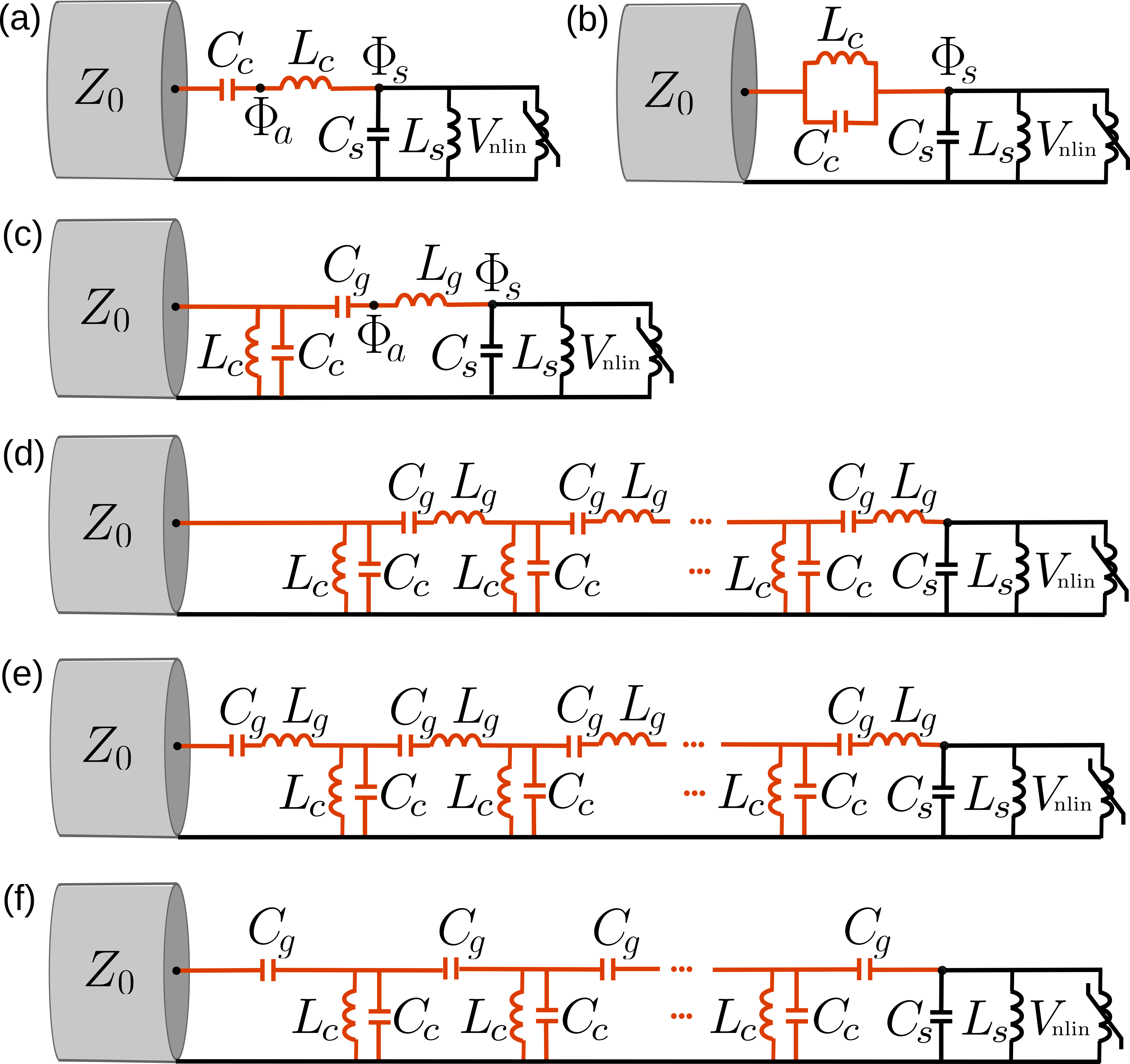}
\caption{
A (nonlinear) lumped element resonator (e.g.~a JPA)
coupled to a transmission line of impedance $Z_0$ via
various coupling networks (red):
(a) a series-$\sLC$ circuit,
(b) a parallel-$\sLC$ circuit,
(c) a simple ladder filter,
(d,e) extended ladder filters and (f) an extended parallel-$\sLC$ train filter.
}
\label{fig:filters}
\end{figure}

We consider a lumped-element system of interest, such as a resonator circuit, 
that is coupled to a TL, e.g. by one of the filters shown
in Fig.~\ref{fig:filters}.
The natural starting point of circuit analysis is usually the Lagrangian.
In the Lagrangian formalism,
the distinction between degrees of freedom of the
TL and the lumped-element system is quite natural,
in particular in a weak-coupling scenario.
For a quantum description of the circuit, however, one needs to perform a Legendre transformation
from the Lagrangian to the Hamiltonian.
As has been noted before~\cite{ParraRodriguezETAL2018} in similar scenarios,
where a subsystem with infinitely many modes (the TL) is coupled
to a finite-dimensional subsystem (the lumped-element system of interest), this transformation is problematic
as it requires the inversion of an infinite-dimensional capacitance matrix.

Our approach to this problem is to diagonalize the `dressed' TL Hamiltonian,
which differs from the `bare' TL Hamiltonian in that it incorporates
the influence of the lumped-element system and the coupling circuit.
As a result, the diagonalized TL modes are subject to a nontrivial boundary condition,
as determined by the coupling circuit. 
\textcolor{black}{These boundary conditions introduce finite characteristic length scales which reflect the dressing effect imposed by the coupling circuit.
In contrast, in the limit where these length scales are not finite (zero or infinite), free TL modes are obtained.}
This approach is comparable to that of Ref.~\cite{ParraRodriguezETAL2018},
where the authors consider (among other scenarios) a parallel-$\sLC$ coupling circuit,
as shown in Fig.~\ref{fig:filters}(b).
Here we analyze other standard coupling circuits in a comparable way,
specifically a series-$\sLC$ circuit, Fig.~\ref{fig:filters}(a),
as well as a simple ladder filter consisting of a series-$\sLC$ circuit
followed by a parallel-$\sLC$ circuit, Fig.~\ref{fig:filters}(c).

With the help of the dressed TL modes, the quantum Hamiltonian
of the coupled system can be formulated without invoking a weak-coupling approximation.
The resulting coupling coefficients \textcolor{black}{have a natural frequency cut-off and are thus not divergent,}
\textcolor{black}{as a result of the finite length scales in the dressed-mode boundary condition~\cite{ParraRodriguezETAL2018}.
(This is in contrast to the approximation of free TL modes, where the coupling coefficients are usually divergent.)
}
\textcolor{black}{For coupling circuits based on $\sLC$ resonators, the coupling coefficients 
have a pronounced resonance structure.}

\subsection{Series-$\sLC$ coupling}
\label{subsec:summary_seriesLC}

Here we consider a series-$\sLC$ circuit which connects a lumped-element resonator
and a TL,
\textcolor{black}{as depicted in Fig.~\ref{fig:filters}(a) (see also Fig.~\ref{fig:LCresonator_seriesLcCccoupled__discreteTL})}.
This section summarizes the main results of our analysis, which is described
in more detail in App.~\ref{app:seriesLC_TLmodes} and App.~\ref{sec:selfenergy_seriesCcLc}.

\subsubsection{Diagonalization of dressed TL Hamiltonian}

The TL is characterized by the flux field $\Phi(x)$ ($x>0$),
and the lumped-element system by one or more flux variables, in particular
the node flux $\Phi_s$ at the connection node to the coupling circuit, as shown in \textcolor{black}{Fig.~\ref{fig:filters}(a)}.
The Lagrangian of the composite system can be written as
\begin{align}
 \mathcal{L} &= \mathcal{L}_{\sys}^{(0)}(\Phi_s, \ldots) + \int_0^\infty \diff x
 \left[ \frac{C_0}{2} \dot \Phi^2 - \frac{1}{2L_0} (\Phi')^2 \right] \nonumber\\
\label{eqQ:Lagrangian_orig}
 &\quad+ \frac{C_c}{2} \left( \dot \Phi_a - \dot \Phi(0) \right)^2
 - \frac{1}{2 L_c} \left( \Phi_s - \Phi_a \right)^2
 \,,
\end{align}
where $L_0, C_0$ are the inductance and capacitance per unit length of the TL,
with wave velocity $v=1/\sqrt{L_0 C_0}$ and impedance $Z_0=\sqrt{L_0/C_0}$.

The Lagrangian in \Eq{eqQ:Lagrangian_orig} is written with an auxiliary
node flux $\Phi_a$ between the two components of the coupling circuit, 
see~\textcolor{black}{Fig.~\ref{fig:filters}(a)}.
However, when written in this form, the composite system is singular, as the Lagrangian EOMs for $\Phi(0)$ and $\Phi_a$ generate linearly dependent rows
in the capacitance matrix.
This singularity is lifted when we eliminate $\Phi_a$
by making use of the constraint from the two Langrangian EOMs,
$(\Phi_s - \Phi_a)/L_c = -\Phi'(0)/L_0$,
which simply manifests the current conservation through the coupling circuit.

For concreteness, we assume that the system node with node flux
$\Phi_s$ connects to ground via a capacitance $C_s$, and there is an (effective) inductance $L_s$ between $\Phi_s$
and the ground, such that the Lagrangian for the system of interest can be
written as
\begin{equation}\label{eqQ:Lagrangian_system}
\mathcal{L}_{\sys}^{(0)}(\Phi_s,\ldots)
= \frac{C_s}{2} \dot \Phi_s^2 - \frac{1}{2L_s} \Phi_s^2 + \ldots
\end{equation}
The omitted part may consist of bilinear contributions from all other
system degrees of freedom $\Phi_i$ ($i\neq s$) and their coupling with $\Phi_s$,
as well as a possible (weakly) nonlinear potential $V_\nlin(\Phi_s, \ldots)$.
For simplicity, we assume there are no other degrees of freedom besides $\Phi_s$, such that the bare system Hamiltonian can be obtained through a simple Legendre transformation of \Eq{eqQ:Lagrangian_system},
\begin{equation}\label{eqQ:Hsys_bare}
 H_{\sys}^{(0)} = \frac{1}{2 C_s} Q_s^2 + \frac{1}{2L_s} \Phi_s^2 + V_\nlin(\Phi_s)
 \,
\end{equation}
where $Q_s = C_s \dot \Phi_s$ is the conjugate charge of the system node flux $\Phi_s$.

Under the assumption \Eq{eqQ:Hsys_bare}, we can perform the Legendre transformation on the composite system (\Eq{eqQ:Lagrangian_orig} after eliminating $\Phi_a$); details of this transformation are found in App.~\ref{app:seriesLC_TLmodes}. It leads to a nonperturbative Hamiltonian of the form
\begin{equation}\label{eqQ:ham0}
    H = H_{\sys}+H_{\mathrm{TL}}+H_{\mathrm{cpl}},
\end{equation}
where $H_{\sys}$ describes the system degrees of freedom, $H_{\mathrm{TL}}$ describes the dressed TL degrees of freedom, and $H_{\mathrm{cpl}}$ describes the (bilinear) coupling between the dressed TL and the system (see \Eq{eqQ:Hamiltonian_definitionHtl}). Here, in case of the series-$\sLC$ coupling, there is no coupling-induced capacitive or inductive shift, such that 
\begin{equation}\label{eqQ:Hsys}
    H_{\sys} = H_{\sys}^{(0)} \,.
\end{equation}
(In contrast, for a purely inductive coupling or parallel-$\sLC$ coupling
the system Hamiltonian is shifted by $H_{\sys} - H_{\sys}^{(0)} = \Phi_s^2/(2 L_c)$, cf.~\Eq{eqP:Hamiltonian_definitionHtl}.)

A crucial step in our treatment is to diagonalize the dressed TL Hamiltonian $H_{\mathrm{TL}}$. One straightforward way to achieve this is to diagonalize the dressed TL Lagrangian $\mathcal{L}_{\mathrm{TL}}$, which is obtained through an inverse Legendre transformation of $H_{\mathrm{TL}}$, within the TL phase space alone. 
As a result, $\mathcal{L}_{\mathrm{TL}}$ depends on $\Phi(x)$, but not on $\Phi_s$ (see \Eq{eqQ:Lagrangian_TL_continuous}): 
\begin{align}\label{eqQ:Lagrangian_TL_continuous_main}
    \mathcal{L}_{\mathrm{TL}} &= \int_{0}^{\infty} \diff x \left[ \frac{C_0 }{2} \dot \Phi^2- \frac{1}{2 L_0} (\Phi')^2 \right]  \nonumber\\
 &\quad +\frac{\alpha C_0}{2} \left( \dot \Phi(0) - \beta \dot{\Phi}'(0) \right)^2 - \frac{\beta}{2 L_0} (\Phi'(0))^2
 \,,
\end{align}
with the coupling length parameters $\alpha, \beta$
defined as (using similar notation as in Ref.~\cite{ParraRodriguezETAL2018})
\begin{align}\label{eqQ:alpha_beta}
 \alpha = \frac{C_c C_s}{C_0 (C_c + C_s)} ,\ 
  \beta = \frac{L_c}{L_0}
 \;.
\end{align}
Equation~\eqref{eqQ:Lagrangian_TL_continuous_main} leads to the following boundary condition for the TL flux,
\begin{equation}\label{eqQ:boundarycondition}
 \alpha \left( \ddot \Phi(0) - \beta \ddot{\Phi}'(0) \right) - v^2 \Phi'(0) = 0 \,.
\end{equation}
The bulk dynamics determined by \Eq{eqQ:Lagrangian_TL_continuous_main}
is as usual treated by
expanding $\Phi(x)$ in a set of dressed TL modes,

\begin{equation}\label{eqAB:Phix_TLmodeexpansion}
 \Phi(x) = \int_0^\infty \diff k u_k(x) \Phi_k
 \,,
\end{equation}
thus separating the spatial mode dynamics $u_k''(x)=-k^2 u_k(x)$
from the equations of motion
$\ddot \Phi_k = -\omega_k^2 \Phi_k$ for the mode fluxes $\Phi_k$, 
with the mode frequencies $\omega_k = v k$.
Inserting expansion \Eq{eqAB:Phix_TLmodeexpansion} into the boundary condition \Eq{eqQ:boundarycondition} leads to the individual boundary conditions for the TL modes
\begin{equation}\label{eqQ:dressedTLmodes_boundarycondition}
 u_k'(0) + \alpha k^2  u_k(0) - \alpha \beta k^2 u'_k(0) = 0
 \,.
\end{equation}
From these boundary conditions, it is straightforward to show\textcolor{black}{~\cite{walter1973regular} } that the modes $u_k(x)$ 
\textcolor{black}{satisfy the orthogonality conditions
(\Eqs{eqQ2:orthogonality_alpha}--\eqref{eqQ2:orthogonality_beta}) 
that are needed to diagonalize $\mathcal{L}_{\mathrm{TL}}$ (see \Eq{eqQ:Lagrangian_TL}).} 
The corresponding Hamiltonian (again following from the Legendre transformation within the TL phase space alone) is 
\begin{align}
 H_{\mathrm{TL}} =
  \int_0^\infty \diff k \left( \frac{1}{2C_0} q_k^2 + \frac{k^2}{2 L_0} \Phi_k^2 \right) \,,
\end{align}
where $q_k = C_0 \dot \Phi_k$ are the canonical TL charges.
Inserting the ansatz $u_k(x) = \sqrt{2/\pi} \cos(k x + \varphi_k)$ into \Eq{eqQ:dressedTLmodes_boundarycondition},
we find the dressed TL modes as
\begin{align}
\label{eqQ:TLmode}
 u_k(x)
 &= u_k(0) \left( \cos kx + \left(\beta k - \frac{1}{\alpha k}\right)^{-1} \sin kx \right) \\
\label{eqQ:TLmodeamplitudes}
 u_k(0) &= \sqrt{\frac{2}{\pi}} \frac{\beta k - 1/(\alpha k)}{\sqrt{1+\left(\beta k - 1/(\alpha k)\right)^2}} \\
 \label{eqQ:TLmodephase}
\tan(\varphi_k) &= -\left(\beta k - \frac{1}{\alpha k} \right)^{-1}
\,.
\end{align}
Notably, for arbitrary $\alpha$ and $\beta$, there is always a resonant wave vector $k=1/\sqrt{\alpha\beta}$ where the dressed TL mode has a strong-coupling boundary condition such that $\varphi_k \to \pm \pi/2$. However, at any given $k$ of practical interest, we can achieve the weak-coupling limit by taking $C_c \to 0$ or $L_c \to \infty$ so that $k$ is far from the resonance. In this case, we recover vanishing phase shifts $\varphi_k\to 0$ corresponding to free TL modes $u_k(x) \to \sqrt{2/\pi} \cos(k x )$.

With the help of $u_k(x)$, we can also express the coupling Hamiltonian $H_{\mathrm{cpl}}$ in terms of the TL modes (see \Eq{eqQ:Hamiltonian_final}). Eventually, we arrive at the Hamiltonian of the composite system \Eq{eqQ:ham0} in the series-$\sLC$ case,
\begin{align}
 H &= H_{\sys}^{(0)}
 + \int_0^\infty \diff k
 \left( \frac{1}{2 C_0} q_k^2 + \frac{k^2}{2 L_0} \Phi_k^2  \right) \nonumber\\
\label{eqQ:Ham}
 &\quad + \frac{\alpha}{C_s} Q_s \int_0^\infty \diff k \left(u_k(0) - \beta u_k'(0)\right) q_k \,.
\end{align}
Importantly, by taking into account the proper dressing of the TL, the Hamiltonian \Eq{eqQ:Ham} is not limited to a weak-coupling situation.

\subsubsection{Quantum Hamiltonian and coupling coefficients}\label{sec:coef_selfenergy_seriesCcLc}

We now proceed to quantize the Hamiltonian \Eq{eqQ:Ham} and investigate the frequency dependence of the self-energy due to the coupling.
We introduce the amplitude $a_s$ for the system,
\begin{align}\label{eqQq:Phis_Qs_fromas}
 a_s =  \frac{1}{\sqrt{2 \hbar Z_s}} \Phi_s + \ui  \sqrt{\frac{Z_s}{2 \hbar}} Q_s
\end{align}
and 
the corresponding system frequency and impedance are
\begin{align}
\label{eq:ws_seriescoupled}
 \omega_s &= (C_s L_s)^{-1/2} \\
\label{eq:Zs_seriescoupled}
 Z_s &= (L_s/C_s)^{1/2}
 \,,
\end{align}
as determined by $H_{\sys}$, \Eq{eqQ:Hsys} (see App.~\ref{app:RF_fluxpumpedSQUID}).
Similarly, the TL mode amplitude is
\begin{align}
\label{eqQq:qk_phik_fromak}
a_k = \sqrt{\frac{\omega_k C_0}{2 \hbar}} \Phi_k + \frac{\ui}{\sqrt{2 \hbar \omega_k C_0}} q_k
\,.
\end{align}
The Hamiltonian \Eq{eqQ:Ham} then reads
\begin{align}
 H&= H_{\sys} + \int_0^\infty \diff k \hbar \omega_k a_k^\dag a_k \nonumber \\
\label{eqQq:Ham_orig}
 &\quad + \hbar \int_0^\infty \diff k f_k \left( a_k^\dag a_s + a_s^\dag a_k - a_k^\dag a_s^\dag - a_k a_s \right)
\end{align}
where the coupling coefficients are defined as
\begin{align}
 f_k &:= \sqrt{\frac{\omega_k C_0}{Z_s}} \frac{\alpha}{2C_s} \left(u_k(0) -\beta  u_k'(0)\right) \nonumber\\
\label{eqQq:def_fk}
&\quad = -\sqrt{\frac{ Z_s}{2\pi Z_0 }}\frac{\omega_s}{\sqrt{k}}\frac{1}{\sqrt{1+(\beta k - 1/(\alpha k))^2}}
\end{align}
with the dressed mode amplitudes $u_k(0)$ from \Eq{eqQ:TLmodeamplitudes}.

\subsection{Simple ladder filter
}
\label{subsec:summary_parallelLCbeforeseriesLC}

We now consider a coupling network that
consists of a series-$\sLC$ circuit (with parameters $C_g, L_g$)
followed by a parallel-$\sLC$ circuit (with parameters $C_c, L_c$),
as seen from the system side. \textcolor{black}{Figure~\ref{fig:filters}(c) shows a sketch of this network (see also Fig.~\ref{fig:LCresonator_parallelLcCc_beforeseriesLCcoupled__discreteTL}), which we refer to} as the `simple ladder filter'.
This section summarizes the main results of our analysis, which is described
in more detail in App.~\ref{app:parallelLC_beforeseriesLC_TLmodes}
and \ref{sec:selfenergy_parallelCcLc_beforeseriesLC}.
It is based on analysis of the parallel-$\sLC$ circuit
(Fig.~\ref{fig:filters}(b))
which is presented in App.~\ref{app:parallelLC_TLmodes}.

\subsubsection{Diagonalization of dressed TL Hamiltonian}

The Lagrangian of the composite system reads
\begin{align}
 \mathcal{L} &= \mathcal{L}_{\sys}^{(0)}(\Phi_s, \ldots)
 + \int_0^\infty \diff x
 \left[\frac{C_0}{2} \dot{\Phi}^2 - \frac{1}{2 L_0} (\Phi')^2 \right] \nonumber\\
 &\quad + \frac{C_c}{2} (\dot \Phi(0))^2 - \frac{1}{2 L_c} (\Phi(0))^2
 + \frac{C_g}{2} \left( \dot \Phi_a - \dot \Phi(0) \right)^2 \nonumber\\
\label{eqQ:Lagrangian_filter}
 &\quad - \frac{1}{2 L_g} \left( \Phi_s - \Phi_a \right)^2
 \,,
\end{align}
with the node fluxes as indicated in \textcolor{black}{Fig.~\ref{fig:filters}(c)}.
Here, in contrast to the original Lagrangian of the series-$\sLC$ coupling, \Eq{eqQ:Lagrangian_orig},
the composite system is not singular, i.e.~the auxiliary node flux $\Phi_a$
cannot be removed from the set of system coordinates.
Furthermore, in contrast to the purely series-$\sLC$ and the purely parallel-$\sLC$ cases, the simple ladder filter circuit remains closed even if the lumped-element system is replaced by an open circuit. As a result, we can cleanly divide the composite system into a dressed TL part, and a system part which contains both $\Phi_s$ and $\Phi_a$. The dressing of the TL is due to the parallel-$\sLC$ part
of the filter (with parameters $C_c, L_c$) alone, and is formally identical to
the dressing for the purely parallel-$\sLC$ coupling circuit of Fig.~\ref{fig:filters}(b),
as analyzed in App.~\ref{app:parallelLC_TLmodes}.
The only difference lies in the different
values of the coupling length parameter $\alpha$.
While for purely parallel-$\sLC$ coupling\textcolor{black}{,} $\alpha$ takes the same value as for the purely series-$\sLC$ coupling, \Eq{eqQ:alpha_beta}, 
here it is given by
\begin{align}\label{eqQ:alpha_beta__filter}
 \alpha = \frac{C_c}{C_0}
 ,\, \textcolor{black}{\text{and } } \beta = \frac{L_c}{L_0}
 \;.
\end{align}
This is due to the separation between the boundary TL flux $\Phi(0)$ and $\Phi_s$
brought about by the additional circuit elements $C_g, L_g$.
The coupling length $\beta$ is the same for all three cases.

As detailed in App.~\ref{app:parallelLC_beforeseriesLC_TLmodes}, we can again perform a Legendre transformation on the composite system to obtain a Hamiltonian \Eq{eqF:Hamiltonian_definitionHtl}, which is of the form \Eq{eqQ:ham0}. 
For the simple ladder filter, the system part of the Hamiltonian $H_{\sys}$ is modified by the additional $\Phi_a$ degree of freedom (see \Eq{eqF:Hsysext}):
\begin{equation}\label{eqQ:Hsys_orig__filter}
    H_{\sys}=H_{\sys}^{(0)}+ \frac{C_c + C_g}{2 C_c C_g} Q_a^2+  \frac{1}{2 L_g} \left(\Phi_s - \Phi_a\right)^2\, ,
\end{equation}
where the bare system Hamiltonian $H_{\sys}^{(0)}$ is given by \Eq{eqQ:Hsys_bare}. To diagonalize the dressed TL Hamiltonian $H_{\mathrm{TL}}$, we transform back to a Lagrangian
\begin{align}
  \mathcal{L}_{\mathrm{TL}} =&
  \int_{0}^{\infty} \diff x \left[\frac{C_0 }{2} \dot \Phi^2- \frac{1}{2 L_0} (\Phi')^2 \right]\nonumber\\
  &+ \frac{\alpha C_0}{2} \left( \dot \Phi(0) \right)^2 
 - \frac{1}{2 \beta L_0} (\Phi(0))^2\, ,
\end{align}
see \Eq{eqF:Lagrangian_TL_continuous}. This leads to the boundary condition for the TL flux $\Phi(x)$
\begin{equation}
 \alpha \ddot \Phi(0) + \frac{v^2}{\beta} \Phi(0) - v^2 \Phi'(0) = 0 \,,
\end{equation}
and, by virtue of the TL mode expansion \Eq{eqAB:Phix_TLmodeexpansion}, the boundary condition for the TL modes
\begin{equation}\label{eqQ:dressedTLmodes_boundarycondition__filter}
 u_k'(0) + \alpha k^2  u_k(0) - \frac{1}{\beta} u_k(0) = 0
 \,.
\end{equation}
\textcolor{black}{Equation~\eqref{eqQ:dressedTLmodes_boundarycondition__filter} leads to the orthogonality conditions, \Eqs{eqP2:orthogonality_alpha}--\eqref{eqP2:orthogonality_beta},
}which allow us to diagonalize $\mathcal{L}_{\mathrm{TL}}$ and thus $H_{\mathrm{TL}}$ in terms of $u_k$.

Inserting the ansatz $u_k(x) = \sqrt{2/\pi} \cos(k x + \varphi_k)$ into \Eq{eqQ:dressedTLmodes_boundarycondition__filter},
we find the dressed TL modes as
\begin{align}
\label{eqP:TLmode}
 u_k(x)
 &= u_k(0) \left( \cos kx - \left(\alpha k - \frac{1}{\beta k}\right) \sin kx \right) \\
\label{eqP:TLmodeamplitudes}
 u_k(0) &= \sqrt{\frac{2}{\pi}} \frac{1}{\sqrt{1+\left(\alpha k - 1/(\beta k)\right)^2}} \\
\label{eqP:TLmodephase}
 \tan(\varphi_k) &= \alpha k - \frac{1}{\beta k}
 \,.
\end{align}
Using $u_k$, we express the system-TL coupling Hamiltonian in terms of the dressed modes, and finally arrive at the non-perturbative Hamiltonian of the composite system:
\begin{align}
  H &= H_{\sys}
 + \int_0^\infty \diff k \left( \frac{1}{2C_0} q_k^2 + \frac{k^2}{2 L_0} \Phi_k^2 \right) \nonumber \\
\label{eqQ:Ham__filter}
 &\quad + \frac{\alpha}{C_c} Q_a \int_0^\infty \diff k u_k(0) q_k\, ,
\end{align}
where the modified system Hamiltonian $H_{\sys}$ is given in \Eq{eqQ:Hsys_orig__filter}. Note that the coupling Hamiltonian in \Eq{eqQ:Ham__filter}
only indirectly depends on the filter parameters via
the dressed mode amplitudes $u_k(0)$,
whereas the prefactor evaluates to $\alpha/C_c = 1/C_0$.
This is in contrast to \Eq{eqQ:Ham} (purely series-$\sLC$ coupling) or
\Eq{eqP:Hsys} (purely parallel-$\sLC$ coupling)
where the prefactors of the coupling Hamiltonian are explicitly coupling-dependent.

\subsubsection{Quantum Hamiltonian and coupling coefficients}\label{sec:coef_selfenergy_simpleladder}

We introduce the amplitudes $a_s$ and $a_a$ of the extended system
as in \Eq{eqQq:Phis_Qs_fromas},
where now the system frequencies and impedances are
\begin{align}
\label{eqF:ws}
 \omega_s &= \sqrt{\frac{L_s + L_g}{C_s L_s L_g}} \\
\label{eqF:Zs}
 Z_s &= \sqrt{\frac{L_s L_g}{C_s (L_s + L_g)}} \\
\label{eqF:wa}
 \omega_a &= \sqrt{\frac{C_c + C_g}{C_c C_g L_g}} \\
\label{eqF:Za}
 Z_a &= \sqrt{\frac{L_g (C_c + C_g)}{C_c C_g}}
 \,,
\end{align}
as determined by $H_{\sys}$, \Eq{eqQ:Hsys_orig__filter}.
Using also the TL mode amplitudes $a_k$ from \Eq{eqQq:qk_phik_fromak},
the Hamiltonian, \Eq{eqQ:Ham__filter}, 
then reads
\begin{align}
H
 &= H_{\sys} + \int_0^\infty \diff k \hbar \omega_k a_k^\dag a_k \nonumber \\
\label{eqFq:Ham_orig}
 &\quad- \hbar \int_0^\infty \diff k f_k (a_a - a_a^\dag)(a_k - a_k^\dag) \\
 H_{\sys}
 &= \hbar \omega_s a_s^\dag a_s + \hbar \omega_a a_a^\dag a_a
 - \hbar g (a_s + a_s^\dag) (a_a + a_a^\dag) \nonumber \\
\label{eqFq:Hsysext}
 &\quad+ V_{\nlin}(\Phi_s)
 \,,
\end{align}
where we have defined the coupling parameter between the amplitudes of the extended system,
\begin{equation}\label{eqFq:def_g}
 g := \frac{\sqrt{Z_s Z_a}}{2 L_g}
 = \frac{1}{2} \left( \frac{L_s}{(L_s + L_g) L_g^2} \frac{C_c + C_g}{C_s C_c C_g} \right)^{1/4}
 \,,
\end{equation}
and the coupling coefficients between the extended system and the TL,
\begin{align}\label{eqFq:def_fk_0}
 f_k &:= \sqrt{\frac{\omega_k C_0}{Z_a}} \frac{\alpha}{2C_c} u_k(0)
\end{align}
with the dressed mode amplitudes $u_k(0)$ from \Eq{eqP:TLmodeamplitudes}.

Both Hamiltonians \Eqs{eqQq:Ham_orig} and \eqref{eqFq:Ham_orig} describe a lumped-element system coupled to a dressed TL through an explicitly frequency-dependent coupling $f_k$. In the rest of this paper, we will analyze these Hamiltonians with a focus on the non-Markovian aspects of their dynamics.

\section{Non-Markovian analysis for frequency-dependent coupling}\label{sec:frequencydependentcoupling}

In cQED analysis, the dynamics of a system circuit of interest coupled to a TL
is usually formulated by the Langevin equation of motion for the mode operators
of the system \cite{GarZol_book}.
It allows to quantify the internal system response and --
in combination with an input-output relation
(which itself is a consequence of the Langevin dynamics) --
also the output field in response to an input signal.
Often, the frequency dependence of the coupling is neglected (Markov approximation),
under the assumption that correlations in the system's environment (the TL and coupling circuit) decay much faster than the time scale on which the system changes.

In Secs.~\ref{subsec:wdependentcoupling_systresponse}--\ref{subsec:wdependentcoupling_IOrelation}, with the frequency dependence of the coupling retained, we re-derive both the Langevin equation and the input-output relation. Our results highlight the self-energy due to the coupling to the TL, whose frequency dependence plays a key role beyond the Markov approximation. In Sec.~\ref{subsec:wdependentcoupling_unitarity}, we proceed to show that the usual commutation relations of the input and output fields are also modified by the frequency-dependent coupling.
In Sec.~\ref{subsec:wdependentcoupling_Markov} we show how the
system response, input-output relations, and commutation relations
reduce to their familiar form in the frequency-independent Markov approximation.

The analysis in Secs.~\ref{subsec:wdependentcoupling_systresponse}--\ref{subsec:wdependentcoupling_Markov} does not assume a particular form of the coupling \textcolor{black}{coefficient} $f_k$.
In Sec.~\ref{subsec:self_energy_in_sec3}, we calculate the self-energy for the specific examples of the series-$\sLC$ and simple ladder filter discussed in Sec.~\ref{sec:summary_seriesLC_andfilter}, and analyze its frequency dependence both in the exact form and in resonant approximations.
In App.~\ref{app:selfenergy_variouscouplings} we list the expressions of $f_k$ and the corresponding self-energies
for simple passive coupling circuits: parallel-$\sLC$, purely inductive
and purely capacitive,
series-$\sLC$, and the simple ladder filter case.

\subsection{System response: quantum Langevin equation}\label{subsec:wdependentcoupling_systresponse}

To obtain closed-form equations of motion from the Hamiltonian, \Eq{eqQq:Ham_orig},
we make the usual rotating wave approximation (RWA),
which assumes that only near-resonant TL frequencies $\omega_k \approx \omega_s$
will significantly contribute to the coupling integrals in the equations of motion \cite{GarZol_book}.
One therefore can neglect the non-resonant (counter-rotating) contributions
$a_s a_k$ and $a_s^\dag a_k^\dag$ in the coupling Hamiltonian,
\begin{align}
\label{eqQq:Ham_RWA}
 H &= H_{\sys} + \int_0^\infty \diff k \hbar \omega_k a_k^\dag a_k
 + \hbar \int_{0}^\infty \diff k f_k \left( a_k^\dag a_s + a_s^\dag a_k\right)
\end{align}
The resulting Heisenberg equations of motion for the $a_k$ are
\begin{align}
 \dot a_k &= -\ui \omega_k a_k -\ui f_k a_s
\end{align}
and at times $t$ larger than an initial time $t_0$, this equation formally integrates to
\begin{align}\label{eq:ak_t(t0)}
 a_k(t) &= a_k(t_0) e^{-\iexp \omega_k (t-t_0)}
 -\ui f_k\int_{t_0}^t \diff t' a_s(t') e^{-\iexp \omega_k (t-t')}  %
\end{align}
Subsequently, for an arbitrary system operator $W$ the Heisenberg equations of motion read:
\begin{align}
\dot W & - \frac{\ui}{\hbar}\left[H_\sys, W\right]
 =  \ui \int_0^\infty \diff k f_k \left\{ a_k^\dag [a_s, W]
 + [a_s^\dag, W] a_k \right\} \nonumber \\
&=
 \sqrt{2\Gamma_E} \left\{ b_{\In}^\dag \left[ a_s, W \right] - \left[ a_s^\dag, W \right] b_{\In} \right\} \nonumber \\
 &\quad- \int_{0}^\infty \diff k f_k^2 \int_{t_0}^t \diff t'
 \Bigl\{ a_s^\dag(t') e^{\iexp \omega_k(t-t')} [a_s, W] \Bigr. \nonumber \\
\label{eq:EOM_W_orig}
& \hspace*{3.3cm}\Bigl. - [a_s^\dag, W] a_s(t') e^{-\iexp \omega_k(t-t')} \Bigr\}
\end{align}
Here we have defined amplitudes of the input and output fields \cite{GarZol_book, Yurke_DruFic2004}:
\begin{align}\label{eq:bInOut}
 b_{\In,\Out}(t) &:=
 \frac{-\ui}{\sqrt{2 \Gamma_E}} \int_0^\infty \diff k f_k a_k(t_{0,1}) e^{-\iexp \omega_k (t-t_{0,1})} 
\end{align}
where the time $t_0 < t$ ($t_1 > t$) lies in the distant past (future),
before (after) which the system is assumed decoupled from the TL.
We have factored out in \Eq{eq:bInOut} a parameter $\Gamma_E$
which characterizes the effective strength of the coupling-induced damping.
While it is not uniquely defined in \Eq{eq:bInOut},
to be specific, we here define it as
\begin{equation}\label{eq:def_GammaE}
 \Gamma_E := \frac{\pi}{v} f_{\wref/v}^2
\end{equation}
This form is suitable for comparison with the Markov approximation,
as discussed below in Sec.~\ref{subsec:wdependentcoupling_Markov}.

The equation of motion for the system amplitude,
$W = a_s$, follows from \Eq{eq:EOM_W_orig},
\begin{align}
\dot a_s &= \frac{\ui}{\hbar} \left[H_\sys, a_s\right] - \Gamma_I a_s
 + \sqrt{2\Gamma_E}  b_\In \nonumber \\
\label{eq:EOM_as_preMarkov}
&\quad- \int_0^\infty \diff k f_k^2 \int_{t_0}^t \diff t' a_s(t') e^{-\iexp \omega_k(t-t')}
\end{align}
where we have now also included a term which reflects the
additional damping due to internal losses with the damping rate $\Gamma_I$.
This term can easily be derived \cite{RosenthalETAL2021}
from the circuit equations of motion if the
system (with capacitance $C_s$) is shunted by a resistor $R_s$,
such that e.g.~$2\Gamma_I = 1/(R_s C_s)$ for an inductively coupled system
or $2\Gamma_I = 1/(R_s (C_s+C_c))$ for a capacitively coupled system.
Alternatively, internal losses can be modeled within the Hamiltonian framework, by
assuming the (frequency-independent) coupling to an additional virtual TL.

Now we express operators $a_s$, $b_{\In,\Out}$ in a frame rotating at a reference
frequency $\wref \approx \omega_s$,
\begin{align}
 \label{eq:as_As_wref}
 a_s(t) &= A_s(t) e^{-\iexp \wref t} \\
 \label{eq:binout_Binout_wref}
 b_{\In,\Out}(t) &= B_{\In,\Out}(t) e^{-\iexp \wref t}
\end{align}
The system Hamiltonian transforms accordingly under the unitary transformation
\begin{align}\label{eq:unitarytrafo2RF}
 U & = e^{-\ui \wref a_s^\dag a_s t} \\
 H_\sys \to \tilde{H}_\sys &= U^\dag H_\sys U -\ui\hbar U^\dag \dot U \\
 & 
 = U^\dag H_\sys U  - \hbar \wref A_s^\dag A_s \nonumber
\end{align}
Then,  \Eq{eq:EOM_as_preMarkov} becomes
\begin{align}
\dot A_s &= \frac{\ui}{\hbar}\left[\tilde{H}_\sys, A_s\right] - \Gamma_I A_s
 + \sqrt{2\Gamma_E}  B_\In \nonumber \\
\label{eq:EOM_As_preMarkov}
 &\quad - \int_0^\infty \diff k f_k^2 \int_{t_0}^t \diff t' A_s(t') e^{-\iexp \Delta_k (t-t')}
\end{align}
with
\begin{equation}\label{eq:def_Delta_k}
\Delta_k := \omega_k -\wref\,.
\end{equation}
Usually, the coherent part of the evolution is averaged on short time scales,
$\left\langle \left[\tilde{H}_\sys, A_s\right] \right\rangle_t$,
such that fast oscillating terms with frequencies $\gg \wref$ cancel.
Looking at small detunings $\Delta$ from $\wref$, 
we write the amplitudes in the form
\begin{align}
 \label{eq:As_decompose_general}
 A_s(t) &= \int \frac{\diff \Delta}{2\pi} A_s(\Delta) e^{-\iexp \Delta t}\\
 \label{eq:Binout_decompose_general}
 B_{\In,\Out}(t) &= \int \frac{\diff \Delta}{2\pi} B_{\In,\Out}(\Delta) e^{-\iexp \Delta t}
\end{align}
The time integration on the rhs of \Eq{eq:EOM_As_preMarkov} can then be evaluated
in the long-time limit, setting $t_0 \to -\infty$.
As a result, after separating the contribution rotating at $\Delta$,
one obtains
\begin{align}
-\ui \Delta A_s(\Delta)
 &= \frac{\ui}{\hbar} \left\langle \left[\tilde{H}_\sys, A_s\right] \right\rangle_t(\Delta)
- \Gamma_I A_s(\Delta) \nonumber \\
\label{eq:As_wdependent_general_version1}
&\quad+ \sqrt{2\Gamma_E} B_\In(\Delta)
- \ui \Sigma_E(\Delta) A_s(\Delta) \,.
\end{align}
Here $\Sigma_E(\Delta)$ is the self-energy related to the coupling with TL, determined by
the coupling coefficient $f_k$ which appears in \Eqs{eqQq:Ham_orig} and \eqref{eqFq:Ham_orig}:
\begin{align}\label{eq:def_selfenergy}
 \Sigma_E(\Delta) &:= \int_0^\infty \diff k \frac{f_k^2}{\Delta - \Delta_k + \ui \epsilon} \\
 &= \int_0^\infty \diff k \frac{f_k^2}{\omega - \omega_k + \ui \epsilon}
 =:  \hat{\Sigma}_E(\omega=\wref + \Delta)
 \,. \nonumber
\end{align}
It can be combined with the contribution from the internal damping,
\begin{equation}\label{eq:def_selfenergy_tot}
 \Sigma(\Delta) = \Sigma_E(\Delta) - \ui \Gamma_I \;,
\end{equation}
such that
\begin{align}
-\ui \Delta A_s(\Delta)
 &= \frac{\ui}{\hbar} \left\langle \left[\tilde{H}_\sys, A_s\right] \right\rangle_t(\Delta) \nonumber \\
\label{eq:As_wdependent_general}
&\quad+ \sqrt{2\Gamma_E} B_\In(\Delta)
- \ui \Sigma(\Delta) A_s(\Delta) \,.
\end{align}
The commutator term on the rhs
denotes the contribution from the detuning $\Delta$ to the coherent system evolution
in the rotating frame: it is evaluated by first computing the commutator
and its time-average, and then inserting \Eq{eq:As_decompose_general}. 

\subsection{Input-output relation}\label{subsec:wdependentcoupling_IOrelation}

The input-output relations can also be obtained for
general coupling coefficients $f_k$, suitable for the quantum Langevin equations
of Sec.~\ref{subsec:wdependentcoupling_systresponse}.
To this end, consider the integrated TL mode amplitudes,
integrated either at $t>t_0$, see~\Eq{eq:ak_t(t0)}, or at $t<t_1$:
\begin{align}
 a_k(t) &= a_k(t_0) e^{-\iexp \omega_k (t-t_0)}
 - \ui f_k\int_{t_0}^t \diff t' a_s(t') e^{-\iexp \omega_k (t-t')}  \\ 
%
 a_k(t) &= a_k(t_1) e^{-\iexp \omega_k (t-t_1)}
 + \ui f_k \int_t^{t_1} \diff t' a_s(t') e^{-\iexp \omega_k (t-t')}  
\end{align}
The integral $\int \diff k f_k a_k(t)$ can thus be expressed in two alternative ways,
\begin{align}
 &\int_0^\infty \diff k f_k a_k(t) \nonumber \\
 \label{eq:int_fk_ak_t(t0)}
 &= \ui \sqrt{2\Gamma_E} b_{\In}(t)
 -\ui \int_0^\infty \diff k f_k^2 \int_{t_0}^t \diff t' a_s(t') e^{-\iexp \omega_k (t-t')}
 \\ 
 \label{eq:int_fk_ak_t(t1)}
  &= \ui \sqrt{2\Gamma_E} b_{\Out}(t)
 +\ui \int_0^\infty \diff k f_k^2 \int_t^{t_1} \diff t' a_s(t') e^{-\iexp \omega_k (t-t')}
\end{align}
with the definitions of the input and output amplitudes, \Eq{eq:bInOut}.
In general, it is nontrivial to analytically evaluate the $k$-integration on the right side of
\Eqs{eq:int_fk_ak_t(t0)}--\eqref{eq:int_fk_ak_t(t1)}
for a given form of the coupling \textcolor{black}{coefficient} $f_k$ \footnote{%
Because of the oscillating factor, the contour integral along
either the upper or lower semicircle with radius $\to \infty$ diverges.
Although one can calculate the integral $\int_{-\infty}^{\infty}$
using the non-diverging semicircle, this is in general of little help to
determine $\int_{0}^{\infty}$.
While a branch cut trick would in principle allow to calculate $\int_{0}^{\infty}$,
it also leads to divergence because it uses the full circle contour integral.%
}.

We proceed as in Sec.~\ref{subsec:wdependentcoupling_systresponse},
transforming the amplitudes in \Eqs{eq:int_fk_ak_t(t0)} and \eqref{eq:int_fk_ak_t(t1)}
to a frame rotating at reference frequency $\wref$,
as defined in \Eqs{eq:as_As_wref} and \eqref{eq:binout_Binout_wref},
\begin{align}
& \int_0^\infty \diff k f_k a_k(t) e^{\iexp \wref t} \nonumber \\
\label{eq:int_fk_ak_t(t0)_version2}
 &= \ui \sqrt{2\Gamma_E} B_{\In}(t)  \:
 -\ui \int_0^\infty \diff k f_k^2 \int_{t_0}^t \diff t' A_s(t') e^{-\iexp \Delta_k (t-t')}
\\ 
 \label{eq:int_fk_ak_t(t1)_version2}
  &= \ui \sqrt{2\Gamma_E} B_{\Out}(t)
 + \ui \int_0^\infty \diff k f_k^2 \int_t^{t_1} \diff t' A_s(t') e^{-\iexp \Delta_k (t-t')}
\end{align}
with $\Delta_k$ defined in \Eq{eq:def_Delta_k}.
As in Sec.~\ref{subsec:wdependentcoupling_systresponse},
we assume the rotating-frame amplitudes to be of the form given
in \Eqs{eq:As_decompose_general}--\eqref{eq:Binout_decompose_general}
and evaluate the time-integrals on the rhs of \Eqs{eq:int_fk_ak_t(t0)_version2}--\eqref{eq:int_fk_ak_t(t1)_version2} with the usual assumption $t_0 \to -\infty$, $t_1 \to \infty$.
This results in
\begin{align}
 &\int_0^\infty \diff k f_k a_k(t) e^{\iexp \wref t} \nonumber \\
 &= \int \frac{\diff \Delta}{2\pi}\left( \ui \sqrt{2\Gamma_E}
 B_{\In}(\Delta) + \Sigma_E(\Delta) A_s(\Delta)\right) e^{-\iexp \Delta t}
       \\
  &=  \int \frac{\diff \Delta}{2\pi}\left( \ui \sqrt{2\Gamma_E}
 B_{\Out}(\Delta) + \Sigma_E^*(\Delta) A_s(\Delta)\right) e^{-\iexp \Delta t} \,.
\end{align}
Subtracting the two equations
and separating the terms rotating at different frequencies, 
one arrives at
the input-output relation for the frequency-dependent coupling,
\begin{equation}\label{eq:inputoutput_frequencydependent}
B_{\Out}(\Delta) = B_{\In}(\Delta)
+ \frac{2\im\Sigma_E(\Delta)}{\sqrt{2\Gamma_E}} A_s(\Delta)
\,.
\end{equation}

\subsection{Unitarity condition}
\label{subsec:wdependentcoupling_unitarity}

If the system has no other loss channels except through its coupling with the TL,
the system response together with the input-output relations have to guarantee
the unitarity of the input-to-output field transformation.
The invariance of the quantum commutation relations
of the output relative to those of the input field
usually serves as a test of unitarity.
In the case of frequency-dependent coupling, care has to be taken,
since these commutation relations are now different from the familiar case, as we show below.

Assuming canonical commutation relations for the bare TL amplitudes,
$\left[a_k(t_0), a_{k'}^\dag(t_0)\right] = \delta(k-k')$
and $\left[a_k(t_0), a_{k'}(t_0)\right] = 0$,
the input and output amplitudes defined in \Eqs{eq:bInOut}
fulfill the commutation relations,
$\left[ b_{\In}(t), b_{\In}^\dag(t') \right]
 = \frac{1}{2\Gamma_E} \int_0^\infty \diff k f_k^2 e^{-\iexp \omega_k (t-t')}$
and
$\left[ B_{\In}(t), B_{\In}^\dag(t') \right]
 = \frac{1}{2\Gamma_E} \int_0^\infty \diff k f_k^2 e^{-\iexp \Delta_k (t-t')}$,
in the rotating frame.
Similarly,
$\left[ b_{\In}(t), b_{\In}(t') \right] = \left[ B_{\In}(t), B_{\In}(t') \right]  = 0$
holds.
After Fourier transformation,
the resulting commutation relations in frequency space are found to be
\begin{align}\label{eq:commutator1_Bin}
 \left[ B_{\In}(\Delta), B_{\In}^\dag(\Delta') \right]
 &= -\frac{\im\Sigma_E(\Delta)}{\Gamma_E}  \delta(\Delta - \Delta') \\
\label{eq:commutator2_Bin}
 \left[ B_{\In}(\Delta), B_{\In}(\Delta') \right] &= 0
 \,.
\end{align}

The corresponding commutation relations for the output amplitudes $B_{\Out}(\Delta)$
can be found from \Eq{eq:inputoutput_frequencydependent}
and \Eq{eq:As_wdependent_general} for a specific system of interest.
In the absence of internal losses, unitarity implies that the commutation relations
will be preserved,
$\left[ B_{\Out}(\Delta), B_{\Out}^\dag(\Delta') \right] =
\left[ B_{\In}(\Delta), B_{\In}^\dag(\Delta') \right]$
and $\left[ B_{\Out}(\Delta), B_{\Out}(\Delta') \right] =
\left[ B_{\In}(\Delta), B_{\In}(\Delta') \right]$.
This unitarity condition thus can be used to check the consistency of \Eq{eq:inputoutput_frequencydependent} with \Eq{eq:As_wdependent_general}.
In Sec.~\ref{sec:paramp_frequencydependentcoupling} we perform this
unitarity check for the example of a system subject to (degenerate) parametric resonance.

\subsection{Comparison with the Markov approximation}
\label{subsec:wdependentcoupling_Markov}

A frequent assumption in the analysis of open quantum systems
is that changes of the system state happen
on a time scale much larger than the decay time of correlations in the
environment \cite{GarZol_book}.
In the cQED context, where the environment consists of the TL and coupling network,
this condition is fulfilled if the coupling coefficients
are frequency-independent or
do not vary strongly over the range of relevant system frequencies or $k$-values,
$f_k \approx \text{const}$.
This (first) Markov approximation is usually justified in the weak-coupling limit,
i.e. if the coupling coefficients $f_k$ are small compared with the energy scales
of the system such that any $f_k$-variations are automatically small too.

To compare the results of Sec.~\ref{subsec:wdependentcoupling_systresponse}--\ref{subsec:wdependentcoupling_unitarity} with their counterparts under
the Markov approximation,
we set $f_k^2 \to f_{\wref/v}^2 =: v \Gamma_E/\pi$
in the self-energy \Eqs{eq:def_selfenergy} and the equations of motion 
\Eqs{eq:EOM_as_preMarkov} and \eqref{eq:EOM_As_preMarkov} .
The external damping rate $\Gamma_E$
is here defined as in \Eq{eq:def_GammaE}.
Additionally, since only the resonant contributions $a_s^\dag a_k + a_k^\dag a_s$
appear in the RWA Hamiltonian, \Eq{eqQq:Ham_RWA},
the lower integration boundary of $k$ can be set to $-\infty$.
This allows to evaluate the integral
$\int_{-\infty}^\infty \diff k f_k^2 e^{\pm \iexp \omega_k(t-t')} = 2\Gamma_E \delta(t-t')$.
Then, \Eqs{eq:EOM_as_preMarkov} and \eqref{eq:EOM_As_preMarkov} become
\begin{align}
\label{eq:EOM_as_Markov}
\dot a_s &= \frac{\ui}{\hbar}\left[H_\sys, a_s\right]
 - \Gamma_I a_s  + \sqrt{2\Gamma_E}  b_\In - \Gamma_E a_s \\
\label{eq:EOM_As_Markov}
\dot A_s &= \frac{\ui}{\hbar}\left[\tilde{H}_\sys, A_s\right]
 - \Gamma_I A_s  + \sqrt{2\Gamma_E}  B_\In - \Gamma_E A_s
\end{align}
and the contribution of the coupling to the self-energy $\Sigma_E(\Delta)$,
\Eq{eq:def_selfenergy}, reduces to
\begin{equation}\label{eq:SigmaE_Markov}
 \Sigma_E(\Delta) = \frac{\Gamma_E v}{\pi} \int_{-\infty}^\infty \diff k (\omega - v k + \ui \epsilon)^{-1} = - \ui \Gamma_E
\end{equation}
using the Cauchy principal value integral
$\mathcal{P} \int_{-\infty}^{\infty} \diff x\, x/ (x^2 + \epsilon^2) = 0$.
Thus, the quantum Langevin equation \Eq{eq:As_wdependent_general} becomes
\begin{align}\label{eq:As_Markov_general}
-\ui \Delta A_s(\Delta)
 &= \frac{\ui}{\hbar} \left\langle \left[\tilde{H}_\sys, A_s\right] \right\rangle_t(\Delta)\nonumber\\
&\quad + \sqrt{2\Gamma_E} B_\In(\Delta)
- \Gamma A_s(\Delta) \,,
\end{align}
with $\Gamma = \Gamma_E + \Gamma_I$.

Under the Markov approximation,
\Eqs{eq:int_fk_ak_t(t0)}--\eqref{eq:int_fk_ak_t(t1)} reduce to
\begin{align}
 \int_0^\infty \diff k f_k a_k(t)
 &= \ui \sqrt{2\Gamma_E} b_{\In}(t) - \ui \Gamma_E a_s(t) \\ 
 \int_0^\infty \diff k f_k a_k(t)
  &= \ui \sqrt{2\Gamma_E} b_{\Out}(t) + \ui \Gamma_E a_s(t) 
\end{align}
and these determine the input-output relation
\begin{align}%
  b_{\Out} &= b_{\In} - \sqrt{2\Gamma_E} a_s
\end{align}
valid in the time as well as the frequency representation, and equally
in the chosen rotating frame,
\begin{align}\label{eq:inputoutput_Markov}
  B_{\Out} &= B_{\In} - \sqrt{2\Gamma_E} A_s
  \,.
\end{align}
The same result can be obtained from \Eq{eq:inputoutput_frequencydependent}
using \Eq{eq:SigmaE_Markov}.

Finally, the Markov approximation reduces
the commutation relations for the input amplitudes to
$\left[ B_{\In}(t), B_{\In}^\dag(t') \right] = \delta(t-t')$,
cf.~Sec.~\ref{subsec:wdependentcoupling_unitarity},
and the rhs of \Eq{eq:commutator1_Bin} becomes
$\delta(\Delta - \Delta')$.

\subsection{Self-energy}\label{subsec:self_energy_in_sec3}

\textcolor{black}{
As we can see from \Eqs{eq:As_wdependent_general} and \eqref{eq:inputoutput_frequencydependent}, the self-energy characterizes the response of the system of interest to 
an input from the TL at detuning $\Delta$.
It is related to the damping of the system due to the environment, and plays a key role in the understanding of the open system dynamics beyond the Markov approximation. }

\textcolor{black}{
The imaginary part of the self-energy, $\im \Sigma_E(\Delta)$, quantifies the coupling contribution
to the effective damping rate of the system.
In the limit $\epsilon \to 0$, the imaginary part of \Eq{eq:def_selfenergy} evaluates to }

\textcolor{black}{
\begin{equation}\label{eq:def_imSig}
\im \hat{\Sigma}_E(\omega) =-\pi\int_0^\infty \diff k \delta(\omega - \omega_k)f_k^2= -\frac{\pi}{v}  f_{k=\frac{\omega}{v}}^2  \,.
\end{equation}
}

\textcolor{black}{
The real part of the self-energy, $\re \Sigma_E(\Delta)$,
in general needs to be evaluated
by performing the nontrivial integration in \Eq{eq:def_selfenergy}.
In this section, we compute the self-energy for the two coupling circuits discussed in Sec.~\ref{sec:summary_seriesLC_andfilter}, namely series-$\sLC$ and the simple ladder filter, with further analytical details given in App.~\ref{app:selfenergy_variouscouplings}.
}

\subsubsection{Series-$\sLC$ filter case}

\begin{figure}\centering
\includegraphics[width=\columnwidth]{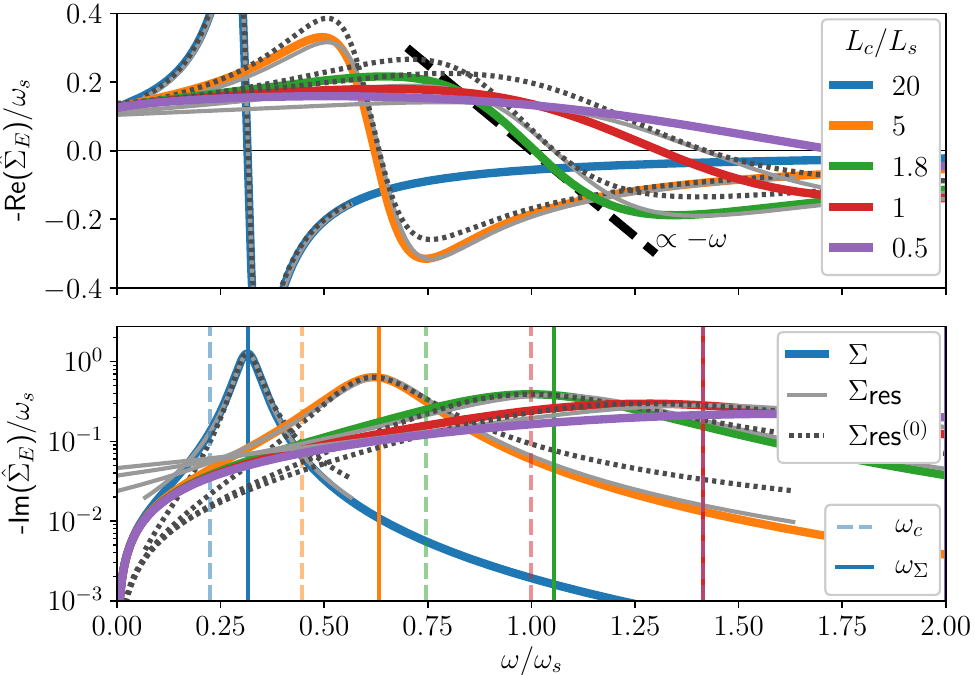}
\caption{
 Self-energy $\Sigma_E(\Delta)$ vs $\omega = \omega_s + \Delta$
 of series-$\sLC$ coupling circuit for $Z_s/Z_0=0.8$, $C_c/C_s = 1.0$
 and various values of $L_c/L_s$.
 All cases belong to the parameter regime $4\beta/\alpha > 1$,
 with $\beta/\alpha = 32 L_c/(25 L_s)$.
 Grey solid and black dotted lines show the resonant approximations
 $\Sigma_{E,\text{res}}(\Delta)$
 and $\Sigma_{E,\text{res},0}(\Delta)$
 of \Eq{eq:selfenergy_seriescoupled_resapprox}
 and \Eq{eqapp:selfenergy_seriescoupled_resapprox_exact}, respectively.
 The vertical lines show the (approximate) resonance position $\omega_\Sigma$ (solid), \Eq{eq:wres_selfenergy_imag_seriescoupled},
 and the bare coupling frequency $\omega_c = 1/\sqrt{C_c L_c}$ (dashed). The black dashed line has slope $-1$, which represents the so-called `slope-matching' condition \Eq{eq:resonantcondition_bandwidthbroadening} (see Sec.~\ref{sec:results_bandwidthbroadening}).
}
\label{fig:selfenergy_seriescoupled}
\end{figure}

\textcolor{black}{For the particular coupling constant $f_k$ of the series-$\sLC$ coupling,
\Eq{eqQq:def_fk}, \Eq{eq:def_imSig} evaluates to }
\begin{align}\label{eq:imSig_seriescoupled}
 \frac{\im \hat{\Sigma}_E(\omega)}{\omega_s}
 &=
 -\frac{Z_s}{2 Z_0} \frac{\omega_s}{\omega}
 \left( 1 + \left( \frac{\beta \omega}{v} - \frac{v}{\alpha \omega}\right)^2 \right)^{-1} \\
 &=  -\frac{Z_s}{2 Z_0} \frac{\omega_s}{\omega}
 \frac{\omega^2}{\omega^2 + \frac{L_c^2}{Z_0^2}\left(\omega^2 - \omega_\Sigma^2\right)^2 }
\nonumber
\end{align}
with
\begin{equation}\label{eq:wres_selfenergy_imag_seriescoupled}
 \omega_{\Sigma}
 = \frac{v}{\sqrt{\alpha \beta}}
 = \frac{\sqrt{1+\frac{C_c}{C_s}}}{\sqrt{L_c C_c}}
 \,.
\end{equation}
\textcolor{black}{For the real part of the self-energy, 
the integral of \Eq{eq:def_selfenergy} can be solved exactly, see~App.~\ref{app:selfenergy_variouscouplings}.}
The result, given in \Eqs{eqapp:reSig1_seriescoupled}--\eqref{eqapp:reSig2_seriescoupled},
takes different forms depending on the ratio of the coupling lengths $\alpha$ and $\beta$.
Figure~\ref{fig:selfenergy_seriescoupled} shows both $\re \Sigma_E(\Delta)$
and $\im \Sigma_E(\Delta)$ from \Eq{eq:imSig_seriescoupled} for various
parameter cases.

While the exact analytical solution for $\re \Sigma_E(\Delta)$
in principle contains all information,
it is not very intuitive.
A simplification for the self-energy can be found in the vicinity of the
(approximate) resonance frequency $\omega_\Sigma$ of the coupling, \Eq{eq:wres_selfenergy_imag_seriescoupled},
near which the damping rate $-\im\Sigma(\Delta)$, \Eq{eq:imSig_seriescoupled}, is maximized.
Approximating $f_k^2$ by Lorentzians centered at $k = \omega_\Sigma/v$,
we find the following Lorentzian approximation for the self-energy,
\begin{equation}\label{eq:selfenergy_seriescoupled_resapprox}
 \frac{\Sigma_{E,\text{res}}(\Delta)}{\omega_s} = \frac{1}{4 C_s L_c \omega_s \omega_{\Sigma}}
 \frac{\omega_s}{\omega - \omega_{\Sigma} + \ui Z_0/(2L_c)}
 \,,
\end{equation}
see~\Eq{eqapp:selfenergy_seriescoupled_resapprox}. 
\textcolor{black}{This resonant approximation 
corresponds to the approximate self-energy used in Ref.~\onlinecite{RoyETAL2015},
see further discussion in App.~\ref{sec:selfenergy_seriesLC_resonant}.
}
\textcolor{black}{This approximate self-energy} is shown by the gray solid lines in Fig.~\ref{fig:selfenergy_seriescoupled}.
For the cases with small $L_c$, the maxima of $-\im\Sigma_{E,\text{res}}(\Delta)$
at $\omega_{\Sigma}$ (vertical solid lines)
start to deviate
from the actual maxima $\omega_\Sigma^{(0)}$ of $-\im\Sigma(\Delta)$.
This is because $\omega_{\Sigma} \geq \omega_\Sigma^{(0)}$ is only an approximation of the
exact resonance position $\omega_\Sigma^{(0)}$, valid in the limit $\beta \gg \alpha$.
In App.~\ref{app:selfenergy_variouscouplings} we derive
the general expression for $\omega_\Sigma^{(0)}$,
\Eq{eqapp:wres_exact_selfenergy_imag_seriescoupled},
and --
using Lorentzian expansions of $f_k^2$ around $k=\omega_\Sigma^{(0)}/v$ --
an alternative resonant approximation $\Sigma_{E,\text{res}}^{(0)}(\Delta)$ for the self-energy, \Eq{eqapp:selfenergy_seriescoupled_resapprox_exact}.
It is shown by the black dotted lines in Fig.~\ref{fig:selfenergy_seriescoupled}.
Despite giving a more accurate approximation for $\Sigma_{E}(\Delta)$
in the closest vicinity to $\omega_\Sigma^{(0)}$, at moderate detuning $|\Delta|>0$
this approximation tends to deviate much faster and more asymmetrically
from $\Sigma_{E}(\Delta)$. This can be seen especially in the local maxima and minima
of $\re\Sigma_E(\Delta)$.
The comparison with the two resonant approximations -- each of which by construction
has perfect alignment between the maximum of $-\im(\Sigma_E(\Delta))$
and the root of $\re\Sigma_E(\Delta)$ -- also demonstrates that this alignment is
not perfect for the full solution, \Eq{eq:imSig_seriescoupled}
and \Eqs{eqapp:reSig1_seriescoupled}--\eqref{eqapp:reSig2_seriescoupled}.

\subsubsection{Simple ladder filter case}
\begin{figure}\centering
\includegraphics[width=\columnwidth]{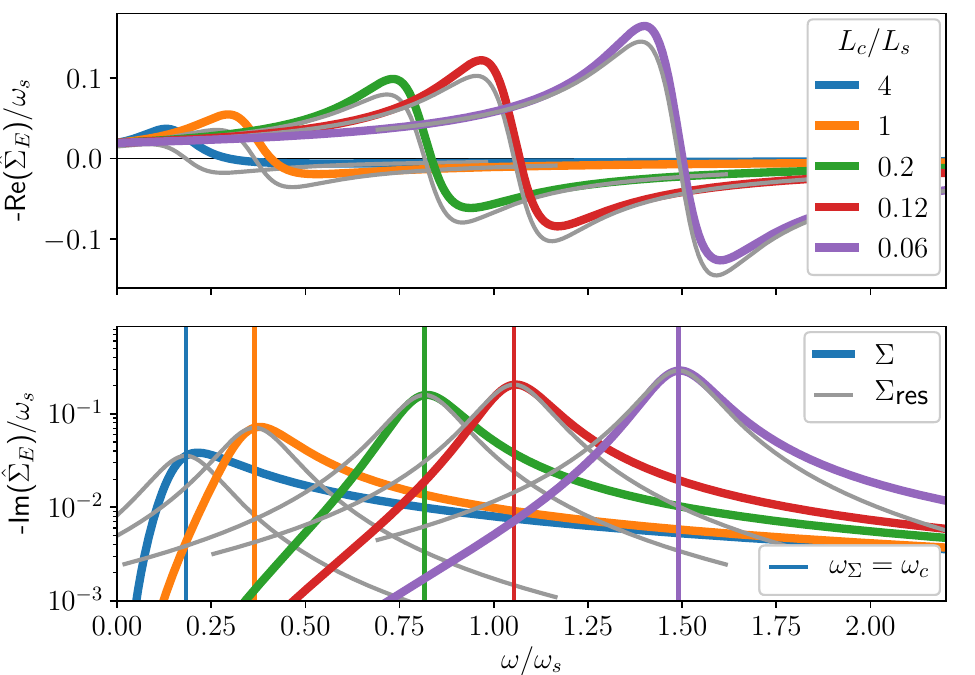}
\caption{
 Self-energy $\Sigma_E(\Delta)$ vs $\omega = \omega_s + \Delta$
 for the simple ladder filter with $Z_s/Z_0=1.0$, $C_g/C_s=0.5$, $L_g/L_s=2.0$, $C_c/C_s = 5.0$
 and various values of $L_c/L_s$.
 All cases belong to the parameter regime $4\alpha/\beta > 1$,
 with $\alpha/\beta = 10 L_s/(3 L_c)$.
 Grey solid lines show the resonant approximation $\Sigma_{E,\text{res}}(\Delta)$
 of \Eq{eq:selfenergy_parallelbeforeseriescoupled_resapprox}.
 The vertical lines show the (approximate) resonance position $\omega_\Sigma$ (solid), \Eq{eq:wres_selfenergy_imag_parallelbeforeseriescoupled},
 which here coincides with the bare coupling frequency $\omega_c = 1/\sqrt{L_c C_c}$.
}
\label{fig:selfenergy_parallelbeforeseriescoupled}
\end{figure}

For the coupling coefficients of the simple ladder filter, \Eq{eqFq:def_fk},
the imaginary part of the self-energy, \Eq{eq:def_imSig}, evaluates to
\begin{align}
 \frac{ \im \hat{\Sigma}_E(\omega) }{\omega_s}
 &=
-\frac{Z_0}{2 Z_a} \frac{\omega}{\omega_s}
\left( 1 + \left( \frac{\alpha \omega}{v} - \frac{v}{\beta \omega}\right)^2 \right)^{-1} \nonumber \\
\label{eq:imSig_filtercoupled}
 &= -\frac{Z_0}{2 Z_a} \frac{\omega}{\omega_s}
 \frac{\omega^2}{\omega^2 + Z_0^2 C_c^2 \left(\omega^2 - \omega_\Sigma^2 \right)^2 }
\end{align}
with
\begin{equation}\label{eq:wres_selfenergy_imag_parallelbeforeseriescoupled}
 \omega_{\Sigma}
 = \frac{v}{\sqrt{\alpha \beta}}
 = \frac{1}{\sqrt{L_c C_c}}
 \,.
\end{equation}
The real part of the self-energy $\re \Sigma_E(\Delta)$ is calculated in
App.~\ref{app:selfenergy_variouscouplings},
resulting in 
\Eqs{eqapp:reSig1_parallelbeforeseriescoupled}--\eqref{eqapp:reSig2_parallelbeforeseriescoupled}.
Again it takes different forms depending
on the ratio of the coupling lengths $\alpha$ and $\beta$.
Figure~\ref{fig:selfenergy_parallelbeforeseriescoupled} shows $\re \Sigma_E(\Delta)$
together with $\im \Sigma_E(\Delta)$ from \Eq{eq:imSig_filtercoupled} for various
parameter cases, here all in the regime $4\alpha/\beta > 1$.
It is worth mentioning that $\im \Sigma_E(\Delta)$
here has a single resonance and does not turn zero at any finite frequency,
in contrast to the purely parallel-$\sLC$ coupling,
cf.~Fig.~\ref{fig:selfenergy_parallelcoupled} and discussion in
App.~\ref{sec:selfenergy_parallelCcLc}.
This difference is due to the absence of the $L_c^{-1}$-proportional
contribution to the coupling Hamiltonian in \Eq{eqQ:Ham__filter}
compared with \Eq{eqP:Hamiltonian_final},
and ultimately originates from the different structures of the
underlying circuit Lagrangians, \Eqs{eqF:Lagrangian_discretized} and \eqref{eqP:Lagrangian_discretized}.
The coupling characteristics of the simple ladder filter are thus similar to those of the
purely series-$\sLC$ coupling.
However, due to the additional degree of freedom $a_a$ of the system, \Eq{eqFq:Hsysext},
the spectral characteristics of the system response becomes richer,
as will be discussed in Sec.~\ref{subsec:gains_filtercoupling}.

The damping rate $-\im\Sigma_E(\Delta)$ is maximized near the (approximate)
resonance frequency $\omega_\Sigma$ of the coupling, cf. \Eq{eq:wres_selfenergy_imag_parallelbeforeseriescoupled}.
Approximating $f_k^2$ by Lorentzians centered at $k = \omega_\Sigma/v$,
we find the following Lorentzian approximation for the self-energy,
\begin{equation}\label{eq:selfenergy_parallelbeforeseriescoupled_resapprox}
 \frac{ \hat{\Sigma}_{E,\text{res}} }{\omega_s}
= \frac{\omega_{\Sigma}}{4 C_c Z_a \omega_s^2}
  \frac{\omega_s}{\omega - \omega_{\Sigma} + \ui/(2 Z_0 C_c)}
  \,,
\end{equation}
see~\Eq{eqapp:selfenergy_parallelbeforeseriescoupled_resapprox}.
It is shown by the gray solid lines in Fig.~\ref{fig:selfenergy_parallelbeforeseriescoupled}.

\section{Spectral properties of parametric resonance
beyond Markov approximation
}
\label{sec:paramp_frequencydependentcoupling}

The frequency-dependent versions of the system response equation, \Eq{eq:As_wdependent_general},
and the input-output relation, \Eq{eq:inputoutput_frequencydependent},
are distinct from their Markovian counterparts, \Eqs{eq:As_Markov_general}
and \eqref{eq:inputoutput_Markov}. This has implications for the spectral properties of the open system under consideration.
Of special interest are the spectral consequences for systems driven in a parametric
resonance regime where the pump frequency coincides with a combination of internal
system frequencies.
The resulting dynamics is slow relative to the fast oscillations of the unperturbed
system and can exhibit large amplitudes at only moderate pump strength \cite{NayfehMook_book}.

Here we focus on the case of a lumped-element system, such as a single JJ
or a SQUID, characterized by a linear system frequency $\omega_s$.
Degenerate parametric resonance occurs if the pump frequency is close to
twice the system frequency, $\omega_p \approx 2\omega_s$.
If we choose the reference frequency to be $\wref = \omega_p/2$, in the rotating frame, the effective (time-averaged) system Hamiltonian becomes
\begin{equation}\label{eq:Hsys_paramp_RF}
 \tilde{H}_{\sys} = -\hbar \delta_p A_s^\dag A_s
 - \frac{\hbar}{2} \left( \epsilon_p A_s^\dag A_s^\dag + \epsilon_p^\ast A_s A_s \right)
\end{equation}
where $\delta_p = \wref - \omega_s$ is the pump detuning and $\epsilon_p$
is the effective pump strength. With this Hamiltonian $\tilde{H}_{sys}$, we anticipate the mixing of frequency components at $+\Delta$ and $-\Delta$, such that the commutator term in \Eq{eq:As_wdependent_general} has contributions from both $A_s(+\Delta)$ and $A_s(-\Delta)$.
In App.~\ref{app:RF_fluxpumpedSQUID} we provide a derivation of $H_{\sys}$ for the
particular case of a capacitively shunted, flux-tunable SQUID pumped with frequency
$\omega_p = 2\omega_s + 2\delta_p \approx 2\omega_s$ and effective pump strength $\epsilon_p$, cf.~\Eq{eq:pump_strength}.

It should be pointed out that many other cQED realizations for the same Hamiltonian exist.
These include lumped-element or cavity systems,
and they may either be directly flux-pumped \cite{RosenthalETAL2021, LinETAL2013_JPA, YamamotoETAL2014, KrantzETAL2016_NatComm, WusShu2013}
via a 3-wave mixing process with $\omega_p \approx 2\omega_s$,
or indirectly `current-pumped' via a 4-wave mixing process \cite{VijayETAL2011, BergealETAL2010b}
with $2\omega_p \approx 2\omega_s$.
Other extended or coupled systems with several linear system frequencies
allow for non-degenerate parametric resonance \cite{BergealETAL2010b, SimoenETAL2015, WusShu2017, BengtssonETAL2018}
where $\omega_p \approx \omega_{s,1} + \omega_{s,2}$.
The spectral properties of the non-degenerate resonance are closely related
to those of the degenerate case, and generalizations of the results below
can be obtained. %

\subsection{Parametric gains in the Markovian limit}

Let us first recall the familiar description of parametric resonance
and parametric amplification in the Markov approximation \cite{GarZol_book,ClerkETAL_review2010}.
For the Hamiltonian~\Eq{eq:Hsys_paramp_RF}, the Markovian quantum Langevin equation \Eq{eq:EOM_As_Markov} has the form
\begin{align}\label{eq:EOM_As_Markov__paramp}
\dot A_s &= \ui \delta_p A_s + \ui \epsilon_p A_s^\dag - \Gamma A_s
 + \sqrt{2\Gamma_E} B_\In
\end{align}
with $\Gamma = \Gamma_E + \Gamma_I$. 
\textcolor{black}{Once the amplitudes $A_s$, $B_{\mathrm{in}}$ and $B_{\mathrm{out}}$ are expressed in the form
of \Eqs{eq:As_decompose_general}--\eqref{eq:Binout_decompose_general} as functions of the input detuning $\Delta$ ,}
the quantum Langevin equation in the frequency domain \Eq{eq:As_Markov_general} becomes
\begin{align}\label{eq:As_Delta_Markov__paramp}
(\delta_p + \Delta) A_s(\Delta)  + \epsilon_p A_s^\dag(-\Delta)
+ \ui \Gamma A_s(\Delta)
&= \ui \sqrt{2\Gamma_E} B_\In(\Delta)
\,.
\end{align}
Using \Eq{eq:As_Delta_Markov__paramp} together with the corresponding equation
for $A_s^\dag(-\Delta)$,
we can express
$A_s$ in terms of $B_\In(\Delta)$ and $B^\dagger_\In(-\Delta)$. Inserting $A_s$ into the Markovian input-output relation, \Eq{eq:inputoutput_Markov},
one obtains the relation between the input and output amplitudes
\begin{eqnarray}\label{eq:Bin_Bout_Markov}
B_{\Out}(\Delta)  = u(\Delta) B_{\In}(\Delta) + v(\Delta) B_{\In}^\dag(-\Delta)
\end{eqnarray}
with the gain (Bogoliubov) coefficients
\begin{align}
 \label{eq:u_Delta__Markov}
 u(\Delta) &= \displaystyle 1 -  \frac{2 \ui\Gamma_E }{D(\Delta)}  (\delta_p - \Delta -\ui\Gamma)\\
 \label{eq:v_Delta__Markov}
 v(\Delta) &= \displaystyle -\frac{2 \ui\epsilon_p\Gamma_E}{D(\Delta)}
\end{align}
where the parametric resonance determinant is
\begin{equation}\label{eq:determinant_Markov}
D(\Delta) = (\delta_p+\Delta+\ui\Gamma)(\delta_p-\Delta-\ui\Gamma) - |\epsilon_p|^2
\end{equation}
and fulfills $D^\ast(\Delta) = D(-\Delta)$.
\textcolor{black}{
Equations~\eqref{eq:As_Delta_Markov__paramp}--\eqref{eq:determinant_Markov} are similarly obtained within the coupled-mode formalism~\cite{NaamanAumentado2022}. In particular, this formalism makes use of the Markov approximation, after approximating the frequency-dependent environment as a chain of (nearest-neighbor-coupled) parallel $\sLC$ resonators.}

In \Eqs{eq:As_Delta_Markov__paramp} and \eqref {eq:Bin_Bout_Markov}
one observes the effect of the parametric system Hamiltonian $H_{\text{sys}}$,
\Eq{eq:Hsys_paramp_RF}, to mix a signal tone at detuning $\Delta$ with an idler tone
at detuning $-\Delta$.
As a result, even if the input consists only of a contribution $B_\In(\Delta)$
at $\Delta$, the output will contain both contributions,
$B_\Out(\Delta) = u(\Delta) B_\In(\Delta)$, and
$B_\Out(-\Delta) = v(-\Delta) B_\In^\dagger(\Delta)$.

If there are no internal losses, $\Gamma_I=0$, the gain coefficients
fulfill the unitarity conditions,
\begin{eqnarray}
\label{eq:Bogoliubov_rel1__Markov}
 |u(\Delta)|^2 - |v(\Delta)|^2  &=& 1   \\
\label{eq:Bogoliubov_rel2__Markov}
 u(\Delta)v(-\Delta)- v(\Delta) u(-\Delta) &=& 0
\end{eqnarray}
These relations ensure unitarity, i.e. that the commutators are preserved,
\begin{align}
 &\left[ B_{\Out}(\Delta), B_{\Out}^\dag(\Delta') \right]
 =\left[ B_{\In}(\Delta), B_{\In}^\dag(\Delta') \right]
 =\delta(\Delta - \Delta') \\
 &\left[ B_{\Out}(\Delta), B_{\Out}(\Delta') \right]
 =\left[ B_{\In}(\Delta), B_{\In}(\Delta') \right]
 =0
\end{align}
cf.~discussion in Secs.~\ref{subsec:wdependentcoupling_unitarity} and \ref{subsec:wdependentcoupling_Markov}.

The oscillations described by the linear dynamical system
\Eq{eq:EOM_As_Markov__paramp}
-- or by \Eq{eq:EOM_As_preMarkov} --
are only stable within a finite parameter regime.
In absence of an input signal, $B_\In=0$, and for sufficiently small pump strength $\epsilon_p$,
the system amplitude evolves towards its single steady state, $A_s = 0$.
The stability of the steady state $A_s = 0$ is determined by the eigenvalues
of the matrix based on \Eq{eq:EOM_As_Markov__paramp} and its conjugation: 
it is stable if its determinant --
which equals $D(\Delta=0)$ from \Eq{eq:determinant_Markov} --
is larger than $0$.
This is the case as long as the pump strength $\epsilon_p$
is below the instability threshold
\begin{equation}\label{eq:h_thresh__Markov}
 \textcolor{black}{\epsilon^2_{p,\thresh} = \delta_{p}^2 + \Gamma^2}
 \,,
\end{equation}
found from the zero of $|D(\Delta=0)|$.
For $|\epsilon_p| \geq \epsilon_{p,\thresh}$,
the trivial steady state turns unstable, leading to exponential increase of $A_s$.
Only by taking into account (higher-order) nonlinear terms in the expansion of
$\tilde H_{\sys}$, the unphysical amplitude growth is lifted and
the system is instead found to evolve towards a finite-amplitude steady state, $A_s > 0$.
This finite-amplitude steady state corresponds to parametric self-oscillations of the system
(i.e. not in response to an input field $B_{\In}$), which lead to measurable
output amplitudes $B_{\Out}$ according to \Eq{eq:inputoutput_Markov}.

When $|\epsilon|$ approaches $\epsilon_{p,\thresh}$ from below,
the gain factors $u(0),v(0)$ diverge, and this behavior is the basis of
all Josephson parametric amplifiers and related devices.
The (classical) parametric gains -- direct (signal) gain and conversion (idler) gain -- are given by
\begin{align}\label{eq:signalgain}
 G_s(\Delta)
 &:= \frac{|B_{\Out}(\Delta)|^2}{|B_{\In}(\Delta)|^2} = |u(\Delta)|^2 \\
 G_i(\Delta)
 &:= \frac{|B_{\Out}(-\Delta)|^2}{|B_{\In}(\Delta)|^2} = |v(-\Delta)|^2
\end{align}

\subsection{Parametric gains and unitarity for frequency-dependent coupling}
\label{subsec:paramp_frequencydependentcoupling_gains_unitarity}

Based on \Eq{eq:EOM_As_preMarkov}
and the frequency-decompositions \eqref{eq:As_decompose_general}--\eqref{eq:Binout_decompose_general}
of the amplitudes, \Eq{eq:As_wdependent_general} becomes
\begin{align}\label{eq:As_Delta_nonMarkov__paramp}
&(\delta_p + \Delta) A_s(\Delta)  + \epsilon_p A_s^\dag(-\Delta)
- \Sigma(\Delta) A_s(\Delta)\nonumber\\
=& \ui \sqrt{2\Gamma_E} B_\In(\Delta),
\end{align}
and can be inverted together with the corresponding equation for $A_s^\dag(-\Delta)$,
\begin{align}\label{eq:As_paramp_matrixform_wdependent}
&
\begin{pmatrix} A_s(\Delta) \\ A_s^\dag(-\Delta) \end{pmatrix}
= \boldsymbol{\chi}
\begin{pmatrix} B_{\In}(\Delta) \\ B_{\In}^\dag(-\Delta) \end{pmatrix} \\
&
\boldsymbol{\chi} = \frac{\ui \sqrt{2\Gamma_E}}{D(\Delta)}
 \begin{pmatrix} \delta_p - \Delta -\Sigma^\ast(-\Delta) & \epsilon_p \\
-\epsilon_p^\ast & -(\delta_p + \Delta -\Sigma(\Delta) ) \end{pmatrix}
\nonumber
\end{align}
Herein, the parametric resonance determinant is
\begin{equation}\label{eq:determinant_wdependent}
D(\Delta) = \left(\delta_p+\Delta - \Sigma(\Delta)\right)\left(\delta_p-\Delta-\Sigma^\ast(-\Delta)\right) - |\epsilon_p|^2
\end{equation}
and fulfills $D^\ast(\Delta) = D(-\Delta)$,
independent of the specific form of $\Sigma(\Delta)$.
Next, by inserting the system response $A_s(\Delta)$ from \Eq{eq:As_paramp_matrixform_wdependent}
into the input-output relation, \Eq{eq:inputoutput_frequencydependent},
one arrives at a relation between the in- and output amplitudes alone,
\begin{equation}\label{eq:Bin_Bout}
 B_{\Out}(\Delta) = u(\Delta) B_{\In}(\Delta) + v(\Delta) B_{\In}^\dag(-\Delta)
\end{equation}
with the gain (Bogoliubov) coefficients
\begin{align}
\label{eq:u_Delta__wdependent}
 u(\Delta) &= \displaystyle 1 + \frac{2 \ui\im \Sigma_E(\Delta)}{D(\Delta)} (\delta_p - \Delta - \Sigma^\ast(-\Delta)) \\
\label{eq:v_Delta__wdependent}
 v(\Delta) &= \displaystyle \frac{2 \ui\epsilon_p \im \Sigma_E(\Delta)}{D(\Delta)}
\end{align}
The (classical) parametric gains
evaluate to
\begin{align}\label{eq:usq_Delta__wdependent}
  |u(\Delta)|^2
  &= 1 + \frac{4 |\epsilon_p|^2 \im\Sigma_E(\Delta) \im\Sigma_E(-\Delta) }{|D(\Delta)|^2} \\
  &\quad - \frac{4 \Gamma_I \im\Sigma_E(\Delta) }{|D(\Delta)|^2}
  \left( |\epsilon_p|^2 - |\delta_p - \Delta - \Sigma(-\Delta)|^2 \right) \nonumber\\
\label{eq:vsq_Delta__wdependent}
  |v(\Delta)|^2 &= \frac{4 |\epsilon_p|^2 \left(\im\Sigma_E(\Delta)\right)^2}{|D(\Delta)|^2}
\end{align}
Recall that $\Sigma(\Delta) = \Sigma_E(\Delta) - \ui \Gamma_I$, where $\Sigma_E(\Delta)$
is the coupling-generated contribution to the self-energy, \Eq{eq:def_selfenergy},
and the purely imaginary contribution $-\ui\Gamma_I$ is due to internal losses.

As in the Markov approximation,
the trivial steady state $A_s=0$ is only stable below a threshold value of the
pump strength, $\epsilon_p < \epsilon_{p,\thresh}$,
above which it turns unstable.
In order to analyze the above-threshold regime,
(higher-order) nonlinear contributions to the system Hamiltonian $\tilde H_{\sys}$
need to be taken into account (not included in \Eq{eq:Hsys_paramp_RF}).
Depending on their form,
the system is then found to saturate to a new, finite-amplitude steady state
with $A_s \neq 0$,
even in absence of input $B_{\In}$ from the TL.

\begin{figure}\centering
\includegraphics[width=\columnwidth]{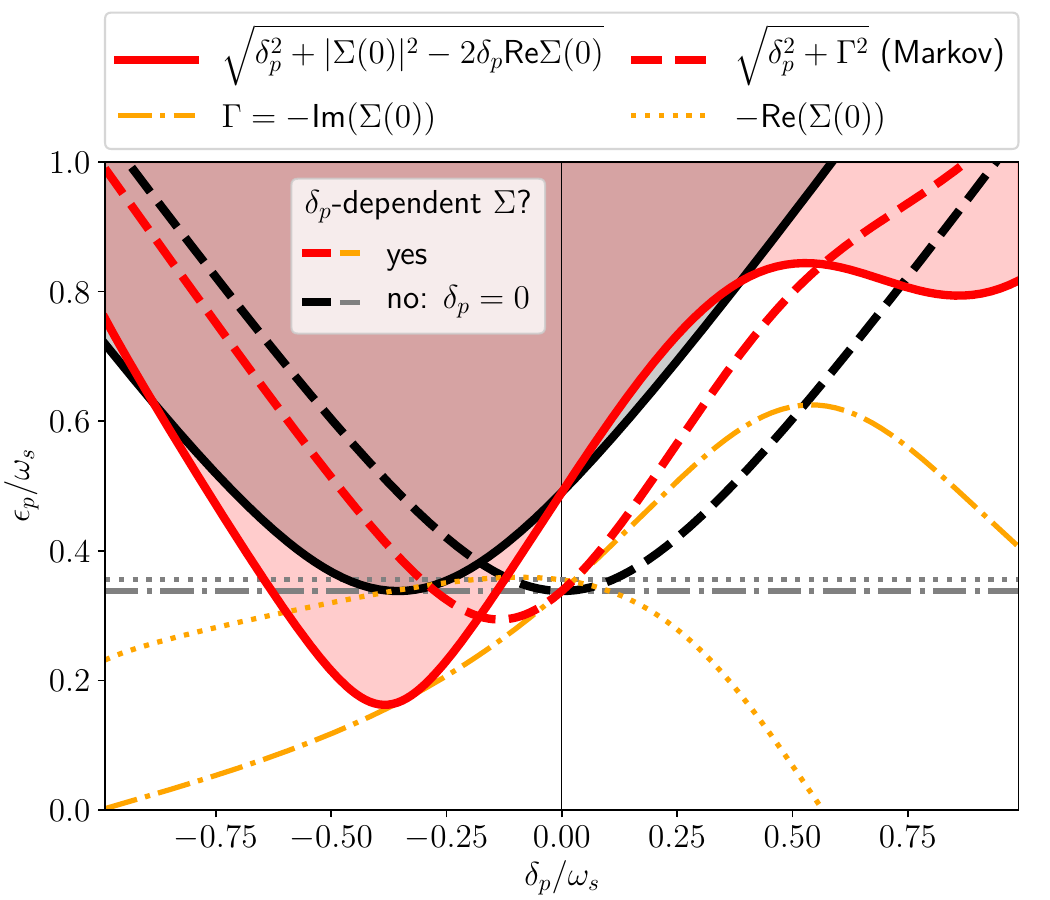}
\caption{
Pump-parameter space $(\delta_p, \epsilon_p)$
showing stable and unstable parameter regions for trivial state $A_s=0$
of undriven resonator ($B_{\In} = 0$).
It is stable for pump strength $\epsilon_p$ below
the pump threshold $\epsilon_{p,\thresh}$ (bold solid and dashed lines),
but for $|\epsilon| \geq \epsilon_{p,\thresh}$ it turns unstable, while
parametric oscillations with finite amplitude $A_s \neq 0$ set in.
The solid and dashed bold lines show $\epsilon_{p,\thresh}$
from \Eq{eq:h_thresh__wdependent} and from
the corresponding Markov approximation, \Eq{eq:h_thresh__Markov}, respectively.
Thin dash-dotted and dotted lines show the underlying $\delta_p$-dependence
of $\Sigma(\Delta=0)=\Sigma_E(\Delta=0)$ for $\Gamma_I=0$:
either calculated for series-$\sLC$ coupling
with $Z_s/Z_0 = 2$, $C_c/C_s = 10$, $L_c/L_s = 0.4$
from \Eqs{eq:imSig_seriescoupled}
and \eqref{eqapp:reSig1_seriescoupled}--\eqref{eqapp:reSig2_seriescoupled}
(yellow, resulting $\epsilon_{p,\thresh}$ in red),
or held fixed at value $\Sigma(\Delta=0)[\delta_p=0] \approx (-0.36 -  0.34\ui)\omega_s$
(gray, resulting $\epsilon_{p,\thresh}$ in black).
}
\label{fig:threshold}
\end{figure}

The stability of the steady state $A_s=0$ is determined by the eigenvalues
of the linear dynamical system \Eq{eq:EOM_As_preMarkov} at $B_{\In}=0$.
We obtain the dynamical matrix from a closed-form approximation of
\Eq{eq:EOM_As_preMarkov}
which omits its dependence on the evolution in the past, $A_s(t' < t)$.
This is done by approximating $A_s(t') \approx A_s(t)$ in the integral
and leads to a dynamical equation of the form of \Eq{eq:EOM_As_Markov__paramp},
with the replacement $\Gamma \to \ui \Sigma(\Delta=0)$.
(This is equivalent to setting $\Delta=0$ in \Eq{eq:As_Delta_nonMarkov__paramp}.)
As discussed in the Markov case,
this dynamical system is stable as long as its determinant --
which equals $D(\Delta=0)$ from \Eq{eq:determinant_wdependent} --
is larger than $0$.
This is fulfilled if the pump strength $\epsilon_p$ is below the instability threshold,
\begin{equation}\label{eq:h_thresh__wdependent}
 |\epsilon_{p,\thresh}|^2 = \delta_{p}^2 + |\Sigma(0)|^2 - 2\delta_{p} \re\Sigma(0)
 \,.
\end{equation}
Assuming that $\Sigma(\Delta=0)$ is independent of $\delta_p$,
the comparison of \Eq{eq:h_thresh__wdependent} with
its Markov approximation counterpart, \Eq{eq:h_thresh__Markov},
shows that the frequency-dependent coupling shifts
the minimum of the pump strength threshold from $\delta_p=0$ to $\delta_p = \re\Sigma(0)$,
while leaving the minimum value invariant at $|\im \Sigma(0)|$.
This is illustrated in Fig.~\ref{fig:threshold}
by the black data lines (solid vs dashed),
which show the respective $\epsilon_{p,\thresh}$-values as functions of $\delta_{p,\thresh}$
for a fixed value of $\Sigma(\Delta=0)$.
In practice, when taking into account the actual $\delta_p$-dependence
of $\Sigma(\Delta=0)$,
the two $\epsilon_{p,\thresh}$-functions as well as their relation can be more complicated,
as illustrated by the red data lines (solid vs dashed).
At $\delta_p=0$, \Eq{eq:h_thresh__wdependent},
always assumes a higher or equal value than \Eq{eq:h_thresh__Markov},
$|\epsilon_{p,\thresh}| = |\Sigma(0)| \geq |\im\Sigma(0)|$,
implying that a higher pump strength is required to reach the threshold
under the frequency-dependent coupling.
Depending on the experimental setup, this may be a disadvantage,
e.g.~if the available pump strength is too limited for the device to be operated
near the threshold.
If this limitation does not exist, the higher pump-threshold
on the other hand enables a more robust tuning of the pump parameters.

\subsubsection{Unitarity conditions}

An important consistency check
(between \Eq{eq:inputoutput_frequencydependent} and \Eq{eq:As_wdependent_general})
is the preservation of the commutation relations for the TL amplitudes
under unitary system evolution, i.e. in absence of internal losses, $\Gamma_I=0$.
We had found the commutation relations, \Eq{eq:commutator1_Bin},
for the input amplitudes under frequency-dependent coupling.
Using  \Eq{eq:Bin_Bout} we can write the commutation relations for the output
amplitudes as
\begin{align}
  &\left[ B_{\Out}(\Delta), B_{\Out}^\dag(\Delta') \right]
  = u(\Delta) u^\ast(\Delta') \left[ B_{\In}(\Delta), B_{\In}^\dag(\Delta') \right] \nonumber \\
  &\hspace{3.2cm}+ v(\Delta) v^\ast(\Delta') \left[ B_{\In}^\dag(-\Delta), B_{\In}(-\Delta') \right] \nonumber \\
\label{eq:commutator1_Bout}
  &= -\left(  |u(\Delta)|^2 \frac{\im\Sigma_E(\Delta)}{\Gamma_E}
           - |v(\Delta)|^2 \frac{\im\Sigma_E(-\Delta)}{\Gamma_E}
   \right) \delta(\Delta-\Delta') \\
  &\left[ B_{\Out}(\Delta), B_{\Out}(\Delta') \right]
  = u(\Delta) v(\Delta') \left[ B_{\In}(\Delta), B_{\In}^\dag(-\Delta') \right] \nonumber \\
  &\hspace{3.2cm}+ v(\Delta) u(\Delta') \left[ B_{\In}^\dag(-\Delta), B_{\In}(\Delta') \right] \nonumber \\
\label{eq:commutator2_Bout}
  &= -\left( u(\Delta)v(-\Delta) \frac{\im\Sigma_E(\Delta)}{\Gamma_E} \right. \nonumber \\
   &\left.\hspace{0.7cm} - v(\Delta) u(-\Delta) \frac{\im\Sigma_E(-\Delta)}{\Gamma_E}
   \right) \delta(\Delta+\Delta')
\end{align}
Setting the rhs of \Eqs{eq:commutator1_Bout} and \eqref{eq:commutator2_Bout} equal to the rhs of \Eqs{eq:commutator1_Bin} and \eqref{eq:commutator2_Bin}, respectively,
we obtain the following unitarity conditions for the gain coefficients,
\begin{eqnarray}
\label{eq:Bogoliubov_rel1_wdependent}
 |u(\Delta)|^2
   - |v(\Delta)|^2  \frac{\im\Sigma_E(-\Delta)}{\im\Sigma_E(\Delta)}
   &=& 1   \\
\label{eq:Bogoliubov_rel2_wdependent}
 u(\Delta)v(-\Delta)
   - v(\Delta) u(-\Delta) \frac{\im\Sigma_E(-\Delta)}{\im\Sigma_E(\Delta)}
   &=& 0
\end{eqnarray}
It can easily be checked that the gain coefficients, \Eq{eq:u_Delta__wdependent} and \Eq{eq:v_Delta__wdependent}
fulfill \Eqs{eq:Bogoliubov_rel1_wdependent}--\eqref{eq:Bogoliubov_rel2_wdependent}
independent of the form of $\Sigma_E(\Delta)$,
provided that no internal damping is present, $\Gamma_I=0$.
In this case, the direct gain simplifies to
\begin{equation}
  |u(\Delta)|^2 = 1 + \frac{4 |\epsilon_p|^2 \im\Sigma_E(\Delta) \im\Sigma_E(-\Delta) }{|D(\Delta)|^2}
\end{equation}

\subsubsection{Quantum noise}

\textcolor{black}{
The system amplitude in the linearized response equation, \Eq{eq:As_Delta_nonMarkov__paramp}, and the input and output amplitudes in the parametric amplifier relation, \Eq{eq:Bin_Bout}, may have noise contributions in addition to their classical contributions, i.e.~the expectation values $\langle A_s \rangle$, $\langle B_{\text{in,out}}\rangle$. 
Under the usual assumption that the classical expectation values dominate, the noise amplitudes $\delta\!A_s$, $\delta\!B_{\text{in,out}}$ may be separated and are then themselves characterized by \Eqs{eq:As_Delta_nonMarkov__paramp} and \eqref{eq:Bin_Bout}. 
Noise characteristics like the output power spectrum, 
\begin{equation}
n_{\text{out}}(\Delta)=\int_{-\infty}^{\infty} d\Delta' \langle \delta\!B_{\text{out}}^\dagger(\Delta) \delta\!B_{\text{out}}(\Delta') \rangle
\end{equation}
or the output squeezing spectrum are therefore determined by the coefficients $u(\Delta)$ and $v(\Delta)$, as are the classical gains \cite{ClerkETAL_review2010, WusShu2013}. For example, if the input noise consists of vacuum noise alone, then $n_{\text{out}}(\Delta)=|v(\Delta)|^2$.
(In our analysis we have neglected nonlinear contributions to \Eq{eq:As_Delta_nonMarkov__paramp}. If these were taken into account, then the coefficients $u(\Delta)$ and $v(\Delta)$ in the parametric relation for the noise amplitudes, \Eq{eq:Bin_Bout}, become functions of the classical system amplitude $\langle A_s \rangle$.)
}

\subsection{Parametric gains for coupling with simple ladder filter}
\label{subsec:gains_filtercoupling}

Here we calculate the response and gain of a system under parametric resonance,
when coupled to the input and output TL via the simple ladder filter
discussed in Sec.~\ref{subsec:summary_parallelLCbeforeseriesLC}.
In this case, we choose not to proceed directly with the results of Sec.~\ref{subsec:paramp_frequencydependentcoupling_gains_unitarity},
since the system here consists of more degrees of freedom: it is extended by the
flux $\Phi_a$, and it is also $\Phi_a$ (or rather its conjugate charge $Q_a$)
instead of $\Phi_s$ that is involved in the coupling of the extended system
with the TL.

Starting with \Eqs{eqFq:Ham_orig} -- \eqref{eqFq:Hsysext},
where in \Eqs{eqFq:Ham_orig} we apply the RWA to the contribution describing
the coupling of the extended system with the TL.
Continuing as in Sec.~\ref{subsec:wdependentcoupling_systresponse}
we obtain, instead of \Eq{eq:EOM_as_preMarkov},
\begin{align}
\label{eqF:EOM_as_preMarkov}
\dot a_s &= \frac{\ui}{\hbar} \left[H_{\sys}, a_s\right] - \Gamma_I a_s \\
\dot a_a &= \frac{\ui}{\hbar} \left[H_{\sys}, a_a\right]
 + \sqrt{2\Gamma_E}  b_\In \nonumber \\
\label{eqF:EOM_aa_preMarkov}
&\quad- \int_0^\infty \diff k f_k^2 \int_{t_0}^t \diff t' a_a(t') e^{-\iexp \omega_k(t-t')}
\end{align}
with the coupling coefficients $f_k$ from \Eq{eqFq:def_fk}.
The unitary transformation to the frame rotating at the reference frequency
$\wref$, \Eq{eq:unitarytrafo2RF},
needs to be modified as $U = e^{-\ui \wref (a_s^\dag a_s + a_a^\dag a_a) t}$
and gives
$\tilde H_{\sys} = U^\dag H_{\sys} U  - \hbar \wref (A_s^\dag A_s + A_a^\dag A_a)$.
Finally, instead of \Eq{eq:As_wdependent_general}, we obtain
\begin{align}\label{eqF:As_wdependent_general}
-\ui \Delta A_s(\Delta)
 &= \frac{\ui}{\hbar} \left\langle \left[\tilde{H}_{\sys}, A_s\right] \right\rangle_t(\Delta)
 - \Gamma_I A_s(\Delta) \\
-\ui \Delta A_a(\Delta)
 &= \frac{\ui}{\hbar} \left\langle \left[\tilde{H}_{\sys}, A_a\right] \right\rangle_t(\Delta)
  \nonumber  \\
\label{eqF:Aa_wdependent_general}
&\quad + \sqrt{2\Gamma_E} B_\In(\Delta) - \ui \Sigma_E(\Delta) A_a(\Delta) \,,
\end{align}
and, instead of \Eq{eq:inputoutput_frequencydependent}, the input-output relation
\begin{equation}\label{eqF:inputoutput_frequencydependent}
B_{\Out}(\Delta) = B_{\In}(\Delta)
+ \frac{2\im\Sigma_E(\Delta)}{\sqrt{2\Gamma_E}} A_a(\Delta)
\,.
\end{equation}
Here, $\Sigma_E(\Delta)$ is given in \Eq{eq:imSig_filtercoupled}
and \Eqs{eqapp:reSig1_parallelbeforeseriescoupled}--\eqref{eqapp:reSig2_parallelbeforeseriescoupled}.

For the parametric resonance,
after applying the unitary transformation $U$ with $\wref = \omega_p/2$
to the extended system Hamiltonian, \Eq{eqFq:Hsysext}, and averaging over fast oscillations,
one obtains
\begin{align}
 \tilde H_{\sys}
 &= -\hbar \delta_{p,s} A_s^\dag A_s - \hbar \delta_{p,a} A_a^\dag A_a
 - \hbar g (A_s^\dag A_a + A_a^\dag A_s) \nonumber \\
 &\quad - \frac{\hbar}{2} \left(\epsilon_p A_s^\dag A_s^\dag + \epsilon_p^\ast A_s A_s\right)
\end{align}
where $\delta_{p,s} = \omega_p/2 - \omega_s$
and $\delta_{p,a} = \omega_p/2 - \omega_a = \delta_{p,s} + \omega_s - \omega_a$
are the pump detunings, $\epsilon_p$ is the pump strength, cf.~\Eq{eq:pump_strength},
and the coupling constant $g$ is defined in \Eq{eqFq:def_g}.
Here, we have also assumed that $\tilde{V}_{\nlin} = U^\dag V_{\nlin} U$
takes the form as in \Eq{eqG0:Vnlin_rotatingframe}, cf.~\Eq{eq:Hsys_paramp_RF}.
Thus, \Eqs{eqF:As_wdependent_general}--\eqref{eqF:Aa_wdependent_general} become
\begin{align}\label{eqF:As_Delta_nonMarkov__paramp}
&\!\!\!(\delta_{p,s} + \Delta + \ui \Gamma_I) A_s(\Delta)
+ g A_a(\Delta) + \epsilon_p A_s^\dag(-\Delta) = 0 \\
\label{eqF:Aa_Delta_nonMarkov__paramp}
&\!\!\!(\delta_{p,a} + \Delta - \Sigma_E(\Delta)) A_a(\Delta)
+ g A_s(\Delta) = \ui \sqrt{2\Gamma_E} B_{\In}(\Delta)
\,,
\end{align}

Extracting $A_a(\Delta)$ from the 2nd equation,
\begin{align}
A_a(\Delta) &= \frac{-g}{\delta_{p,a} + \Delta - \Sigma_E(\Delta)} A_s(\Delta) \nonumber \\
\label{eqF:AaDelta_AsDelta_BinDelta}
&\quad + \frac{\ui \sqrt{2\Gamma_E}}{\delta_{p,a} + \Delta - \Sigma_E(\Delta)}  B_{\In}(\Delta)
\,,
\end{align}
inserting into the first one, and inverting the corresponding matrix equation
one obtains a relation for the response of the resonator amplitude $A_s$ alone,
\begin{align}\label{eqF:As_paramp_matrixform_wdependent}
& \begin{pmatrix} A_s(\Delta) \\ A_s^\dag(-\Delta) \end{pmatrix}
= \boldsymbol{\chi}
\begin{pmatrix} B_{\In}(\Delta) \\ B_{\In}^\dag(-\Delta) \end{pmatrix} \\
& \boldsymbol{\chi} =
- \frac{\ui g \sqrt{2\Gamma_E}}{D(\Delta)}
 \begin{pmatrix}
 \frac{ \delta_{p,s} - \Delta - \Sigma^\ast(-\Delta) }{\delta_{p,a} + \Delta - \Sigma_E(\Delta)} & \frac{\epsilon_p}{\delta_{p,a} - \Delta - \Sigma^\ast_E(-\Delta)} \\
 \frac{-\epsilon_p^\ast}{\delta_{p,a} + \Delta - \Sigma_E(\Delta)} &
 \frac{\delta_{p,s} + \Delta - \Sigma(\Delta)}{\delta_{p,a} - \Delta - \Sigma^\ast_E(-\Delta)}
 \end{pmatrix}
\nonumber
\end{align}
where we have defined the total self-energy
\begin{equation}\label{eqF:selfenergy_tot}
 \Sigma(\Delta) := \frac{g^2}{\delta_{p,a} + \Delta - \Sigma_E(\Delta)} -\ui \Gamma_I
\end{equation}
which combines the effects of the external coupling and the internal loss.
The parametric resonance determinant 
$D(\Delta)$ has the same form as in \Eq{eq:determinant_wdependent},
with the (formal) replacement $\delta_p \to \delta_{p,s}$,
\begin{equation}\label{eq:determinant_wdependent_filter}
D(\Delta) = \left(\delta_{p,s}+\Delta - \Sigma(\Delta)\right)\left(\delta_{p,s}-\Delta-\Sigma^\ast(-\Delta)\right) - |\epsilon_p|^2
\end{equation}
Inserting the solution $A_s(\Delta)$ of \Eq{eqF:As_paramp_matrixform_wdependent} into
\Eq{eqF:AaDelta_AsDelta_BinDelta} one obtains $A_a(\Delta)$ as function of $B_\In(\Delta)$.
Finally, inserting this solution into
the input-output relation, \Eq{eqF:inputoutput_frequencydependent},
one arrives again at the amplifier relation, \Eq{eq:Bin_Bout}, where
here the gain coefficients read
\begin{align}
&u(\Delta) - 1 \nonumber\\
\label{eqF:u_Delta__wdependent}
 &= \frac{2 \ui g^2 \im \Sigma_E(\Delta)}{D(\Delta) d_a(\Delta)} \biggl(\frac{D(\Delta)}{g^2} +\frac{\delta_{p,s} - \Delta - \Sigma^\ast(-\Delta)}{d_a(\Delta)}  \biggr) \\
 &= \frac{2 \ui \im \Sigma_E(\Delta)}{D(\Delta) d_a(\Delta)}
 \Bigl[ \phantom{\cdot}\left(\delta_{p,s} - \Delta - \Sigma^\ast(-\Delta) \right)\Bigr. \nonumber\\
 &\hspace*{2.5cm} \Bigl.\cdot \left(\delta_{p,s} + \Delta + \ui \Gamma_I\right) -  |\epsilon_p|^2 \Bigr] \\
\label{eqF:v_Delta__wdependent}
 &v(\Delta) =  \frac{2 \ui \epsilon_p g^2 \im \Sigma_E(\Delta)}{D(\Delta)d_a(\Delta)d_a^\ast(-\Delta)} \\
 &\hspace*{0.8cm} = \frac{2 \ui \epsilon_p \left(\Sigma(\Delta) + \ui \Gamma_I \right) \im \Sigma_E(\Delta)}{D(\Delta)d_a^\ast(-\Delta)}
\end{align}
defining the short-hand notation
 $d_a(\Delta) = \delta_{p,a} + \Delta - \Sigma_E(\Delta)$.

It can be checked that \Eqs{eqF:u_Delta__wdependent}--\eqref{eqF:v_Delta__wdependent}
fulfill the unitarity conditions,
\Eqs{eq:Bogoliubov_rel1_wdependent}--\eqref{eq:Bogoliubov_rel2_wdependent},
independent of the form of $\Sigma_E(\Delta)$,
provided that there is no internal damping, $\Gamma_I=0$.

\begin{figure}\centering
\includegraphics[width=\columnwidth]{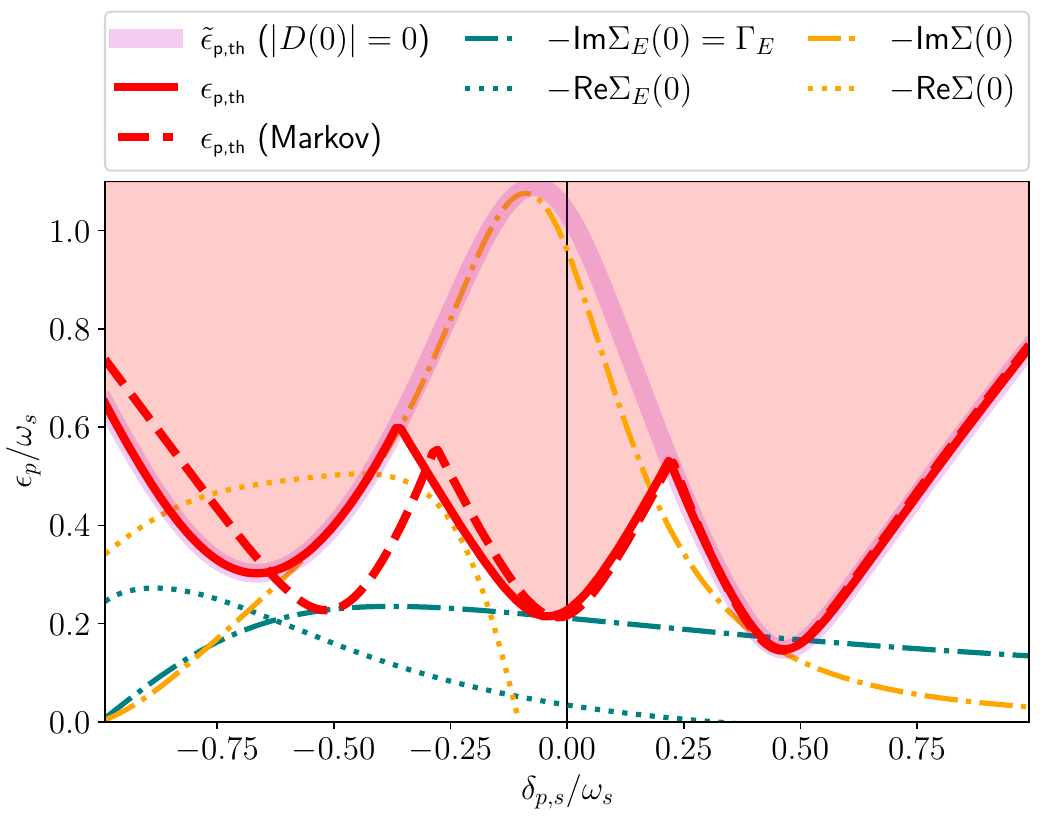}
\caption{
Pump-parameter space $(\delta_{p,s}, \epsilon_p)$
showing stable and unstable parameter regions for trivial state $A_s = A_a =0$
of undriven resonator ($B_{\In} = 0$) coupled to the TL via the simple ladder filter.
It is stable for pump strength $\epsilon_p$ below
the pump threshold $\epsilon_{p,\thresh}$ (red solid line),
where all eigenvalues of \Eq{eq:dynamicalmatrix_filter} have negative real parts.
For the outer two lobes of the instability region (shaded),
the threshold $\epsilon_{p,\thresh}$
(where the real part of one or more eigenvalues turns zero)
is equivalent to the value $\tilde{\epsilon}_{p,\thresh}$ (bold magenta)
defined by the condition $|\det(\boldsymbol{\mathcal{A}})| = |D(0)| = 0$.
The red dashed line shows $\epsilon_{p,\thresh}$
according to the Markov approximation for the threshold,
obtained by setting $\re\Sigma_E(0) = 0$ in \Eq{eq:dynamicalmatrix_filter}.
Thin dash-dotted and dotted lines show the underlying $\delta_{p,s}$-dependence
of $\Sigma(\Delta=0)$ (yellow) and $\Sigma_E(\Delta=0)$ (teal),
calculated for parameters
$Z_s/Z_0 = 0.7$, $C_g/C_s = 120$, $L_g/L_s = 0.002$,
$C_c/C_s = 1.12$, $L_c/L_s = 4.0$
and $\Gamma_I=0$ from \Eqs{eq:imSig_filtercoupled},
\Eqs{eqapp:reSig1_parallelbeforeseriescoupled}--\eqref{eqapp:reSig2_parallelbeforeseriescoupled},
and \eqref{eqF:selfenergy_tot}.
}
\label{fig:threshold_filtercoupling}
\end{figure}

As discussed in Sec.~\ref{subsec:paramp_frequencydependentcoupling_gains_unitarity},
the threshold for parametric instability
can be found from the eigenvalues of the linear dynamical system,
\Eqs{eqF:EOM_as_preMarkov}--\eqref{eqF:EOM_aa_preMarkov}.
Again, we bring \Eqs{eqF:EOM_as_preMarkov}--\eqref{eqF:EOM_aa_preMarkov}
into closed form with the approximation $a_a(t') \approx a_a(t)$.
The resulting dynamical matrix in the vector space of
$(\re A_s, \im A_s, \re A_a, \im A_a)$ is
\begin{equation}\label{eq:dynamicalmatrix_filter}
 \boldsymbol{\mathcal{A}} = \begin{pmatrix}
 -\Gamma_I & \epsilon_p - \delta_{p,s} & 0 & -g \\
  \epsilon_p + \delta_{p,s} & -\Gamma_I & g & 0 \\
  0 & -g & -\Gamma_E & -\delta_{p,a}' \\
  g &  0 & \delta_{p,a}' & -\Gamma_E
 \end{pmatrix}
\end{equation}
with $\delta_{p,a}' = \delta_{p,a} - \re\Sigma_E(0)$ and $\Gamma_E = -\im\Sigma_E(0)$.
The dynamical system is stable if all four eigenvalues of $\boldsymbol{\mathcal{A}}$
have negative real parts.
The resulting instability region is shown in Fig.~\ref{fig:threshold_filtercoupling}
in the pump-parameter space $(\delta_{p,s},\epsilon_p)$
by the shaded area, confined from below by the threshold pump strength $\epsilon_{p,\thresh}$
(solid red line).
In contrast to the $2\times2$-dynamical system of Sec.~\ref{subsec:paramp_frequencydependentcoupling_gains_unitarity},
the stability criterion is here no longer equivalent to the condition that
the determinant $\det(\boldsymbol{\mathcal{A}})$ is positive.
The parameter line $\tilde \epsilon_{p,\thresh}$ where $\det(\boldsymbol{\mathcal{A}})$ changes sign
is shown in Fig.~\ref{fig:threshold_filtercoupling} (bold magenta line),
and this is also the line where $|D(0)| = 0$,
because of the relation $\det(\boldsymbol{\mathcal{A}}) = |D(0)|\cdot\left((\delta_{p,a}')^2 + \Gamma_E^2 \right)$,
with $D(0)$ from \Eq{eq:determinant_wdependent_filter}.
Of the three lobes (red solid line) formed by $\epsilon_{p,\thresh}$,
the two outer ones are also determined by $\det(\boldsymbol{\mathcal{A}}) = |D(0)| = 0$,
whereas the inner lobe is not.
In the $\delta_{p,s}$-range of the outer lobes near the
threshold, $\epsilon_p \approx \epsilon_{p,\thresh}$,
two of the eigenvalues form a complex conjugate pair,
whereas the other two are purely real and $<0$ at $\epsilon_p < \epsilon_{p,\thresh}$.
As one of them changes its sign to $>0$ at $\epsilon_p > \epsilon_{p,\thresh}$,
this transition then also gives rise to a sign-change of $\det(\boldsymbol{\mathcal{A}}) = \Pi_{i=1-4} \lambda_i$
from $>0$ to $<0$.
In the $\delta_{p,s}$-range of the inner lobe,
there are two complex conjugate eigenvalue pairs and one of the pairs
changes its real part from $<0$ for $\epsilon_p < \epsilon_{p,\thresh}$
to $>0$ for $\epsilon_p > \epsilon_{p,\thresh}$.
Since this happens simultaneously to two eigenvalues, this transition is here
not reflected in a sign-change of $\det(\boldsymbol{\mathcal{A}})$.
The multi-lobed character of the instability regime differs from
the behavior for simple, frequency-independent coupling, cf.~\Eq{eq:h_thresh__Markov},
and may offer applications analogous to e.g. threshold-sensitive qubit-state readout
\cite{KrantzETAL2016_NatComm}.

The red dashed line in Fig.~\ref{fig:threshold_filtercoupling} shows the
Markov approximation for the threshold, obtained by setting $\re\Sigma_E(0) = 0$
in \Eq{eq:dynamicalmatrix_filter}.
This is very close to $\epsilon_{p,\thresh}$ at $\delta_{p,s}>0$,
where $|\re\Sigma_E(0)| \ll |\im\Sigma_E(0)|$ ($|\re\Sigma_E(0)|$ is the dotted teal)
is small in our parameter example,
but substantial deviations occur for $\delta_{p,s}<0$
where $\re\Sigma_E(0)$ is of the same order as $\im\Sigma_E(0)$ (dash-dotted teal).
For comparison, we also show the total self-energy $\Sigma(0)$.

\section{Bandwidth broadening in resonant and non-resonant coupling regimes}
\label{sec:results_bandwidthbroadening}

A lumped-element parametric device directly coupled to a TL via
a frequency-independent coupling rate $\Gamma_E$
has a fixed relation between its maximum gain $G_{s,\max} = G_s(\Delta=0)$ and its bandwidth
$\sigma$,
\begin{align}\label{eq:gain_bandwidth_tradeoff}
 &\sigma \sqrt{G_{s,\max}}
 = \frac{|\Gamma (\Gamma_E - \Gamma_I) + |\epsilon_p|^2|}{\sqrt{\Gamma^2 - 2\Gamma_I^2/G_{s,\max} }} 
\end{align}
as follows from \Eq{eq:u_Delta__Markov} for $\delta_p=0$ and moderate to large gain.
In the limit of large gain,
$\epsilon_p \to \Gamma$, this relation can be further simplified to read
\begin{align}\label{eq:gain_bandwidth_tradeoff_approx}
 &\sigma \sqrt{G_{s,\max}}
 \approx \frac{2\Gamma \Gamma_E}{\sqrt{\Gamma^2 - 2\Gamma_I^2/G_{s,\max}}} 
\end{align}
Herein, the bandwidth is defined as the full-width half-maximum of the
(approximate) gain profile $G_s(\Delta) = |u(\Delta)|^2$.

The trade-off \textcolor{black}{\Eq{eq:gain_bandwidth_tradeoff}}
between maximum gain and bandwidth critically limits
applications of the device even over a moderate bandwidth
and excludes e.g.~multiplexed readout \cite{ChenETAL2012, JergerETAL2012, ChapmanETAL2017}.
Frequency-dependent environments, on the other hand, allow to overcome
this limitation and to broaden the bandwidth at fixed gain,
as has been shown in many demonstrations \cite{MutusETAL2014, RoyETAL2015, GrebelETAL2021, DuanETAL2021, WhiteETAL2023}.
The idea was originally established in Ref.~\onlinecite{RoyETAL2015},
where the authors proposed to engineer the impedance of the environment
such that the real part of the self-energy approximates to
$\re \Sigma(\Delta) \approx 0 + \Delta$ in the vicinity of the reference
frequency $\wref$.
This removes the explicit occurrence of $\Delta$ in the susceptibility equation
for $A_s(\Delta)$, \Eq{eq:As_Delta_nonMarkov__paramp}.
The resulting determinant in the denominator of \Eq{eq:usq_Delta__wdependent}
is then only weakly $\Delta$-dependent,
$|D(\Delta)|^2 \approx \delta_p^2 + (\im\Sigma(\Delta))^2 - |\epsilon_p|^2 $,
where one assumes that $\im\Sigma(\Delta)$ is even about $\Delta=0$.
Usually, it is assumed that the coupling is resonant, i.e.
 $|\im\Sigma|$ has its maximum at $\Delta=0$, while $\re\Sigma(0) = 0$:
\begin{align}
  &\frac{\diff |\im\Sigma|}{\diff \Delta}(\Delta=0) = 0 \,, \nonumber \\
\label{eq:resonantcondition_bandwidthbroadening}
  &\re\Sigma(0) = 0 \,,\qquad
  \frac{\diff |\re\Sigma|}{\diff \Delta}(\Delta=0) = 1
\end{align}
Below, we refer to this condition for bandwidth-broadening
as `resonant slope-matching'.

\subsection{Series $\sLC$-coupling}\label{subsec:results_seriesLC}

\begin{figure}[bh]\centering
\includegraphics[width=\columnwidth]{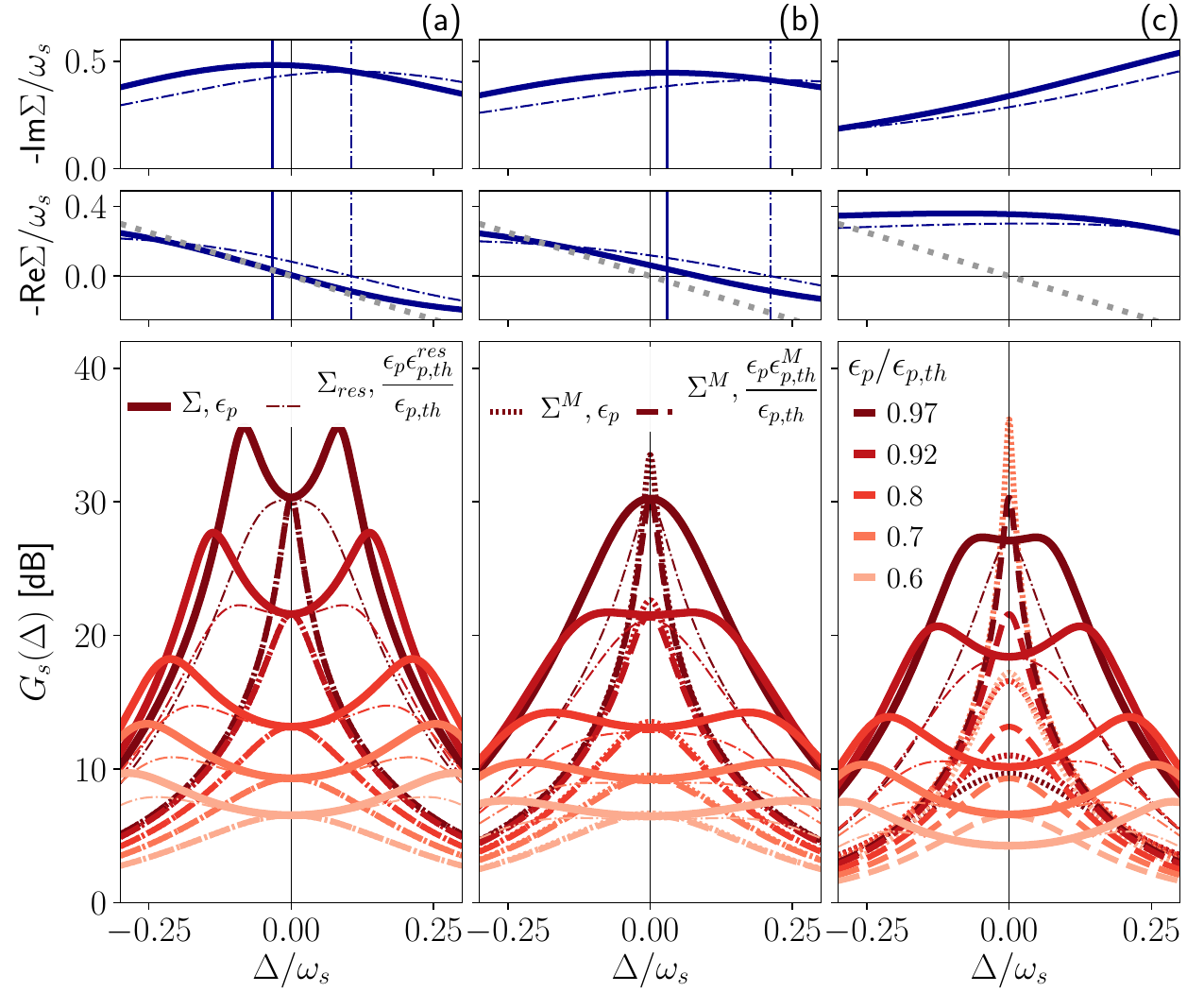}
\caption{
\textcolor{black}{
Coupling of parametric resonator via a series-$\sLC$ circuit,} 
Fig.~\ref{fig:filters}(a),
with parameters $(Z_s/Z_0, C_c/C_s, L_c/L_s) =$
(a) (1,10,0.9), (b) (1,10,0.75), and (c) (2,10,0.4).
The two upper panels show the self-energy $\Sigma(\Delta) = \Sigma_E(\Delta)$
from the exact solution, \Eqs{eq:imSig_seriescoupled}
and \eqref{eqapp:reSig1_seriescoupled}--\eqref{eqapp:reSig2_seriescoupled} (solid),
and its resonant approximation $\Sigma_{E,\text{res}}(\Delta)$,
\Eq{eq:selfenergy_seriescoupled_resapprox} (thin dash-dotted).
The gray dashed lines in the $\re\Sigma$-panel indicates the condition
$\re\Sigma(\Delta) = -\Delta$. The solid and dash-dotted vertical lines mark $\omega_{\Sigma,0}$
and $\omega_\Sigma$, respectively.
\textcolor{black}{
The lowest panels show the resulting parametric gain profiles $G_s(\Delta)$ 
for various parametric pump strengths $\epsilon_p = r \epsilon_{p,\thresh}$ (distinguished by colors)
and different models of the self-energy $\Sigma$ (distinguished by line styles). 
While $r$ is fixed for each color, 
$\epsilon_{p,\thresh}$ and thus $\epsilon_p$ may differ, as indicated by the legend entries:
}
the solid \textcolor{black}{(marked $\Sigma, \epsilon_p$)}
and thin dash-dotted \textcolor{black}{(marked $\Sigma_{\text{res}}, \epsilon_p \epsilon^{\text{res}}_{p,\thresh}/\epsilon_{p,\thresh}$)} 
lines show $G_s(\Delta)$ from \Eq{eq:usq_Delta__wdependent}
with $\epsilon_{p,\thresh}$ from \Eq{eq:h_thresh__wdependent},
and where $\Sigma_E(\Delta)$ (in the evaluation of both \Eqs{eq:usq_Delta__wdependent} and \eqref{eq:h_thresh__wdependent}) is based on either the exact solution,
\Eqs{eq:imSig_seriescoupled}
and \eqref{eqapp:reSig1_seriescoupled}--\eqref{eqapp:reSig2_seriescoupled} \textcolor{black}{(solid)},
or the resonant approximation, \Eq{eq:selfenergy_seriescoupled_resapprox} (\textcolor{black}{thin }dash-dotted).
The dotted \textcolor{black}{(marked $\Sigma^M,\epsilon_p$)} and dashed \textcolor{black}{(marked $\Sigma^M,\epsilon_p\epsilon^M_{p,\thresh}/\epsilon_{p,\thresh}$)} lines show $G_s(\Delta)$ based on the Markov approximation, \Eq{eq:u_Delta__Markov},
using $\epsilon_{p,\thresh}$ from \Eq{eq:h_thresh__wdependent} (dotted)
and from \Eq{eq:h_thresh__Markov} (dashed), respectively.
(Thus, in comparison with the solid line, the dotted uses the same $\epsilon_p$-value, whereas the dashed uses the same $\epsilon_p/\epsilon_{p,\thresh}$-ratio.%
)
In (a,b), \textcolor{black}{the  values of $\epsilon_{p,\thresh}$ and $\epsilon^M_{p,\thresh}$} are nearly equal
because of $|\re\Sigma_E(\Delta=0)| \ll |\im\Sigma(\Delta=0)|$ \textcolor{black}{-- leading to the coincidence of both Markov gains (dotted and dashed curves in lowest panels)},
whereas in (c) the Markov approximation gives a $30\%$ smaller
threshold value, cf.~Fig.~\ref{fig:threshold}
(red dashed vs red solid line at $\delta_p=0$).
The Markov cases use $\Gamma_E = -\im\Sigma_E(\Delta=0)$, \Eq{eq:def_GammaE},
with $\im\Sigma_E$ from \Eq{eq:imSig_seriescoupled}.
Other parameters are $\delta_p = 0$, $\Gamma_I = 0$.
}
\label{fig:gains_wdependentcoupling}
\end{figure}

Figure \ref{fig:gains_wdependentcoupling} shows the parametric gain profiles
$G_s(\Delta)$ for several constellations of coupling and system parameters,
distributed in panels (a-c).
In all cases, the parametric system is coupled to the TL via series-$\sLC$ circuits,
with coupling parameters $L_c, C_c$.
The upper panels show the resulting real and imaginary parts of the self-energy
$\Sigma(\Delta) = \Sigma_E(\Delta)$,
based on the exact solution $\Sigma_E(\Delta)$, \Eqs{eq:imSig_seriescoupled}
and \eqref{eqapp:reSig1_seriescoupled}--\eqref{eqapp:reSig2_seriescoupled} (solid),
and its resonant approximation $\Sigma_{E,\text{res}}(\Delta)$, \Eq{eq:selfenergy_seriescoupled_resapprox} (dash-dotted).
In the bottom panels,
$G_s(\Delta)$ is shown for a sequence of pump strengths $\epsilon_p$.
Each color refers to a fixed ratio
$r = \epsilon_p/\epsilon_{p,\thresh}$ to the threshold value,
where $\epsilon_{p,\thresh}$ is from \Eq{eq:h_thresh__wdependent}
except stated otherwise.
The thick solid and thin dash-dotted lines show $G_s(\Delta) = |u(\Delta)|^2$
from \Eq{eq:usq_Delta__wdependent}.
It is based on the frequency-dependent self-energy $\Sigma_E(\Delta)$,
which is taken either from the exact solution,
\Eqs{eq:imSig_seriescoupled}
and \eqref{eqapp:reSig1_seriescoupled}--\eqref{eqapp:reSig2_seriescoupled} (solid),
or from the resonant approximation, \Eq{eq:selfenergy_seriescoupled_resapprox} (dash-dotted).
Note, that in the latter case, the pump strength is adapted to the threshold for this resonant approximation, $\epsilon_p = r \epsilon_{p,\thresh}^{\text{res}}$,
where $\epsilon_{p,\thresh}^{\text{res}}$ follows from \Eq{eq:h_thresh__wdependent}
with $\Sigma_{E,\text{res}}$ replacing $\Sigma_{E}$.
By this choice, the two gain profiles based on either $\Sigma_E$ or $\Sigma_{E,\text{res}}$
are (approximately) equal at $\Delta=0$.
The partly strong deviation between the two gain profiles is mostly due to the
deviations between $\Sigma_E$ and $\Sigma_{E,\text{res}}$
as seen in the upper panels.
These deviations, in turn, are largely caused because the
resonant approximation $\Sigma_{E,\text{res}}$ uses
an expansion around $\omega_\Sigma$, \Eq{eq:wres_selfenergy_imag_seriescoupled},
which is only an approximate value for the actual position $\omega_{\Sigma,0}$, \Eq{eqapp:wres_exact_selfenergy_imag_seriescoupled},
of the coupling resonance, i.e. of the maximum of $\im\Sigma(\Delta)$.
In the upper panels, the solid and dash-dotted vertical lines mark $\omega_{\Sigma,0}$
and $\omega_\Sigma$, respectively.
As described in App.~\ref{app:selfenergy_variouscouplings},
an alternative expansion $\Sigma_{E,\text{res},0}$, \Eq{eqapp:selfenergy_seriescoupled_resapprox_exact},
can be obtained using the accurate resonance position $\omega_{\Sigma,0}$,
\Eq{eqapp:wres_exact_selfenergy_imag_seriescoupled}.
This expansion leads to much better agreement (not shown) with the exact gain profile
in panel (a),
where $\Delta_{\Sigma,0} = \omega_{\Sigma,0} - \wref \approx 0$.
In panel (b), this resonant approximation (thin dash-dotted line) is still not optimal.
Note in both cases how $\omega_{\Sigma,0}$
is not exactly aligned with the location where $\re\Sigma(\Delta)=0$.
In case (c), where $\Delta_{\Sigma,0} \approx \omega_s/2$ is comparably large,
both resonant approximations deviate significantly from the exact gain profile.

Next we compare the gain $G_s(\Delta)$ with the gain according to the Markov approximation.
Dashed and dotted lines in the bottom panels of Fig.~\ref{fig:gains_wdependentcoupling}
show $G_s^M$ based on the Markov approximation,
$\Sigma_E(\Delta) = -\ui \Gamma_E = -\ui \Sigma_E(0)$, cf.~\Eq{eq:def_GammaE},
leading to \Eq{eq:u_Delta__Markov}.
The pump strength in these Markov cases
is either chosen identical to the frequency-dependent case,
$\epsilon_p = r \epsilon_{p,\thresh}$ (dotted),
or adapted to the Markov threshold,
$\epsilon_p = r \epsilon_{p,\thresh}^M$
(dashed),
with $\epsilon_{p,\thresh}^M$ from \Eq{eq:h_thresh__Markov}.
From $|\im \Sigma(0)| \leq |\Sigma(0)|$ follows that
$\epsilon_{p,\thresh}^M < \epsilon_{p,\thresh}$ at $\delta_p=0$,
and hence the latter choice of pump strength (dashed) gives smaller or equal values of $G_s^M$
compared to the former (dotted).
In turn, at $\Delta=0$ the latter are always larger or equal to the gains
based on the frequency-dependent self-energy $\Sigma_E(\Delta)$.
This is because of the contributions of $(\re \Sigma_E)^2$ in the denominator
of $G_s$ %
compared with its absence in the denominator of $G_s^M$. %
For example, at $\delta_p=\Gamma_I=0$,
the gains evaluate to
$G_s^M(0) - 1 = 4 |\epsilon_p^2|\Gamma_E^2 (\Gamma_E^2 - |\epsilon_p^2|)^{-2}$
and
$G_s(0) - 1 = 4 |\epsilon_p^2|\Gamma_E^2 (\Gamma_E^2 + (\re\Sigma(0))^2 - |\epsilon_p^2|)^{-2}$,
respectively.
Panel (a) illustrates the `resonant slope-matching' scenario described above,
where $|\im\Sigma|$ has its maximum near $\Delta=0$
and $\re \Sigma(\Delta) \approx 0 + \Delta$, as indicated by the gray dashed line.
This constellation indeed leads to significant broadening of the gain $G_s(\Delta)$ (solid)
compared with the Markov approximation (dotted and dashed).
In this constellation, due to $\re\Sigma(0)=\im\Sigma(0) = 0$, the thesholds
$\epsilon_{p,\thresh}$ and $\epsilon_{p,\thresh}^M$ are equal,
cf. \Eqs{eq:h_thresh__wdependent} and \eqref{eq:h_thresh__Markov},
such that both Markov approximations (dotted and dashed curves) coincide.
Moreover, because $\re\Sigma(0) = 0$ and $\Gamma_E = -\im\Sigma_E(0)$,
at $\Delta=0$ the gains $G_s(0)$ and $G_s^M(0)$ are equal.
In panel (b), the two gains are still approximately equal at $\Delta=0$,
provided that $\epsilon_p = r \epsilon_{p,\thresh}^M$
is adapted to be the Markov threshold $\epsilon_p^M$ (dashed).
This is because $|\re\Sigma(0)| \ll |\im\Sigma(0)|$ is fulfilled in panel (b).
Although the behavior of $\Sigma(\Delta)$ in (b)
noticeably deviates from the optimal `resonant slope-matching' condition,
bandwidth broadening is still observed.
Even in panel (c), where the self-energy deviates much more substantially
from the `resonant slope-matching' condition
-- the coupling resonance $\omega_\Sigma$ is far detuned from the reference
frequency and $|\re\Sigma(0)| \approx |\im\Sigma(0)|$ --
a significant bandwidth broadening is nevertheless observed.

\textcolor{black}{For the cases of Fig.~\ref{fig:gains_wdependentcoupling},} the comparison with the Markovian counterparts \textcolor{black}{-- which always have a single-peaked gain profile at zero pump detuning, $\delta_p=0$ --} shows that the double-peaked gain profile (resonance splitting) and the resulting bandwidth broadening result from the frequency-dependent coupling. 
\textcolor{black}{This frequency-dependence produces significant deviations from the Markov approximation whenever the self-energy has a large slope of either the real part (cases a, b) or the imaginary part at $\omega_s$, or if the real part itself is of comparable magnitude to the imaginary part there (case c). 
(For example, for the case of Fig.~\ref{fig:gains_wdependentcoupling}(b) with $L_c/L_s=0.75$ the slope of $\re\Sigma(0)$ is big enough to invalidate the Markov approximation in the parameter region where $Z_{0}/Z_{s} \lesssim 2$ and $C_c/C_s \gtrsim 1.5$.)
On the other hand, if neither of these criteria is fulfilled, the exact gain profile approximately agrees with the Markov approximation (not shown here).
}
It is worth mentioning, however,
that even a frequency-independent, Markovian coupling
can give rise to a double-peaked gain profile \cite{WusShu2013},
under the condition that $\delta_p^2 > \Gamma_E^2 + |\epsilon_p|^2$,
and the maxima of the Markovian gain profile are then located at the minima of $|D(\Delta)|^2$.
In Fig.~\ref{fig:gains_wdependentcoupling}, no such \textcolor{black}{resonance} splitting
can be observed for the Markovian gains due to $\delta_p = 0$.
These two mechanisms for resonance splitting are independent,
such that e.g. resonance splitting under Markovian coupling does not necessarily
entail also its presence under comparable frequency-dependent coupling.
For a uniform, frequency-independent amplification,
strong rippling is of course undesirable,
i.e. the large-amplitude enhancement of $G_s(\Delta)$ at $\Delta \neq 0$
due to the resonance splitting.
Usually, an optimal value for the pump strength can be found where the split gain
maxima merge and thus provide a flat gain profile over a relatively wide bandwidth.

\begin{figure}\centering
\includegraphics[width=\columnwidth]{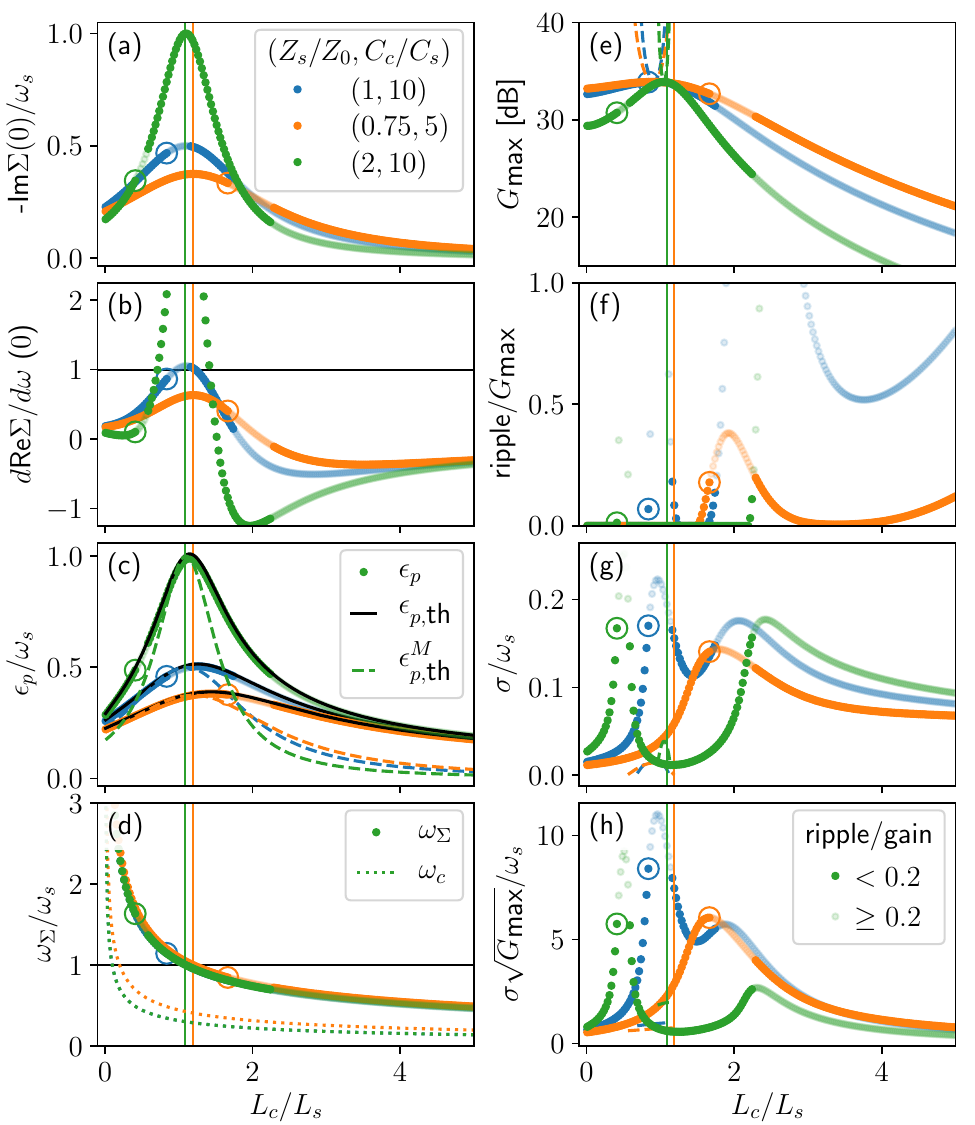}
\caption{
Performance of degenerate parametric amplifier
coupled to a TL via a series-$\sLC$ circuit,
here as function of coupling parameter $L_c$
and for several combinations of ($Z_s$, $C_c$),
with $\Gamma_I = 0$, $\delta_p=0$.
The self-energy $\Sigma(\Delta)=\Sigma_E(\Delta)$ is calculated from
the exact solution,
\Eqs{eq:imSig_seriescoupled}
and \eqref{eqapp:reSig1_seriescoupled}--\eqref{eqapp:reSig2_seriescoupled}.
(a) $\im\Sigma(\Delta=0)$;
(b) $(\diff \re\Sigma/\diff \Delta)(\Delta=0)$;
(c) pump strength $\epsilon_p$, set to $0.98 \epsilon_{p,\thresh}$,
with $\epsilon_{p,\thresh}$ from \Eq{eq:h_thresh__wdependent} (black lines);
(d) approximate coupling resonance frequency $\omega_\Sigma$,
\Eq{eq:wres_selfenergy_imag_seriescoupled},
and bare coupling frequency $\omega_c = (L_c C_c)^{-1/2}$ (dotted);
(e) maximum gain $G_{\max} = G_s(\Delta=0) = |u(0)|^2$;
(f) ripple-to-gain ratio;
(g) bandwidth $\sigma$ of $G_s(\Delta) = |u(\Delta)|^2$;
(h) gain-bandwidth product $\sigma \sqrt{G_{\max}}$.
The respective Markov approximations (dashed lines) are shown in (c,e,g), see
discussion in text.
The intervals where the ripple-to-gain amplitude ratio is above $20\%$ are indicated by transparent markers.
Special data points (large circles) are marked where
the gain-bandwidth product is maximized within these intervals (at their edges).
Vertical lines in all panels mark where $\omega_\Sigma = \omega_s$ is fulfilled.
}
\label{fig:gaincharacteristics_Lc}
\end{figure}

Figure \ref{fig:gaincharacteristics_Lc}
illustrates the performance of parametric amplifiers
coupled to a TL via a series-$\sLC$ circuit,
as function of the coupling parameter $L_c/L_s$ 
for $\Gamma_I=0$:
(e) the `maximum' gain $G_{\max} = G_s(\Delta=0) = |u(0)|^2$,
(f) the ratio of the ripple amplitude $\max(G_s) - G_s(\Delta=0)$ to gain $G_s(\Delta=0)$
in the parameter ranges where rippling occurs,
(g) the bandwidth $\sigma$ of $G_s(\Delta) = |u(\Delta)|^2$,
and
(h) the resulting bandwidth-gain product $\sigma \sqrt{G_{\max}}$.
We note that the bandwidth $\sigma$ is defined as usual as full-width-half-maximum
with respect to the `maximum' gain $G_s(\Delta=0)$ at $\Delta=0$, rather than the
larger actual maximum of $G_s(\Delta)$ in case of rippling.
These quantities are based on $u(\Delta)$ from \Eq{eq:usq_Delta__wdependent},
where the self-energy $\Sigma(0) = \Sigma_E(0)$ (a,b) is calculated from
the exact solution, \Eqs{eq:imSig_seriescoupled} and
\eqref{eqapp:reSig1_seriescoupled}--\eqref{eqapp:reSig2_seriescoupled}.
The pump strength $\epsilon_p$ (c)
is set to $\epsilon_p = 0.98 \epsilon_{p,\thresh}$,
with $\epsilon_{p,\thresh}$ from \Eq{eq:h_thresh__wdependent} (black lines).
The pump detuning is $\delta_p=0$, such that $\wref = \omega_p/2 = \omega_s$.

For comparison, the respective Markov approximations are also shown (dashed lines)
in Fig.~\ref{fig:gaincharacteristics_Lc}:
(c) $\epsilon_{p,\thresh}^M$ from \Eq{eq:h_thresh__Markov}
and (e,g,h) $G_{\max}^M$ and $\sigma^M$ from $G_s(\Delta) = |u(\Delta)|^2$ of \Eq{eq:u_Delta__Markov}.
Due to the sole contribution of $|\im \Sigma(0)|^2$ to $\epsilon_{p,\thresh}^M$
-- compared with the combined contributions of $|\Sigma(0)|^2 = (\im \Sigma(0))^2 + (\re \Sigma(0))^2$
to $\epsilon_{p,\thresh}$ --
the Markov threshold at $\delta_p=0$ is generally lower than the frequency-dependent one,
$\epsilon_{p,\thresh}^M \leq \epsilon_{p,\thresh}$,
cf.~Fig.~\ref{fig:threshold}.
In the Markov case, stable parametric amplification
is thus limited to the narrow parameter range defined by
the condition $\epsilon_{p,\thresh}^M \geq \epsilon_p = 0.98 \epsilon_{p,\thresh} \geq 0.98 \epsilon_{p,\thresh}^M$, cf.~panel (c).
In this interval, the Markov case has here by construction a larger relative pump strength
than the frequency-dependent coupling case,
$\epsilon_p/\epsilon_{p,\thresh}^M \geq \epsilon_p/\epsilon_{p,\thresh}$,
and thus features a larger maximum gain in panel (e).

Non-transparent markers in some panels indicate the intervals
where the ripple-to-gain amplitude ratio is below $20\%$,
set here as a criterion of acceptably small rippling.
Large circles mark special data points, where
the gain-bandwidth product is maximized within these intervals, typically at their edges.
At those points, the parametric device simultaneously achieves
large maximum gain $G_{\max}$,
moderately high bandwidth $\sigma$ ,
and no or only moderate rippling.
The achievable values for these metrics depend on the parameters
$Z_s/Z_0$, $L_c/L_s$, and $C_c/C_s$.

The three different parameter cases (blue, orange, green) shown in Fig.~\ref{fig:gaincharacteristics_Lc}
serve as an illustration that large bandwidth (or a large gain-bandwidth product)
can be achieved both under the established criterion of `resonant slope-matching',
\Eq{eq:resonantcondition_bandwidthbroadening},
but similarly also for `off-resonant' coupling.
In the former case, which is approximately fulfilled for parameters
$Z_s/Z_0=1$ and $C_c/C_s=10$ (blue) and $L_c/L_s \approx 0.84$ \textcolor{black}{(large circle mark)},
the (approximate) coupling resonance
$\omega_\Sigma$, \Eq{eq:wres_selfenergy_imag_seriescoupled},
is close to $\omega_s$ and $\diff \re \Sigma/\diff \Delta (\Delta=0) \approx 1$,
see panels (d) and (b), respectively.
However, other parameter constellations can reach similarly good results,
such as $Z_s/Z_0 = 0.75$, $C_c/C_s=4.9$ and $L_c/L_s \approx 1.66$ (orange)
or $Z_s/Z_0 = 2.0$, $C_c/C_s=10$ and $L_c/L_s \approx 0.41$ (green).
In these `off-resonant' coupling cases, $|\omega_\Sigma - \omega_s|$
is not small and the criterion $\diff \re\Sigma/\diff \omega \approx 1$
is not fulfilled. Nevertheless, they achieve bandwidth broadening and
gain-bandwidth products comparable to the `resonant slope-matching' case.



\subsection{Simple ladder filter}

\begin{figure}\centering
\includegraphics[width=\columnwidth]{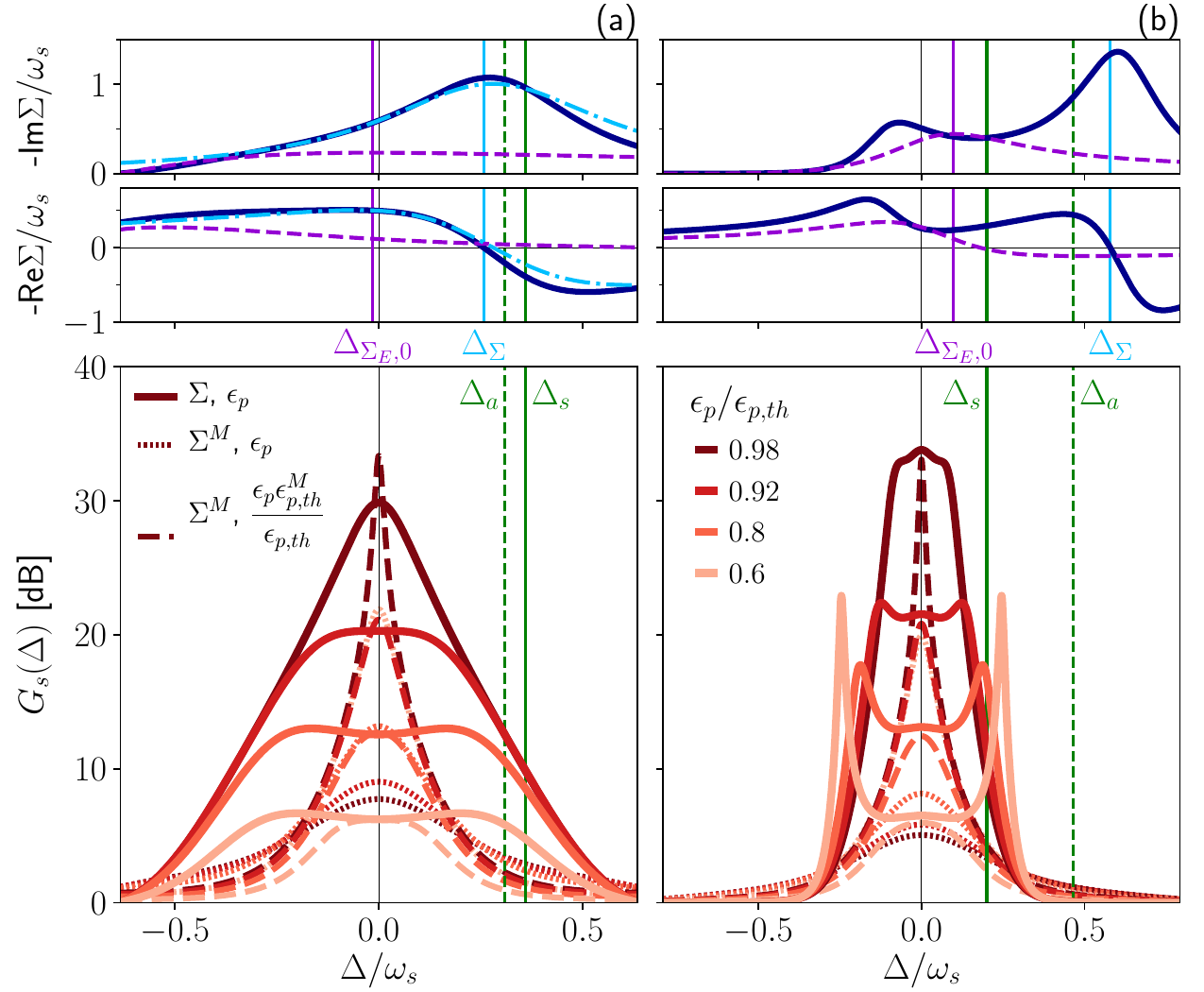}
\caption{
Self-energy $\Sigma(\Delta)$ (upper panels) and parametric gain profiles $G_s(\Delta)$
(bottom panels)
for various parametric pump strengths $\epsilon_p = r \epsilon_{p,\thresh}$
and parameters $(Z_s/Z_0, C_g/C_s, L_g/L_s, C_c/C_s, L_c/L_s, \delta_{p,s}/\omega_s) =$
(a) ($0.7$, $120$, $0.002$, $1.12$, $4.0$, $-0.36$)
and (b) ($0.6$, $0.8$, $0.3$, $1.2$, $0.26$, $-0.2$)
for a lumped-element parametric resonator coupled to the TL via the simple ladder filter
of Fig.~\ref{fig:filters}(c).
The upper panels show the self-energy $\Sigma(\Delta)$, \Eq{eqF:selfenergy_tot} (blue solid),
and the approximation for $\Sigma(\Delta)$ given in \Eq{eqF:selfenergy_tot_approx} (light blue dash-dotted),
and the self-energy $\Sigma_E(\Delta)$ of the coupling,
\Eq{eq:imSig_filtercoupled}
and \Eqs{eqapp:reSig1_parallelbeforeseriescoupled}--\eqref{eqapp:reSig2_parallelbeforeseriescoupled} (purple dashed).
In the bottom panels, the resulting gain
$G_s(\Delta) = |u(\Delta)|^2$ from \Eq{eqF:u_Delta__wdependent} is shown (solid lines),
with $\epsilon_{p,\thresh}$ calculated from the eigenvalues of \Eq{eq:dynamicalmatrix_filter},
cf.~Fig.~\ref{fig:threshold_filtercoupling}.
The dashed lines show $G_s(\Delta)$ based on the Markov approximation
of \Eq{eqF:u_Delta__wdependent}, i.e. under the approximations $\re\Sigma(\Delta) = 0$
and $\im\Sigma(\Delta) = \im\Sigma(0)$.
Consistently, the threshold pump strength $\epsilon_{p,\thresh}$
is calculated from the eigenvalues of \Eq{eq:dynamicalmatrix_filter}
under the approximation $\re\Sigma(\Delta) = 0$.
Other parameters are $\Gamma_I = 0$.
}
\label{fig:gainprofiles_filter}
\end{figure}

Figure \ref{fig:gainprofiles_filter} shows the gain profiles for two
parameter cases where large bandwidth broadening at moderately large gain
can be achieved with the simple ladder filter of Sec.~\ref{subsec:summary_parallelLCbeforeseriesLC}.
As discussed there and in Sec.~\ref{subsec:gains_filtercoupling},
the parallel components of the filter (with parameters $C_c, L_c$)
determine the boundary condition for the TL dressed modes,
whereas the series components of the filter (with parameters $C_g, L_g$)
introduce an additional degree of freedom that hybridizes with the resonator.
The resonator is thus characterized by two frequencies $\omega_s$ and $\omega_a$.
The total self-energy $\Sigma(\Delta)$, given in \Eq{eqF:selfenergy_tot},
is shown in the upper panels (blue solid)
and is compared with the self-energy $\Sigma_E(\Delta)$ of the coupling alone (purple dashed).
For these parameter cases, especially of panel (a), $\Sigma_E(\Delta)$ is
very broad in frequency space.
In an approximation for $\Sigma(\Delta)$,
we may therefore neglect $\diff \im\Sigma_E/\diff \Delta \approx 0$
and set
$\diff \re\Sigma_E/\diff \Delta \approx \left(\diff \re\Sigma_E/\diff \Delta \right)(\Delta=0) \cdot \Delta$.
The resulting approximation,
\begin{align}\label{eqF:selfenergy_tot_approx}
 \Sigma(\Delta) \approx  \frac{g^2}{\delta_{p,a} + \Delta \left(1 - \frac{\diff \re\Sigma_E}{\diff \Delta}(0)\right) - \Sigma_E(0)} - \ui \Gamma_I
 \,,
\end{align}
is shown by the light-blue dash-dotted line.

A criterion for the large bandwidth broadening observed in the example of
Fig.~\ref{fig:gainprofiles_filter}(a) may be given as follows:
In order for the amplification of an incoming signal to be strong,
the effective coupling between the TL and the pumped system needs to be large
over the range covering, on the one hand $\omega=\omega_p/2$ ($\Delta = 0$) where the parametric
effect takes place, and on the other hand $\omega=\omega_{s}$
($\Delta=\Delta_s = \omega_s-\omega_p/2$) where the system resonates.
Ideally, $-\im \Sigma(\Delta)$ thus assumes its maximum
within the $\Delta$-interval $[0,\Delta_s]$ (or $[\Delta_s,0]$ if $\Delta_s < 0$),
and is sufficiently large over this interval,
as is the case in Fig.~\ref{fig:gainprofiles_filter}(a).
Note that the frequency axis is $\Delta := \omega - \wref = \omega - \omega_p/2$,
for which the system frequencies translate to
$\Delta_{i} = \omega_{i} - \omega_p/2 = -\delta_{p,i}$ ($i=s,a$).
The position of the maximum of $-\im\Sigma(\Delta)$ may be approximated by the
minimum of the denominator's absolute value,
which in turn is approximated by
\begin{equation}\label{eq:rootcondition_maximSig_filter}
 \Delta - \re \Sigma_E(\Delta) + \delta_{p,a} = 0
 \,.
\end{equation}
In Fig.~\ref{fig:gainprofiles_filter}(a) the value $\Delta_\Sigma$
resulting from this root condition is marked by the vertical light-blue line.
Since $\re \Sigma_E(\Delta)$ is small around $\Delta \approx 0$,
the solution of \Eq{eq:rootcondition_maximSig_filter}
may even be further approximated as $\Delta_\Sigma \approx -\delta_{p,a} = \Delta_a$,
(vertical green dashed line).
The width of $\Sigma(\Delta)$ in \Eq{eqF:selfenergy_tot_approx},
is approximately $2 |\im\Sigma_E(0)|$.
According to the criterion for large bandwidth broadening,
the pump detunings should therefore fulfill
$-\delta_{p,a} = \Delta_a < 2|\im \Sigma_E(0)|$ and
$-\delta_{p,s} = \Delta_s \lesssim 2|\im \Sigma_E(0)|$, respectively.
In the example of Fig.~\ref{fig:gainprofiles_filter}(a) these are
$\delta_{p,s} = -0.36$, $\delta_{p,a} \approx -0.31$
and $2|\im \Sigma_E(0)| \approx 0.47$.

The gain profiles of Fig.~\ref{fig:gainprofiles_filter}
show varying numbers of gain maxima: two (one) for medium (large) $\epsilon_p$-values of (a)
and three (one) for medium (large) $\epsilon_p$-values of (b).
The gain profiles of the corresponding Markov approximations
-- when setting $\Sigma_E(\Delta) \to \ui \im \Sigma_E(0) = -\ui \Gamma_E$
in \Eqs{eqF:selfenergy_tot}--\eqref{eqF:v_Delta__wdependent} --
have only a single maximum, except for the smallest $\epsilon_p$-values of (a)
where there are two maxima at $\Delta \neq 0$.
In general, the maxima of the Markovian gain profile are located at the minima of
$|D(\Delta)|^2 |\delta_a + \Delta + \ui \Gamma_E|^2 |\delta_a - \Delta - \ui \Gamma_E|^2$
and there may be up to four of them.
(This is similar to the simpler coupling discussed in Sec.~\ref{subsec:results_seriesLC},
where the Markovian gain profile can have up to two maxima.)
Bandwidth broadening thus may either be inherited from the Markovian limit,
or generated by the frequency-dependent coupling, or be a combination of the two mechanisms.
For the cases of Fig.~\ref{fig:gainprofiles_filter}
we conclude from the comparison that only the effect of the frequency-dependent coupling
contributes.

\section{Summary and discussion}

We have analyzed the non-Markovian, i.e. frequency-dependent, coupling
between a superconducting resonator (such as a JPD) and a TL with the help of
a dressed-TL formalism.
The dressed-mode analysis is done specifically for a series-$\sLC$ resonator
and for a simple filter composed of a series-$\sLC$ resonator in series with a parallel-$\sLC$ resonator, see Fig.~\ref{fig:filters}(a) and (c).

With the help of the dressed-TL formalism we have calculated the exact self-energies
for these coupling circuits and their resonant approximations.
These frequency-dependent self-energies are then included
in the non-Markovian Langevin dynamics,
specifically that of a degenerate parametric amplifier.
We have identified the generalized input-output conditions consistent with this
non-Markovian dynamics, as well as the resulting generalized unitarity conditions.
Further spectral analysis yields the gain factors and
the instability thresholds for the frequency-dependent coupling.

In comparison with their Markovian counterparts,
the frequency-dependent couplings can show significant
bandwidth broadening and enhanced gain-bandwidth products.
One favorable coupling regime is the `resonant slope-matching' situation,
where the coupling resonance $\omega_\Sigma$ is near the amplifier resonance $\omega_s$
and where the real part of the self-energy $\re \Sigma(\Delta)$ has unity slope there.
This regime has been identified before \cite{RoyETAL2015} in terms of a generic
impedance of the amplifier's environment, without making use of a dressed-TL mode formalism.
We believe that the self-energy obtained through the non-perturbative formalism of dressed TL modes
provides a more accurate description of the coupling
and can help with the identification of optimal filter parameters.
Furthermore, we have identified other favorable coupling regimes
which despite being `off-resonant' show good amplifier performance
in terms of bandwidth and gain-bandwidth product.
In one of these regimes, $\omega_\Sigma$ is far detuned from $\omega_s$
but has $|\re \Sigma(0)| \approx |\im \Sigma(0)|$ there.
In case of the simple ladder filter another favorable regime exists for
a finite pump detuning together with a finite detuning between the two
resonances of the extended amplifier system.
We believe that the existence of these new coupling regimes
widens the parameter scope for practical devices and are worth further investigation.

\section*{Acknowledgements}

R. Yang and Z.R. Lin were supported by the National Key Research and Development Program of China (Grant No. 2023YFB4404904). \textcolor{black}{Z. Shi acknowledges the Canada First Research Excellence Fund (CFREF), NSERC of Canada, the Canadian Foundation for Innovation, the Ontario Ministry of Research and Innovation, and Industry Canada for financial support.} W. Wustmann was supported by {\it Quantum Technologies Aotearoa},
a research program funded by the New Zealand Ministry of Business,
Innovation and Employment, contract number UOO2347.

\onecolumngrid
\begin{appendix}

\section{Rotating-frame Hamiltonian of capacitively shunted, flux-pumped SQUID}
\label{app:RF_fluxpumpedSQUID}
For a capacitively shunted effective JJ the system Hamiltonian is
\begin{equation}\label{eqG0:Hsys__JJres}
 H^{(0)}_{\sys} = \frac{1}{2 C_s} Q_s^2 + \flqu I_{c,s} \left( 1 - \cos(2\pi \Phi_s/\Phi_0) \right)
\end{equation}
where 
\begin{equation}\label{eqG0:Ics}
I_{c,s} = \bar I_{c,s} + \delta\!I_{c,s}(t)
 = \bar I_{c,s} - \widehat{\delta\!I_{c,s}} \cos(\omega_p t - \theta_p)
\end{equation}
is the effective critical current of the JJ which may be flux-tuned and 
pumped.
For example, if the effective JJ consists of a flux-tuned and/or flux-pumped 
dc-SQUID, with external flux
$\bar \Phi_{e} + \widehat{\delta\!\Phi_{e}} \cos(\omega_p t - \theta_p)$, 
one may approximate
\begin{align}
 \bar I_{c,s} &\approx 2 I_c \cos(\pi\bar \Phi_{e}/\Phi_0) \\
 \delta\!I_{c,s}(t) 
  &= - \widehat{\delta\!I_{c,s}} \cos(\omega_p t - \theta_p)
\end{align}
with $\widehat{\delta\!I_{c,s}} = I_c \sin(\pi\bar \Phi_{e}/\Phi_0) \iflqu \widehat{\delta\!\Phi_{e}}$.
Below we keep the more general notation, \Eq{eqG0:Ics},
without specifying the origin of parameters $\bar I_{c,s}$ and $\widehat{\delta\!I_{c,s}}$. 

We can formally rewrite \Eq{eqG0:Hsys__JJres}
such that the harmonic, time-independent contribution to the Hamiltonian 
is separated from the remaining nonlinear and/or time-dependent terms:
\begin{align}\label{eqG0:Hsys_bare}
 H^{(0)}_{\sys} &= \hbar \omega_s^{(0)} (a_s^\dag a_s)^{(0)} + V_{\nlin} \\
 \hbar \omega_s^{(0)} (a_s^\dag a_s)^{(0)} &= \frac{1}{2 C_s} Q_s^2 + \frac{1}{2} \flqu \bar I_{c,s} \left(\iflqu \Phi_s \right)^2 \\
\label{eqG0:Vnlin}
 V_{\nlin} &= \flqu I_{c,s} \left( 1 - \cos(2\pi \Phi_s/\Phi_0) \right) - \frac{1}{2} \flqu \bar I_{c,s} \left(\iflqu \Phi_s \right)^2
\end{align}
where the system frequency is $\omega_s^{(0)} = (C_s L_s)^{-1/2}$,
with $L_s = \Phi_0/(2\pi \bar I_{c,s})$.
Depending on the type of coupling, the system Hamiltonian is modified
when taking into account the coupling to the TL,
\begin{align}\label{eqG0:Hsys_inclcoupling_0}
 H_{\sys} &= \hbar \omega_s a_s^\dag a_s + V_{\nlin} \,,
\end{align}
where $\omega_s = (C_s L_\eff)^{-1/2}$ is the (potentially coupling-shifted) system frequency.
Specifically, $L_\eff = (1/L_s + 1/L_c)^{-1}$ for parallel-$\sLC$ coupling
(cf.~\Eq{eqP:Hamiltonian_final})
or for purely inductive coupling (a special case of \Eq{eqP:Hamiltonian_final}),
whereas $L_\eff = L_s$ for series-$\sLC$ coupling
(cf.~\Eq{eqQ:Hamiltonian_final})
or for purely capacitive coupling (a special case of \Eq{eqP:Hamiltonian_final}).
These coupling-induced frequency shifts $\omega_s^{(0)} \to \omega_s$
follow from the combined system-TL Hamiltonians after identifying the correct
TL-modes subject to the boundary condition from the respective coupling~\footnote{%
If the combined Hamiltonians are instead approximated in the usual weak-coupling
approximation and with coupling-independent TL modes, then the coupling induced
frequency shifts may take other forms, e.g.~with
$C_\eff = C_s + C_c$ for the purely capacitive coupling.}.
The system amplitudes $a_s$ are then defined through
\begin{equation*}
 \Phi_s = \sqrt{\frac{\hbar Z_s}{2}} (a_s + a_s^\dag) 
 \hspace*{2cm}
    Q_s = -\ui \sqrt{\frac{\hbar}{2 Z_s}} (a_s - a_s^\dag) 
\end{equation*}
with the system impedance $Z_s = (L_\eff/C_s)^{1/2}$.

The nonlinear potential can be expanded,
\begin{align}
 V_{\nlin}
 &= \flqu I_{c,s} \left( 1 - \cos(2\pi \Phi_s/\Phi_0) \right) - \frac{1}{2} \flqu \bar I_{c,s} \left(\iflqu \Phi_s \right)^2 \\
 &\approx \iflqu \delta\!I_{c,s}(t) \frac{\hbar Z_s}{4} \left(a_s + a_s^\dag\right)^2
 -\left(\iflqu\right)^3 \bar I_{c,s} \frac{\hbar^2 Z_s^2}{96} \left(a_s + a_s^\dag\right)^4
\end{align}
For flux-pumping of the form in \Eq{eqG0:Ics} with $\omega_p \approx 2\omega_s$
we can define slow amplitudes $A_s$ in the frame rotating with $\wref = \omega_p/2$,
by defining rotating amplitudes $a_s  = A_s e^{-\iexp \omega_p t/2}$, cf.~\Eq{eq:as_As_wref},
and then average over fast oscillations on the order of or larger than $\omega_p/2$,
\begin{align}\label{eqG0:Vnlin_rotatingframe}
 \tilde{V}_{\nlin}
 &\approx
 - \frac{\hbar}{2} \left(\epsilon_p A_s^\dag A_s^\dag + \epsilon_p^\ast A_s A_s \right)
 - \frac{\hbar \alpha_{\nlin}}{2} A_s^\dag A_s^\dag A_s A_s
\end{align}
with the pump strength
\begin{equation}\label{eq:pump_strength}
 \epsilon_p = \iflqu \frac{Z_{s}}{4} \widehat{\delta\!I_{c,s}} e^{\iexp \theta_p}
\end{equation}
and the nonlinearity coefficient
\begin{equation}\label{eq:alpha_JJnonlinearity}
 \alpha_{nl} = \left(\iflqu\right)^3 \hbar Z_{s}^2 \frac{\bar I_{c,s}}{8}
\end{equation}
Formally, the transformation to the rotating frame is done with the unitary
transformation \eqref{eq:unitarytrafo2RF}, which entails also a
shift of the linear contribution in \Eq{eqG0:Hsys_inclcoupling},
\begin{align}\label{eqG0:Hsys_inclcoupling}
 \tilde{H}_{\sys} &= -\hbar \delta_p A_s^\dag A_s
 - \frac{\hbar}{2} \left(\epsilon_p A_s^\dag A_s^\dag + \epsilon_p^\ast A_s A_s \right)
 - \frac{\hbar \alpha_{\nlin}}{2} A_s^\dag A_s^\dag A_s A_s
 \end{align}
with the pump detuning
\begin{equation}\label{eq:pump_detuning}
 \delta_{p} = \frac{\omega_p}{2} - \omega_{s}
\end{equation}
(As usual, we have neglected a $\alpha_\nlin$-proportional shift of $\delta_p$.)

\section{Series-$\sLC$ coupling circuit: derivation of dressed TL modes and coupling constants}
\label{app:seriesLC_TLmodes}

\begin{figure}[h]\centering
\includegraphics[height=1.7cm]{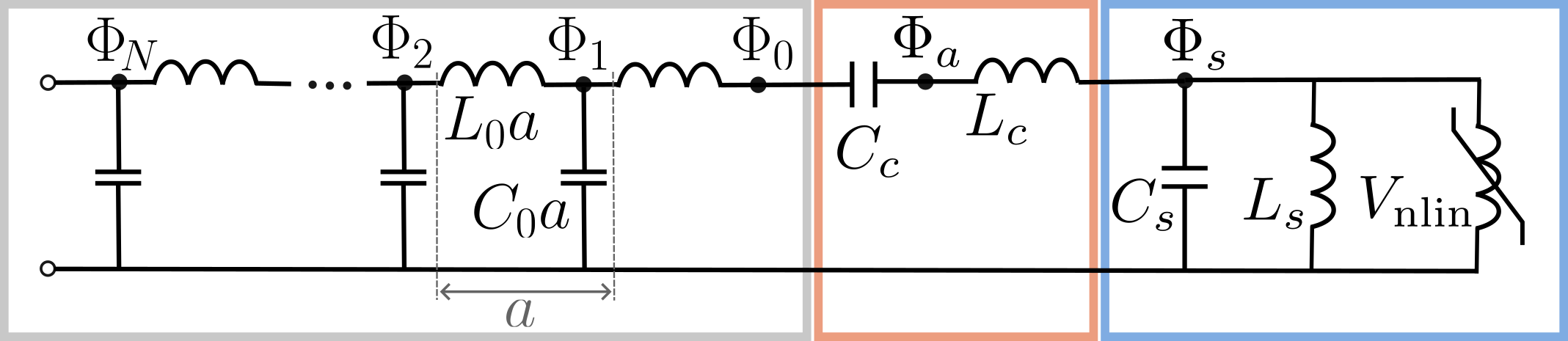}
\caption{
A resonator coupled to the end of a TL (one-port setup) via a series-$\sLC$ network,
cf.~Fig.~\ref{fig:filters}(a).
The TL is here represented in discretized form, where the unit cell has a length $a$
such that the capacitors and inductors of the discrete TL are of size $C_0 a$
and $L_0 a$, respectively.
The resonator is assumed to consist of a parallel circuit of capacitance $C_s$,
inductance $L_s$,
and a nonlinear potential $V_{\nlin}(\Phi_s)$, here symbolized as a nonlinear inductor.
}
\label{fig:LCresonator_seriesLcCccoupled__discreteTL}
\end{figure}

\subsection{Diagonalization of dressed TL Hamiltonian}

Here we derive the Hamiltonian \Eq{eqQ:Ham} for the system
coupled via a series-$\sLC$ circuit to a TL, as shown in Fig.~\ref{fig:filters}(a).
In this analyis, we diagonalize the TL part of the coupled system,
and as a result we obtain
the dressed TL modes \Eqs{eqQ:TLmode}--\eqref{eqQ:TLmodephase}.
This analysis is analogous to a procedure employed in Ref.~\cite{ParraRodriguezETAL2018},
which, however, does not cover the case of the series-$\sLC$ coupling.

In a discretized representation of the TL, as shown in Fig.~\ref{fig:LCresonator_seriesLcCccoupled__discreteTL},
the system Lagrangian \eqref{eqQ:Lagrangian_orig} reads
\begin{equation}\label{eqApp:Lagrangian_orig_discretized}
 \mathcal{L} = \mathcal{L}_{\sys}^{(0)}
 + \sum_{n \geq 1} \left[\frac{C_0 a}{2} \dot{\Phi}_n^2
 - \frac{1}{2 L_0 a} \left(\Phi_{n+1} - \Phi_{n} \right)^2 \right]
 -  \frac{1}{2 L_0 a} \left(\Phi_1 - \Phi_0 \right)^2
 + \frac{C_c}{2} \left( \dot \Phi_a - \dot \Phi_0 \right)^2
 - \frac{1}{2 L_c} \left( \Phi_s - \Phi_a \right)^2
\end{equation}
where $a$ is the size of the TL unit cell.
The Lagrangian is here written for the order where the coupling capacitor
directly faces the TL whereas the coupling inductor faces the system.

Written in the form of \Eq{eqApp:Lagrangian_orig_discretized},
the system is overdetermined, as is apparent on
inspection of the Lagrangian EOMs for $\Phi_0$ and $\Phi_a$,
\begin{align}
 -C_c (\ddot \Phi_a - \ddot \Phi_0) &= - \frac{1}{L_0 a} \left(\Phi_0 - \Phi_1 \right) \\
 C_c (\ddot \Phi_a - \ddot \Phi_0) &= \frac{1}{L_c} (\Phi_s - \Phi_a) \,.
\end{align}
These produce two linearly dependent rows in the capacitance matrix
and thus make the system singular.
We therefore need to make use of the constraint defined by the rhs:
\begin{equation}\label{eqApp:constraint_Phia}
 \frac{1}{L_c} (\Phi_s - \Phi_a) = \frac{1}{L_0 a} \left(\Phi_0 - \Phi_1 \right)
\end{equation}
This simply manifests the current conservation along the coupling axis
and the fact that we have discretized the TL in such a way that it ends with
an inductor of size $L_0 a$~\footnote{Terminating the TL with a capacitor of size $C_0 a$ will apparently remove the singularity. However, the resulting Hamiltonian becomes pathological in the continuum limit $a\to 0$, unless $\Phi_a$ is integrated into the TL part of the Hamiltonian which is equivalent to our treatment in this appendix.}.
Using the constraint \eqref{eqApp:constraint_Phia}, the actual system Lagrangian reads
\begin{equation}\label{eqApp:Lagrangian_discretized}
 \mathcal{L} = \mathcal{L}_{\sys}^{(0)}
 + \sum_{n \geq 1} \left[\frac{C_0 a}{2} \dot{\Phi}_n^2
 - \frac{1}{2 L_0 a} \left(\Phi_{n+1} - \Phi_{n} \right)^2 \right]
 -  \frac{1 + l}{2 L_0 a} \left(\Phi_1 - \Phi_0 \right)^2
 + \frac{C_c}{2} \left( \dot \Phi_s - \dot \Phi_0 - l (\dot \Phi_0 - \dot \Phi_1)  \right)^2
\end{equation}
where we introduce the shorthand notation (partly only used later)
\begin{align}
 l := \frac{L_c}{L_0 a} = \frac{\beta}{a} &&
 l_s := \frac{L_s}{L_0 a} &&
 c := \frac{C_c}{C_0 a} &&
 c_s := \frac{C_s}{C_0 a} && \frac{1}{c} + \frac{1}{c_s} = \frac{a}{\alpha}
\end{align}

Assuming that the system of interest can be represented as a linear oscillator, plus
a nonlinear potential, see Fig.~\ref{fig:LCresonator_seriesLcCccoupled__discreteTL},
\begin{equation}
 \mathcal{L}_{\sys}^{(0)}(\Phi_s, \dot \Phi_s) = \frac{C_s}{2} \dot{\Phi}_s^2
 - \frac{1}{2 L_s} \Phi_s^2 - V_{\nlin}(\Phi_s)
\end{equation}
\begin{equation}
 \mathcal{L} = \frac{1}{2} \dot{\boldsymbol{\Phi}}^T \uuline{C} \dot{\boldsymbol{\Phi}}
 - \frac{1}{2} \boldsymbol{\Phi}^T \uuline{L^{-1}} \boldsymbol{\Phi}
 - V_{\nlin}(\Phi_s)
\end{equation}
the matrices for the capacitance and inverse inductance of the full system,
with flux vector $\boldsymbol{\Phi}^T = (\Phi_s, \Phi_0, \Phi_1, \Phi_2, \ldots)$,
can be written as
\begin{align}
 \uuline{C} = C_0 a \begin{pmatrix}
  c_s + c & -(1+l) c & lc \\
  -(1+l) c & (1+l)^2 c & -l(1+l)c \\
  lc & -l(1+l)c & 1+l^2 c \\
  &&& 1 & & \\
  &&&& 1 & \\
  &&&&& \ddots
 \end{pmatrix}
 &&
 \uuline{L^{-1}} = \frac{1}{L_0 a} \begin{pmatrix}
  l_s^{-1} & 0 & 0 \\
  0 & (1+l) & -(1+l) \\
  0 & -(1+l) & 2+l & -1 \\
  &&-1& 2 & -1 & \\
  &&&-1& 2 & \\
  &&&&&&\ddots \\
 \end{pmatrix}
\end{align}
The Legendre transformation from the Lagrangian to the Hamiltonian
involves the inverse capacitance matrix,
\begin{align}
 \uuline{C}^{-1} = \frac{1}{C_0 a} \begin{pmatrix}
  \frac{1}{c_s} & \frac{1}{c_s (1+l)} & 0 \\
  \frac{1}{c_s (1+l)} & \frac{1}{c_s(1+l)^2} + \frac{1 + c l^2}{c(1+l)^2} & \frac{l}{1+l} \\
  0 & \frac{l}{1+l} & 1 \\
  &&& 1 & & \\
  &&&& 1 & \\
  &&&&& \ddots
 \end{pmatrix}
\end{align}
using the determinant $(C_0 a)^3 c_s c (1+l)^2$ of the upper $(3\times 3)$-capacitance matrix.
The resulting Hamiltonian is
\begin{align}
 H &= \frac{1}{2} \boldsymbol{Q}^T \uuline{C}^{-1} \boldsymbol{Q}
 + \frac{1}{2} \boldsymbol{\Phi}^T \uuline{L^{-1}} \boldsymbol{\Phi}
 + V_{\nlin}(\Phi_s) \\
\label{eqQ:Hamiltonian_definitionHtl}
 H &= \frac{1}{2 C_s} Q_s^2 + \frac{1}{2L_s} \Phi_s^2 + V_{\nlin}(\Phi_s)
 + \frac{1}{C_s (1 + l)} Q_s Q_0 + H_{TL}
\end{align}
where we have defined the dressed TL Hamiltonian
\begin{align}
 H_{TL} &=
 \frac{1}{2} \boldsymbol{Q}_{TL}^T \left[\uuline{C}^{-1}\right]_{1:,1:} \boldsymbol{Q}_{TL}
 + \frac{1}{2} \boldsymbol{\Phi}_{TL}^T \left[\uuline{L^{-1}}\right]_{1:,1:} \boldsymbol{\Phi}_{TL} \\
 &= \frac{1}{2 (1+l)^2} \left[\frac{1}{C_s} + \frac{1}{C_c} + \frac{l^2}{C_0 a}\right] Q_0^2
 + \frac{1}{C_0 a} \frac{l}{1+l} Q_0 Q_1 + \frac{1}{2C_0 a} \sum_{n\geq 1} Q_n^2 \nonumber \\
 &\qquad + \frac{1+l}{2 L_0 a} \Phi_0^2 - \frac{1+l}{L_0 a} \Phi_0 \Phi_1 +  \frac{1+l}{2 L_0 a} \Phi_1^2
 + \frac{1}{2L_0 a} \sum_{n\geq 1} (\Phi_{n+1} - \Phi_n)^2 \\
 \label{eqQ:H_TL_orig}
 &= \frac{1}{2 (1+l)^2 \alpha C_0} Q_0^2
  +  \frac{1}{2C_0 a} \left( \frac{l}{1+l} Q_0 + Q_1 \right)^2 + \frac{1}{2C_0 a} \sum_{n\geq 2} Q_n^2
 + \frac{1+l}{2 L_0 a} (\Phi_0 - \Phi_1)^2
 + \frac{1}{2L_0 a} \sum_{n\geq 1} (\Phi_{n+1} - \Phi_n)^2
\end{align}
The TL flux and charge vectors are
$\boldsymbol{\Phi}_{TL}^T = (\Phi_0, \Phi_1, \Phi_2, \ldots)$
and $\boldsymbol{Q}_{TL}^T = (Q_0, Q_1, Q_2, \ldots)$,
and we have here used the coupling length parameter $\alpha$,
\Eq{eqQ:alpha_beta}.
Because of the property
\begin{equation}\label{eqQ:property_invC}
 \left[\uuline{C}^{-1}\right]_{1:,1:}
\neq \left[\uuline{C}_{1:,1:} \right]^{-1}
\end{equation}
the `dressed' TL-Hamiltonian
differs from the `bare' TL-Hamiltonian which would be obtained
by performing the Legendre transformation 
while setting $\Phi_s=0$ in \Eq{eqApp:Lagrangian_discretized}.

We now want to perform a continuum limit $a\to 0$,
and to this end transform back to the Lagrangian representation.
(In the Hamiltonian \Eq{eqQ:H_TL_orig} the continuum limit appears
somewhat unclear.)
The Legendre transformation back to the Lagrangian involves the inverted
relation between the $Q_n$ and $\dot\Phi_n$
\begin{align}\label{eqQ:inv_invCmat_TL}
 \left[\left[\uuline{C}^{-1}\right]_{1:,1:}\right]^{-1} = \begin{pmatrix}
  \alpha C_0 (1+l)^2 & -\alpha C_0 l (1+l)  \\
  -\alpha C_0 l (1+l) & C_0 a + \alpha C_0 l^2 &  \\
  && C_0 a & & \\
  &&& C_0 a & \\
  &&&& \ddots
 \end{pmatrix}
\end{align}
and gives the dressed TL Lagrangian,
\begin{align}
 \mathcal{L}_{TL} &=
 \frac{1}{2} \dot{\boldsymbol{\Phi}}_{TL}^T \left[\left[\uuline{C}^{-1}\right]_{1:,1:}\right]^{-1} \dot{\boldsymbol{\Phi}}_{TL}
 - \frac{1}{2} \boldsymbol{\Phi}_{TL}^T \left[\uuline{L^{-1}}\right]_{1:,1:} \boldsymbol{\Phi}_{TL} \\
\label{eqQ:Lagrangian_TL_discrete}
 &= \frac{\alpha C_0}{2} \left( (1+l)\dot \Phi_0 - l \dot \Phi_1 \right)^2 + \frac{C_0 a}{2} \sum_{n\geq 1} \dot \Phi_n^2
 - \frac{1+l}{2 L_0 a} (\Phi_0 - \Phi_1)^2
 - \frac{1}{2L_0 a} \sum_{n\geq 1} (\Phi_{n+1} - \Phi_n)^2 \\
\label{eqQ:Lagrangian_TL_continuous}
 &\to \frac{\alpha C_0}{2} \left( \dot \Phi(0) - \beta \dot{\Phi}'(0) \right)^2 + \frac{C_0 }{2} \int_{0}^{\infty} \diff x \dot \Phi^2
 - \frac{\beta}{2 L_0} (\Phi'(0))^2
 - \frac{1}{2 L_0} \int_{0}^{\infty} \diff x (\Phi')^2
\end{align}
This \textcolor{black}{inverse} Legendre transformation from $H_{TL}$ to $\mathcal{L}_{TL}$
is only with respect to the TL subspace
and due to the property \eqref{eqQ:property_invC} does not agree
with the Legendre transformation of the full system.
In \Eq{eqQ:Lagrangian_TL_continuous}
we have associated the coupling point with $x=0$ and the far end of the TL as $x=+\infty$,
such that $(\Phi_0 - \Phi_1)/a \to -\Phi'(0)$
(other than seen from left to right in the geometry of Fig.~\ref{fig:LCresonator_seriesLcCccoupled__discreteTL}).

We can now diagonalize the TL contribution $\mathcal{L}_{TL}$,
instead of diagonalizing the full system.
To this end, we expand the TL flux with respect to the still unknown TL modes $u_k(x)$,
\begin{align}
 \Phi(x) &= \int_0^{\infty} \diff k u_k(x) \Phi_k \,,
\end{align}
introducing the mode fluxes $\Phi_k$.
The TL Lagrangian then reads
\begin{align}
 \mathcal{L}_{TL} &=
 \frac{\alpha C_0}{2} \left( \int \diff k \left[u_k(0) - \beta u'_k(0)\right] \dot \Phi_k \right)^2 + \frac{C_0 }{2} \int_{0}^{\infty} \diff x \iint \diff k \diff k' u_k(x) u_{k'}(x) \dot \Phi_k \dot \Phi_{k'} \nonumber \\
 \label{eqQ:Lagrangian_TL}
 &\quad - \frac{\beta}{2 L_0} \left(\int \diff k u_k'(0) \Phi_k\right)^2
 - \frac{1}{2 L_0} \int_{0}^{\infty} \diff x \int \diff k \diff k' u'_{k}(x) u'_{k'}(x) \Phi_k \Phi_{k'}
\end{align}
The TL modes diagonalize $\mathcal{L}_{TL}$ if they fulfill the following
orthogonality relations \textcolor{black}{(which are derived~\cite{walter1973regular} from the boundary equation, cf.~\Eqs{eqQ:boundarycondition}--\eqref{eqQ:boundarycondition_TLmodes}),}

\begin{align}\label{eqQ2:orthogonality_alpha}
 \int_0^\infty \diff x u_k(x) u_{k'}(x) + \alpha \left(u_k(0) - \beta u'_k(0) \right) \left(u_{k'}(0) - \beta u'_{k'}(0) \right) &= \delta(k-k') \\
 \label{eqQ2:orthogonality_beta}
 \int_{0}^{\infty} \diff x u'_{k}(x) u'_{k'}(x) + \beta u'_k(0) u'_{k'}(0) &= k^2 \delta(k-k') \,.
\end{align}
Under these conditions, the TL Lagrangian becomes
\begin{align}
 \mathcal{L}_{TL} &=
\int \diff k \left( \frac{C_0}{2} \dot \Phi_k^2 - \frac{k^2}{2 L_0} \Phi_k^2 \right)
\end{align}
and its Lagrangian EOMs, $\ddot \Phi_k = -\frac{k^2}{L_0 C_0} \Phi_k$,
as usual define the mode frequencies
\begin{align}
 \omega_k^2 = \frac{k^2}{L_0 C_0} = v^2 k^2
 \;.
\end{align}
The corresponding Hamiltonian is
\begin{align}
 H_{TL} &=
 \int \diff k \left( \frac{1}{2C_0} q_k^2 + \frac{k^2}{2 L_0} \Phi_k^2 \right)
\end{align}
with the mode charges $q_k = C_0 \dot \Phi_k$.

In the coupling term between the system and the TL Hamiltonian
in \Eq{eqQ:Hamiltonian_definitionHtl}
we still need to evaluate the boundary node charge $Q_0$ in terms of the
TL modes.
Using \Eq{eqQ:inv_invCmat_TL} and performing the continuum limit,
we find
\begin{align}
 Q_0
 &= \alpha C_0 (1+l)^2 \dot \Phi_0 - \alpha C_0 l (1+l) \dot \Phi_1
 = \alpha C_0 (1+l) \left[ \dot \Phi_0 + l (\dot \Phi_0 - \dot \Phi_1) \right] \\
 &\to \alpha C_0 (1+l) \left[  \dot \Phi(0) - \beta  \dot \Phi'(0) \right]
 = \alpha C_0 (1+l) \int \diff k \left( u_k(0) - \beta u'_k(0) \right) \dot \Phi_k \\
 &= \alpha (1+l) \int \diff k \left( u_k(0) - \beta u'_k(0) \right) q_k
\end{align}

Using the TL mode representation,
the full Hamiltonian, \Eq{eqQ:Hamiltonian_definitionHtl}, now reads
\begin{align}
  H &= \frac{1}{2 C_s} Q_s^2 + \frac{1}{2L_s} \Phi_s^2 + V_{\nlin}(\Phi_s)
 + \frac{\alpha}{C_s} Q_s \int_0^\infty \diff k \left( u_k(0) - \beta u'_k(0) \right) q_k
 + \int_0^\infty \diff k \left( \frac{1}{2C_0} q_k^2 + \frac{k^2}{2 L_0} \Phi_k^2 \right) \\
\label{eqQ:Hamiltonian_final}
   &= H_{\sys}^{(0)}
 + \frac{\alpha}{C_s} Q_s \int_0^\infty \diff k \left( u_k(0) - \beta u_k'(0) \right) q_k
 + \int_0^\infty \diff k  \left( \frac{1}{2 C_0} q_k^2 + \frac{k^2}{2 L_0} \Phi_k^2  \right)
\end{align}
cf.~\Eq{eqQ:Ham}.

It remains to evaluate the form of the TL modes $u_k(x)$.
To this end, we consider the boundary condition
that follows from the TL Langrangian, \Eq{eqQ:Lagrangian_TL_discrete},
\begin{equation}\label{eqQ:boundarycondition}
 \alpha C_0 \left( \ddot \Phi(0) - \beta \ddot{\Phi}'(0) \right)
 - \frac{1}{L_0} \Phi'(0) = 0
\end{equation}
Using $\ddot \Phi_k = -v^2 k^2 \Phi_k$ this can be cast into a
boundary condition for the modes,
\begin{equation}\label{eqQ:boundarycondition_TLmodes}
 \alpha k^2  \Bigl( u_k(0) - \beta u'_k(0) \Bigr) + u'_k(0) = 0
\end{equation}
Note how this differs from a Sturm-Liouville problem, for which the boundary condition
does not depend on the value $k$.
Using the ansatz $u_k(x) = a \cos(kx) + b \sin(kx) = \sqrt{2/\pi} \cos(k x + \varphi_k)$,
we then determine the modes
from this boundary condition, together with the normalization condition.
This results in the secular equation
\begin{equation}
\tan(\varphi_k) = -\left(\beta k - \frac{1}{\alpha k}\right)^{-1}
\end{equation}
and the amplitudes of \Eqs{eqQ:TLmode}--\eqref{eqQ:TLmodephase}.

\subsection{Coupling constants}

We briefly comment on the coupling coefficient \Eq{eqQq:def_fk} and its weak-coupling limit. Focusing on the system frequency $\omega_k = \omega_s$, we can express $f_{\omega_s/v}$ in terms of the wave function $u_{\omega_s/v}(0)$ at the end of the TL:
\begin{align}\label{eqQq:fk_@ws}
 f_{\omega_s/v}
 & = -\sqrt{\frac{v\Gamma_E^{\eff}}{2}}u_{\omega_s/v}(0) \nonumber\\
 & = -\sqrt{\frac{v\Gamma_E^{\eff}}{\pi}}
 \left( 1 + \frac{2 \Gamma_E^\eff Z_0}{\omega_s Z_s} \right)^{-1/2}\,.
\end{align}
where we have defined an effective coupling strength at the system frequency,
\begin{equation}
    \Gamma_E^\eff = \frac{\omega_s Z_s}{2Z_0}\tan^2(\varphi_{k_s})\,.
\end{equation}
In particular, in the weak-coupling limit $C_c \to 0$ or $L_c \to \infty$, $u_{\omega_s/v}(0) \to \sqrt{2/\pi}$, and
\begin{equation}
    \Gamma_E^\eff \approx \frac{\pi}{v} f_{\omega_s/v}^2\, .
\end{equation}

\section{Parallel-$\sLC$ coupling circuit: derivation of dressed TL modes and coupling constants}
\label{app:parallelLC_TLmodes}

\begin{figure}[h]\centering
\includegraphics[height=1.7cm]{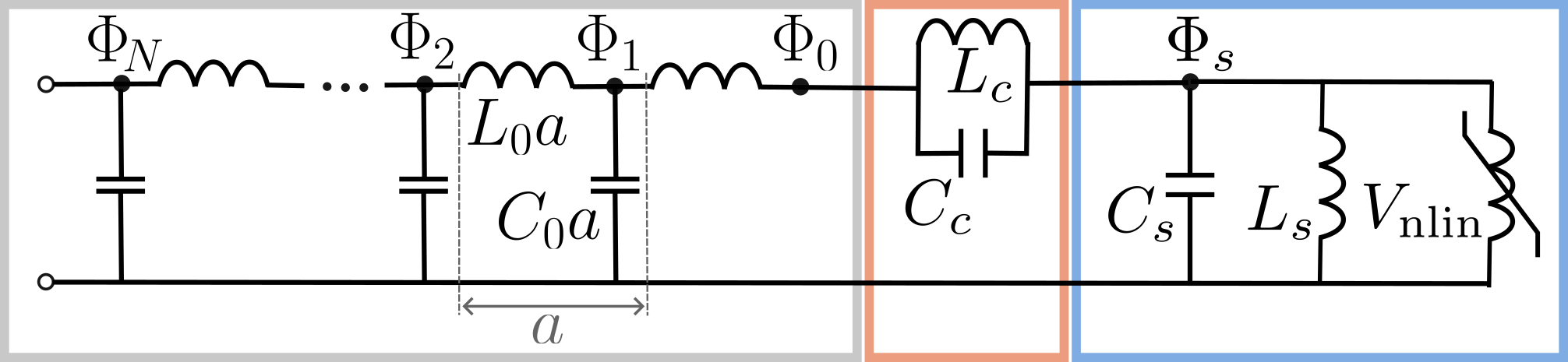}
\caption{
A resonator coupled to the end of a TL (one-port setup) via a parallel-$\sLC$ network,
cf.~Fig.~\ref{fig:filters}(b).
The TL is here represented in discretized form, where the unit cell has a length $a$
such that the capacitors and inductors of the discrete TL are of size $C_0 a$
and $L_0 a$, respectively.
The resonator is assumed to consist of a parallel circuit of capacitance $C_s$,
inductance $L_s$,
and a nonlinear potential $V_{\nlin}(\Phi_s)$, here symbolized as a nonlinear inductor.
}
\label{fig:LCresonator_parallelLcCccoupled__discreteTL}
\end{figure}

\subsection{Diagonalization of dressed TL Hamiltonian}

Here we derive the Hamiltonian for the system
coupled via a parallel-$\sLC$ circuit to a TL, as shown in Fig.~\ref{fig:filters}(b).
In this analyis, we diagonalize the TL part of the coupled system,
and as a result we obtain
the dressed TL modes \Eqs{eqP:TLmode}--\eqref{eqP:TLmodephase}.
This analysis is analogous to a procedure employed in Ref.~\cite{ParraRodriguezETAL2018}.

In a discretized representation of the TL, as shown in Fig.~\ref{fig:LCresonator_parallelLcCccoupled__discreteTL},
the system Lagrangian reads
\begin{equation}\label{eqP:Lagrangian_discretized}
 \mathcal{L} = \mathcal{L}_{\sys}^{(0)}
 + \sum_{n \geq 1} \left[\frac{C_0 a}{2} \dot{\Phi}_n^2
 - \frac{1}{2 L_0 a} \left(\Phi_{n+1} - \Phi_{n} \right)^2 \right]
 -  \frac{1}{2 L_0 a} \left(\Phi_1 - \Phi_0 \right)^2
 + \frac{C_c}{2} \left( \dot \Phi_s - \dot \Phi_0 \right)^2
 - \frac{1}{2 L_c} \left( \Phi_s - \Phi_0 \right)^2
\end{equation}
where $a$ is the size of the TL unit cell.

Assuming that the system of interest can be represented as a linear oscillator, plus
a nonlinear potential, see Fig.~\ref{fig:LCresonator_parallelLcCccoupled__discreteTL},
\begin{equation}
 \mathcal{L}_{\sys}^{(0)}(\Phi_s, \dot \Phi_s) = \frac{C_s}{2} \dot{\Phi}_s^2
 - \frac{1}{2 L_s} \Phi_s^2 - V_{\nlin}(\Phi_s) \,,
\end{equation}
\begin{equation}
 \mathcal{L} = \frac{1}{2} \dot{\boldsymbol{\Phi}}^T \uuline{C} \dot{\boldsymbol{\Phi}}
 - \frac{1}{2} \boldsymbol{\Phi}^T \uuline{L^{-1}} \boldsymbol{\Phi}
 - V_{\nlin}(\Phi_s)
 \,,
\end{equation}
the matrices for the capacitance and inverse inductance of the full system,
with flux vector $\boldsymbol{\Phi}^T = (\Phi_s, \Phi_0, \Phi_1, \Phi_2, \ldots)$,
can be written as
\begin{align}
 \uuline{C} = \begin{pmatrix}
  C_s + C_c & -C_c & 0 \\
  -C_c & C_c & 0 \\
  0 & 0 & C_0 a \\
  &&& C_0 a &  \\
  &&&& \ddots
 \end{pmatrix}
 &&
 \uuline{L^{-1}} = \frac{1}{L_0 a} \begin{pmatrix}
  l_s^{-1} + l^{-1} & -l^{-1} & 0 \\
 -l^{-1} & 1 + l^{-1} & -1 \\
  0 & -1 & 2 & -1 \\
  &&-1& 2 & \\
  &&&&\ddots \\
 \end{pmatrix}
\end{align}
with parameters
\begin{align}
 l := \frac{L_c}{L_0 a} 
 &&
 l_s := \frac{L_s}{L_0 a}
\end{align}
The Legendre transformation from the Lagrangian to the Hamiltonian
involves the inverse capacitance matrix,
\begin{align}
 \uuline{C}^{-1} = \begin{pmatrix}
  \frac{1}{C_s} & \frac{1}{C_s} & 0 \\
  \frac{1}{C_s} & \frac{1}{C_s} + \frac{1}{C_c} & 0 \\
  0 & 0 & \frac{1}{C_0 a} \\
  &&& \frac{1}{C_0 a} & \\
  &&&& \ddots
 \end{pmatrix}
\end{align}
and gives the Hamiltonian
\begin{align}
  H &= \frac{1}{2} \boldsymbol{Q}^T \uuline{C}^{-1} \boldsymbol{Q}
 + \frac{1}{2} \boldsymbol{\Phi}^T \uuline{L^{-1}} \boldsymbol{\Phi}
 + V_{\nlin}(\Phi_s) \\
\label{eqP:Hamiltonian_definitionHtl}
 &= \frac{1}{2 C_s} Q_s^2 + \frac{1}{2}\left(\frac{1}{L_s} + \frac{1}{L_c}\right) \Phi_s^2 + V_{\nlin}(\Phi_s)
 + \frac{1}{C_s} Q_s Q_0 - \frac{1}{L_c} \Phi_s \Phi_0 + H_{TL}
\end{align}
where we have defined the dressed TL Hamiltonian
\begin{align}
 H_{TL} &=
 \frac{1}{2} \boldsymbol{Q}_{TL}^T \left[\uuline{C}^{-1}\right]_{1:,1:} \boldsymbol{Q}_{TL}
 + \frac{1}{2} \boldsymbol{\Phi}_{TL}^T \left[\uuline{L^{-1}}\right]_{1:,1:} \boldsymbol{\Phi}_{TL} \\
 &= \frac{1}{2} \left[\frac{1}{C_s} + \frac{1}{C_c}\right] Q_0^2
 + \frac{1}{2C_0 a} \sum_{n\geq 1} Q_n^2
 + \frac{1+l^{-1}}{2 L_0 a} \Phi_0^2 - \frac{1}{L_0 a} \Phi_0 \Phi_1 + \frac{1}{2 L_0 a} \Phi_1^2
 + \frac{1}{2L_0 a} \sum_{n\geq 1} (\Phi_{n+1} - \Phi_n)^2 \\
 \label{eqP:H_TL_orig}
 &= \frac{1}{2\alpha C_0} Q_0^2 + \frac{1}{2C_0 a} \sum_{n\geq 2} Q_n^2
 + \frac{1}{2 \beta L_0} \Phi_0^2
 + \frac{1}{2 L_0 a} (\Phi_0 - \Phi_1)^2
 + \frac{1}{2L_0 a} \sum_{n\geq 1} (\Phi_{n+1} - \Phi_n)^2
\end{align}
The TL flux and charge vectors are
$\boldsymbol{\Phi}_{TL}^T = (\Phi_0, \Phi_1, \Phi_2, \ldots)$
and $\boldsymbol{Q}_{TL}^T = (Q_0, Q_1, Q_2, \ldots)$,
and we have here used the coupling length parameters $\alpha,\beta$,
\Eq{eqQ:alpha_beta},
which are of the same form here as in the series $\sLC$ coupling.
Because of the property
\begin{equation}\label{eqP:property_invC}
 \left[\uuline{C}^{-1}\right]_{1:,1:}
\neq \left[\uuline{C}_{1:,1:} \right]^{-1}
\end{equation}
the `dressed' TL-Hamiltonian
differs from the `bare' TL-Hamiltonian which would be obtained
by performing the Legendre transformation 
while setting $\Phi_s=0$ in \Eq{eqP:Lagrangian_discretized}.

We now want to perform a continuum limit $a\to 0$,
and to this end transform back to the Lagrangian representation.
The \textcolor{black}{inverse} Legendre transformation back to the Lagrangian involves the inverted
relation between the $Q_n$ and $\dot\Phi_n$
\begin{align}\label{eqP:inv_invCmat_TL}
 \left[\left[\uuline{C}^{-1}\right]_{1:,1:}\right]^{-1}
 = C_0 \cdot \text{diag}\left(\alpha, a, a, \ldots \right)
\end{align}
and gives the dressed TL Lagrangian,
\begin{align}
 \mathcal{L}_{TL} &=
 \frac{1}{2} \dot{\boldsymbol{\Phi}}_{TL}^T \left[\left[\uuline{C}^{-1}\right]_{1:,1:}\right]^{-1} \dot{\boldsymbol{\Phi}}_{TL}
 - \frac{1}{2} \boldsymbol{\Phi}_{TL}^T \left[\uuline{L^{-1}}\right]_{1:,1:} \boldsymbol{\Phi}_{TL} \\
\label{eqP:Lagrangian_TL_discrete}
 &= \frac{\alpha C_0}{2} \left(\dot \Phi_0\right)^2 + \frac{C_0 a}{2} \sum_{n\geq 1} \dot \Phi_n^2
 - \frac{1}{2 \beta L_0}  \Phi_0^2
 - \frac{1}{2 L_0 a} (\Phi_0 - \Phi_1)^2
 - \frac{1}{2L_0 a} \sum_{n\geq 1} (\Phi_{n+1} - \Phi_n)^2 \\
\label{eqP:Lagrangian_TL_continuous}
 &\to \frac{\alpha C_0}{2} \left( \dot \Phi(0) \right)^2 + \frac{C_0 }{2} \int_{0}^{\infty} \diff x \dot \Phi^2
 - \frac{1}{2 \beta L_0} (\Phi(0))^2
 - \frac{1}{2 L_0} \int_{0}^{\infty} \diff x (\Phi')^2
\end{align}
This \textcolor{black}{inverse} Legendre transformation from $H_{TL}$ to $\mathcal{L}_{TL}$
is only with respect to the TL subspace
and due to the property \eqref{eqP:property_invC} does not agree
with the Legendre transformation of the full system.
In \Eq{eqP:Lagrangian_TL_continuous}
we have associated the coupling point with $x=0$ and the far end of the TL as $x=+\infty$,
such that $(\Phi_0 - \Phi_1)/a \to -\Phi'(0)$
(other than seen in the geometry of Fig.~\ref{fig:LCresonator_parallelLcCccoupled__discreteTL}).

We can now diagonalize the TL contribution $\mathcal{L}_{TL}$,
instead of diagonalizing the full system.
To this end, we expand the TL flux with respect to the still unknown TL modes $u_k(x)$,
\begin{align}\label{eqP:mode_expansion}
 \Phi(x) &= \int_0^{\infty} \diff k u_k(x) \Phi_k \,,
\end{align}
introducing the mode fluxes $\Phi_k$.
The TL Lagrangian then reads
\begin{align}
 \mathcal{L}_{TL} &=
 \frac{\alpha C_0}{2} \left( \int \diff k u_k(0) \dot \Phi_k \right)^2 + \frac{C_0 }{2} \int_{0}^{\infty} \diff x \iint \diff k \diff k' u_k(x) u_{k'}(x) \dot \Phi_k \dot \Phi_{k'} \nonumber \\
 \label{eqP:Lagrangian_TL}
 &\quad - \frac{1}{2 \beta L_0} \left(\int \diff k u_k(0) \Phi_k\right)^2
 - \frac{1}{2 L_0} \int_{0}^{\infty} \diff x \int \diff k \diff k' u'_{k}(x) u'_{k'}(x) \Phi_k \Phi_{k'}
\end{align}
The TL modes diagonalize $\mathcal{L}_{TL}$ since they fulfill the following
orthogonality relations
\textcolor{black}{(which are derived~\cite{walter1973regular} from the boundary equation, cf.~\Eqs{eqP:boundarycondition}--\eqref{eqP:boundarycondition_TLmodes} ),}
\begin{align}\label{eqP2:orthogonality_alpha}
 \int_0^\infty \diff x u_k(x) u_{k'}(x) + \alpha u_k(0) u_{k'}(0)  &= \delta(k-k') \\
 \label{eqP2:orthogonality_beta}
 \int_{0}^{\infty} \diff x u'_{k}(x) u'_{k'}(x) + \frac{1}{\beta} u_k(0) u_{k'}(0) &= k^2 \delta(k-k') \,.
\end{align}
Under these conditions, the TL Lagrangian becomes
\begin{align}
 \mathcal{L}_{TL} &=
\int \diff k \left( \frac{C_0}{2} \dot \Phi_k^2 - \frac{k^2}{2 L_0} \Phi_k^2 \right)
\end{align}
and its Lagrangian EOMs, $\ddot \Phi_k = -\frac{k^2}{L_0 C_0} \Phi_k$,
as usual define the mode frequencies
\begin{align}
 \omega_k^2 = \frac{k^2}{L_0 C_0} = v^2 k^2
 \;.
\end{align}
The corresponding Hamiltonian is
\begin{align}
 H_{TL} &=
 \int \diff k \left( \frac{1}{2C_0} q_k^2 + \frac{k^2}{2 L_0} \Phi_k^2 \right)
\end{align}
with the mode charges $q_k = C_0 \dot \Phi_k$.

In the coupling term between the system and the TL Hamiltonian
in \Eq{eqP:Hamiltonian_definitionHtl}
we still need to evaluate the boundary node charge $Q_0$ in terms of the
TL modes.
Using \Eq{eqP:inv_invCmat_TL} and performing the continuum limit,
we find
\begin{align}\label{eqP:Q0_boundary}
 Q_0
 &= \alpha C_0 \dot \Phi_0 \to \alpha C_0 \dot \Phi(0)
 = \alpha C_0 \int \diff k u_k(0) \dot \Phi_k
 = \alpha \int \diff k u_k(0) q_k
\;.
\end{align}

Using the TL mode representation,
the full Hamiltonian, \Eq{eqP:Hamiltonian_definitionHtl}, now reads
\begin{align}\label{eqP:Hamiltonian_final}
  H &= H_{\sys}
 + \frac{\alpha}{C_s} Q_s \int_0^\infty \diff k u_k(0) q_k
 - \frac{1}{L_c} \Phi_s \int_0^\infty \diff k u_k(0) \Phi_k
 + \int_0^\infty \diff k \left( \frac{1}{2C_0} q_k^2 + \frac{k^2}{2 L_0} \Phi_k^2 \right) \\
\label{eqP:Hsys}
 H_{\sys}
   &= \frac{1}{2 C_s} Q_s^2 + \frac{1}{2} \left(\frac{1}{L_s} + \frac{1}{L_c}\right) \Phi_s^2 + V_{\nlin}(\Phi_s)
   = H_{\sys}^{(0)} + \frac{1}{2 L_c} \Phi_s^2
\end{align}

It remains to evaluate the form of the TL modes $u_k(x)$.
To this end, we consider the boundary condition
that follows from the TL Langrangian, \Eq{eqP:Lagrangian_TL_discrete},
\begin{equation}\label{eqP:boundarycondition}
 \alpha C_0 \ddot \Phi(0)  - \frac{1}{L_0} \Phi'(0) + \frac{1}{L_c} \Phi(0) = 0
\end{equation}
Using $\ddot \Phi_k = -v^2 k^2 \Phi_k$ this can be cast into a
boundary condition for the modes,
\begin{equation}\label{eqP:boundarycondition_TLmodes}
 \alpha k^2  u_k(0) + u'_k(0) - \frac{1}{\beta} u_k(0) = 0
\end{equation}
Note  \textcolor{black}{again} how this differs from a Sturm-Liouville problem, for which the boundary condition
does not depend on the value $k$.
Using the ansatz $u_k(x) = a \sin(k x) + b \cos(k x) = \sqrt{\frac{2}{\pi}} \cos(k x + \varphi_k)$,
we find the dressed TL modes as
\begin{align}
\label{eqP2:TLmode}
 u_k(x)
 &= u_k(0) \left( \cos kx - \left(\alpha k - \frac{1}{\beta k}\right) \sin kx \right) \\
\label{eqP2:TLmodeamplitudes}
 u_k(0) &= \sqrt{\frac{2}{\pi}} \frac{1}{\sqrt{1+\left(\alpha k - 1/(\beta k)\right)^2}} \\
\label{eqP2:TLmodephase}
 \tan(\varphi_k) &= \alpha k - \frac{1}{\beta k}
\end{align}

\subsection{Coupling constants}

We introduce the system amplitude $a_s$ as in \Eq{eqQq:Phis_Qs_fromas},
where here the system frequency and impedance are
\begin{align}
\label{eqP:ws}
 \omega_s &= \sqrt{\frac{L_s + L_c}{C_s L_s L_c}} \\
\label{eqP:Zs}
 Z_s &= \sqrt{\frac{L_s L_c}{C_s (L_s + L_c)}}
 \,,
\end{align}
as determined by $H_{\sys}$, \Eq{eqP:Hsys}.
Using also the TL mode amplitudes $a_k$ from \Eq{eqQq:qk_phik_fromak},
the Hamiltonian, \Eq{eqP:Hamiltonian_final} then reads
\begin{align}\label{eqPq:Ham_orig}
H &= H_{\sys} + \int_0^\infty \diff k \hbar \omega_k a_k^\dag a_k
 + \hbar \int_0^\infty \diff k \left[ ( f_{k}^{(C_c)} + f_{k}^{(L_c)} ) ( a_k^\dag a_s + a_s^\dag a_k ) -  ( f_{k}^{(C_c)} - f_{k}^{(L_c)} ) ( a_k^\dag a_s^\dag + a_s a_k ) \right]
\end{align}
where the capacitive and inductive coupling coefficients are defined as
\begin{align}
 f_k^{(C_c)} &:= \sqrt{\frac{\omega_k C_0}{Z_s}} \frac{\alpha}{2C_s} u_k(0)
 \\
 f_k^{(L_c)} &:= -\sqrt{\frac{Z_s}{\omega_k C_0}} \frac{1}{2\beta L_0} u_k(0)
\end{align}
As described in Sec.~\ref{subsec:wdependentcoupling_systresponse},
in this work we consider only the RWA-form of the coupled system-TL Hamiltonian,
i.e. neglecting the counter-rotating contributions $a_s a_k$ and $a_s^\dag a_k^\dag$
in \Eq{eqPq:Ham_orig}.
Then, only the sum of capacitive and inductive coupling coefficients remains.
Using the capacitive and inductive coupling rates,

\begin{align}
  \label{eqPq:GammaE_Cc}
 2\Gamma_E^{(C_c)} &:= \frac{C_c^2 \omega_s Z_0}{(C_s+C_c)^2 Z_s}
 = \frac{Z_s \omega_s^3 \alpha^2}{Z_0 v^2} \\
 \label{eqPq:GammaE_Lc}
 2\Gamma_E^{(L_c)} &:= \frac{Z_s Z_0}{L_c^2 \omega_s}
 = \frac{Z_s v^2}{\omega_s Z_0 \beta^2}
\end{align}
and defining an effective coupling rate,
\begin{align}
\label{eqPq:def_GammaE_eff}
\sqrt{2\Gamma_E^{\eff}}
&:= \left| \sqrt{2\Gamma_E^{(C_c)}} - \sqrt{2\Gamma_E^{(L_c)}} \right|
= \sqrt{\frac{\omega_s Z_s}{Z_0}}
\left| \frac{\alpha \omega_s}{v} - \frac{v}{\beta \omega_s} \right| \,,
\end{align}
we can cast the resulting coupling constant into the form
\begin{align}
\label{eqPq:def_fk}
 f_k := f_k^{(C_c)} + f_k^{(L_c)}
 &= \sqrt{2\Gamma_E^{\eff} v}
 \left(
 \sqrt{\frac{2\Gamma_E^{(C_c)}}{2\Gamma_E^{\eff}} } \sqrt{\frac{\omega_k}{\omega_s} }
 -
 \sqrt{\frac{2\Gamma_E^{(L_c)}}{2\Gamma_E^{\eff}} } \sqrt{\frac{\omega_s}{\omega_k} }
 \right) \frac{u_k(0)}{2}
\end{align}
At $k=\omega_s/v$,
using $u_{k=\omega_s/v}(0) = (2/\pi)^{1/2} ( 1 + 2 \Gamma_E^\eff Z_0/(\omega_s Z_s))^{-1/2}$,
this becomes
\begin{align}
 f_{k = \omega_s/v}
 = -\sqrt{\frac{v\Gamma_E^{\eff}}{\pi}}
 \left( 1 + \frac{2 \Gamma_E^\eff Z_0}{\omega_s Z_s} \right)^{-1/2}
\,.
\end{align}

\section{Simple ladder filter:
derivation of dressed TL modes and coupling constants}
\label{app:parallelLC_beforeseriesLC_TLmodes}

\begin{figure}[h]\centering
\includegraphics[height=1.75cm]{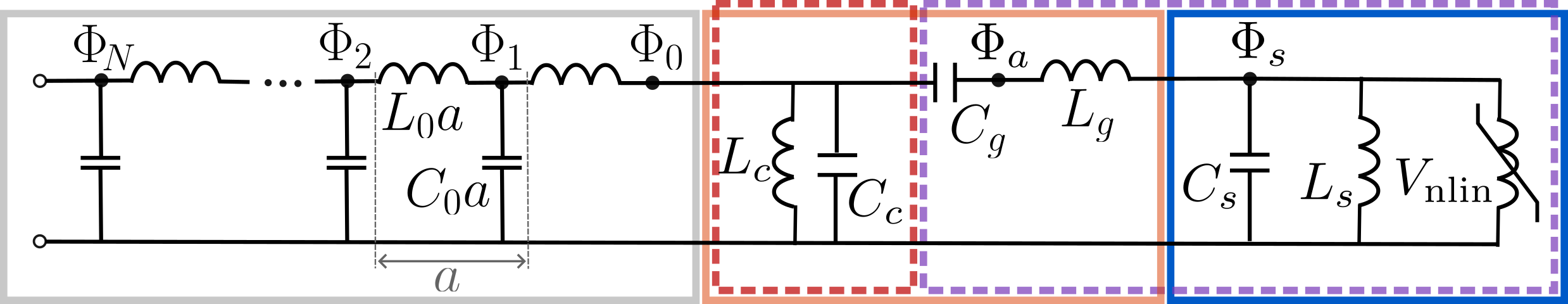}
\caption{
A resonator coupled to the end of a TL (one-port setup) via a simple ladder filter,
consisting of a series-$\sLC$ resonator (with parameters $C_g, L_g$)
followed by a parallel-$\sLC$ resonator (with parameters $C_c, L_c$),
cf.~Fig.~\ref{fig:filters}(c).
The TL is here represented in discretized form, where the unit cell has a length $a$
such that the capacitors and inductors of the discrete TL are of size $C_0 a$
and $L_0 a$, respectively.
The system resonator is assumed to consist of a parallel circuit of capacitance $C_s$,
inductance $L_s$,
and a nonlinear potential $V_{\nlin}(\Phi_s)$, here symbolized as a nonlinear inductor.
The dressed TL modes are subject to the boundary condition set by the
parallel coupling network (with parameters $C_c, L_c$),
whereas the series coupling network (with parameters $C_g, L_g$)
together with the bare resonator form the extended resonator system
characterized by two degrees of freedom.
}
\label{fig:LCresonator_parallelLcCc_beforeseriesLCcoupled__discreteTL}
\end{figure}

Here we derive the Hamiltonian for the system
coupled via a simple ladder filter to a TL, as shown in Fig.~\ref{fig:filters}(c).
The filter consists of a series-$\sLC$ circuit (with parameters $C_g, L_g$)
followed by a parallel-$\sLC$ circuit (with parameters $C_c, L_c$),
as seen from the system side.
In this analyis, we diagonalize the TL part of the coupled system,
and as a result we obtain the same expressions for the dressed TL modes
that were obtained for the
bare parallel-$\sLC$ coupling, \Eqs{eqP2:TLmode}--\eqref{eqP2:TLmodephase}.
The coupling length parameters $\alpha,\beta$ are defined as
\begin{align}\label{eqF:alpha_beta}
 \alpha = \frac{C_c}{C_0} \qquad
  \beta = \frac{L_c}{L_0}
 \;,
\end{align}
where $\alpha$ here differs from its form, \Eq{eqQ:alpha_beta}, in the bare series-$\sLC$
as well as the bare parallel-$\sLC$ coupling, whereas $\beta$ is still the same.

In a discretized representation of the TL, as shown in Fig.~\ref{fig:LCresonator_parallelLcCc_beforeseriesLCcoupled__discreteTL},
the system Lagrangian reads
\begin{equation}\label{eqF:Lagrangian_discretized}
 \mathcal{L} = \mathcal{L}_{\sys}^{(0)}
 + \sum_{n \geq 1} \left[\frac{C_0 a}{2} \dot{\Phi}_n^2
 - \frac{1}{2 L_0 a} \left(\Phi_{n+1} - \Phi_{n} \right)^2 \right]
 -  \frac{1}{2 L_0 a} \left(\Phi_1 - \Phi_0 \right)^2
 + \frac{C_c}{2} \dot \Phi_0^2 - \frac{1}{2 L_c} \Phi_0^2
 + \frac{C_g}{2} \left( \dot \Phi_a - \dot \Phi_0 \right)^2
 - \frac{1}{2 L_g} \left( \Phi_s - \Phi_a \right)^2
\end{equation}
where $a$ is the size of the TL unit cell.
In contrast to the pure series coupling, \Eq{eqApp:Lagrangian_orig_discretized},
the system is here not overdetermined or singular,
i.e. $\Phi_a$ is a genuine degree of freedom needed to completely describe the system.
We treat $\Phi_a$ as a coordinate of the `extended' system $(\Phi_s, \Phi_a)$.

Assuming that the system of interest can be represented as a linear oscillator, plus
a nonlinear potential, see Fig.~\ref{fig:LCresonator_parallelLcCc_beforeseriesLCcoupled__discreteTL},
\begin{equation}
 \mathcal{L}_{\sys}^{(0)}(\Phi_s, \dot \Phi_s) = \frac{C_s}{2} \dot{\Phi}_s^2
 - \frac{1}{2 L_s} \Phi_s^2 - V_{\nlin}(\Phi_s) \,,
\end{equation}
\begin{equation}
 \mathcal{L} = \frac{1}{2} \dot{\boldsymbol{\Phi}}^T \uuline{C} \dot{\boldsymbol{\Phi}}
 - \frac{1}{2} \boldsymbol{\Phi}^T \uuline{L^{-1}} \boldsymbol{\Phi}
 - V_{\nlin}(\Phi_s)
 \,,
\end{equation}
the matrices for the capacitance and inverse inductance of the full system,
with flux vector $\boldsymbol{\Phi}^T = (\Phi_s, \Phi_a, \Phi_0, \Phi_1, \Phi_2, \ldots)$,
can be written as
\begin{align}
 \uuline{C} = \begin{pmatrix}
 C_s & 0 & 0 & \\
  0 & C_g & -C_g &  \\
  0 & -C_g & C_g + C_c &  \\
  &&& C_0 a & & \\
  &&&& C_0 a& \\
  &&&&& \ddots
 \end{pmatrix}
 &&
 \uuline{L^{-1}} = \frac{1}{L_0 a} \begin{pmatrix}
  l_s^{-1} + l_g^{-1} & -l_g^{-1} & 0 \\
  -l_g^{-1} & l_g^{-1} & 0 \\
  0 & 0 & 1 + l_c^{-1} & -1 \\
  &&-1 & 2 & -1 & \\
  &&&-1 & 2 & \\
  &&&&&&\ddots \\
 \end{pmatrix}
\end{align}
with parameters
\begin{align}
 l_c := \frac{L_c}{L_0 a} 
 &&
 l_g := \frac{L_g}{L_0 a}
 &&
 l_s := \frac{L_s}{L_0 a}
\end{align}
The Legendre transformation from the Lagrangian to the Hamiltonian
involves the inverse capacitance matrix,
\begin{align}
 \uuline{C}^{-1} = \begin{pmatrix}
  \frac{1}{C_s} & 0 & 0 & \\
  0& \frac{C_c + C_g}{C_c C_g} & \frac{1}{C_c} \\
  0 & \frac{1}{C_c} & \frac{1}{C_c} \\
  &&& \frac{1}{C_0 a} \\
  &&&& \frac{1}{C_0 a} & \\
  &&&&& \ddots
 \end{pmatrix}
\end{align}
and gives the Hamiltonian
\begin{align}
  H &= \frac{1}{2} \boldsymbol{Q}^T \uuline{C}^{-1} \boldsymbol{Q}
 + \frac{1}{2} \boldsymbol{\Phi}^T \uuline{L^{-1}} \boldsymbol{\Phi}
 + V_{\nlin}(\Phi_s) \\
\label{eqF:Hamiltonian_definitionHtl}
 &= H_{\sys} + H_{TL} + \frac{1}{C_c} Q_a Q_0
\end{align}
with the extended system Hamiltonian
\begin{align}\label{eqF:Hsysext}
 H_{\sys}
 &=
 \frac{1}{2 C_s} Q_s^2 + \frac{1}{2}\left(\frac{1}{L_s} + \frac{1}{L_g}\right) \Phi_s^2 + V_{\nlin}(\Phi_s)
 + \frac{1}{2L_g} \Phi_a^2 - \frac{1}{L_g}\Phi_a \Phi_s
 + \frac{1}{2} \left(\frac{1}{C_c} + \frac{1}{C_g}\right) Q_a^2 \\
 &= H_{\sys}^{(0)}
 + \frac{1}{2} \left(\frac{1}{C_c} + \frac{1}{C_g}\right) Q_a^2
 +  \frac{1}{2 L_g} \left(\Phi_s - \Phi_a\right)^2
\end{align}
and where we have defined the dressed TL Hamiltonian
\begin{align}
 H_{TL} &=
 \frac{1}{2} \boldsymbol{Q}_{TL}^T \left[\uuline{C}^{-1}\right]_{2:,2:} \boldsymbol{Q}_{TL}
 + \frac{1}{2} \boldsymbol{\Phi}_{TL}^T \left[\uuline{L^{-1}}\right]_{2:,2:} \boldsymbol{\Phi}_{TL} \\
 &= \frac{1}{2 C_c}  Q_0^2 + \frac{1}{2C_0 a} \sum_{n\geq 1} Q_n^2
 + \frac{1+l_c^{-1}}{2 L_0 a} \Phi_0^2 - \frac{1}{L_0 a} \Phi_0 \Phi_1 +  \frac{1}{2 L_0 a} \Phi_1^2
 + \frac{1}{2L_0 a} \sum_{n\geq 1} (\Phi_{n+1} - \Phi_n)^2 \\
 \label{eqF:H_TL_orig}
 &= \frac{1}{2 \alpha C_0}  Q_0^2 + \frac{1}{2C_0 a} \sum_{n\geq 1} Q_n^2
 + \frac{1}{2 \beta L_0} \Phi_0^2 + \frac{1}{2 L_0 a}  (\Phi_0 - \Phi_1)^2
 + \frac{1}{2L_0 a} \sum_{n\geq 1} (\Phi_{n+1} - \Phi_n)^2
\end{align}
The TL flux and charge vectors are
$\boldsymbol{\Phi}_{TL}^T = (\Phi_0, \Phi_1, \Phi_2, \ldots)$
and $\boldsymbol{Q}_{TL}^T = (Q_0, Q_1, Q_2, \ldots)$,
and we have here used the coupling length parameters $\alpha,\beta$,
\Eq{eqF:alpha_beta}.
Because of the property
\begin{equation}\label{eqF:property_invC}
 \left[\uuline{C}^{-1}\right]_{2:,2:}
\neq \left[\uuline{C}_{2:,2:} \right]^{-1}
\end{equation}
the `dressed' TL-Hamiltonian
differs from the `bare' TL-Hamiltonian which would be obtained
by performing the Legendre transformation 
while setting $\Phi_s=0$ in \Eq{eqF:Lagrangian_discretized}.

We now want to perform a continuum limit $a\to 0$,
and to this end transform back to the Lagrangian representation.
The Legendre transformation back to the Lagrangian involves the inverted
relation between the $Q_n$ and $\dot\Phi_n$,
\begin{align}\label{eqF:inv_invCmat_TL}
 \left[\left[\uuline{C}^{-1}\right]_{2:,2:}\right]^{-1} = C_0 \cdot \text{diag}\left(\alpha, a, a, \ldots \right)
\end{align}
and gives the dressed TL Lagrangian,
\begin{align}
 \mathcal{L}_{TL} &=
 \frac{1}{2} \dot{\boldsymbol{\Phi}}_{TL}^T \left[\left[\uuline{C}^{-1}\right]_{2:,2:}\right]^{-1} \dot{\boldsymbol{\Phi}}_{TL}
 - \frac{1}{2} \boldsymbol{\Phi}_{TL}^T \left[\uuline{L^{-1}}\right]_{2:,2:} \boldsymbol{\Phi}_{TL} \\
\label{eqF:Lagrangian_TL_discrete}
 &= \frac{\alpha C_0}{2} \left( \dot \Phi_0 \right)^2 + \frac{C_0 a}{2} \sum_{n\geq 1} \dot \Phi_n^2
 - \frac{1}{2 L_c} \Phi_0^2 - \frac{1}{2 L_0 a}  (\Phi_0 - \Phi_1)^2
 - \frac{1}{2L_0 a} \sum_{n\geq 1} (\Phi_{n+1} - \Phi_n)^2 \\
\label{eqF:Lagrangian_TL_continuous}
 &\to \frac{\alpha C_0}{2} \left( \dot \Phi(0) \right)^2 + \frac{C_0 }{2} \int_{0}^{\infty} \diff x \dot \Phi^2
 - \frac{1}{2 \beta L_0} (\Phi(0))^2
 - \frac{1}{2 L_0} \int_{0}^{\infty} \diff x (\Phi')^2
\end{align}
This Legendre transformation from $H_{TL}$ to $\mathcal{L}_{TL}$
is only with respect to the TL subspace
and due to the property \eqref{eqF:property_invC} does not agree
with the Legendre transformation of the full system.
In \Eq{eqF:Lagrangian_TL_continuous}
we have associated the coupling point with $x=0$ and the far end of the TL as $x=+\infty$,
such that $(\Phi_0 - \Phi_1)/a \to -\Phi'(0)$
(other than seen in the geometry of Fig.~\ref{fig:LCresonator_parallelLcCc_beforeseriesLCcoupled__discreteTL}).

The TL Hamiltonian, \Eq{eqF:H_TL_orig},
and Lagrangian, \Eqs{eqF:Lagrangian_TL_discrete}--\eqref{eqF:Lagrangian_TL_continuous},
are formally identical to their counterparts
in the purely parallel-$\sLC$ coupling case, \Eq{eqP:H_TL_orig} and \Eqs{eqP:Lagrangian_TL_discrete}--\eqref{eqP:Lagrangian_TL_continuous}.
The only difference lies in their respective definitions of the coupling length $\alpha$,
\Eq{eqF:alpha_beta} compared with \Eq{eqQ:alpha_beta}.
Thus also the orthogonality relations
\eqref{eqP2:orthogonality_alpha}--\eqref{eqP2:orthogonality_beta},
boundary conditions
\eqref{eqP:boundarycondition}--\eqref{eqP:boundarycondition_TLmodes},
and solutions for the dressed TL modes, \Eqs{eqP2:TLmode}--\eqref{eqP2:TLmodephase},
can be taken over from the purely parallel-$\sLC$ coupling case,
keeping in mind the different definitions of $\alpha$.
Using \Eq{eqF:inv_invCmat_TL} we also find the same expression for the
boundary node charge $Q_0$ as in \Eq{eqP:Q0_boundary}.
Using the TL mode representation,
the full Hamiltonian, \Eq{eqF:Hamiltonian_definitionHtl}, now reads
\begin{align}\label{eqF:Hamiltonian_final}
  H &= H_{\sys}
 + \frac{\alpha}{C_c} Q_a \int_0^\infty \diff k u_k(0) q_k
 + \int_0^\infty \diff k \left( \frac{1}{2C_0} q_k^2 + \frac{k^2}{2 L_0} \Phi_k^2 \right)
\end{align}
with $H_{\sys}$ from \Eq{eqF:Hsysext}
and $u_k(0)$ from \Eq{eqP2:TLmodeamplitudes}.

The quantization of the Hamiltonian \Eq{eqF:Hamiltonian_final} proceeds as in Sec.~\ref{sec:coef_selfenergy_simpleladder}, and results in \Eq{eqFq:Ham_orig}. Defining an effective coupling rate,
\begin{align}
\label{eqFq:def_GammaE_eff}
\sqrt{2\Gamma_E^{\eff}} &:= \sqrt{\frac{\omega_s Z_0}{Z_a}}
 \,,
\end{align}
we can cast the coupling constant $f_k$ into the form
\begin{align}
 f_k
 &= \sqrt{2\Gamma_E^{\eff} v}\, \sqrt{\frac{\omega_k}{\omega_s}} \frac{u_k(0)}{2} \nonumber\\
\label{eqFq:def_fk}
 & = \sqrt{\frac{\Gamma_E^{\eff} v}{\pi}} \sqrt{\frac{\omega_k}{\omega_s}} \left(1 + \left(\frac{\alpha \omega_k}{v} - \frac{v}{\beta \omega_k} \right)^2 \right)^{-1/2}
\end{align}
At $k=\omega_s/v$,
this becomes
\begin{align}
 f_{k = \omega_s/v}
 = \sqrt{\frac{v\Gamma_E^{\eff}}{\pi}}
 \left( 1 + \left(\frac{\alpha \omega_s}{v} - \frac{v}{\beta \omega_s} \right)^2 \right)^{-1/2}
\,.
\end{align}

\section{Evaluation of self-energy $\Sigma_E(\Delta)$ for passive coupling networks}
\label{app:selfenergy_variouscouplings}

The self-energy of the system-bath coupling is defined as
\begin{align}\label{eqapp:def_selfenergy}
\Sigma_E(\Delta)
:=& \int_0^\infty \diff k \frac{f_k^2}{\Delta - \Delta_k + \ui\epsilon}
&&= \int_0^\infty \diff k \frac{f_k^2}{\omega - \omega_k + \ui\epsilon}
&&=: \hat \Sigma_E(\omega) 
\end{align}
where $\Delta_{(k)} = \omega_{(k)} - \wref$ is the detuning
from the reference frequency $\wref$ of the rotating frame in which the
slow system dynamics takes place, i.e. where the Langevin equation
is not explicitly time-dependent.
For an autonomous system, this frame is given by the relevant system frequency,
$\wref = \omega_s$. For a driven system, it is usually determined by the drive frequency,
e.g. $\wref = \omega_p/2 \approx \omega_s$ in case of a system
driven into degenerate parametric resonance with pump frequency $\omega_p \approx 2\omega_s$.
The imaginary part of the self-energy 
quantifies the coupling contribution
to the effective damping rate of the system.
With the help of the nascent delta function
$f_\epsilon(x) = \pi^{-1} \epsilon (x^2 + \epsilon^2)^{-1}$,
it can be evaluated as
\begin{equation}\label{eqapp:def_imSig}
 \im \hat{\Sigma}_E(\omega) = \left.-\frac{\pi}{v} f_k^2\right|_{k=\omega/v}
 \,,
\end{equation}
cf.~\Eq{eq:def_imSig}.
The real part remains to be evaluated from the integration,
\begin{equation}\label{eqapp:def_reSig}
 \re \hat{\Sigma}_E(\omega)
 =  \re \int_0^{\infty} \diff k \frac{ f_k^2 }{\omega - vk + \ui \epsilon}
 =  - \int_0^{\infty} \diff k \frac{ f_k^2 (vk - \omega) }{(vk - \omega)^2 + \epsilon^2}
 \,.
\end{equation}

Here we evaluate the self-energy integral for the simplest coupling circuits:
firstly a parallel-$\sLC$ circuit in App.~\ref{sec:selfenergy_parallelCcLc}
and its limiting cases, the purely capacitive and inductive coupling,
in App.~\ref{sec:selfenergy_Cc} and \ref{sec:selfenergy_Lc}, respectively.
Finally, the series-$\sLC$ circuit is treated in App.~\ref{sec:selfenergy_seriesCcLc}.
We will make use of the following dimensionless parameters,
\begin{align}
 x:= \frac{vk}{\omega}  
  && \alpha_\omega := \frac{\alpha \omega}{v}
  && \beta_\omega := \frac{\beta \omega}{v} 
  \,.
\end{align}

%

\textcolor{black}{In Appendices \ref{app:seriesLC_TLmodes}--\ref{app:parallelLC_beforeseriesLC_TLmodes} we have derived the dressed TL modes for series $\sLC$ coupling, parallel $\sLC$ coupling, and the simple ladder filter, respectively.
The characteristic length scales $\alpha$ and $\beta$ appearing in the boundary condition and dressed system frequencies for these cases are different. Table~\ref{table:circuit_paras_alpha_beta_etc} lists the different definitions of these length scales, together with the system frequency and impedance.}

\begin{table}
\begin{center}
\begin{tabular}{ | m{3cm} | m{2cm}| m{2cm} |m{2cm}|m{3cm}| } 
  \hline
  Coupling circuit& $\alpha$ & $\beta$ & $Z_s$ & $\omega_s$ \\ 
  \hline
  Series-$\sLC$ & $\frac{C_cC_s}{C_0(C_c+C_s)}$ & $\frac{L_c}{L_0}$ & $\sqrt{\frac{L_s}{C_s}}$ & $\sqrt{\frac{1}{L_sC_s}}$ \\ 
  \hline
  Ladder & $\frac{C_c}{C_0}$ & $\frac{L_c}{L_0}$ & $\sqrt{\frac{L_sL_g}{C_s(L_s+L_g)}}$  & $\sqrt{\frac{L_s+L_g}{C_sL_sL_g}}$\\ 
  \hline
Parallel-$\sLC$ & $\frac{C_cC_s}{C_0(C_c+C_s)}$ & $\frac{L_c}{L_0}$ & $\sqrt{\frac{L_sL_c}{C_s(L_s+L_c)}}$ & $\sqrt{\frac{1}{C_s}(\frac{1}{L_s}+\frac{1}{L_c})}$ \\
\hline
\end{tabular}
\caption{\textcolor{black}{Parameters for different coupling circuits.}}    
\label{table:circuit_paras_alpha_beta_etc}
\end{center}
\end{table}


\subsection{Parallel $C_c-L_c$ coupling}\label{sec:selfenergy_parallelCcLc}

In App.~\ref{app:parallelLC_TLmodes} we had derived the coupling coefficients
for the case of parallel-$\sLC$ coupling, cf.~\Eqs{eqPq:def_fk} and \eqref{eqP2:TLmodeamplitudes},
\begin{align}\label{eqapp:fksquared_parallelcoupled}
 f_k^2 = \frac{v}{2\pi} 
 \frac{
  \left(\sqrt{2\Gamma_E^{(C_c)}} \sqrt{\frac{k v}{\omega_s}} - \sqrt{2\Gamma_E^{(L_c)}} \sqrt{\frac{\omega_s}{kv}}\right)^2
  }{ 1 + \left[\alpha k - 1/(\beta k) \right]^2 }
\end{align}
where $\alpha = \frac{C_c C_s}{(C_s+C_c) C_0}$,
$\beta = \frac{L_c}{L_0}$,
and where $\omega_s$, $Z_s$ and $\Gamma_E^{(C_c)}$, $\Gamma_E^{(L_c)}$
are defined in \Eqs{eqP:ws}--\eqref{eqP:Zs}
and \Eqs{eqPq:GammaE_Cc}--\eqref{eqPq:GammaE_Lc}.

The coupling constants $|f_k|$ are bounded and thus divergence free,
as seen in $\im \hat{\Sigma}(\omega) \propto |f_{k=\omega/v}|^2$
shown in Fig.~\ref{fig:selfenergy_parallelcoupled}.
The position $k_{\Sigma,0}$ of the maxima of $f_k$ is determined by the equation
\begin{align}\label{eqapp:k_maxfk_parallelcoupled}
 0 &= \frac{k v}{\omega_s} \left[ \left(\frac{\beta}{\alpha} - 2\right) \left(\frac{kv}{\omega_s}\right)^2
 - \frac{\omega_s}{\omega_0} \left(\frac{kv}{\omega_s}\right)^4
 + 3 \frac{\omega_0}{\omega_s} \right]
 + \frac{\omega_0}{\omega_s} \left[ \left(\frac{\beta}{\alpha} - 2\right) \left(\frac{kv}{\omega_s}\right)^2
 + 3 \frac{\omega_s}{\omega_0} \left(\frac{kv}{\omega_s}\right)^4
 - \frac{\omega_0}{\omega_s} \right]
 \;, \\
\label{eqapp:w0_selfenergy_imag_parallelcoupled}
& \text{with}\quad \frac{\omega_0}{\omega_s}
 = \sqrt{\frac{2\Gamma_E^{(L_c)}}{2\Gamma_E^{(C_c)}}}
 = \frac{C_s + C_c}{C_c \left(1+\frac{L_c}{L_s}\right)}
 = \frac{v^2}{\alpha \beta \omega_s^2}
\end{align}
There are in general two maxima,
cf.~Fig.~\ref{fig:selfenergy_parallelcoupled};
only the dominant one is relevant here and will be used in our further discussion
after \Eq{eqapp:wres_exact_selfenergy_imag_parallelcoupled}.

According to \Eq{eqapp:def_imSig} one obtains the imaginary part of the self-energy,
\begin{align}\label{eqapp:selfenergy_imag_parallelcoupled}
 \frac{\im \hat{\Sigma}_E}{\omega_s}
 = -\frac{1}{2 \omega_s} \left(\sqrt{2\Gamma_E^{(C_c)}} \sqrt{\frac{\omega}{\omega_s}} - \sqrt{2\Gamma_E^{(L_c)}} \sqrt{\frac{\omega_s}{\omega}}\right)^2 \frac{1}{1 + \left( \frac{\alpha \omega}{v} - \frac{v}{\beta \omega}\right)^2}
\end{align}
A resonant enhancement of the damping rate $-\im \hat{\Sigma}_E(\omega)$
is expected near $\omega \approx \omega_{\Sigma}$, with
\begin{align}\label{eqapp:wres_selfenergy_imag_parallelcoupled}
 \omega_{\Sigma}
 &= \frac{v}{\sqrt{\alpha \beta}} 
 = \frac{\sqrt{1+\frac{C_c}{C_s}}}{\sqrt{L_c C_c}} 
 \,.
\end{align}
However, at the same time $-\im \hat{\Sigma}_E(\omega)$ turns zero at $\omega_0$,
\Eq{eqapp:w0_selfenergy_imag_parallelcoupled},
which is closely related to $\omega_\Sigma$
through the property $\omega_0/\omega_s = (\omega_\Sigma/\omega_s)^2$.
As a result, under variation of the coupling ($\alpha$ or $\beta$)
the two functions $\omega_\Sigma$ and $\omega_0$
intersect exactly at $\omega_s$:
(i) if $\omega_\Sigma=\omega_s$ then also $\omega_0 = \omega_s$.
Futhermore,
(ii) if $\omega_\Sigma>\omega_s$ then $\omega_s < \omega_0$, and
(iii) if $\omega_\Sigma<\omega_s$ then $\omega_s > \omega_0$ .
Examples for these cases can be seen in Fig.~\ref{fig:selfenergy_parallelcoupled}.
Thus we see that the naively expected condition for strong system-TL-coupling,
$\omega_\Sigma \approx \omega_s$,
implies at the same time $\omega_0 \approx \omega_s$,
such that the resonant enhancement is not realized
and instead $-\im \hat{\Sigma}_E(\omega_s) \approx 0$.
The distance between $\omega_0$ and $\omega_{\Sigma}$ is
\begin{eqnarray}
 \omega_0 - \omega_{\Sigma} &=& \frac{\sqrt{1 + C_c/C_s}}{\sqrt{L_c C_c}}
 \left( \sqrt{\frac{1 + C_c/C_s}{1 + L_c/L_s} }\sqrt{ \frac{C_s}{C_c} } - 1\right)
\end{eqnarray}
and in order for the resonance to be resolved, this distance
has to be large compared with the resonance width $\gamma_\Sigma$,
\begin{eqnarray}
 \gamma_\Sigma &=& \frac{1}{Z_0 C_c} \left(1 + \frac{C_c}{C_s}\right) 
\,.
\end{eqnarray}
In principle, this condition can be met with $Z_s \ll Z_0$,
independent of the resonance locations.

Note that $\omega_{\Sigma}$ is only an approximation for the
location of the resonance, while in general
the maximum of $-\im \hat{\Sigma}_E(\omega)$ is somewhat shifted.
The exact resonance location is determined by the condition
$\diff(\im \hat{\Sigma}_E)/\diff(\omega) = 0$.
Owing to the relation between $\im \hat{\Sigma}_E$ and $f_k$,
\Eq{eqapp:def_imSig}, it is is given by the maximum position $k_{\Sigma,0}$ of $|f_k|$,
\Eq{eqapp:k_maxfk_parallelcoupled},
\begin{equation}\label{eqapp:wres_exact_selfenergy_imag_parallelcoupled}
 \omega_{\Sigma,0} = v k_{\Sigma,0}
\end{equation}
In general, \Eq{eqapp:k_maxfk_parallelcoupled} determines
two maxima, separated by a zero of $-\im \hat{\Sigma}_E(\omega)$ at $\omega_0$,
cf.~\Eq{eqapp:w0_selfenergy_imag_parallelcoupled} and Fig.~\ref{fig:selfenergy_parallelcoupled}.
Our definition of
$k_{\Sigma,0}$ and $\omega_{\Sigma,0}$ refers to the dominant one of the two maxima.
In the limit $\alpha/\beta \gg 1$ the approximate resonance position $\omega_\Sigma$
is a good approximation for $\omega_{\Sigma,0}$.
In the limit $\alpha/\beta \ll 1$, $\omega_{\Sigma,0}$ approaches 0, but only
if $\omega_\Sigma > \omega_s$ (case (ii)),
and in this case $\omega_{\Sigma,0}$ is the lower-frequency maximum of $-\im \hat{\Sigma}_E(\omega)$.
Otherwise, if $\omega_\Sigma < \omega_s$ (case (iii)),
then $\omega_{\Sigma,0}$ is the higher-frequency maximum of $-\im \hat{\Sigma}_E(\omega)$
and it can be large.

From \Eq{eqapp:def_reSig} we obtain the real part of the self-energy,
\begin{align}
  \frac{\re \hat{\Sigma}_E}{\omega_s}
  &= -\re \frac{v}{2\pi \omega_s} \int_0^\infty \diff k \frac{
  \left(\sqrt{2\Gamma_E^{(C_c)}} \sqrt{\frac{k v}{\omega_s}} - \sqrt{2\Gamma_E^{(L_c)}} \sqrt{\frac{\omega_s}{kv}}\right)^2
  }{(vk -\omega -\ui\epsilon)(1+\left[\alpha k -1/(\beta k)\right]^2)} \\
\label{eqapp:selfenergy_real_parallelcoupled_preintegration}
  &=  -\frac{1}{2\pi} \int_0^\infty \diff x \frac{
  x (x-1) (g_C x - g_L)^2
  }{( [x-1]^2 + \epsilon^2)(x^2 + [ \alpha_\omega x^2 - 1/\beta_\omega ]^2)} 
\end{align}
with the parameters
\begin{align}
g_C &:= \sqrt{\frac{2\Gamma_E^{(C_c)}}{\omega_s}} \sqrt{\frac{\omega}{\omega_s}} \\
g_L &:= \sqrt{\frac{2\Gamma_E^{(L_c)}}{\omega_s}} \sqrt{\frac{\omega_s}{\omega}}
\,.
\end{align}

For the integration of \Eq{eqapp:selfenergy_real_parallelcoupled_preintegration}
the poles of the integrand are needed.
Poles are located at $x = 1 \pm \ui \epsilon$
and at the complex roots of the polynomial 
$p(x) = x^2 + [\alpha_\omega x^2 - 1/\beta_\omega]^2$.
These roots take the form
\begin{IEEEeqnarray}{llCl}\label{eqapp:roots_denom_selfenergy_real_parallelcoupled}
\IEEEyesnumber
 \uuline{\frac{4 \alpha}{\beta} < 1}:\hspace*{1cm}
 &x_{1,2} &=& \pm \frac{\ui}{\sqrt{2} \alpha_\omega} \sqrt{\left(1- \frac{2\alpha}{\beta}\right) - \sqrt{1 - \frac{4\alpha}{\beta}} } \IEEEyessubnumber\\
 &x_{3,4} &=& \pm \frac{\ui}{\sqrt{2} \alpha_\omega} \sqrt{\left(1- \frac{2\alpha}{\beta}\right) + \sqrt{1 - \frac{4\alpha}{\beta}} } \IEEEyessubnumber\\[2ex]
 \uuline{\frac{4 \alpha}{\beta} > 1}:\hspace*{1cm} 
 &x_{i} &=& r e^{\iexp \varphi_i} 
 \qquad r = \frac{1}{\sqrt{\alpha_\omega \beta_\omega}}
 \qquad (i=1,2,3,4) \IEEEyessubnumber\\
 &\tan(2\varphi_1) &=& \frac{\sqrt{\frac{4\alpha}{\beta} - 1}}{\frac{2\alpha}{\beta} - 1} 
 \hspace*{1cm} 2\varphi_1 \in \left\{ \begin{array}{ll} 
        \left[0, \frac{\pi}{2}\right) & \text{if}\left(\frac{4 \alpha}{\beta} > \frac{2 \alpha}{\beta} > 1 \right) \IEEEyessubnumber\\
        \left[\frac{\pi}{2}, \pi\right) & \text{if}\left(\frac{4 \alpha}{\beta} > 1 > \frac{2 \alpha}{\beta}\right)
      \end{array}\right. \IEEEyessubnumber\\
 &&&\varphi_2 = \pi + \varphi_1,\qquad 
    \varphi_3 = \pi - \varphi_1,\qquad 
    \varphi_4 = \pi + \varphi_3 = - \varphi_1 \nonumber\\
 &&& \text{i.e.}\quad x_4 = x_1^\ast \qquad x_2 = x_3^\ast \nonumber
\end{IEEEeqnarray}

Using a keyhole contour 
for the integration of \Eq{eqapp:selfenergy_real_parallelcoupled_preintegration}
on the complex plane \cite{complexanalysis_book}
we find the real part of the self-energy.
For $4\alpha/\beta < 1$ it takes the form
\begin{align}\label{eqapp:reSig1_parallelcoupled}
\frac{\re \hat{\Sigma}_E}{\omega_s}
&= -\frac{1}{4 \alpha_\omega^2}
 \frac{g_C^2 (|x_1|^2  + |x_3|^2 + |x_1||x_3| + |x_1|^2|x_3|^2) - g_L^2(1- |x_1||x_3|) + 2 g_C g_L (1 - |x_1||x_3|)
 }{(|x_3| + |x_1|)(1 + |x_1|^2)(1 + |x_3|^2)} \\
&- \frac{1}{2\pi \alpha_\omega^2}
 \frac{ [g_L^2 - g_C (g_C-2 g_L) |x_1|^2] (1 + |x_3|^2) \ln(|x_1|)
       -[g_L^2 - g_C (g_C-2 g_L) |x_3|^2] (1 + |x_1|^2) \ln(|x_3|)
 }{(|x_3|^2 - |x_1|^2)(1 + |x_1|^2)(1 + |x_3|^2)} \nonumber
\end{align}
and for $4\alpha/\beta > 1$ it takes the form
\begin{align}\label{eqapp:reSig2_parallelcoupled}
\frac{\re \Sigma_E}{\omega_s}
  &= \frac{1}{4\pi \alpha_\omega^2} \frac{1}{ r^2 \sin(2\varphi_1)
                [1 + r^4 - 2 r^2 \cos(2\varphi_1)]} \\
&   \times
  \biggl\{  \left[g_L^2 - g_C(g_C-2 g_L) r^4\right] (\pi - 2\varphi_1)
  - \pi r \cos(\varphi_1) \left[ g_C^2 r^4 + g_L(g_L - 2g_C) (r^2-1) \right]  \biggr.  \nonumber \\
&
  \hspace*{0.5cm}+\biggl.
  \left[ g_C^2 - 2 g_C g_L - g_L^2 \right] r^2 \cos(2\varphi_1) (\pi - 2\varphi_1)
  + \pi g_C^2 r^3 \cos(3\varphi_1)
  - 2(g_C - g_L)^2 r^2 \ln(r) \sin(2\varphi_1)
\biggr\} \nonumber
\end{align}
For purely capacitive coupling, i.e. setting $\beta \to \infty$ and $g_L = 0$,
only the case $4 \alpha/\beta < 1$ is relevant and 
it can be checked that \Eq{eqapp:reSig1_parallelcoupled}
then reduces to
\Eq{eqapp:selfenergy_real_Cccoupled},
using that $|x_1| \to 0$ and $|x_3| \to 1/\alpha_\omega$.

\begin{figure}\centering
\includegraphics[width=8.6cm]{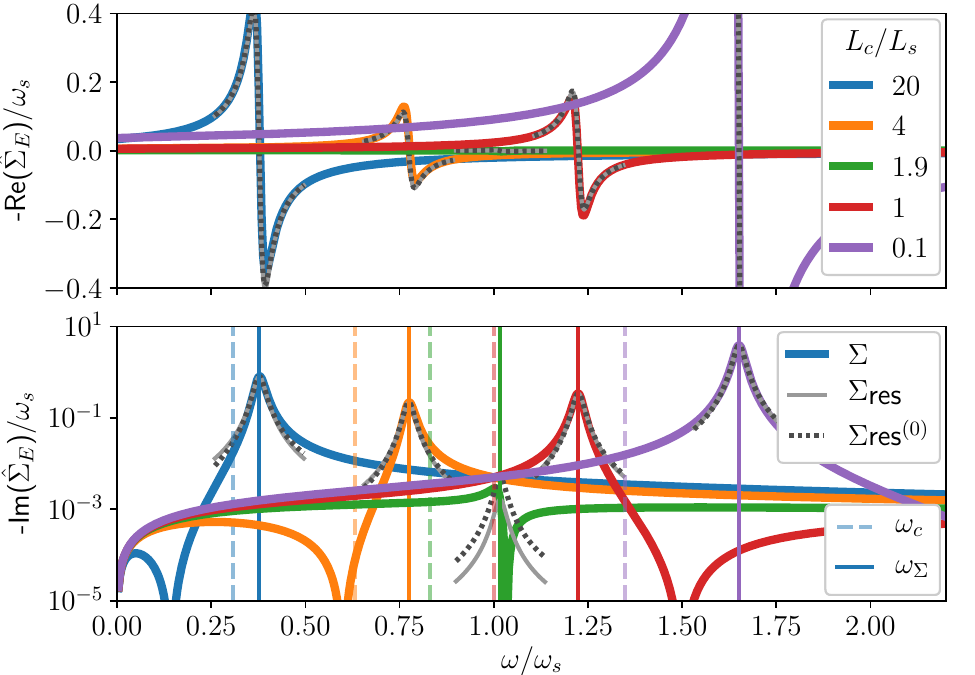}
\caption{
 Self energy $\Sigma_E(\Delta)$ vs $\omega = \omega_s + \Delta$
 for parallel-$\sLC$ coupling circuit with $Z_s/Z_0=0.01$, $C_c/C_s = 0.5$
 and various values of $L_c/L_s$.
 All cases belong to the parameter regime $4\alpha/\beta > 1$,
 with $\beta/\alpha = 3 \cdot 10^{-4} (1 + L_c/L_s)$.
 Grey solid and black dotted lines show the resonant approximations $\Sigma_{E,\text{res}}(\Delta)$
 and $\Sigma_{E,\text{res},0}(\Delta)$
 of \Eq{eqapp:selfenergy_parallelcoupled_resapprox} and \Eq{eqapp:selfenergy_parallelcoupled_resapprox_exact}, respectively.
 The vertical lines show the (approximate) resonance position $\omega_\Sigma$ (solid), \Eq{eqapp:wres_selfenergy_imag_parallelcoupled},
 and the bare coupling frequency
 $\omega_c = \sqrt{(L_c+L_s)/(C_c L_c L_s)}$ (dashed).
}
\label{fig:selfenergy_parallelcoupled}
\end{figure}

\subsubsection{Resonant approximation}\label{sec:selfenergy_parallelcoupled_resapprox}

In order to simplify the expression for the self-energy,
one may approximate the coupling coefficients, \Eq{eqapp:fksquared_parallelcoupled},
near the (approximate) resonance position
$\omega_{\Sigma} = v k_{\Sigma}$,
\Eq{eqapp:wres_selfenergy_imag_parallelcoupled}, as Lorentzians:
\begin{eqnarray}
 f_k^2
\label{eqapp:fksquared_parallelcoupled_v2}
 &=& \frac{v k}{4\pi} 
   \left(\sqrt{2\Gamma_E^{(C_c)}} \sqrt{\frac{k v}{\omega_s}} - \sqrt{2\Gamma_E^{(L_c)}} \sqrt{\frac{\omega_s}{kv}}\right)^2
 \left( \frac{1}{k + \ui \alpha (k^2 - k_{\Sigma}^2)} + \frac{1}{k - \ui \alpha (k^2 - k_{\Sigma}^2)} \right) \\
 &\approx& 
 \frac{v k_\Sigma}{4\pi}
   \left(\sqrt{2\Gamma_E^{(C_c)}} \sqrt{\frac{\omega_\Sigma}{\omega_s}} - \sqrt{2\Gamma_E^{(L_c)}} \sqrt{\frac{\omega_s}{\omega_\Sigma}}\right)^2
 \left( \frac{1}{k_{\Sigma} + \ui 2\alpha k_{\Sigma} (k - k_{\Sigma})} + \frac{1}{k_{\Sigma} - \ui 2 \alpha (k - k_{\Sigma})} \right) \\
 &=& 
 -\frac{\ui v}{8\pi \alpha}
 \left(\sqrt{2\Gamma_E^{(C_c)}} \sqrt{\frac{\omega_\Sigma}{\omega_s}} - \sqrt{2\Gamma_E^{(L_c)}} \sqrt{\frac{\omega_s}{\omega_\Sigma}}\right)^2
 \left( \frac{1}{k - k_{\Sigma} - \ui/(2\alpha)} - \frac{1}{k - k_{\Sigma} +  \ui/(2 \alpha)} \right)
\end{eqnarray}
In line with this resonant approximation, the integration boundaries in \Eq{eqapp:def_selfenergy}
can be extended,
\begin{eqnarray*}
\hat{\Sigma}_{E,\text{res}}
&\approx& \int_{-\infty}^\infty \diff k \frac{f_k^2}{\omega - \omega_k + \ui\epsilon} \\
&\approx& 
-\frac{\ui v}{8\pi \alpha} 
\left(\sqrt{2\Gamma_E^{(C_c)}} \sqrt{\frac{\omega_\Sigma}{\omega_s}} - \sqrt{2\Gamma_E^{(L_c)}} \sqrt{\frac{\omega_s}{\omega_\Sigma}}\right)^2
\int_{-\infty}^\infty \diff k 
 \frac{1}{\omega - v k + \ui\epsilon}
 \left( \frac{1}{k - k_{\Sigma} - \ui/(2\alpha)} - \frac{1}{k - k_{\Sigma} +  \ui/(2\alpha)} \right) \\
&=& 
\frac{\ui}{8\pi \alpha} 
\left(\sqrt{2\Gamma_E^{(C_c)}} \sqrt{\frac{\omega_\Sigma}{\omega_s}} - \sqrt{2\Gamma_E^{(L_c)}} \sqrt{\frac{\omega_s}{\omega_\Sigma}}\right)^2
\int_{-\infty}^\infty \diff k 
 \frac{1}{k - \omega/v - \ui\epsilon/v}
 \left( \frac{1}{k - k_{\Sigma} - \ui/(2\alpha)} - \frac{1}{k - k_{\Sigma} +  \ui/(2 \alpha)} \right)
\end{eqnarray*}
With integration closed over the upper half of the complex plane, the integrals amount to
\begin{eqnarray}
\label{eqapp:integral1_selfenergy_resapprox_parallelcoupled}
 \int_{-\infty}^{\infty} \diff k \frac{1}{k - \omega/v -\ui \epsilon} \frac{1}{k-k_{\Sigma}-\ui/(2\alpha)} &=& 0 \\
\label{eqapp:integral2_selfenergy_resapprox_parallelcoupled}
 \int_{-\infty}^{\infty} \diff k \frac{1}{k - \omega/v -\ui \epsilon} \frac{1}{k-k_{\Sigma}+\ui/(2\alpha)} &=& \frac{2\pi \ui}{\omega/v + \ui \epsilon - k_{\Sigma} + \ui/(2\alpha) }
\end{eqnarray}
and one obtains
\begin{eqnarray}
%
\label{eqapp:selfenergy_parallelcoupled_resapprox}
 \frac{\hat{\Sigma}_{E,\text{res}}}{\omega_s}
&=& \frac{v}{4 \alpha \omega_s}
\left(\sqrt{2\Gamma_E^{(C_c)}} \sqrt{\frac{\omega_\Sigma}{\omega_s}} - \sqrt{2\Gamma_E^{(L_c)}} \sqrt{\frac{\omega_s}{\omega_\Sigma}}\right)^2
 \cdot \frac{1}{\omega - \omega_{\Sigma} + \ui v/(2\alpha)} \\
 &=& \frac{1}{4} \frac{ 2\Gamma_E^{(C_c)} }{\omega_s} \frac{\omega_\Sigma}{\omega_s}
\left( 1 - \frac{\omega_0}{\omega_\Sigma} \right)^2
 \cdot \frac{v/\alpha}{\omega - \omega_{\Sigma} + \ui v/(2\alpha)} \nonumber
\end{eqnarray}
where in the last step we have evaluated the prefactor
from \Eqs{eqPq:GammaE_Cc}--\eqref{eqPq:GammaE_Lc}.
As mentioned before, $\omega_{\Sigma}$ is only an approximation for the resonance position,
valid for $\alpha \gg \beta$.
Since \Eq{eqapp:selfenergy_parallelcoupled_resapprox}
is based on an expansion around $\omega_{\Sigma}$,
this resonant approximation is only good under the same condition.
An alternative resonant approximation is derived in the following section.

\subsubsection{Resonant approximation 2}
\label{sec:selfenergy_parallelcoupled_resapproxE}

As discussed above, the approximation $\omega_\Sigma$ in general deviates
from the exact maximum position $\omega_{\Sigma,0}$ of $-\im\hat{\Sigma}(\omega)$.
Here we derive a similar resonant approximation as in Sec.~\ref{sec:selfenergy_parallelcoupled_resapprox},
using the Lorentzian expansion of the coupling coefficients
around $k_{\Sigma,0} = \omega_{\Sigma,0}/v$ instead of $k_\Sigma$.
Starting point is again \Eq{eqapp:fksquared_parallelcoupled_v2}, where we now
replace $k = k_{\Sigma,0} + (k-k_{\Sigma,0})$
and neglect contributions $\mathcal{O}((k-k_{\Sigma,0})^2)$.
\begin{align*}
 f_k^2 &\approx
   \frac{B_k}{1 + \ui 2\alpha k_{\Sigma,0}} \frac{1}{k - k_{\Sigma,0} + k_{\Sigma,0} a }
 + \frac{B_k}{1 - \ui 2\alpha k_{\Sigma,0}} \frac{1}{k - k_{\Sigma,0} + k_{\Sigma,0} a^\ast }  \\
 &\approx
 \frac{B}{1 + \ui 2\alpha k_{\Sigma,0}} \frac{1}{k - k_{\Sigma,0} + k_{\Sigma,0} a }
   + \frac{B}{1 - \ui 2\alpha k_{\Sigma,0}}
 \frac{1}{k - k_{\Sigma,0} + k_{\Sigma,0} a^\ast }
\end{align*}
with parameters
\begin{align}
B_k &= \frac{v k}{4\pi}
   \left(\sqrt{2\Gamma_E^{(C_c)}} \sqrt{\frac{k v}{\omega_s}} - \sqrt{2\Gamma_E^{(L_c)}} \sqrt{\frac{\omega_s}{kv}}\right)^2
   = \frac{1}{4\pi}
   \left(\sqrt{2\Gamma_E^{(C_c)}} \frac{k v}{\sqrt{\omega_s}} - \sqrt{2\Gamma_E^{(L_c)}} \sqrt{\omega_s}\right)^2 \\
   \approx B &=  \frac{1}{4\pi}
   \left(\sqrt{2\Gamma_E^{(C_c)}} \frac{\omega_{\Sigma,0}}{\sqrt{\omega_s}} - \sqrt{2\Gamma_E^{(L_c)}} \sqrt{\omega_s} \right)^2 \\
  a &= \frac{1 - \ui \alpha (k_\Sigma^2 - k_{\Sigma,0}^2)/k_{\Sigma,0}
            }{ 1 + \ui 2\alpha k_{\Sigma,0}  }
\end{align}
Thus,
\begin{align}
 \hat{\Sigma}_{E,\text{res},0}
 &\approx -\frac{1}{v} \int_{-\infty}^{\infty} \diff k \frac{f_k^2}{k-\omega/v -\ui \epsilon/v} \\
\label{eqapp:selfenergy_parallelcoupled_resapproxE_pre}
 &=
 -\frac{B}{v (1 + \ui 2\alpha k_{\Sigma,0})}
 \int_{-\infty}^{\infty} \diff k \frac{1}{k-\omega/v -\ui \epsilon/v} \frac{1}{k-k_{\Sigma,0} + k_{\Sigma,0} a} \\
 &\quad
 -\frac{B}{v (1 - \ui 2\alpha k_{\Sigma,0})}
 \int_{-\infty}^{\infty} \diff k \frac{1}{k-\omega/v -\ui \epsilon/v} \frac{1}{k-k_{\Sigma,0} + k_{\Sigma,0} a^\ast} \nonumber
\end{align}
In order to solve the integrals, we need to know whether parameter $a$
lies in the upper or lower half of the complex plane.
Numerical inspection of the solutions of \Eq{eqapp:k_maxfk_parallelcoupled}
shows that $k_\Sigma$ can assume values on either side of $k_{\Sigma,0}$, such that
the sign of $k_\Sigma^2 - k_{\Sigma,0}^2$ is not fixed.
For $k_\Sigma > k_{\Sigma,0}$ it is clear
that $a$ always lies in the lower half of the complex plane.
This turns out to be true even in the opposite case, $k_\Sigma < k_{\Sigma,0}$,
because of the property $|k_\Sigma^2 - k_{\Sigma,0}^2| \leq 2 k_\Sigma^2$
.
For integration closed over the upper half of the complex plane,
we therefore find the integrals
\begin{eqnarray}
\label{eqapp:integral1_selfenergy_resapproxE_parallelcoupled}
 \int_{-\infty}^{\infty} \diff k \frac{1}{k - \omega/v -\ui \epsilon} \frac{1}{k-k_{\Sigma,0} + k_{\Sigma,0} a} &=& 0 \\
\label{eqapp:integral2_selfenergy_resapproxE_parallelcoupled}
 \int_{-\infty}^{\infty} \diff k \frac{1}{k - \omega/v -\ui \epsilon} \frac{1}{k-k_{\Sigma,0}+ k_{\Sigma,0} a^\ast} &=& \frac{2\pi \ui}{\omega/v + \ui \epsilon - k_{\Sigma,0} + k_{\Sigma,0} a^\ast}
\end{eqnarray}
Thus one obtains

\begin{equation}\label{eqapp:selfenergy_parallelcoupled_resapprox_exact}
 \frac{\hat{\Sigma}_{E,\text{res},0}}{\omega_s} =
 -\ui \frac{ \left(
   \sqrt{\frac{2\Gamma_E^{(C_c)}}{\omega_s}} \frac{\omega_{\Sigma,0}}{\omega_s}
 - \sqrt{\frac{2\Gamma_E^{(L_c)}}{\omega_s}} \right)^2
 }{2(1 - \ui 2\alpha k_{\Sigma,0})}
\frac{\omega_s}{\omega - \omega_{\Sigma,0} + \omega_{\Sigma,0} a^\ast}
\end{equation}

\subsection{Capacitive coupling}\label{sec:selfenergy_Cc}

The purely capacitive coupling is a special case of the parallel-$\sLC$ coupling,
obtained in the limit $L_c, \beta \to \infty$.
In this limit, $\omega_s$ and $Z_s$ from \Eqs{eqP:ws},
\eqref{eqP:Zs}
reduce to $\omega_s := 1/\sqrt{L_s C_s}$ and  $Z_s = \sqrt{L_s/C_s}$,
while $\Gamma_E^{(L_c)} \to 0$
and $\Gamma_E^{\eff} \to \Gamma_E^{(C_c)}$, cf.~\Eqs{eqPq:GammaE_Cc}--\eqref{eqPq:def_GammaE_eff}.
The coupling coefficients of \Eq{eqapp:fksquared_parallelcoupled}
then become
\begin{align}
 f_k^2 = 2\Gamma_E^{\eff} \frac{v^2 k}{2\pi\omega_s} \frac{1}{1 + \alpha^2 k^2}
\end{align}
They are bounded and thus divergence free,
assuming a maximum at $k_{\Sigma,0} = \alpha^{-1}$.

According to \Eq{eqapp:def_imSig} one obtains the imaginary part of the self-energy,
\begin{align}\label{eqapp:selfenergy_imag_Cccoupled}
 \frac{\im \hat{\Sigma}_E}{\omega_s}
 = -\frac{\Gamma_E^{\eff} \omega}{\omega_s^2} \frac{1}{1 + (\alpha \omega/v)^2 }
 \,.
\end{align}
At $\omega=\omega_s$ and in the weak-coupling limit, $\alpha \omega_s/v \to 0$,
the coupling-induced damping of the system becomes $\im \Sigma_E = -\Gamma_E^{\eff}$.
If the reference frame is chosen
as $\wref=\omega_s$, the effective damping rate $\Gamma_E^{\eff}$
equals the Markov damping rate $\Gamma_E$
according to \Eq{eq:def_GammaE}. 

From \Eq{eqapp:def_reSig} we obtain the real part of the self-energy,
\begin{align}
\label{eqapp:selfenergy_real_Cccoupled_preintegration}
  \frac{\re \hat{\Sigma}_E}{\omega_s}
  &= \frac{\Gamma_E^\eff v^2}{\pi \omega_s^2} \re \int_0^\infty \diff k \frac{k}{(\omega -vk + \ui \epsilon)(1+\alpha^2 k^2)}
  = -\frac{\Gamma_E^\eff \omega}{\pi \omega_s^2} \int_0^\infty \diff x \frac{x (x-1)}{([x-1]^2 + \epsilon^2)(1+\alpha_\omega^2 x^2)}
\end{align}
Using a keyhole contour for the integration over the semi-infinite axis,
the integral is found to be
\begin{equation}
 \int_0^\infty \diff x \frac{x}{(x-1)(1+\alpha_\omega^2 x^2)}
 = \frac{1}{2 \alpha_\omega} \frac{\pi - 2 \alpha_\omega \ln(\alpha_\omega)}{1 + \alpha_\omega^2 }
\end{equation}
and thus
\begin{align}\label{eqapp:selfenergy_real_Cccoupled}
  \frac{\re \hat{\Sigma}_E}{\omega_s}
  &=  -\frac{\Gamma_E^\eff v}{2 \pi \omega_s^2 \alpha}
  \frac{\pi - 2 (\alpha \omega/v) \ln(\alpha \omega/v)}{1 + (\alpha \omega/v)^2 }
  & &=- \frac{C_c}{4 C_s}
  \frac{1 - \frac{2 \alpha \omega}{\pi v} \ln(\alpha\omega/v)}{1 + (\alpha \omega/v)^2 }
\end{align}

\subsection{Inductive coupling}\label{sec:selfenergy_Lc}

The purely inductive coupling is a special case of the parallel-$\sLC$ coupling,
obtained in the limit $C_c, \alpha \to 0$.
In this limit,
$\omega_s$ and $Z_s$
from \Eqs{eqP:ws}, \eqref{eqP:Zs}
remain unchanged, while $\Gamma_E^{(C_c)} \to 0$
and $\Gamma_E^{\eff} \to \Gamma_E^{(L_c)}$, cf.~\Eqs{eqPq:GammaE_Cc}--\eqref{eqPq:def_GammaE_eff}.

The coupling coefficients of \Eq{eqapp:fksquared_parallelcoupled} then read
\begin{align}
 f_k^2 = 2\Gamma_E^\eff \frac{\omega_s}{2\pi k } \frac{1}{1 + 1/(\beta^2 k^2)}
\end{align}
They are bounded and thus divergence free,
assuming a maximum at $k_{\Sigma,0} = \beta^{-1}$.

According to \Eq{eqapp:def_imSig} one obtains the imaginary part of the self-energy,
\begin{align}\label{eqapp:selfenergy_imag_Lccoupled}
 \frac{\im \hat{\Sigma}_E}{\omega_s}
 = -\frac{\Gamma_E^\eff}{\omega} \frac{1}{1 + v^2/(\beta \omega)^2 }
 \,.
\end{align}
At $\omega=\omega_s$ and in the weak-coupling limit, $v/(\beta \omega_s) \to 0$,
the coupling-induced damping of the system becomes $\im \Sigma_E = -\Gamma_E^{\eff}$.
If the reference frame is chosen
as $\wref=\omega_s$, the effective damping rate $\Gamma_E^{\eff}$
equals the Markov damping rate $\Gamma_E$
according to \Eq{eq:def_GammaE}. 

From \Eq{eqapp:def_reSig} we obtain the real part of the self-energy,
\begin{align}
  \frac{\re \hat{\Sigma}_E}{\omega_s}
  &= \frac{\Gamma_E^\eff \beta^2}{\pi} \re \int_0^\infty \diff k \frac{k}{(\omega - vk + \ui \epsilon)(1+\beta^2 k^2)}
  = -\frac{\Gamma_E^\eff \beta^2 \omega}{\pi v^2} \int_0^\infty \diff x \frac{x (x-1)}{([x-1]^2+\epsilon^2)(1+\beta_\omega^2 x^2)}
\end{align}
The structure of the integral here is the same as in  \Eq{eqapp:selfenergy_real_Cccoupled_preintegration},
so we readily obtain
\begin{equation}
 \int_0^\infty \diff x \frac{x}{(x-1)(1+\beta_\omega^2 x^2)}
 = \frac{1}{2 \beta_\omega} \frac{\pi - 2 \beta_\omega \ln(\beta_\omega)}{1 + \beta_\omega^2 }
\end{equation}
and thus
\begin{align}\label{eqapp:selfenergy_real_Lccoupled}
  \frac{\re \hat{\Sigma}_E}{\omega_s}
  &=  -\frac{\Gamma_E^\eff \beta}{2\pi v}
  \frac{\pi - 2 (\beta \omega/v) \ln(\beta \omega/v)}{1 + (\beta \omega/v)^2 }
  =- \frac{L_s}{4(L_s + L_c)}
  \frac{1 - \frac{2 \beta \omega}{\pi v} \ln(\beta\omega/v)}{1 + (\beta \omega/v)^2 }
\end{align}

\FloatBarrier
\subsection{Series $C_c-L_c$ coupling}\label{sec:selfenergy_seriesCcLc}

In Sec.~\ref{sec:coef_selfenergy_seriesCcLc} we had derived the coupling coefficients
for the case of series-$\sLC$ coupling,
\begin{align}\label{eqapp:fksquared_seriescoupled}
 f_k^2 = \frac{1}{C_s Z_0} \frac{\omega_s}{2\pi k}
 \frac{1}{1 + (\beta k - 1/(\alpha k))^2}
 \,,
\end{align}
cf.~\Eqs{eqQq:def_fk} and \eqref{eqQ:TLmodeamplitudes}.
Here, $\alpha = \frac{C_c C_s}{(C_s+C_c) C_0}$,
$\beta = \frac{L_c}{L_0}$, and 
and $\omega_s$, $Z_s$ are given in \Eqs{eq:ws_seriescoupled}--\eqref{eq:Zs_seriescoupled}.

The coupling constants $f_k$ are bounded and thus divergence free,
as seen in $\im \hat{\Sigma}(\omega) \propto |f_{k=\omega/v}|^2$
shown in Fig.~\ref{fig:selfenergy_seriescoupled}.
Their maximum is located at
\begin{equation}\label{eqapp:k_maxfk_seriescoupled}
 k_{\Sigma,0}
= \frac{1}{\sqrt{6}\beta} \sqrt{  2\frac{\beta}{\alpha} - 1 + \sqrt{1 - 4\frac{\beta}{\alpha} + 16\frac{\beta^2}{\alpha^2}} }
\end{equation}

According to \Eq{eqapp:def_imSig} one obtains the imaginary part of the self-energy,
\begin{align}\label{eqapp:selfenergy_imag_seriescoupled}
 \frac{\im \hat{\Sigma}_E}{\omega_s}
 \:=\: -\frac{Z_s}{2 Z_0} \frac{\omega_s}{\omega} \frac{1}{1 + \left( \frac{\beta \omega}{v} - \frac{v}{\alpha \omega}\right)^2}
 \,.
\end{align}
For purely capacitive coupling, corresponding here to the limit $L_c, \beta \to 0$
and the approximation $Z_s C_s^2 \omega_s/(C_s+C_c) \approx Z_s C_s \omega_s = 1$,
\Eq{eqapp:selfenergy_imag_seriescoupled} becomes equal to \Eq{eqapp:selfenergy_imag_Cccoupled}.

A resonant enhancement of the damping rate $-\im \Sigma_E$
occurs near $\omega \approx \omega_{\Sigma}$, where
\begin{align}\label{eqapp:wres_selfenergy_imag_seriescoupled}
 \omega_{\Sigma}
 &= \frac{v}{\sqrt{\alpha \beta}} 
 = \frac{\sqrt{1+\frac{C_c}{C_s}}}{\sqrt{L_c C_c}} 
\,.
\end{align}
Note that $\omega_{\Sigma}$ is only an approximation for the
location of the resonance, valid in the limit that $\beta \gg \alpha$.
In general, the maximum of $\im \Sigma_E(\omega)$ may be somewhat shifted.
The exact resonance location is determined by the condition
$\diff(\im \Sigma_E)/\diff(\omega) = 0$.
Owing to the relation between $\im \hat{\Sigma}_E(\omega)$ and $f_k$,
\Eq{eqapp:def_imSig},
it is given by the maximum position $k_{\Sigma,0}$ of $|f_k|$,
\Eq{eqapp:k_maxfk_seriescoupled},
\begin{equation}\label{eqapp:wres_exact_selfenergy_imag_seriescoupled}
 \omega_{\Sigma,0} = v k_{\Sigma,0} \leq \omega_{\Sigma} \,.
\end{equation}
In the limit $\beta/\alpha \gg 1$ this reduces to $\omega_{\Sigma}$
and in the limit $\beta/\alpha \ll 1$ this reduces to $0$.

From \Eq{eqapp:def_reSig} we obtain the real part of the self-energy,
\begin{align}
  \frac{\re \hat{\Sigma}_E}{\omega_s}
  &= -\re \frac{1}{2\pi} \int_0^\infty \diff k \frac{ 1/(C_s Z_0 k)
  }{(vk -\omega - \ui \epsilon)(1+\left[\beta k -1/(\alpha k)\right]^2)} \\
\label{eqapp:selfenergy_real_seriescoupled_preintegration}
  &=  -\frac{1}{2\pi} \int_0^\infty \diff x \frac{
  x (x-1) g^2
  }{([x -1]^2 + \epsilon^2)(x^2 + [\beta_\omega x^2 - 1/\alpha_\omega ]^2)} 
\end{align}
with the parameter
\begin{align}
g &:= \frac{1}{\sqrt{C_s Z_0 \omega_s}} \sqrt{\frac{\omega_s}{\omega}}
\,.
\end{align}
The integral in \Eq{eqapp:selfenergy_real_seriescoupled_preintegration}
is a special form of that in \Eq{eqapp:selfenergy_real_parallelcoupled_preintegration},
here with $g_C=0$ and $g_L \to g$, 
and with swapped roles of the coupling parameters $\alpha \leftrightarrow \beta$.
We can therefore take over from the solution,
\Eqs{eqapp:reSig1_parallelcoupled}--\eqref{eqapp:reSig2_parallelcoupled},
and similarly the roots of the polynomial in the denominator,
$p(x) = x^2 + [\beta_\omega x^2 - 1/\alpha_\omega]^2$ are obtained from
\Eq{eqapp:roots_denom_selfenergy_real_parallelcoupled}.
For $4\beta/\alpha < 1$ the solution takes the form
\begin{align}\label{eqapp:reSig1_seriescoupled}
\frac{\re \hat{\Sigma}_E}{\omega_s}
&=
-\frac{g^2 v^2}{4 \beta^2 \omega^2} \frac{ - (1- |x_1||x_3|) }{(|x_3| + |x_1|)(1 + |x_1|^2)(1 + |x_3|^2)} \\
& - \frac{g^2 v^2}{2\pi \beta^2 \omega^2}
 \frac{(1 + |x_3|^2) \ln(|x_1|) - (1 + |x_1|^2) \ln(|x_3|)
 }{(|x_3|^2 - |x_1|^2)(1 + |x_1|^2)(1 + |x_3|^2)} \nonumber \\
x_{1,2} &= \pm \frac{\ui v}{\sqrt{2} \beta \omega} \sqrt{\left(1- \frac{2\beta}{\alpha}\right) - \sqrt{1 - \frac{4\beta}{\alpha}} } \nonumber \\
x_{3,4} &= \pm \frac{\ui v}{\sqrt{2} \beta \omega} \sqrt{\left(1- \frac{2\beta}{\alpha}\right) + \sqrt{1 - \frac{4\beta}{\alpha}} } \nonumber
\end{align}
and for $4\beta/\alpha > 1$ it takes the form
\begin{align}\label{eqapp:reSig2_seriescoupled}
\frac{\re \Sigma_E}{\omega_s}
  &= \frac{g^2 v^2}{4\pi \beta^2 \omega^2}
  \frac{ (\pi - 2\varphi_1) \left[ 1 - r^2 \cos(2\varphi_1) \right]
  - \pi r (r^2-1) \cos(\varphi_1)
  - 2 r^2 \ln(r) \sin(2\varphi_1) }{ r^2 \sin(2\varphi_1)
                [1 + r^4 - 2 r^2 \cos(2\varphi_1)]} \\
r &= \frac{v}{\sqrt{\alpha \beta} \omega} \nonumber\\
 \tan(2\varphi_1) &= \frac{\sqrt{\frac{4\beta}{\alpha} - 1}}{\frac{2\beta}{\alpha} - 1}
 \hspace*{1cm} 2\varphi_1 \in \left\{ \begin{array}{ll}
        \left[0, \frac{\pi}{2}\right) & \text{if}\left(\frac{4 \beta}{\alpha} > \frac{2 \beta}{\alpha} > 1 \right) \\
          \left[\frac{\pi}{2}, \pi\right) & \text{if}\left(\frac{4 \beta}{\alpha} > 1 > \frac{2\beta}{\alpha} \right)
      \end{array}\right. \nonumber
\end{align}

In the limit of purely capacitive coupling, $\beta \to 0$, only the case $4\beta/\alpha < 1$ is relevant, and the roots can be approximated 
with $|x_1| \to 1/\alpha_\omega$ 
and $|x_3| \to 1/\beta_\omega \to \infty$, respectively. 
From \Eq{eqapp:reSig1_seriescoupled} one then obtains
\begin{align*}
 \frac{\re \Sigma_E}{\omega_s} \to -\frac{C_c}{4(C_s + C_c)} \frac{1 - \frac{2 \alpha \omega}{\pi v}\ln(\alpha \omega/v)}{1 + (\alpha \omega/v)^2 }
\end{align*}
which is equivalent to \Eq{eqapp:selfenergy_real_Cccoupled}.

\subsubsection{Resonant approximation}
\label{sec:selfenergy_seriesLC_resonant}

In order to simplify the expression for the self-energy,
one may approximate the coupling coefficients, \Eq{eqapp:fksquared_seriescoupled},
near the (approximate) resonance position
$\omega_{\Sigma} = v k_{\Sigma}$,
\Eq{eqapp:wres_selfenergy_imag_seriescoupled}, as Lorentzians:
\begin{eqnarray}
 f_k^2
 \label{eqapp:fksquared_seriescoupled_v2}
 &=& \frac{\omega_s}{4\pi C_s Z_0} 
 \left( \frac{1}{k + \ui \beta (k^2 - k_{\Sigma}^2)} + \frac{1}{k - \ui \beta (k^2 - k_{\Sigma}^2)} \right) \\
 &\approx& \frac{\omega_s}{4\pi C_s Z_0}
 \left( \frac{1}{k_{\Sigma} + \ui 2\beta k_{\Sigma} (k - k_{\Sigma})} + \frac{1}{k_{\Sigma} - \ui 2 \beta (k - k_{\Sigma})} \right) \\
 &=& 
 -\frac{\ui \omega_s}{8\pi C_s Z_0 \beta k_{\Sigma}}
 \left( \frac{1}{k - k_{\Sigma} - \ui/(2\beta)} - \frac{1}{k - k_{\Sigma} +  \ui/(2 \beta)} \right)
\end{eqnarray}
In line with this resonance approximation, the integration boundaries in \Eq{eqapp:def_selfenergy}
can be extended,
\begin{eqnarray*}
\hat{\Sigma}_{E,\text{res}}
&\approx& \int_{-\infty}^\infty \diff k \frac{f_k^2}{\omega - \omega_k + \ui\epsilon} \\
&\approx& 
-\frac{\ui \omega_s}{8\pi C_s Z_0 \beta k_{\Sigma}}
\int_{-\infty}^\infty \diff k 
 \frac{1}{\omega - v k + \ui\epsilon}
 \left( \frac{1}{k - k_{\Sigma} - \ui/(2\beta)} - \frac{1}{k - k_{\Sigma} +  \ui/(2 \beta)} \right) \\
&=& 
\frac{\ui \omega_s}{8\pi C_s Z_0 \beta \omega_{\Sigma}}
\int_{-\infty}^\infty \diff k 
 \frac{1}{k - \omega/v - \ui\epsilon/v}
 \left( \frac{1}{k - k_{\Sigma} - \ui/(2\beta)} - \frac{1}{k - k_{\Sigma} +  \ui/(2 \beta)} \right)
\end{eqnarray*}
The integrals are equivalent to those in
\Eqs{eqapp:integral1_selfenergy_resapprox_parallelcoupled}--\eqref{eqapp:integral2_selfenergy_resapprox_parallelcoupled}
under the replacement $\alpha\to\beta$, and one obtains
\begin{equation}\label{eqapp:selfenergy_seriescoupled_resapprox}
\frac{\hat{\Sigma}_{E,\text{res}}}{\omega_s}
= \frac{1}{4 C_s L_c \omega_{\Sigma}}
 \cdot \frac{1}{\omega - \omega_{\Sigma} + \ui Z_0/(2L_c)}
 {\color{black}
 =\frac{-\ui}{2C_{s}\omega_{\Sigma}}\cdot\frac{1}{-\ui\left(\omega-\omega_{\Sigma}\right)2L_{c}+Z_{0}}
 }
\end{equation} 
As mentioned before, $\omega_{\Sigma}$ is only an approximation for the resonance position,
valid for $\beta \gg \alpha$.
Since \Eq{eqapp:selfenergy_seriescoupled_resapprox}
is based on an expansion around $\omega_{\Sigma}$,
this resonant approximation is only good under the same condition.
An alternative resonant approximation is derived in the following section. \textcolor{black}{It is worth mentioning that, from \Eq{eqapp:selfenergy_seriescoupled_resapprox}, one obtains the approximation $\hat{\Sigma}_{E,\text{res}} \approx -\ui/(2C_{s} Z_{\text{in}})$, where $Z_{\text{in}}=-\ui\left(\omega-\omega_{\Sigma}\right)2L_{c}+Z_{0}$ is the environment impedance. This approximation corresponds to the self-energy formula used in Ref.~\onlinecite{RoyETAL2015}.}

\subsubsection{Resonant approximation 2}
\label{sec:selfenergy_seriesLC_resonantE}

As discussed above, the approximation $\omega_\Sigma$ in general deviates
from the exact maximum position $\omega_{\Sigma,0}$ of $-\im\hat{\Sigma}(\omega)$.
Here we derive a similar resonant approximation as in Sec.~\ref{sec:selfenergy_seriesLC_resonant},
using the Lorentzian expansion of the coupling coefficients
around $k_{\Sigma,0} = \omega_{\Sigma,0}/v$ instead of $k_\Sigma$.
Starting point is \Eq{eqapp:fksquared_seriescoupled_v2}, where we now
replace $k = k_{\Sigma,0} + (k-k_{\Sigma,0})$
and neglect contributions $\mathcal{O}((k-k_{\Sigma,0})^2)$.
\begin{align*}
 f_k^2 &\approx
 \frac{\omega_s}{4\pi C_s Z_0 (1 + \ui 2\beta k_{\Sigma,0})}
 \frac{1}{k - k_{\Sigma,0} + k_{\Sigma,0} a }
 + \frac{\omega_s}{4\pi C_s Z_0 (1 - \ui 2\beta k_{\Sigma,0})}
 \frac{1}{k - k_{\Sigma,0} + k_{\Sigma,0} a^\ast }
\end{align*}
with parameter
\begin{equation}
 a = \frac{1 - \ui \beta (k_\Sigma^2 - k_{\Sigma,0}^2)/k_{\Sigma,0}
 }{ 1 + \ui 2\beta k_{\Sigma,0}  }
\end{equation}
Thus,
\begin{align}
 \hat{\Sigma}_{E,\text{res},0}
 &\approx -\frac{1}{v} \int_{-\infty}^{\infty} \diff k \frac{f_k^2}{k-\omega/v -\ui \epsilon/v} \\
 &=
 -\frac{\omega_s}{4\pi C_s Z_0 v (1 + \ui 2\beta k_{\Sigma,0})}
 \int_{-\infty}^{\infty} \diff k \frac{1}{k-\omega/v -\ui \epsilon/v} \frac{1}{k-k_{\Sigma,0} + k_{\Sigma,0} a} \\
 &\quad-\frac{\omega_s}{4\pi C_s Z_0 v (1 - \ui 2\beta k_{\Sigma,0})}
 \int_{-\infty}^{\infty} \diff k \frac{1}{k-\omega/v -\ui \epsilon/v} \frac{1}{k-k_{\Sigma,0} + k_{\Sigma,0} a^\ast} \nonumber
\end{align}
The integrals are exactly the same as in \Eq{eqapp:selfenergy_parallelcoupled_resapproxE_pre}
(for the parallel-$\sLC$ coupling)
and due to the relation $k_\Sigma \geq k_{\Sigma,0}$, cf.~\Eq{eqapp:wres_exact_selfenergy_imag_seriescoupled},
the parameter $a$ again always lies in the lower half of the complex plane.
We can thus directly take over integrations from \Eqs{eqapp:integral1_selfenergy_resapproxE_parallelcoupled}--\eqref{eqapp:integral2_selfenergy_resapproxE_parallelcoupled}
and obtain
\begin{equation}\label{eqapp:selfenergy_seriescoupled_resapprox_exact}
 \frac{\hat{\Sigma}_{E,\text{res},0}}{\omega_s} =
-\frac{\ui Z_s}{2 Z_0 (1 - \ui 2\beta k_{\Sigma,0})}
\frac{\omega_s}{\omega - \omega_{\Sigma,0} + \omega_{\Sigma,0} a^\ast}
\,.
\end{equation}

\subsection{Simple ladder filter: series-$C_g-L_g$ plus parallel-$C_c-L_c$ circuit}\label{sec:selfenergy_parallelCcLc_beforeseriesLC}

In App.~\ref{app:parallelLC_beforeseriesLC_TLmodes} we had derived the coupling coefficients $f_k$
for the case of the simple ladder filter, consisting of a series circuit (with parameters $C_g, L_g$),
followed by a parallel circuit (with parameters $C_c, L_c$),
as seen from the system side,
\begin{align}\label{eqapp:fksquared_parallelbeforeseriescoupled}
 f_k^2 = \frac{v Z_0}{2\pi Z_a}
 \frac{v k}{ 1 + \left[\alpha k - 1/(\beta k) \right]^2 }
\end{align}
cf.~\Eq{eqFq:def_fk}.
Here $\alpha = \frac{C_c}{C_0}$,
$\beta = \frac{L_c}{L_0}$,
and where $\omega_s$, $Z_s$, $\omega_a$, $Z_a$, and $\Gamma_E^{\eff}$
are defined in \Eqs{eqF:ws}--\eqref{eqF:Za} and \Eq{eqFq:def_GammaE_eff}.

The coupling constants $|f_k|$ are bounded and thus divergence free,
as seen in $\im \hat{\Sigma}(\omega) \propto |f_{k=\omega/v}|^2$
shown in Fig.~\ref{fig:selfenergy_parallelbeforeseriescoupled}.
Their maximum is located at
\begin{equation}\label{eqapp:k_maxfk_parallelbeforeseriescoupled}
 k_{\Sigma,0}
= \frac{1}{\sqrt{2}\alpha} \sqrt{ 1 - 2\frac{\alpha}{\beta} + \sqrt{1 - 4\frac{\alpha}{\beta} + 16\frac{\alpha^2}{\beta^2}} }
\,.
\end{equation}

According to \Eq{eqapp:def_imSig} one obtains the imaginary part of the self-energy,
\begin{align}\label{eqapp:selfenergy_imag_parallelbeforeseriescoupled}
 \frac{\im \hat{\Sigma}_E}{\omega_s}
 \:=\: -\frac{Z_0}{2 Z_a} \frac{\omega}{\omega_s} \frac{1}{1 + \left( \frac{\alpha \omega}{v} - \frac{v}{\beta \omega}\right)^2}
 \,.
\end{align}

A resonant enhancement of the damping rate $-\im \hat{\Sigma}_E(\omega)$
is expected near $\omega \approx \omega_{\Sigma}$, with
\begin{align}\label{eqapp:wres_selfenergy_imag_parallelbeforeseriescoupled}
 \omega_{\Sigma}
 &= \frac{v}{\sqrt{\alpha \beta}}
 = \frac{1}{\sqrt{L_c C_c}} = \omega_c
 \,.
\end{align}
In contrast to the bare parallel or bare series-$\sLC$ coupling, $\omega_\Sigma$
coincides here with the bare coupling frequency $\omega_c$.
Also in contrast to the bare parallel-$\sLC$ coupling, $\im \hat{\Sigma}_E(\omega)$
does not turn zero at any finite $\omega$-value.
Note that $\omega_{\Sigma}$ is only an approximation for the
location of the resonance, while in general
the maximum of $-\im \hat{\Sigma}_E(\omega)$ is somewhat shifted.
The exact resonance location is determined by the condition
$\diff(\im \hat{\Sigma}_E)/\diff(\omega) = 0$.
Owing to the relation between $\im \hat{\Sigma}_E$ and $f_k$,
\Eq{eqapp:def_imSig}, it is given by the maximum position $k_{\Sigma,0}$ of $|f_k|$,
\Eq{eqapp:k_maxfk_parallelbeforeseriescoupled},
\begin{equation}\label{eqapp:wres_exact_selfenergy_imag_parallelbeforeseriescoupled}
 \omega_{\Sigma,0} = v k_{\Sigma,0}
\end{equation}
In the limit $\alpha/\beta \gg 1$ this reduces to $\omega_\Sigma$ and in the limit
$\alpha/\beta \ll 1$ this behaves as $v/\alpha$.

From \Eq{eqapp:def_reSig} we obtain the real part of the self-energy,
\begin{align}
  \frac{\re \hat{\Sigma}_E}{\omega_s}
  &= -\re \frac{1}{2\pi} \int_0^\infty \diff k \frac{ v^2 Z_0 k /(Z_a \omega_s)
  }{(vk -\omega - \ui \epsilon)(1+\left[\alpha k -1/(\beta k)\right]^2)} \\
\label{eqapp:selfenergy_real_parallelbeforeseriescoupled_preintegration}
  &=  -\frac{1}{2\pi} \int_0^\infty \diff x \frac{
  x^3 (x-1) g^2
  }{([x-1]^2 + \epsilon^2)(x^2 + [\alpha_\omega x^2 - 1/\beta_\omega ]^2)} 
\end{align}
with the parameter
\begin{equation}
 g := \sqrt{\frac{\omega Z_0}{\omega_s Z_a}}
 \,.
\end{equation}
The integral in \Eq{eqapp:selfenergy_real_parallelbeforeseriescoupled_preintegration}
is a special form of its counterpart for the purely parallel coupling,
\Eq{eqapp:selfenergy_real_parallelcoupled_preintegration},
here with $g_L=0$ and $g_C \to g$.
We can therefore take over from the solution,
\Eqs{eqapp:reSig1_parallelcoupled}--\eqref{eqapp:reSig2_parallelcoupled},
and obtain for $4\alpha/\beta < 1$
\begin{align}\label{eqapp:reSig1_parallelbeforeseriescoupled}
 \frac{\re \hat{\Sigma}_E}{\omega_s}
&= -\frac{g^2}{4 \alpha_\omega^2}
 \frac{|x_1|^2  + |x_3|^2 + |x_1||x_3| + |x_1|^2|x_3|^2
 }{(|x_3| + |x_1|)(1 + |x_1|^2)(1 + |x_3|^2)} \\
 &+ \frac{g^2}{2\pi \alpha_\omega^2}
 \frac{ |x_1|^2 (1 + |x_3|^2) \ln(|x_1|) - |x_3|^2 (1 + |x_1|^2) \ln(|x_3|)
 }{(|x_3|^2 - |x_1|^2)(1 + |x_1|^2)(1 + |x_3|^2)} \nonumber
\end{align}
and for $4\alpha/\beta > 1$
\begin{align}\label{eqapp:reSig2_parallelbeforeseriescoupled}
\frac{\re \hat{\Sigma}_E}{\omega_s}
  &= -\frac{g^2}{4\pi \alpha_\omega^2} \frac{%
   r^2 (\pi - 2\varphi_1) \left[r^2 - \cos(2\varphi_1)\right]
  + \pi r^5 \cos(\varphi_1) 
  - \pi r^3 \cos(3\varphi_1)
  + 2 r^2 \ln(r) \sin(2\varphi_1)
  }{ r^2 \sin(2\varphi_1)
                [1 + r^4 - 2 r^2 \cos(2\varphi_1)]}
\end{align}
Here, the roots of the polynomial in the denominator,
$p(x) = x^2 + [\alpha_\omega x^2 - 1/\beta_\omega]^2$,
are given by \Eq{eqapp:roots_denom_selfenergy_real_parallelcoupled}.

\subsubsection{Resonant approximation}
\label{sec:selfenergy_parallelLC_beforeseriesLC_resonant}

In order to simplify the expression for the self-energy,
one may approximate the coupling coefficients, \Eq{eqapp:fksquared_parallelbeforeseriescoupled},
near the (approximate) resonance position
$\omega_{\Sigma} = v k_{\Sigma}$,
\Eq{eqapp:wres_selfenergy_imag_parallelbeforeseriescoupled}, as Lorentzians:
\begin{eqnarray}
 f_k^2
 \label{eqapp:fksquared_parallelbeforeseriescoupled_v2}
 &=& \frac{v^2 k^2 Z_0}{4\pi Z_a}
 \left( \frac{1}{k + \ui \alpha (k^2 - k_{\Sigma}^2)} + \frac{1}{k - \ui \alpha (k^2 - k_{\Sigma}^2)} \right) \\
 &\approx& \frac{v^2 k_\Sigma^2 Z_0}{4\pi Z_a}
 \left( \frac{1}{k_{\Sigma} + \ui 2\alpha k_{\Sigma} (k - k_{\Sigma})} + \frac{1}{k_{\Sigma} - \ui 2 \alpha (k - k_{\Sigma})} \right) \\
 &=&
 -\frac{\ui v^2 k_\Sigma Z_0}{8\pi Z_a \alpha}
 \left( \frac{1}{k - k_{\Sigma} - \ui/(2\alpha)} - \frac{1}{k - k_{\Sigma} +  \ui/(2 \alpha)} \right)
\end{eqnarray}
In line with this resonance approximation, the integration boundaries in \Eq{eqapp:def_selfenergy}
can be extended,
\begin{eqnarray*}
\hat{\Sigma}_{E,\text{res}}
&\approx& \int_{-\infty}^\infty \diff k \frac{f_k^2}{\omega - \omega_k + \ui\epsilon} \\
&\approx&
-\frac{\ui v^2 k_\Sigma Z_0}{8\pi Z_a \alpha}
\int_{-\infty}^\infty \diff k
 \frac{1}{\omega - v k + \ui\epsilon}
 \left( \frac{1}{k - k_{\Sigma} - \ui/(2\alpha)} - \frac{1}{k - k_{\Sigma} +  \ui/(2 \alpha)} \right) \\
&=&
\frac{\ui v k_\Sigma Z_0}{8\pi Z_a \alpha}
\int_{-\infty}^\infty \diff k
 \frac{1}{k - \omega/v - \ui\epsilon/v}
 \left( \frac{1}{k - k_{\Sigma} - \ui/(2\alpha)} - \frac{1}{k - k_{\Sigma} +  \ui/(2 \alpha)} \right)
\end{eqnarray*}
The integrals are equal to those in
Sec.~\ref{sec:selfenergy_parallelcoupled_resapprox}, \Eqs{eqapp:integral1_selfenergy_resapprox_parallelcoupled}--\eqref{eqapp:integral2_selfenergy_resapprox_parallelcoupled}
and one obtains
\begin{equation}\label{eqapp:selfenergy_parallelbeforeseriescoupled_resapprox}
\frac{\hat{\Sigma}_{E,\text{res}}}{\omega_s}
= \frac{\omega_{\Sigma}}{4 C_c Z_a \omega_s}
 \cdot \frac{1}{\omega - \omega_{\Sigma} + \ui/(2 Z_0 C_c)}
\end{equation}
As mentioned before, $\omega_{\Sigma}$ is only an approximation for the resonance position,
valid for $\alpha \gg \beta$.
Since \Eq{eqapp:selfenergy_parallelbeforeseriescoupled_resapprox}
is based on an expansion around $\omega_{\Sigma}$,
this resonant approximation is only suitable under the same condition.


\end{appendix}
\twocolumngrid


\begin{thebibliography}{72}%
\makeatletter
\providecommand \@ifxundefined [1]{%
 \@ifx{#1\undefined}
}%
\providecommand \@ifnum [1]{%
 \ifnum #1\expandafter \@firstoftwo
 \else \expandafter \@secondoftwo
 \fi
}%
\providecommand \@ifx [1]{%
 \ifx #1\expandafter \@firstoftwo
 \else \expandafter \@secondoftwo
 \fi
}%
\providecommand \natexlab [1]{#1}%
\providecommand \enquote  [1]{``#1''}%
\providecommand \bibnamefont  [1]{#1}%
\providecommand \bibfnamefont [1]{#1}%
\providecommand \citenamefont [1]{#1}%
\providecommand \href@noop [0]{\@secondoftwo}%
\providecommand \href [0]{\begingroup \@sanitize@url \@href}%
\providecommand \@href[1]{\@@startlink{#1}\@@href}%
\providecommand \@@href[1]{\endgroup#1\@@endlink}%
\providecommand \@sanitize@url [0]{\catcode `\\12\catcode `\$12\catcode `\&12\catcode `\#12\catcode `\^12\catcode `\_12\catcode `\%12\relax}%
\providecommand \@@startlink[1]{}%
\providecommand \@@endlink[0]{}%
\providecommand \url  [0]{\begingroup\@sanitize@url \@url }%
\providecommand \@url [1]{\endgroup\@href {#1}{\urlprefix }}%
\providecommand \urlprefix  [0]{URL }%
\providecommand \Eprint [0]{\href }%
\providecommand \doibase [0]{https://doi.org/}%
\providecommand \selectlanguage [0]{\@gobble}%
\providecommand \bibinfo  [0]{\@secondoftwo}%
\providecommand \bibfield  [0]{\@secondoftwo}%
\providecommand \translation [1]{[#1]}%
\providecommand \BibitemOpen [0]{}%
\providecommand \bibitemStop [0]{}%
\providecommand \bibitemNoStop [0]{.\EOS\space}%
\providecommand \EOS [0]{\spacefactor3000\relax}%
\providecommand \BibitemShut  [1]{\csname bibitem#1\endcsname}%
\let\auto@bib@innerbib\@empty
\bibitem [{\citenamefont {Aumentado}(2020)}]{JPA_StateoftheArt}%
  \BibitemOpen
  \bibfield  {author} {\bibinfo {author} {\bibfnamefont {J.}~\bibnamefont {Aumentado}},\ }\bibfield  {title} {\bibinfo {title} {Superconducting parametric amplifiers: The state of the art in josephson parametric amplifiers},\ }\href {https://doi.org/10.1109/MMM.2020.2993476} {\bibfield  {journal} {\bibinfo  {journal} {IEEE Microw. Mag.}\ }\textbf {\bibinfo {volume} {21}},\ \bibinfo {pages} {45} (\bibinfo {year} {2020})}\BibitemShut {NoStop}%
\bibitem [{\citenamefont {Clerk}\ \emph {et~al.}(2010)\citenamefont {Clerk}, \citenamefont {Devoret}, \citenamefont {Girvin}, \citenamefont {Marquardt},\ and\ \citenamefont {Schoelkopf}}]{ClerkETAL_review2010}%
  \BibitemOpen
  \bibfield  {author} {\bibinfo {author} {\bibfnamefont {A.~A.}\ \bibnamefont {Clerk}}, \bibinfo {author} {\bibfnamefont {M.~H.}\ \bibnamefont {Devoret}}, \bibinfo {author} {\bibfnamefont {S.~M.}\ \bibnamefont {Girvin}}, \bibinfo {author} {\bibfnamefont {F.}~\bibnamefont {Marquardt}},\ and\ \bibinfo {author} {\bibfnamefont {R.~J.}\ \bibnamefont {Schoelkopf}},\ }\bibfield  {title} {\bibinfo {title} {Introduction to quantum noise, measurement, and amplification},\ }\href {https://doi.org/10.1103/RevModPhys.82.1155} {\bibfield  {journal} {\bibinfo  {journal} {Rev. Mod. Phys.}\ }\textbf {\bibinfo {volume} {82}},\ \bibinfo {pages} {1155} (\bibinfo {year} {2010})}\BibitemShut {NoStop}%
\bibitem [{\citenamefont {Krantz}\ \emph {et~al.}(2019)\citenamefont {Krantz}, \citenamefont {Kjaergaard}, \citenamefont {Yan}, \citenamefont {Orlando}, \citenamefont {Gustavsson},\ and\ \citenamefont {Oliver}}]{KrantzETAL2019_review}%
  \BibitemOpen
  \bibfield  {author} {\bibinfo {author} {\bibfnamefont {P.}~\bibnamefont {Krantz}}, \bibinfo {author} {\bibfnamefont {M.}~\bibnamefont {Kjaergaard}}, \bibinfo {author} {\bibfnamefont {F.}~\bibnamefont {Yan}}, \bibinfo {author} {\bibfnamefont {T.~P.}\ \bibnamefont {Orlando}}, \bibinfo {author} {\bibfnamefont {S.}~\bibnamefont {Gustavsson}},\ and\ \bibinfo {author} {\bibfnamefont {W.~D.}\ \bibnamefont {Oliver}},\ }\bibfield  {title} {\bibinfo {title} {A quantum engineer's guide to superconducting qubits},\ }\href {https://doi.org/10.1063/1.5089550} {\bibfield  {journal} {\bibinfo  {journal} {Appl. Phys. Rev.}\ }\textbf {\bibinfo {volume} {6}},\ \bibinfo {pages} {021318} (\bibinfo {year} {2019})}\BibitemShut {NoStop}%
\bibitem [{\citenamefont {Castellanos-Beltran}\ and\ \citenamefont {Lehnert}(2007)}]{LehnertETAL2007}%
  \BibitemOpen
  \bibfield  {author} {\bibinfo {author} {\bibfnamefont {M.~A.}\ \bibnamefont {Castellanos-Beltran}}\ and\ \bibinfo {author} {\bibfnamefont {K.~W.}\ \bibnamefont {Lehnert}},\ }\bibfield  {title} {\bibinfo {title} {Widely tunable parametric amplifier based on a superconducting quantum interference device array resonator},\ }\href {https://doi.org/10.1063/1.2773988} {\bibfield  {journal} {\bibinfo  {journal} {Appl. Phys. Lett.}\ }\textbf {\bibinfo {volume} {91}},\ \bibinfo {pages} {083509} (\bibinfo {year} {2007})}\BibitemShut {NoStop}%
\bibitem [{\citenamefont {Castellanos-Beltran}\ \emph {et~al.}(2008)\citenamefont {Castellanos-Beltran}, \citenamefont {Irwin}, \citenamefont {Hilton}, \citenamefont {Vale},\ and\ \citenamefont {Lehnert}}]{LehnertETAL2008}%
  \BibitemOpen
  \bibfield  {author} {\bibinfo {author} {\bibfnamefont {M.~A.}\ \bibnamefont {Castellanos-Beltran}}, \bibinfo {author} {\bibfnamefont {K.}~\bibnamefont {Irwin}}, \bibinfo {author} {\bibfnamefont {G.}~\bibnamefont {Hilton}}, \bibinfo {author} {\bibfnamefont {L.}~\bibnamefont {Vale}},\ and\ \bibinfo {author} {\bibfnamefont {K.}~\bibnamefont {Lehnert}},\ }\bibfield  {title} {\bibinfo {title} {Amplification and squeezing of quantum noise with a tunable josephson metamaterial},\ }\href@noop {} {\bibfield  {journal} {\bibinfo  {journal} {Nat. Phys.}\ }\textbf {\bibinfo {volume} {4}},\ \bibinfo {pages} {929} (\bibinfo {year} {2008})}\BibitemShut {NoStop}%
\bibitem [{\citenamefont {Yamamoto}\ \emph {et~al.}(2008)\citenamefont {Yamamoto}, \citenamefont {Inomata}, \citenamefont {Watanabe}, \citenamefont {Matsuba}, \citenamefont {Miyazaki}, \citenamefont {Oliver}, \citenamefont {Nakamura},\ and\ \citenamefont {Tsai}}]{YamamotoETAL2008}%
  \BibitemOpen
  \bibfield  {author} {\bibinfo {author} {\bibfnamefont {T.}~\bibnamefont {Yamamoto}}, \bibinfo {author} {\bibfnamefont {K.}~\bibnamefont {Inomata}}, \bibinfo {author} {\bibfnamefont {M.}~\bibnamefont {Watanabe}}, \bibinfo {author} {\bibfnamefont {K.}~\bibnamefont {Matsuba}}, \bibinfo {author} {\bibfnamefont {T.}~\bibnamefont {Miyazaki}}, \bibinfo {author} {\bibfnamefont {W.~D.}\ \bibnamefont {Oliver}}, \bibinfo {author} {\bibfnamefont {Y.}~\bibnamefont {Nakamura}},\ and\ \bibinfo {author} {\bibfnamefont {J.~S.}\ \bibnamefont {Tsai}},\ }\bibfield  {title} {\bibinfo {title} {Flux-driven josephson parametric amplifier},\ }\href {https://doi.org/10.1063/1.2964182} {\bibfield  {journal} {\bibinfo  {journal} {Appl. Phys. Lett.}\ }\textbf {\bibinfo {volume} {93}},\ \bibinfo {pages} {042510} (\bibinfo {year} {2008})}\BibitemShut {NoStop}%
\bibitem [{\citenamefont {Walter}\ \emph {et~al.}(2017)\citenamefont {Walter}, \citenamefont {Kurpiers}, \citenamefont {Gasparinetti}, \citenamefont {Magnard}, \citenamefont {Poto\ifmmode~\check{c}\else \v{c}\fi{}nik}, \citenamefont {Salath\'e}, \citenamefont {Pechal}, \citenamefont {Mondal}, \citenamefont {Oppliger}, \citenamefont {Eichler},\ and\ \citenamefont {Wallraff}}]{WalterETAL2017}%
  \BibitemOpen
  \bibfield  {author} {\bibinfo {author} {\bibfnamefont {T.}~\bibnamefont {Walter}}, \bibinfo {author} {\bibfnamefont {P.}~\bibnamefont {Kurpiers}}, \bibinfo {author} {\bibfnamefont {S.}~\bibnamefont {Gasparinetti}}, \bibinfo {author} {\bibfnamefont {P.}~\bibnamefont {Magnard}}, \bibinfo {author} {\bibfnamefont {A.}~\bibnamefont {Poto\ifmmode~\check{c}\else \v{c}\fi{}nik}}, \bibinfo {author} {\bibfnamefont {Y.}~\bibnamefont {Salath\'e}}, \bibinfo {author} {\bibfnamefont {M.}~\bibnamefont {Pechal}}, \bibinfo {author} {\bibfnamefont {M.}~\bibnamefont {Mondal}}, \bibinfo {author} {\bibfnamefont {M.}~\bibnamefont {Oppliger}}, \bibinfo {author} {\bibfnamefont {C.}~\bibnamefont {Eichler}},\ and\ \bibinfo {author} {\bibfnamefont {A.}~\bibnamefont {Wallraff}},\ }\bibfield  {title} {\bibinfo {title} {Rapid high-fidelity single-shot dispersive readout of superconducting qubits},\ }\href {https://doi.org/10.1103/PhysRevApplied.7.054020} {\bibfield  {journal} {\bibinfo  {journal} {Phys. Rev. Appl.}\ }\textbf {\bibinfo
  {volume} {7}},\ \bibinfo {pages} {054020} (\bibinfo {year} {2017})}\BibitemShut {NoStop}%
\bibitem [{\citenamefont {Sliwa}\ \emph {et~al.}(2015)\citenamefont {Sliwa}, \citenamefont {Hatridge}, \citenamefont {Narla}, \citenamefont {Shankar}, \citenamefont {Frunzio}, \citenamefont {Schoelkopf},\ and\ \citenamefont {Devoret}}]{SliwaETAL2015}%
  \BibitemOpen
  \bibfield  {author} {\bibinfo {author} {\bibfnamefont {K.~M.}\ \bibnamefont {Sliwa}}, \bibinfo {author} {\bibfnamefont {M.}~\bibnamefont {Hatridge}}, \bibinfo {author} {\bibfnamefont {A.}~\bibnamefont {Narla}}, \bibinfo {author} {\bibfnamefont {S.}~\bibnamefont {Shankar}}, \bibinfo {author} {\bibfnamefont {L.}~\bibnamefont {Frunzio}}, \bibinfo {author} {\bibfnamefont {R.~J.}\ \bibnamefont {Schoelkopf}},\ and\ \bibinfo {author} {\bibfnamefont {M.~H.}\ \bibnamefont {Devoret}},\ }\bibfield  {title} {\bibinfo {title} {Reconfigurable josephson circulator/directional amplifier},\ }\href {https://doi.org/10.1103/PhysRevX.5.041020} {\bibfield  {journal} {\bibinfo  {journal} {Phys. Rev. X}\ }\textbf {\bibinfo {volume} {5}},\ \bibinfo {pages} {041020} (\bibinfo {year} {2015})}\BibitemShut {NoStop}%
\bibitem [{\citenamefont {Chapman}\ \emph {et~al.}(2017{\natexlab{a}})\citenamefont {Chapman}, \citenamefont {Rosenthal}, \citenamefont {Kerckhoff}, \citenamefont {Moores}, \citenamefont {Vale}, \citenamefont {Mates}, \citenamefont {Hilton}, \citenamefont {Lalumi\`ere}, \citenamefont {Blais},\ and\ \citenamefont {Lehnert}}]{ChapmanETAL2017b}%
  \BibitemOpen
  \bibfield  {author} {\bibinfo {author} {\bibfnamefont {B.~J.}\ \bibnamefont {Chapman}}, \bibinfo {author} {\bibfnamefont {E.~I.}\ \bibnamefont {Rosenthal}}, \bibinfo {author} {\bibfnamefont {J.}~\bibnamefont {Kerckhoff}}, \bibinfo {author} {\bibfnamefont {B.~A.}\ \bibnamefont {Moores}}, \bibinfo {author} {\bibfnamefont {L.~R.}\ \bibnamefont {Vale}}, \bibinfo {author} {\bibfnamefont {J.~A.~B.}\ \bibnamefont {Mates}}, \bibinfo {author} {\bibfnamefont {G.~C.}\ \bibnamefont {Hilton}}, \bibinfo {author} {\bibfnamefont {K.}~\bibnamefont {Lalumi\`ere}}, \bibinfo {author} {\bibfnamefont {A.}~\bibnamefont {Blais}},\ and\ \bibinfo {author} {\bibfnamefont {K.~W.}\ \bibnamefont {Lehnert}},\ }\bibfield  {title} {\bibinfo {title} {Widely tunable on-chip microwave circulator for superconducting quantum circuits},\ }\href {https://doi.org/10.1103/PhysRevX.7.041043} {\bibfield  {journal} {\bibinfo  {journal} {Phys. Rev. X}\ }\textbf {\bibinfo {volume} {7}},\ \bibinfo {pages} {041043} (\bibinfo {year}
  {2017}{\natexlab{a}})}\BibitemShut {NoStop}%
\bibitem [{\citenamefont {Lecocq}\ \emph {et~al.}(2017)\citenamefont {Lecocq}, \citenamefont {Ranzani}, \citenamefont {Peterson}, \citenamefont {Cicak}, \citenamefont {Simmonds}, \citenamefont {Teufel},\ and\ \citenamefont {Aumentado}}]{LecocqETAL2017}%
  \BibitemOpen
  \bibfield  {author} {\bibinfo {author} {\bibfnamefont {F.}~\bibnamefont {Lecocq}}, \bibinfo {author} {\bibfnamefont {L.}~\bibnamefont {Ranzani}}, \bibinfo {author} {\bibfnamefont {G.~A.}\ \bibnamefont {Peterson}}, \bibinfo {author} {\bibfnamefont {K.}~\bibnamefont {Cicak}}, \bibinfo {author} {\bibfnamefont {R.~W.}\ \bibnamefont {Simmonds}}, \bibinfo {author} {\bibfnamefont {J.~D.}\ \bibnamefont {Teufel}},\ and\ \bibinfo {author} {\bibfnamefont {J.}~\bibnamefont {Aumentado}},\ }\bibfield  {title} {\bibinfo {title} {Nonreciprocal microwave signal processing with a field-programmable josephson amplifier},\ }\href {https://doi.org/10.1103/PhysRevApplied.7.024028} {\bibfield  {journal} {\bibinfo  {journal} {Phys. Rev. Appl.}\ }\textbf {\bibinfo {volume} {7}},\ \bibinfo {pages} {024028} (\bibinfo {year} {2017})}\BibitemShut {NoStop}%
\bibitem [{\citenamefont {Lecocq}\ \emph {et~al.}(2021)\citenamefont {Lecocq}, \citenamefont {Ranzani}, \citenamefont {Peterson}, \citenamefont {Cicak}, \citenamefont {Jin}, \citenamefont {Simmonds}, \citenamefont {Teufel},\ and\ \citenamefont {Aumentado}}]{LecocqETAL2021}%
  \BibitemOpen
  \bibfield  {author} {\bibinfo {author} {\bibfnamefont {F.}~\bibnamefont {Lecocq}}, \bibinfo {author} {\bibfnamefont {L.}~\bibnamefont {Ranzani}}, \bibinfo {author} {\bibfnamefont {G.~A.}\ \bibnamefont {Peterson}}, \bibinfo {author} {\bibfnamefont {K.}~\bibnamefont {Cicak}}, \bibinfo {author} {\bibfnamefont {X.~Y.}\ \bibnamefont {Jin}}, \bibinfo {author} {\bibfnamefont {R.~W.}\ \bibnamefont {Simmonds}}, \bibinfo {author} {\bibfnamefont {J.~D.}\ \bibnamefont {Teufel}},\ and\ \bibinfo {author} {\bibfnamefont {J.}~\bibnamefont {Aumentado}},\ }\bibfield  {title} {\bibinfo {title} {Efficient qubit measurement with a nonreciprocal microwave amplifier},\ }\href {https://doi.org/10.1103/PhysRevLett.126.020502} {\bibfield  {journal} {\bibinfo  {journal} {Phys. Rev. Lett.}\ }\textbf {\bibinfo {volume} {126}},\ \bibinfo {pages} {020502} (\bibinfo {year} {2021})}\BibitemShut {NoStop}%
\bibitem [{\citenamefont {Abdo}\ \emph {et~al.}(2021)\citenamefont {Abdo}, \citenamefont {Jinka}, \citenamefont {Bronn}, \citenamefont {Olivadese},\ and\ \citenamefont {Brink}}]{AbdoETAL2021}%
  \BibitemOpen
  \bibfield  {author} {\bibinfo {author} {\bibfnamefont {B.}~\bibnamefont {Abdo}}, \bibinfo {author} {\bibfnamefont {O.}~\bibnamefont {Jinka}}, \bibinfo {author} {\bibfnamefont {N.~T.}\ \bibnamefont {Bronn}}, \bibinfo {author} {\bibfnamefont {S.}~\bibnamefont {Olivadese}},\ and\ \bibinfo {author} {\bibfnamefont {M.}~\bibnamefont {Brink}},\ }\bibfield  {title} {\bibinfo {title} {High-fidelity qubit readout using interferometric directional josephson devices},\ }\href {https://doi.org/10.1103/PRXQuantum.2.040360} {\bibfield  {journal} {\bibinfo  {journal} {PRX Quantum}\ }\textbf {\bibinfo {volume} {2}},\ \bibinfo {pages} {040360} (\bibinfo {year} {2021})}\BibitemShut {NoStop}%
\bibitem [{\citenamefont {Yurke}\ \emph {et~al.}(1988)\citenamefont {Yurke}, \citenamefont {Kaminsky}, \citenamefont {Miller}, \citenamefont {Whittaker}, \citenamefont {Smith}, \citenamefont {Silver},\ and\ \citenamefont {Simon}}]{YurkeETAL1988}%
  \BibitemOpen
  \bibfield  {author} {\bibinfo {author} {\bibfnamefont {B.}~\bibnamefont {Yurke}}, \bibinfo {author} {\bibfnamefont {P.~G.}\ \bibnamefont {Kaminsky}}, \bibinfo {author} {\bibfnamefont {R.~E.}\ \bibnamefont {Miller}}, \bibinfo {author} {\bibfnamefont {E.~A.}\ \bibnamefont {Whittaker}}, \bibinfo {author} {\bibfnamefont {A.~D.}\ \bibnamefont {Smith}}, \bibinfo {author} {\bibfnamefont {A.~H.}\ \bibnamefont {Silver}},\ and\ \bibinfo {author} {\bibfnamefont {R.~W.}\ \bibnamefont {Simon}},\ }\bibfield  {title} {\bibinfo {title} {Observation of 4.2-k equilibrium-noise squeezing via a josephson-parametric amplifier},\ }\href {https://doi.org/10.1103/PhysRevLett.60.764} {\bibfield  {journal} {\bibinfo  {journal} {Phys. Rev. Lett.}\ }\textbf {\bibinfo {volume} {60}},\ \bibinfo {pages} {764} (\bibinfo {year} {1988})}\BibitemShut {NoStop}%
\bibitem [{\citenamefont {Eichler}\ \emph {et~al.}(2011)\citenamefont {Eichler}, \citenamefont {Bozyigit}, \citenamefont {Lang}, \citenamefont {Baur}, \citenamefont {Steffen}, \citenamefont {Fink}, \citenamefont {Filipp},\ and\ \citenamefont {Wallraff}}]{EichlerETAL2011}%
  \BibitemOpen
  \bibfield  {author} {\bibinfo {author} {\bibfnamefont {C.}~\bibnamefont {Eichler}}, \bibinfo {author} {\bibfnamefont {D.}~\bibnamefont {Bozyigit}}, \bibinfo {author} {\bibfnamefont {C.}~\bibnamefont {Lang}}, \bibinfo {author} {\bibfnamefont {M.}~\bibnamefont {Baur}}, \bibinfo {author} {\bibfnamefont {L.}~\bibnamefont {Steffen}}, \bibinfo {author} {\bibfnamefont {J.~M.}\ \bibnamefont {Fink}}, \bibinfo {author} {\bibfnamefont {S.}~\bibnamefont {Filipp}},\ and\ \bibinfo {author} {\bibfnamefont {A.}~\bibnamefont {Wallraff}},\ }\bibfield  {title} {\bibinfo {title} {Observation of two-mode squeezing in the microwave frequency domain},\ }\href {https://doi.org/10.1103/PhysRevLett.107.113601} {\bibfield  {journal} {\bibinfo  {journal} {Phys. Rev. Lett.}\ }\textbf {\bibinfo {volume} {107}},\ \bibinfo {pages} {113601} (\bibinfo {year} {2011})}\BibitemShut {NoStop}%
\bibitem [{\citenamefont {Qiu}\ \emph {et~al.}(2023)\citenamefont {Qiu}, \citenamefont {Grimsmo}, \citenamefont {Peng}, \citenamefont {Kannan}, \citenamefont {Lienhard}, \citenamefont {Sung}, \citenamefont {Krantz}, \citenamefont {Bolkhovsky}, \citenamefont {Calusine}, \citenamefont {Kim}, \citenamefont {Melville}, \citenamefont {Niedzielski}, \citenamefont {Yoder}, \citenamefont {Schwartz}, \citenamefont {Orlando}, \citenamefont {Siddiqi}, \citenamefont {Gustavsson}, \citenamefont {O’Brien},\ and\ \citenamefont {Oliver}}]{OBrienETAL2023}%
  \BibitemOpen
  \bibfield  {author} {\bibinfo {author} {\bibfnamefont {J.~Y.}\ \bibnamefont {Qiu}}, \bibinfo {author} {\bibfnamefont {A.}~\bibnamefont {Grimsmo}}, \bibinfo {author} {\bibfnamefont {K.}~\bibnamefont {Peng}}, \bibinfo {author} {\bibfnamefont {B.}~\bibnamefont {Kannan}}, \bibinfo {author} {\bibfnamefont {B.}~\bibnamefont {Lienhard}}, \bibinfo {author} {\bibfnamefont {Y.}~\bibnamefont {Sung}}, \bibinfo {author} {\bibfnamefont {P.}~\bibnamefont {Krantz}}, \bibinfo {author} {\bibfnamefont {V.}~\bibnamefont {Bolkhovsky}}, \bibinfo {author} {\bibfnamefont {G.}~\bibnamefont {Calusine}}, \bibinfo {author} {\bibfnamefont {D.}~\bibnamefont {Kim}}, \bibinfo {author} {\bibfnamefont {A.}~\bibnamefont {Melville}}, \bibinfo {author} {\bibfnamefont {B.~M.}\ \bibnamefont {Niedzielski}}, \bibinfo {author} {\bibfnamefont {J.}~\bibnamefont {Yoder}}, \bibinfo {author} {\bibfnamefont {M.~E.}\ \bibnamefont {Schwartz}}, \bibinfo {author} {\bibfnamefont {T.~P.}\ \bibnamefont {Orlando}}, \bibinfo {author} {\bibfnamefont
  {I.}~\bibnamefont {Siddiqi}}, \bibinfo {author} {\bibfnamefont {S.}~\bibnamefont {Gustavsson}}, \bibinfo {author} {\bibfnamefont {K.~P.}\ \bibnamefont {O’Brien}},\ and\ \bibinfo {author} {\bibfnamefont {W.~D.}\ \bibnamefont {Oliver}},\ }\bibfield  {title} {\bibinfo {title} {Broadband squeezed microwaves and amplification with a josephson travelling-wave parametric amplifier},\ }\href {https://doi.org/10.1038/s41567-022-01929-w} {\bibfield  {journal} {\bibinfo  {journal} {Nat. Phys.}\ }\textbf {\bibinfo {volume} {19}},\ \bibinfo {pages} {706} (\bibinfo {year} {2023})}\BibitemShut {NoStop}%
\bibitem [{\citenamefont {Bergeal}\ \emph {et~al.}(2010)\citenamefont {Bergeal}, \citenamefont {Schackert}, \citenamefont {Metcalfe}, \citenamefont {Vijay}, \citenamefont {Manucharyan}, \citenamefont {Frunzio}, \citenamefont {Prober}, \citenamefont {Schoelkopf}, \citenamefont {Girvin},\ and\ \citenamefont {Devoret}}]{BergealETAL2010b}%
  \BibitemOpen
  \bibfield  {author} {\bibinfo {author} {\bibfnamefont {N.}~\bibnamefont {Bergeal}}, \bibinfo {author} {\bibfnamefont {F.}~\bibnamefont {Schackert}}, \bibinfo {author} {\bibfnamefont {M.}~\bibnamefont {Metcalfe}}, \bibinfo {author} {\bibfnamefont {R.}~\bibnamefont {Vijay}}, \bibinfo {author} {\bibfnamefont {V.}~\bibnamefont {Manucharyan}}, \bibinfo {author} {\bibfnamefont {L.}~\bibnamefont {Frunzio}}, \bibinfo {author} {\bibfnamefont {D.}~\bibnamefont {Prober}}, \bibinfo {author} {\bibfnamefont {R.}~\bibnamefont {Schoelkopf}}, \bibinfo {author} {\bibfnamefont {S.}~\bibnamefont {Girvin}},\ and\ \bibinfo {author} {\bibfnamefont {M.}~\bibnamefont {Devoret}},\ }\bibfield  {title} {\bibinfo {title} {Phase-preserving amplification near the quantum limit with a josephson ring modulator},\ }\href {https://doi.org/10.1038/nature09035} {\bibfield  {journal} {\bibinfo  {journal} {Nature}\ }\textbf {\bibinfo {volume} {465}},\ \bibinfo {pages} {64} (\bibinfo {year} {2010})}\BibitemShut {NoStop}%
\bibitem [{\citenamefont {Naaman}\ and\ \citenamefont {Aumentado}(2022)}]{NaamanAumentado2022}%
  \BibitemOpen
  \bibfield  {author} {\bibinfo {author} {\bibfnamefont {O.}~\bibnamefont {Naaman}}\ and\ \bibinfo {author} {\bibfnamefont {J.}~\bibnamefont {Aumentado}},\ }\bibfield  {title} {\bibinfo {title} {Synthesis of parametrically coupled networks},\ }\href {https://doi.org/10.1103/PRXQuantum.3.020201} {\bibfield  {journal} {\bibinfo  {journal} {PRX Quantum}\ }\textbf {\bibinfo {volume} {3}},\ \bibinfo {pages} {020201} (\bibinfo {year} {2022})}\BibitemShut {NoStop}%
\bibitem [{\citenamefont {Braunstein}\ and\ \citenamefont {van Loock}(2005)}]{BraunsteinETAL2005}%
  \BibitemOpen
  \bibfield  {author} {\bibinfo {author} {\bibfnamefont {S.~L.}\ \bibnamefont {Braunstein}}\ and\ \bibinfo {author} {\bibfnamefont {P.}~\bibnamefont {van Loock}},\ }\bibfield  {title} {\bibinfo {title} {Quantum information with continuous variables},\ }\href {https://doi.org/10.1103/RevModPhys.77.513} {\bibfield  {journal} {\bibinfo  {journal} {Rev. Mod. Phys.}\ }\textbf {\bibinfo {volume} {77}},\ \bibinfo {pages} {513} (\bibinfo {year} {2005})}\BibitemShut {NoStop}%
\bibitem [{\citenamefont {Flurin}\ \emph {et~al.}(2012)\citenamefont {Flurin}, \citenamefont {Roch}, \citenamefont {Mallet}, \citenamefont {Devoret},\ and\ \citenamefont {Huard}}]{FlurinETAL2012}%
  \BibitemOpen
  \bibfield  {author} {\bibinfo {author} {\bibfnamefont {E.}~\bibnamefont {Flurin}}, \bibinfo {author} {\bibfnamefont {N.}~\bibnamefont {Roch}}, \bibinfo {author} {\bibfnamefont {F.}~\bibnamefont {Mallet}}, \bibinfo {author} {\bibfnamefont {M.~H.}\ \bibnamefont {Devoret}},\ and\ \bibinfo {author} {\bibfnamefont {B.}~\bibnamefont {Huard}},\ }\bibfield  {title} {\bibinfo {title} {Generating entangled microwave radiation over two transmission lines},\ }\href {https://doi.org/10.1103/PhysRevLett.109.183901} {\bibfield  {journal} {\bibinfo  {journal} {Phys. Rev. Lett.}\ }\textbf {\bibinfo {volume} {109}},\ \bibinfo {pages} {183901} (\bibinfo {year} {2012})}\BibitemShut {NoStop}%
\bibitem [{\citenamefont {Petrovnin}\ \emph {et~al.}(2023)\citenamefont {Petrovnin}, \citenamefont {Perelshtein}, \citenamefont {Korkalainen}, \citenamefont {Vesterinen}, \citenamefont {Lilja}, \citenamefont {Paraoanu},\ and\ \citenamefont {Hakonen}}]{PetrovninETAL2022}%
  \BibitemOpen
  \bibfield  {author} {\bibinfo {author} {\bibfnamefont {K.~V.}\ \bibnamefont {Petrovnin}}, \bibinfo {author} {\bibfnamefont {M.~R.}\ \bibnamefont {Perelshtein}}, \bibinfo {author} {\bibfnamefont {T.}~\bibnamefont {Korkalainen}}, \bibinfo {author} {\bibfnamefont {V.}~\bibnamefont {Vesterinen}}, \bibinfo {author} {\bibfnamefont {I.}~\bibnamefont {Lilja}}, \bibinfo {author} {\bibfnamefont {G.~S.}\ \bibnamefont {Paraoanu}},\ and\ \bibinfo {author} {\bibfnamefont {P.~J.}\ \bibnamefont {Hakonen}},\ }\bibfield  {title} {\bibinfo {title} {Generation and structuring of multipartite entanglement in a josephson parametric system},\ }\href@noop {} {\bibfield  {journal} {\bibinfo  {journal} {Adv. Quantum Technol.}\ }\textbf {\bibinfo {volume} {6}},\ \bibinfo {pages} {2200031} (\bibinfo {year} {2023})}\BibitemShut {NoStop}%
\bibitem [{\citenamefont {Wustmann}\ and\ \citenamefont {Shumeiko}(2013)}]{WusShu2013}%
  \BibitemOpen
  \bibfield  {author} {\bibinfo {author} {\bibfnamefont {W.}~\bibnamefont {Wustmann}}\ and\ \bibinfo {author} {\bibfnamefont {V.}~\bibnamefont {Shumeiko}},\ }\bibfield  {title} {\bibinfo {title} {Parametric resonance in tunable superconducting cavities},\ }\href {https://doi.org/10.1103/PhysRevB.87.184501} {\bibfield  {journal} {\bibinfo  {journal} {Phys. Rev. B}\ }\textbf {\bibinfo {volume} {87}},\ \bibinfo {pages} {184501} (\bibinfo {year} {2013})}\BibitemShut {NoStop}%
\bibitem [{\citenamefont {Wustmann}\ and\ \citenamefont {Shumeiko}(2017)}]{WusShu2017}%
  \BibitemOpen
  \bibfield  {author} {\bibinfo {author} {\bibfnamefont {W.}~\bibnamefont {Wustmann}}\ and\ \bibinfo {author} {\bibfnamefont {V.}~\bibnamefont {Shumeiko}},\ }\bibfield  {title} {\bibinfo {title} {Nondegenerate parametric resonance in a tunable superconducting cavity},\ }\href {https://doi.org/10.1103/PhysRevApplied.8.024018} {\bibfield  {journal} {\bibinfo  {journal} {Phys. Rev. Appl.}\ }\textbf {\bibinfo {volume} {8}},\ \bibinfo {pages} {024018} (\bibinfo {year} {2017})}\BibitemShut {NoStop}%
\bibitem [{\citenamefont {Svensson}\ \emph {et~al.}(2017)\citenamefont {Svensson}, \citenamefont {Bengtsson}, \citenamefont {Krantz}, \citenamefont {Bylander}, \citenamefont {Shumeiko},\ and\ \citenamefont {Delsing}}]{SvenssonETAL2017}%
  \BibitemOpen
  \bibfield  {author} {\bibinfo {author} {\bibfnamefont {I.-M.}\ \bibnamefont {Svensson}}, \bibinfo {author} {\bibfnamefont {A.}~\bibnamefont {Bengtsson}}, \bibinfo {author} {\bibfnamefont {P.}~\bibnamefont {Krantz}}, \bibinfo {author} {\bibfnamefont {J.}~\bibnamefont {Bylander}}, \bibinfo {author} {\bibfnamefont {V.}~\bibnamefont {Shumeiko}},\ and\ \bibinfo {author} {\bibfnamefont {P.}~\bibnamefont {Delsing}},\ }\bibfield  {title} {\bibinfo {title} {Period-tripling subharmonic oscillations in a driven superconducting resonator},\ }\href {https://doi.org/10.1103/PhysRevB.96.174503} {\bibfield  {journal} {\bibinfo  {journal} {Phys. Rev. B}\ }\textbf {\bibinfo {volume} {96}},\ \bibinfo {pages} {174503} (\bibinfo {year} {2017})}\BibitemShut {NoStop}%
\bibitem [{\citenamefont {Svensson}\ \emph {et~al.}(2018)\citenamefont {Svensson}, \citenamefont {Bengtsson}, \citenamefont {Bylander}, \citenamefont {Shumeiko},\ and\ \citenamefont {Delsing}}]{SvenssonETAL2018}%
  \BibitemOpen
  \bibfield  {author} {\bibinfo {author} {\bibfnamefont {I.-M.}\ \bibnamefont {Svensson}}, \bibinfo {author} {\bibfnamefont {A.}~\bibnamefont {Bengtsson}}, \bibinfo {author} {\bibfnamefont {J.}~\bibnamefont {Bylander}}, \bibinfo {author} {\bibfnamefont {V.}~\bibnamefont {Shumeiko}},\ and\ \bibinfo {author} {\bibfnamefont {P.}~\bibnamefont {Delsing}},\ }\bibfield  {title} {\bibinfo {title} {Period multiplication in a parametrically driven superconducting resonator},\ }\href {https://doi.org/10.1063/1.5026974} {\bibfield  {journal} {\bibinfo  {journal} {Appl. Phys. Lett.}\ }\textbf {\bibinfo {volume} {113}},\ \bibinfo {pages} {022602} (\bibinfo {year} {2018})}\BibitemShut {NoStop}%
\bibitem [{\citenamefont {Lin}\ \emph {et~al.}(2014)\citenamefont {Lin}, \citenamefont {Inomata}, \citenamefont {Koshino}, \citenamefont {Oliver}, \citenamefont {Nakamura}, \citenamefont {Tsai},\ and\ \citenamefont {Yamamoto}}]{YamamotoETAL2014}%
  \BibitemOpen
  \bibfield  {author} {\bibinfo {author} {\bibfnamefont {Z.}~\bibnamefont {Lin}}, \bibinfo {author} {\bibfnamefont {K.}~\bibnamefont {Inomata}}, \bibinfo {author} {\bibfnamefont {K.}~\bibnamefont {Koshino}}, \bibinfo {author} {\bibfnamefont {W.}~\bibnamefont {Oliver}}, \bibinfo {author} {\bibfnamefont {Y.}~\bibnamefont {Nakamura}}, \bibinfo {author} {\bibfnamefont {J.-S.}\ \bibnamefont {Tsai}},\ and\ \bibinfo {author} {\bibfnamefont {T.}~\bibnamefont {Yamamoto}},\ }\bibfield  {title} {\bibinfo {title} {Josephson parametric phase-locked oscillator and its application to dispersive readout of superconducting qubits},\ }\href@noop {} {\bibfield  {journal} {\bibinfo  {journal} {Nat. Commun.}\ }\textbf {\bibinfo {volume} {5}},\ \bibinfo {pages} {4480} (\bibinfo {year} {2014})}\BibitemShut {NoStop}%
\bibitem [{\citenamefont {Rosenthal}\ \emph {et~al.}(2021)\citenamefont {Rosenthal}, \citenamefont {Schneider}, \citenamefont {Malnou}, \citenamefont {Zhao}, \citenamefont {Leditzky}, \citenamefont {Chapman}, \citenamefont {Wustmann}, \citenamefont {Ma}, \citenamefont {Palken}, \citenamefont {Zanner}, \citenamefont {Vale}, \citenamefont {Hilton}, \citenamefont {Gao}, \citenamefont {Smith}, \citenamefont {Kirchmair},\ and\ \citenamefont {Lehnert}}]{RosenthalETAL2021}%
  \BibitemOpen
  \bibfield  {author} {\bibinfo {author} {\bibfnamefont {E.~I.}\ \bibnamefont {Rosenthal}}, \bibinfo {author} {\bibfnamefont {C.~M.~F.}\ \bibnamefont {Schneider}}, \bibinfo {author} {\bibfnamefont {M.}~\bibnamefont {Malnou}}, \bibinfo {author} {\bibfnamefont {Z.}~\bibnamefont {Zhao}}, \bibinfo {author} {\bibfnamefont {F.}~\bibnamefont {Leditzky}}, \bibinfo {author} {\bibfnamefont {B.~J.}\ \bibnamefont {Chapman}}, \bibinfo {author} {\bibfnamefont {W.}~\bibnamefont {Wustmann}}, \bibinfo {author} {\bibfnamefont {X.}~\bibnamefont {Ma}}, \bibinfo {author} {\bibfnamefont {D.~A.}\ \bibnamefont {Palken}}, \bibinfo {author} {\bibfnamefont {M.~F.}\ \bibnamefont {Zanner}}, \bibinfo {author} {\bibfnamefont {L.~R.}\ \bibnamefont {Vale}}, \bibinfo {author} {\bibfnamefont {G.~C.}\ \bibnamefont {Hilton}}, \bibinfo {author} {\bibfnamefont {J.}~\bibnamefont {Gao}}, \bibinfo {author} {\bibfnamefont {G.}~\bibnamefont {Smith}}, \bibinfo {author} {\bibfnamefont {G.}~\bibnamefont {Kirchmair}},\ and\ \bibinfo {author} {\bibfnamefont
  {K.~W.}\ \bibnamefont {Lehnert}},\ }\bibfield  {title} {\bibinfo {title} {Efficient and low-backaction quantum measurement using a chip-scale detector},\ }\href {https://doi.org/10.1103/PhysRevLett.126.090503} {\bibfield  {journal} {\bibinfo  {journal} {Phys. Rev. Lett.}\ }\textbf {\bibinfo {volume} {126}},\ \bibinfo {pages} {090503} (\bibinfo {year} {2021})}\BibitemShut {NoStop}%
\bibitem [{\citenamefont {Siddiqi}\ \emph {et~al.}(2004)\citenamefont {Siddiqi}, \citenamefont {Vijay}, \citenamefont {Pierre}, \citenamefont {Wilson}, \citenamefont {Metcalfe}, \citenamefont {Rigetti}, \citenamefont {Frunzio},\ and\ \citenamefont {Devoret}}]{JBA}%
  \BibitemOpen
  \bibfield  {author} {\bibinfo {author} {\bibfnamefont {I.}~\bibnamefont {Siddiqi}}, \bibinfo {author} {\bibfnamefont {R.}~\bibnamefont {Vijay}}, \bibinfo {author} {\bibfnamefont {F.}~\bibnamefont {Pierre}}, \bibinfo {author} {\bibfnamefont {C.~M.}\ \bibnamefont {Wilson}}, \bibinfo {author} {\bibfnamefont {M.}~\bibnamefont {Metcalfe}}, \bibinfo {author} {\bibfnamefont {C.}~\bibnamefont {Rigetti}}, \bibinfo {author} {\bibfnamefont {L.}~\bibnamefont {Frunzio}},\ and\ \bibinfo {author} {\bibfnamefont {M.~H.}\ \bibnamefont {Devoret}},\ }\bibfield  {title} {\bibinfo {title} {Rf-driven josephson bifurcation amplifier for quantum measurement},\ }\href {https://doi.org/10.1103/PhysRevLett.93.207002} {\bibfield  {journal} {\bibinfo  {journal} {Phys. Rev. Lett.}\ }\textbf {\bibinfo {volume} {93}},\ \bibinfo {pages} {207002} (\bibinfo {year} {2004})}\BibitemShut {NoStop}%
\bibitem [{\citenamefont {Mallet}\ \emph {et~al.}(2009)\citenamefont {Mallet}, \citenamefont {Ong}, \citenamefont {Palacios-Laloy}, \citenamefont {Nguyen}, \citenamefont {Bertet}, \citenamefont {Vion},\ and\ \citenamefont {Esteve}}]{JBA2}%
  \BibitemOpen
  \bibfield  {author} {\bibinfo {author} {\bibfnamefont {F.}~\bibnamefont {Mallet}}, \bibinfo {author} {\bibfnamefont {F.~R.}\ \bibnamefont {Ong}}, \bibinfo {author} {\bibfnamefont {A.}~\bibnamefont {Palacios-Laloy}}, \bibinfo {author} {\bibfnamefont {F.}~\bibnamefont {Nguyen}}, \bibinfo {author} {\bibfnamefont {P.}~\bibnamefont {Bertet}}, \bibinfo {author} {\bibfnamefont {D.}~\bibnamefont {Vion}},\ and\ \bibinfo {author} {\bibfnamefont {D.}~\bibnamefont {Esteve}},\ }\bibfield  {title} {\bibinfo {title} {Single-shot qubit readout in circuit quantum electrodynamics},\ }\href {https://doi.org/10.1038/nphys1400} {\bibfield  {journal} {\bibinfo  {journal} {Nat. Phys.}\ }\textbf {\bibinfo {volume} {5}},\ \bibinfo {pages} {791} (\bibinfo {year} {2009})}\BibitemShut {NoStop}%
\bibitem [{\citenamefont {Krantz}\ \emph {et~al.}(2016)\citenamefont {Krantz}, \citenamefont {Bengtsson}, \citenamefont {Simoen}, \citenamefont {Gustavsson}, \citenamefont {Shumeiko}, \citenamefont {Oliver}, \citenamefont {Wilson}, \citenamefont {Delsing},\ and\ \citenamefont {Bylander}}]{KrantzETAL2016_NatComm}%
  \BibitemOpen
  \bibfield  {author} {\bibinfo {author} {\bibfnamefont {P.}~\bibnamefont {Krantz}}, \bibinfo {author} {\bibfnamefont {A.}~\bibnamefont {Bengtsson}}, \bibinfo {author} {\bibfnamefont {M.}~\bibnamefont {Simoen}}, \bibinfo {author} {\bibfnamefont {S.}~\bibnamefont {Gustavsson}}, \bibinfo {author} {\bibfnamefont {V.}~\bibnamefont {Shumeiko}}, \bibinfo {author} {\bibfnamefont {W.}~\bibnamefont {Oliver}}, \bibinfo {author} {\bibfnamefont {C.}~\bibnamefont {Wilson}}, \bibinfo {author} {\bibfnamefont {P.}~\bibnamefont {Delsing}},\ and\ \bibinfo {author} {\bibfnamefont {J.}~\bibnamefont {Bylander}},\ }\bibfield  {title} {\bibinfo {title} {Single-shot read-out of a superconducting qubit using a josephson parametric oscillator},\ }\href {https://doi.org/10.1038/ncomms11417} {\bibfield  {journal} {\bibinfo  {journal} {Nat. Commun.}\ }\textbf {\bibinfo {volume} {7}},\ \bibinfo {pages} {11417} (\bibinfo {year} {2016})}\BibitemShut {NoStop}%
\bibitem [{\citenamefont {Grimm}\ \emph {et~al.}(2020)\citenamefont {Grimm}, \citenamefont {Frattini}, \citenamefont {Puri}, \citenamefont {Mundhada}, \citenamefont {Touzard}, \citenamefont {Mirrahimi}, \citenamefont {Girvin}, \citenamefont {Shankar},\ and\ \citenamefont {Devoret}}]{Kerrcatqubit}%
  \BibitemOpen
  \bibfield  {author} {\bibinfo {author} {\bibfnamefont {A.}~\bibnamefont {Grimm}}, \bibinfo {author} {\bibfnamefont {N.~E.}\ \bibnamefont {Frattini}}, \bibinfo {author} {\bibfnamefont {S.}~\bibnamefont {Puri}}, \bibinfo {author} {\bibfnamefont {S.~O.}\ \bibnamefont {Mundhada}}, \bibinfo {author} {\bibfnamefont {S.}~\bibnamefont {Touzard}}, \bibinfo {author} {\bibfnamefont {M.}~\bibnamefont {Mirrahimi}}, \bibinfo {author} {\bibfnamefont {S.~M.}\ \bibnamefont {Girvin}}, \bibinfo {author} {\bibfnamefont {S.}~\bibnamefont {Shankar}},\ and\ \bibinfo {author} {\bibfnamefont {M.~H.}\ \bibnamefont {Devoret}},\ }\bibfield  {title} {\bibinfo {title} {Stabilization and operation of a kerr-cat qubit},\ }\href@noop {} {\bibfield  {journal} {\bibinfo  {journal} {Nature}\ }\textbf {\bibinfo {volume} {584}},\ \bibinfo {pages} {205} (\bibinfo {year} {2020})}\BibitemShut {NoStop}%
\bibitem [{\citenamefont {He}\ \emph {et~al.}(2023)\citenamefont {He}, \citenamefont {Lu}, \citenamefont {Bao}, \citenamefont {Xue}, \citenamefont {Jiang}, \citenamefont {Wang}, \citenamefont {Roudsari}, \citenamefont {Delsing}, \citenamefont {Tsai},\ and\ \citenamefont {Lin}}]{Kerrcatqubit2}%
  \BibitemOpen
  \bibfield  {author} {\bibinfo {author} {\bibfnamefont {X.}~\bibnamefont {He}}, \bibinfo {author} {\bibfnamefont {Y.}~\bibnamefont {Lu}}, \bibinfo {author} {\bibfnamefont {D.}~\bibnamefont {Bao}}, \bibinfo {author} {\bibfnamefont {H.}~\bibnamefont {Xue}}, \bibinfo {author} {\bibfnamefont {W.}~\bibnamefont {Jiang}}, \bibinfo {author} {\bibfnamefont {Z.}~\bibnamefont {Wang}}, \bibinfo {author} {\bibfnamefont {A.}~\bibnamefont {Roudsari}}, \bibinfo {author} {\bibfnamefont {P.}~\bibnamefont {Delsing}}, \bibinfo {author} {\bibfnamefont {J.}~\bibnamefont {Tsai}},\ and\ \bibinfo {author} {\bibfnamefont {Z.}~\bibnamefont {Lin}},\ }\bibfield  {title} {\bibinfo {title} {Fast generation of schr{\"o}dinger cat states using a kerr-tunable superconducting resonator},\ }\href@noop {} {\bibfield  {journal} {\bibinfo  {journal} {Nat. Commun.}\ }\textbf {\bibinfo {volume} {14}},\ \bibinfo {pages} {6358} (\bibinfo {year} {2023})}\BibitemShut {NoStop}%
\bibitem [{\citenamefont {Puri}\ \emph {et~al.}(2017)\citenamefont {Puri}, \citenamefont {Andersen}, \citenamefont {Grimsmo},\ and\ \citenamefont {Blais}}]{PuriETAL2017_JPO}%
  \BibitemOpen
  \bibfield  {author} {\bibinfo {author} {\bibfnamefont {S.}~\bibnamefont {Puri}}, \bibinfo {author} {\bibfnamefont {C.~K.}\ \bibnamefont {Andersen}}, \bibinfo {author} {\bibfnamefont {A.~L.}\ \bibnamefont {Grimsmo}},\ and\ \bibinfo {author} {\bibfnamefont {A.}~\bibnamefont {Blais}},\ }\bibfield  {title} {\bibinfo {title} {Quantum annealing with all-to-all connected nonlinear oscillators},\ }\href@noop {} {\bibfield  {journal} {\bibinfo  {journal} {Nat. Commun.}\ }\textbf {\bibinfo {volume} {8}},\ \bibinfo {pages} {15785} (\bibinfo {year} {2017})}\BibitemShut {NoStop}%
\bibitem [{\citenamefont {Macklin}\ \emph {et~al.}(2015)\citenamefont {Macklin}, \citenamefont {O’brien}, \citenamefont {Hover}, \citenamefont {Schwartz}, \citenamefont {Bolkhovsky}, \citenamefont {Zhang}, \citenamefont {Oliver},\ and\ \citenamefont {Siddiqi}}]{MacklinETAL2015}%
  \BibitemOpen
  \bibfield  {author} {\bibinfo {author} {\bibfnamefont {C.}~\bibnamefont {Macklin}}, \bibinfo {author} {\bibfnamefont {K.}~\bibnamefont {O’brien}}, \bibinfo {author} {\bibfnamefont {D.}~\bibnamefont {Hover}}, \bibinfo {author} {\bibfnamefont {M.}~\bibnamefont {Schwartz}}, \bibinfo {author} {\bibfnamefont {V.}~\bibnamefont {Bolkhovsky}}, \bibinfo {author} {\bibfnamefont {X.}~\bibnamefont {Zhang}}, \bibinfo {author} {\bibfnamefont {W.}~\bibnamefont {Oliver}},\ and\ \bibinfo {author} {\bibfnamefont {I.}~\bibnamefont {Siddiqi}},\ }\bibfield  {title} {\bibinfo {title} {A near--quantum-limited josephson traveling-wave parametric amplifier},\ }\href {https://doi.org/10.1126/science.aaa8525} {\bibfield  {journal} {\bibinfo  {journal} {Science}\ }\textbf {\bibinfo {volume} {350}},\ \bibinfo {pages} {307} (\bibinfo {year} {2015})}\BibitemShut {NoStop}%
\bibitem [{\citenamefont {White}\ \emph {et~al.}(2015)\citenamefont {White}, \citenamefont {Mutus}, \citenamefont {Hoi}, \citenamefont {Barends}, \citenamefont {Campbell}, \citenamefont {Chen}, \citenamefont {Chen}, \citenamefont {Chiaro}, \citenamefont {Dunsworth}, \citenamefont {Jeffrey}, \citenamefont {Kelly}, \citenamefont {Megrant}, \citenamefont {Neill}, \citenamefont {O'Malley}, \citenamefont {Roushan}, \citenamefont {Sank}, \citenamefont {Vainsencher}, \citenamefont {Wenner}, \citenamefont {Chaudhuri}, \citenamefont {Gao},\ and\ \citenamefont {Martinis}}]{WhiteETAL2015}%
  \BibitemOpen
  \bibfield  {author} {\bibinfo {author} {\bibfnamefont {T.~C.}\ \bibnamefont {White}}, \bibinfo {author} {\bibfnamefont {J.~Y.}\ \bibnamefont {Mutus}}, \bibinfo {author} {\bibfnamefont {I.-C.}\ \bibnamefont {Hoi}}, \bibinfo {author} {\bibfnamefont {R.}~\bibnamefont {Barends}}, \bibinfo {author} {\bibfnamefont {B.}~\bibnamefont {Campbell}}, \bibinfo {author} {\bibfnamefont {Y.}~\bibnamefont {Chen}}, \bibinfo {author} {\bibfnamefont {Z.}~\bibnamefont {Chen}}, \bibinfo {author} {\bibfnamefont {B.}~\bibnamefont {Chiaro}}, \bibinfo {author} {\bibfnamefont {A.}~\bibnamefont {Dunsworth}}, \bibinfo {author} {\bibfnamefont {E.}~\bibnamefont {Jeffrey}}, \bibinfo {author} {\bibfnamefont {J.}~\bibnamefont {Kelly}}, \bibinfo {author} {\bibfnamefont {A.}~\bibnamefont {Megrant}}, \bibinfo {author} {\bibfnamefont {C.}~\bibnamefont {Neill}}, \bibinfo {author} {\bibfnamefont {P.~J.~J.}\ \bibnamefont {O'Malley}}, \bibinfo {author} {\bibfnamefont {P.}~\bibnamefont {Roushan}}, \bibinfo {author} {\bibfnamefont {D.}~\bibnamefont
  {Sank}}, \bibinfo {author} {\bibfnamefont {A.}~\bibnamefont {Vainsencher}}, \bibinfo {author} {\bibfnamefont {J.}~\bibnamefont {Wenner}}, \bibinfo {author} {\bibfnamefont {S.}~\bibnamefont {Chaudhuri}}, \bibinfo {author} {\bibfnamefont {J.}~\bibnamefont {Gao}},\ and\ \bibinfo {author} {\bibfnamefont {J.~M.}\ \bibnamefont {Martinis}},\ }\bibfield  {title} {\bibinfo {title} {Traveling wave parametric amplifier with josephson junctions using minimal resonator phase matching},\ }\href {https://doi.org/10.1063/1.4922348} {\bibfield  {journal} {\bibinfo  {journal} {Appl. Phys. Lett.}\ }\textbf {\bibinfo {volume} {106}},\ \bibinfo {pages} {242601} (\bibinfo {year} {2015})}\BibitemShut {NoStop}%
\bibitem [{\citenamefont {Chen}\ \emph {et~al.}(2012)\citenamefont {Chen}, \citenamefont {Sank}, \citenamefont {O'Malley}, \citenamefont {White}, \citenamefont {Barends}, \citenamefont {Chiaro}, \citenamefont {Kelly}, \citenamefont {Lucero}, \citenamefont {Mariantoni}, \citenamefont {Megrant}, \citenamefont {Neill}, \citenamefont {Vainsencher}, \citenamefont {Wenner}, \citenamefont {Yin}, \citenamefont {Cleland},\ and\ \citenamefont {Martinis}}]{ChenETAL2012}%
  \BibitemOpen
  \bibfield  {author} {\bibinfo {author} {\bibfnamefont {Y.}~\bibnamefont {Chen}}, \bibinfo {author} {\bibfnamefont {D.}~\bibnamefont {Sank}}, \bibinfo {author} {\bibfnamefont {P.}~\bibnamefont {O'Malley}}, \bibinfo {author} {\bibfnamefont {T.}~\bibnamefont {White}}, \bibinfo {author} {\bibfnamefont {R.}~\bibnamefont {Barends}}, \bibinfo {author} {\bibfnamefont {B.}~\bibnamefont {Chiaro}}, \bibinfo {author} {\bibfnamefont {J.}~\bibnamefont {Kelly}}, \bibinfo {author} {\bibfnamefont {E.}~\bibnamefont {Lucero}}, \bibinfo {author} {\bibfnamefont {M.}~\bibnamefont {Mariantoni}}, \bibinfo {author} {\bibfnamefont {A.}~\bibnamefont {Megrant}}, \bibinfo {author} {\bibfnamefont {C.}~\bibnamefont {Neill}}, \bibinfo {author} {\bibfnamefont {A.}~\bibnamefont {Vainsencher}}, \bibinfo {author} {\bibfnamefont {J.}~\bibnamefont {Wenner}}, \bibinfo {author} {\bibfnamefont {Y.}~\bibnamefont {Yin}}, \bibinfo {author} {\bibfnamefont {A.~N.}\ \bibnamefont {Cleland}},\ and\ \bibinfo {author} {\bibfnamefont {J.~M.}\ \bibnamefont
  {Martinis}},\ }\bibfield  {title} {\bibinfo {title} {Multiplexed dispersive readout of superconducting phase qubits},\ }\href {https://doi.org/10.1063/1.4764940} {\bibfield  {journal} {\bibinfo  {journal} {Appl. Phys. Lett.}\ }\textbf {\bibinfo {volume} {101}},\ \bibinfo {pages} {182601} (\bibinfo {year} {2012})}\BibitemShut {NoStop}%
\bibitem [{\citenamefont {Jerger}\ \emph {et~al.}(2012)\citenamefont {Jerger}, \citenamefont {Poletto}, \citenamefont {Macha}, \citenamefont {Hübner}, \citenamefont {Il’ichev},\ and\ \citenamefont {Ustinov}}]{JergerETAL2012}%
  \BibitemOpen
  \bibfield  {author} {\bibinfo {author} {\bibfnamefont {M.}~\bibnamefont {Jerger}}, \bibinfo {author} {\bibfnamefont {S.}~\bibnamefont {Poletto}}, \bibinfo {author} {\bibfnamefont {P.}~\bibnamefont {Macha}}, \bibinfo {author} {\bibfnamefont {U.}~\bibnamefont {Hübner}}, \bibinfo {author} {\bibfnamefont {E.}~\bibnamefont {Il’ichev}},\ and\ \bibinfo {author} {\bibfnamefont {A.~V.}\ \bibnamefont {Ustinov}},\ }\bibfield  {title} {\bibinfo {title} {Frequency division multiplexing readout and simultaneous manipulation of an array of flux qubits},\ }\href {https://doi.org/10.1063/1.4739454} {\bibfield  {journal} {\bibinfo  {journal} {Appl. Phys. Lett.}\ }\textbf {\bibinfo {volume} {101}},\ \bibinfo {pages} {042604} (\bibinfo {year} {2012})}\BibitemShut {NoStop}%
\bibitem [{\citenamefont {Chapman}\ \emph {et~al.}(2017{\natexlab{b}})\citenamefont {Chapman}, \citenamefont {Rosenthal}, \citenamefont {Kerckhoff}, \citenamefont {Vale}, \citenamefont {Hilton},\ and\ \citenamefont {Lehnert}}]{ChapmanETAL2017}%
  \BibitemOpen
  \bibfield  {author} {\bibinfo {author} {\bibfnamefont {B.~J.}\ \bibnamefont {Chapman}}, \bibinfo {author} {\bibfnamefont {E.~I.}\ \bibnamefont {Rosenthal}}, \bibinfo {author} {\bibfnamefont {J.}~\bibnamefont {Kerckhoff}}, \bibinfo {author} {\bibfnamefont {L.~R.}\ \bibnamefont {Vale}}, \bibinfo {author} {\bibfnamefont {G.~C.}\ \bibnamefont {Hilton}},\ and\ \bibinfo {author} {\bibfnamefont {K.~W.}\ \bibnamefont {Lehnert}},\ }\bibfield  {title} {\bibinfo {title} {Single-sideband modulator for frequency domain multiplexing of superconducting qubit readout},\ }\href {https://doi.org/10.1063/1.4981390} {\bibfield  {journal} {\bibinfo  {journal} {Appl. Phys. Lett.}\ }\textbf {\bibinfo {volume} {110}},\ \bibinfo {pages} {162601} (\bibinfo {year} {2017}{\natexlab{b}})}\BibitemShut {NoStop}%
\bibitem [{\citenamefont {Roy}\ \emph {et~al.}(2015)\citenamefont {Roy}, \citenamefont {Kundu}, \citenamefont {Chand}, \citenamefont {Vadiraj}, \citenamefont {Ranadive}, \citenamefont {Nehra}, \citenamefont {Patankar}, \citenamefont {Aumentado}, \citenamefont {Clerk},\ and\ \citenamefont {Vijay}}]{RoyETAL2015}%
  \BibitemOpen
  \bibfield  {author} {\bibinfo {author} {\bibfnamefont {T.}~\bibnamefont {Roy}}, \bibinfo {author} {\bibfnamefont {S.}~\bibnamefont {Kundu}}, \bibinfo {author} {\bibfnamefont {M.}~\bibnamefont {Chand}}, \bibinfo {author} {\bibfnamefont {A.~M.}\ \bibnamefont {Vadiraj}}, \bibinfo {author} {\bibfnamefont {A.}~\bibnamefont {Ranadive}}, \bibinfo {author} {\bibfnamefont {N.}~\bibnamefont {Nehra}}, \bibinfo {author} {\bibfnamefont {M.~P.}\ \bibnamefont {Patankar}}, \bibinfo {author} {\bibfnamefont {J.}~\bibnamefont {Aumentado}}, \bibinfo {author} {\bibfnamefont {A.~A.}\ \bibnamefont {Clerk}},\ and\ \bibinfo {author} {\bibfnamefont {R.}~\bibnamefont {Vijay}},\ }\bibfield  {title} {\bibinfo {title} {Broadband parametric amplification with impedance engineering: Beyond the gain-bandwidth product},\ }\href {https://doi.org/10.1063/1.4939148} {\bibfield  {journal} {\bibinfo  {journal} {Appl. Phys. Lett.}\ }\textbf {\bibinfo {volume} {107}},\ \bibinfo {pages} {262601} (\bibinfo {year} {2015})}\BibitemShut {NoStop}%
\bibitem [{\citenamefont {Mutus}\ \emph {et~al.}(2014)\citenamefont {Mutus}, \citenamefont {White}, \citenamefont {Barends}, \citenamefont {Chen}, \citenamefont {Chen}, \citenamefont {Chiaro}, \citenamefont {Dunsworth}, \citenamefont {Jeffrey}, \citenamefont {Kelly}, \citenamefont {Megrant}, \citenamefont {Neill}, \citenamefont {O'Malley}, \citenamefont {Roushan}, \citenamefont {Sank}, \citenamefont {Vainsencher}, \citenamefont {Wenner}, \citenamefont {Sundqvist}, \citenamefont {Cleland},\ and\ \citenamefont {Martinis}}]{MutusETAL2014}%
  \BibitemOpen
  \bibfield  {author} {\bibinfo {author} {\bibfnamefont {J.~Y.}\ \bibnamefont {Mutus}}, \bibinfo {author} {\bibfnamefont {T.~C.}\ \bibnamefont {White}}, \bibinfo {author} {\bibfnamefont {R.}~\bibnamefont {Barends}}, \bibinfo {author} {\bibfnamefont {Y.}~\bibnamefont {Chen}}, \bibinfo {author} {\bibfnamefont {Z.}~\bibnamefont {Chen}}, \bibinfo {author} {\bibfnamefont {B.}~\bibnamefont {Chiaro}}, \bibinfo {author} {\bibfnamefont {A.}~\bibnamefont {Dunsworth}}, \bibinfo {author} {\bibfnamefont {E.}~\bibnamefont {Jeffrey}}, \bibinfo {author} {\bibfnamefont {J.}~\bibnamefont {Kelly}}, \bibinfo {author} {\bibfnamefont {A.}~\bibnamefont {Megrant}}, \bibinfo {author} {\bibfnamefont {C.}~\bibnamefont {Neill}}, \bibinfo {author} {\bibfnamefont {P.~J.~J.}\ \bibnamefont {O'Malley}}, \bibinfo {author} {\bibfnamefont {P.}~\bibnamefont {Roushan}}, \bibinfo {author} {\bibfnamefont {D.}~\bibnamefont {Sank}}, \bibinfo {author} {\bibfnamefont {A.}~\bibnamefont {Vainsencher}}, \bibinfo {author} {\bibfnamefont {J.}~\bibnamefont
  {Wenner}}, \bibinfo {author} {\bibfnamefont {K.~M.}\ \bibnamefont {Sundqvist}}, \bibinfo {author} {\bibfnamefont {A.~N.}\ \bibnamefont {Cleland}},\ and\ \bibinfo {author} {\bibfnamefont {J.~M.}\ \bibnamefont {Martinis}},\ }\bibfield  {title} {\bibinfo {title} {Strong environmental coupling in a josephson parametric amplifier},\ }\href {https://doi.org/10.1063/1.4886408} {\bibfield  {journal} {\bibinfo  {journal} {Appl. Phys. Lett.}\ }\textbf {\bibinfo {volume} {104}},\ \bibinfo {pages} {263513} (\bibinfo {year} {2014})}\BibitemShut {NoStop}%
\bibitem [{\citenamefont {Grebel}\ \emph {et~al.}(2021)\citenamefont {Grebel}, \citenamefont {Bienfait}, \citenamefont {Dumur}, \citenamefont {Chang}, \citenamefont {Chou}, \citenamefont {Conner}, \citenamefont {Peairs}, \citenamefont {Povey}, \citenamefont {Zhong},\ and\ \citenamefont {Cleland}}]{GrebelETAL2021}%
  \BibitemOpen
  \bibfield  {author} {\bibinfo {author} {\bibfnamefont {J.}~\bibnamefont {Grebel}}, \bibinfo {author} {\bibfnamefont {A.}~\bibnamefont {Bienfait}}, \bibinfo {author} {\bibfnamefont {E.}~\bibnamefont {Dumur}}, \bibinfo {author} {\bibfnamefont {H.-S.}\ \bibnamefont {Chang}}, \bibinfo {author} {\bibfnamefont {M.-H.}\ \bibnamefont {Chou}}, \bibinfo {author} {\bibfnamefont {C.~R.}\ \bibnamefont {Conner}}, \bibinfo {author} {\bibfnamefont {G.~A.}\ \bibnamefont {Peairs}}, \bibinfo {author} {\bibfnamefont {R.~G.}\ \bibnamefont {Povey}}, \bibinfo {author} {\bibfnamefont {Y.~P.}\ \bibnamefont {Zhong}},\ and\ \bibinfo {author} {\bibfnamefont {A.~N.}\ \bibnamefont {Cleland}},\ }\bibfield  {title} {\bibinfo {title} {Flux-pumped impedance-engineered broadband josephson parametric amplifier},\ }\href {https://doi.org/10.1063/5.0035945} {\bibfield  {journal} {\bibinfo  {journal} {Appl. Phys. Lett.}\ }\textbf {\bibinfo {volume} {118}},\ \bibinfo {pages} {142601} (\bibinfo {year} {2021})}\BibitemShut {NoStop}%
\bibitem [{\citenamefont {Duan}\ \emph {et~al.}(2021)\citenamefont {Duan}, \citenamefont {Jia}, \citenamefont {Zhang}, \citenamefont {Du}, \citenamefont {Tao}, \citenamefont {Yang}, \citenamefont {Guo}, \citenamefont {Chen}, \citenamefont {Zhang}, \citenamefont {Peng}, \citenamefont {Kong}, \citenamefont {Li}, \citenamefont {Cao},\ and\ \citenamefont {Guo}}]{DuanETAL2021}%
  \BibitemOpen
  \bibfield  {author} {\bibinfo {author} {\bibfnamefont {P.}~\bibnamefont {Duan}}, \bibinfo {author} {\bibfnamefont {Z.}~\bibnamefont {Jia}}, \bibinfo {author} {\bibfnamefont {C.}~\bibnamefont {Zhang}}, \bibinfo {author} {\bibfnamefont {L.}~\bibnamefont {Du}}, \bibinfo {author} {\bibfnamefont {H.}~\bibnamefont {Tao}}, \bibinfo {author} {\bibfnamefont {X.}~\bibnamefont {Yang}}, \bibinfo {author} {\bibfnamefont {L.}~\bibnamefont {Guo}}, \bibinfo {author} {\bibfnamefont {Y.}~\bibnamefont {Chen}}, \bibinfo {author} {\bibfnamefont {H.}~\bibnamefont {Zhang}}, \bibinfo {author} {\bibfnamefont {Z.}~\bibnamefont {Peng}}, \bibinfo {author} {\bibfnamefont {W.}~\bibnamefont {Kong}}, \bibinfo {author} {\bibfnamefont {H.-O.}\ \bibnamefont {Li}}, \bibinfo {author} {\bibfnamefont {G.}~\bibnamefont {Cao}},\ and\ \bibinfo {author} {\bibfnamefont {G.-P.}\ \bibnamefont {Guo}},\ }\bibfield  {title} {\bibinfo {title} {Broadband flux-pumped josephson parametric amplifier with an on-chip coplanar waveguide impedance transformer},\
  }\href {https://doi.org/10.35848/1882-0786/abf029} {\bibfield  {journal} {\bibinfo  {journal} {Appl. Phys.Express}\ }\textbf {\bibinfo {volume} {14}},\ \bibinfo {pages} {042011} (\bibinfo {year} {2021})}\BibitemShut {NoStop}%
\bibitem [{\citenamefont {White}\ \emph {et~al.}(2023)\citenamefont {White}, \citenamefont {Opremcak}, \citenamefont {Sterling}, \citenamefont {Korotkov}, \citenamefont {Sank}, \citenamefont {Acharya}, \citenamefont {Ansmann}, \citenamefont {Arute}, \citenamefont {Arya}, \citenamefont {Bardin}, \citenamefont {Bengtsson}, \citenamefont {Bourassa}, \citenamefont {Bovaird}, \citenamefont {Brill}, \citenamefont {Buckley}, \citenamefont {Buell}, \citenamefont {Burger}, \citenamefont {Burkett}, \citenamefont {Bushnell}, \citenamefont {Chen}, \citenamefont {Chiaro}, \citenamefont {Cogan}, \citenamefont {Collins}, \citenamefont {Crook}, \citenamefont {Curtin}, \citenamefont {Demura}, \citenamefont {Dunsworth}, \citenamefont {Erickson}, \citenamefont {Fatemi}, \citenamefont {Burgos}, \citenamefont {Forati}, \citenamefont {Foxen}, \citenamefont {Giang}, \citenamefont {Giustina}, \citenamefont {Grajales~Dau}, \citenamefont {Hamilton}, \citenamefont {Harrington}, \citenamefont {Hilton}, \citenamefont {Hoffmann},
  \citenamefont {Hong}, \citenamefont {Huang}, \citenamefont {Huff}, \citenamefont {Iveland}, \citenamefont {Jeffrey}, \citenamefont {Kieferová}, \citenamefont {Kim}, \citenamefont {Klimov}, \citenamefont {Kostritsa}, \citenamefont {Kreikebaum}, \citenamefont {Landhuis}, \citenamefont {Laptev}, \citenamefont {Laws}, \citenamefont {Lee}, \citenamefont {Lester}, \citenamefont {Lill}, \citenamefont {Liu}, \citenamefont {Locharla}, \citenamefont {Lucero}, \citenamefont {McCourt}, \citenamefont {McEwen}, \citenamefont {Mi}, \citenamefont {Miao}, \citenamefont {Montazeri}, \citenamefont {Morvan}, \citenamefont {Neeley}, \citenamefont {Neill}, \citenamefont {Nersisyan}, \citenamefont {Ng}, \citenamefont {Nguyen}, \citenamefont {Nguyen}, \citenamefont {Potter}, \citenamefont {Quintana}, \citenamefont {Roushan}, \citenamefont {Sankaragomathi}, \citenamefont {Satzinger}, \citenamefont {Schuster}, \citenamefont {Shearn}, \citenamefont {Shorter}, \citenamefont {Shvarts}, \citenamefont {Skruzny}, \citenamefont {Smith},
  \citenamefont {Szalay}, \citenamefont {Torres}, \citenamefont {Woo}, \citenamefont {Yao}, \citenamefont {Yeh}, \citenamefont {Yoo}, \citenamefont {Young}, \citenamefont {Zhu}, \citenamefont {Zobrist}, \citenamefont {Chen}, \citenamefont {Megrant}, \citenamefont {Kelly},\ and\ \citenamefont {Naaman}}]{WhiteETAL2023}%
  \BibitemOpen
  \bibfield  {author} {\bibinfo {author} {\bibfnamefont {T.}~\bibnamefont {White}}, \bibinfo {author} {\bibfnamefont {A.}~\bibnamefont {Opremcak}}, \bibinfo {author} {\bibfnamefont {G.}~\bibnamefont {Sterling}}, \bibinfo {author} {\bibfnamefont {A.}~\bibnamefont {Korotkov}}, \bibinfo {author} {\bibfnamefont {D.}~\bibnamefont {Sank}}, \bibinfo {author} {\bibfnamefont {R.}~\bibnamefont {Acharya}}, \bibinfo {author} {\bibfnamefont {M.}~\bibnamefont {Ansmann}}, \bibinfo {author} {\bibfnamefont {F.}~\bibnamefont {Arute}}, \bibinfo {author} {\bibfnamefont {K.}~\bibnamefont {Arya}}, \bibinfo {author} {\bibfnamefont {J.~C.}\ \bibnamefont {Bardin}}, \bibinfo {author} {\bibfnamefont {A.}~\bibnamefont {Bengtsson}}, \bibinfo {author} {\bibfnamefont {A.}~\bibnamefont {Bourassa}}, \bibinfo {author} {\bibfnamefont {J.}~\bibnamefont {Bovaird}}, \bibinfo {author} {\bibfnamefont {L.}~\bibnamefont {Brill}}, \bibinfo {author} {\bibfnamefont {B.~B.}\ \bibnamefont {Buckley}}, \bibinfo {author} {\bibfnamefont {D.~A.}\ \bibnamefont
  {Buell}}, \bibinfo {author} {\bibfnamefont {T.}~\bibnamefont {Burger}}, \bibinfo {author} {\bibfnamefont {B.}~\bibnamefont {Burkett}}, \bibinfo {author} {\bibfnamefont {N.}~\bibnamefont {Bushnell}}, \bibinfo {author} {\bibfnamefont {Z.}~\bibnamefont {Chen}}, \bibinfo {author} {\bibfnamefont {B.}~\bibnamefont {Chiaro}}, \bibinfo {author} {\bibfnamefont {J.}~\bibnamefont {Cogan}}, \bibinfo {author} {\bibfnamefont {R.}~\bibnamefont {Collins}}, \bibinfo {author} {\bibfnamefont {A.~L.}\ \bibnamefont {Crook}}, \bibinfo {author} {\bibfnamefont {B.}~\bibnamefont {Curtin}}, \bibinfo {author} {\bibfnamefont {S.}~\bibnamefont {Demura}}, \bibinfo {author} {\bibfnamefont {A.}~\bibnamefont {Dunsworth}}, \bibinfo {author} {\bibfnamefont {C.}~\bibnamefont {Erickson}}, \bibinfo {author} {\bibfnamefont {R.}~\bibnamefont {Fatemi}}, \bibinfo {author} {\bibfnamefont {L.~F.}\ \bibnamefont {Burgos}}, \bibinfo {author} {\bibfnamefont {E.}~\bibnamefont {Forati}}, \bibinfo {author} {\bibfnamefont {B.}~\bibnamefont {Foxen}}, \bibinfo
  {author} {\bibfnamefont {W.}~\bibnamefont {Giang}}, \bibinfo {author} {\bibfnamefont {M.}~\bibnamefont {Giustina}}, \bibinfo {author} {\bibfnamefont {A.}~\bibnamefont {Grajales~Dau}}, \bibinfo {author} {\bibfnamefont {M.~C.}\ \bibnamefont {Hamilton}}, \bibinfo {author} {\bibfnamefont {S.~D.}\ \bibnamefont {Harrington}}, \bibinfo {author} {\bibfnamefont {J.}~\bibnamefont {Hilton}}, \bibinfo {author} {\bibfnamefont {M.}~\bibnamefont {Hoffmann}}, \bibinfo {author} {\bibfnamefont {S.}~\bibnamefont {Hong}}, \bibinfo {author} {\bibfnamefont {T.}~\bibnamefont {Huang}}, \bibinfo {author} {\bibfnamefont {A.}~\bibnamefont {Huff}}, \bibinfo {author} {\bibfnamefont {J.}~\bibnamefont {Iveland}}, \bibinfo {author} {\bibfnamefont {E.}~\bibnamefont {Jeffrey}}, \bibinfo {author} {\bibfnamefont {M.}~\bibnamefont {Kieferová}}, \bibinfo {author} {\bibfnamefont {S.}~\bibnamefont {Kim}}, \bibinfo {author} {\bibfnamefont {P.~V.}\ \bibnamefont {Klimov}}, \bibinfo {author} {\bibfnamefont {F.}~\bibnamefont {Kostritsa}}, \bibinfo
  {author} {\bibfnamefont {J.~M.}\ \bibnamefont {Kreikebaum}}, \bibinfo {author} {\bibfnamefont {D.}~\bibnamefont {Landhuis}}, \bibinfo {author} {\bibfnamefont {P.}~\bibnamefont {Laptev}}, \bibinfo {author} {\bibfnamefont {L.}~\bibnamefont {Laws}}, \bibinfo {author} {\bibfnamefont {K.}~\bibnamefont {Lee}}, \bibinfo {author} {\bibfnamefont {B.~J.}\ \bibnamefont {Lester}}, \bibinfo {author} {\bibfnamefont {A.}~\bibnamefont {Lill}}, \bibinfo {author} {\bibfnamefont {W.}~\bibnamefont {Liu}}, \bibinfo {author} {\bibfnamefont {A.}~\bibnamefont {Locharla}}, \bibinfo {author} {\bibfnamefont {E.}~\bibnamefont {Lucero}}, \bibinfo {author} {\bibfnamefont {T.}~\bibnamefont {McCourt}}, \bibinfo {author} {\bibfnamefont {M.}~\bibnamefont {McEwen}}, \bibinfo {author} {\bibfnamefont {X.}~\bibnamefont {Mi}}, \bibinfo {author} {\bibfnamefont {K.~C.}\ \bibnamefont {Miao}}, \bibinfo {author} {\bibfnamefont {S.}~\bibnamefont {Montazeri}}, \bibinfo {author} {\bibfnamefont {A.}~\bibnamefont {Morvan}}, \bibinfo {author}
  {\bibfnamefont {M.}~\bibnamefont {Neeley}}, \bibinfo {author} {\bibfnamefont {C.}~\bibnamefont {Neill}}, \bibinfo {author} {\bibfnamefont {A.}~\bibnamefont {Nersisyan}}, \bibinfo {author} {\bibfnamefont {J.~H.}\ \bibnamefont {Ng}}, \bibinfo {author} {\bibfnamefont {A.}~\bibnamefont {Nguyen}}, \bibinfo {author} {\bibfnamefont {M.}~\bibnamefont {Nguyen}}, \bibinfo {author} {\bibfnamefont {R.}~\bibnamefont {Potter}}, \bibinfo {author} {\bibfnamefont {C.}~\bibnamefont {Quintana}}, \bibinfo {author} {\bibfnamefont {P.}~\bibnamefont {Roushan}}, \bibinfo {author} {\bibfnamefont {K.}~\bibnamefont {Sankaragomathi}}, \bibinfo {author} {\bibfnamefont {K.~J.}\ \bibnamefont {Satzinger}}, \bibinfo {author} {\bibfnamefont {C.}~\bibnamefont {Schuster}}, \bibinfo {author} {\bibfnamefont {M.~J.}\ \bibnamefont {Shearn}}, \bibinfo {author} {\bibfnamefont {A.}~\bibnamefont {Shorter}}, \bibinfo {author} {\bibfnamefont {V.}~\bibnamefont {Shvarts}}, \bibinfo {author} {\bibfnamefont {J.}~\bibnamefont {Skruzny}}, \bibinfo {author}
  {\bibfnamefont {W.~C.}\ \bibnamefont {Smith}}, \bibinfo {author} {\bibfnamefont {M.}~\bibnamefont {Szalay}}, \bibinfo {author} {\bibfnamefont {A.}~\bibnamefont {Torres}}, \bibinfo {author} {\bibfnamefont {B.~W.~K.}\ \bibnamefont {Woo}}, \bibinfo {author} {\bibfnamefont {Z.~J.}\ \bibnamefont {Yao}}, \bibinfo {author} {\bibfnamefont {P.}~\bibnamefont {Yeh}}, \bibinfo {author} {\bibfnamefont {J.}~\bibnamefont {Yoo}}, \bibinfo {author} {\bibfnamefont {G.}~\bibnamefont {Young}}, \bibinfo {author} {\bibfnamefont {N.}~\bibnamefont {Zhu}}, \bibinfo {author} {\bibfnamefont {N.}~\bibnamefont {Zobrist}}, \bibinfo {author} {\bibfnamefont {Y.}~\bibnamefont {Chen}}, \bibinfo {author} {\bibfnamefont {A.}~\bibnamefont {Megrant}}, \bibinfo {author} {\bibfnamefont {J.}~\bibnamefont {Kelly}},\ and\ \bibinfo {author} {\bibfnamefont {O.}~\bibnamefont {Naaman}},\ }\bibfield  {title} {\bibinfo {title} {Readout of a quantum processor with high dynamic range josephson parametric amplifiers},\ }\href
  {https://doi.org/10.1063/5.0127375} {\bibfield  {journal} {\bibinfo  {journal} {Appl. Phys. Lett.}\ }\textbf {\bibinfo {volume} {122}},\ \bibinfo {pages} {014001} (\bibinfo {year} {2023})}\BibitemShut {NoStop}%
\bibitem [{\citenamefont {Pozar}(2011)}]{pozar2011microwave}%
  \BibitemOpen
  \bibfield  {author} {\bibinfo {author} {\bibfnamefont {D.}~\bibnamefont {Pozar}},\ }\href {https://books.google.ca/books?id=_YEbGAXCcAMC} {\emph {\bibinfo {title} {Microwave Engineering}}}\ (\bibinfo  {publisher} {Wiley},\ \bibinfo {year} {2011})\BibitemShut {NoStop}%
\bibitem [{\citenamefont {Gardiner}\ and\ \citenamefont {Zoller}(2004)}]{GarZol_book}%
  \BibitemOpen
  \bibfield  {author} {\bibinfo {author} {\bibfnamefont {C.}~\bibnamefont {Gardiner}}\ and\ \bibinfo {author} {\bibfnamefont {P.}~\bibnamefont {Zoller}},\ }\href@noop {} {\emph {\bibinfo {title} {Quantum Noise: A Handbook of Markovian and Non-Markovian Quantum Stochastic Methods with Applications to Quantum Optics}}},\ Springer Series in Synergetics\ (\bibinfo  {publisher} {Springer Berlin},\ \bibinfo {year} {2004})\BibitemShut {NoStop}%
\bibitem [{\citenamefont {Bruus}\ and\ \citenamefont {Flensberg}(2004)}]{BruusFlensberg_book}%
  \BibitemOpen
  \bibfield  {author} {\bibinfo {author} {\bibfnamefont {H.}~\bibnamefont {Bruus}}\ and\ \bibinfo {author} {\bibfnamefont {K.}~\bibnamefont {Flensberg}},\ }\href@noop {} {\emph {\bibinfo {title} {Many-body quantum theory in condensed matter physics: an introduction}}}\ (\bibinfo  {publisher} {Oxford university press},\ \bibinfo {year} {2004})\BibitemShut {NoStop}%
\bibitem [{\citenamefont {Parra-Rodriguez}\ \emph {et~al.}(2018)\citenamefont {Parra-Rodriguez}, \citenamefont {Rico}, \citenamefont {Solano},\ and\ \citenamefont {Egusquiza}}]{ParraRodriguezETAL2018}%
  \BibitemOpen
  \bibfield  {author} {\bibinfo {author} {\bibfnamefont {A.}~\bibnamefont {Parra-Rodriguez}}, \bibinfo {author} {\bibfnamefont {E.}~\bibnamefont {Rico}}, \bibinfo {author} {\bibfnamefont {E.}~\bibnamefont {Solano}},\ and\ \bibinfo {author} {\bibfnamefont {I.~L.}\ \bibnamefont {Egusquiza}},\ }\bibfield  {title} {\bibinfo {title} {Quantum networks in divergence-free circuit qed},\ }\href {https://doi.org/10.1088/2058-9565/aab1ba} {\bibfield  {journal} {\bibinfo  {journal} {Quantum Sci. Technol.}\ }\textbf {\bibinfo {volume} {3}},\ \bibinfo {pages} {024012} (\bibinfo {year} {2018})}\BibitemShut {NoStop}%
\bibitem [{\citenamefont {Matveev}\ \emph {et~al.}(1993)\citenamefont {Matveev}, \citenamefont {Yue},\ and\ \citenamefont {Glazman}}]{MatveevETAL1993}%
  \BibitemOpen
  \bibfield  {author} {\bibinfo {author} {\bibfnamefont {K.~A.}\ \bibnamefont {Matveev}}, \bibinfo {author} {\bibfnamefont {D.}~\bibnamefont {Yue}},\ and\ \bibinfo {author} {\bibfnamefont {L.~I.}\ \bibnamefont {Glazman}},\ }\bibfield  {title} {\bibinfo {title} {Tunneling in one-dimensional non-luttinger electron liquid},\ }\href {https://doi.org/10.1103/PhysRevLett.71.3351} {\bibfield  {journal} {\bibinfo  {journal} {Phys. Rev. Lett.}\ }\textbf {\bibinfo {volume} {71}},\ \bibinfo {pages} {3351} (\bibinfo {year} {1993})}\BibitemShut {NoStop}%
\bibitem [{\citenamefont {Yue}\ \emph {et~al.}(1994)\citenamefont {Yue}, \citenamefont {Glazman},\ and\ \citenamefont {Matveev}}]{YueETAL1994}%
  \BibitemOpen
  \bibfield  {author} {\bibinfo {author} {\bibfnamefont {D.}~\bibnamefont {Yue}}, \bibinfo {author} {\bibfnamefont {L.~I.}\ \bibnamefont {Glazman}},\ and\ \bibinfo {author} {\bibfnamefont {K.~A.}\ \bibnamefont {Matveev}},\ }\bibfield  {title} {\bibinfo {title} {Conduction of a weakly interacting one-dimensional electron gas through a single barrier},\ }\href {https://doi.org/10.1103/PhysRevB.49.1966} {\bibfield  {journal} {\bibinfo  {journal} {Phys. Rev. B}\ }\textbf {\bibinfo {volume} {49}},\ \bibinfo {pages} {1966} (\bibinfo {year} {1994})}\BibitemShut {NoStop}%
\bibitem [{\citenamefont {Lal}\ \emph {et~al.}(2002)\citenamefont {Lal}, \citenamefont {Rao},\ and\ \citenamefont {Sen}}]{LalETAL2002}%
  \BibitemOpen
  \bibfield  {author} {\bibinfo {author} {\bibfnamefont {S.}~\bibnamefont {Lal}}, \bibinfo {author} {\bibfnamefont {S.}~\bibnamefont {Rao}},\ and\ \bibinfo {author} {\bibfnamefont {D.}~\bibnamefont {Sen}},\ }\bibfield  {title} {\bibinfo {title} {Junction of several weakly interacting quantum wires: A renormalization group study},\ }\href {https://doi.org/10.1103/PhysRevB.66.165327} {\bibfield  {journal} {\bibinfo  {journal} {Phys. Rev. B}\ }\textbf {\bibinfo {volume} {66}},\ \bibinfo {pages} {165327} (\bibinfo {year} {2002})}\BibitemShut {NoStop}%
\bibitem [{\citenamefont {Das}\ \emph {et~al.}(2004)\citenamefont {Das}, \citenamefont {Rao},\ and\ \citenamefont {Sen}}]{DasETAL2004}%
  \BibitemOpen
  \bibfield  {author} {\bibinfo {author} {\bibfnamefont {S.}~\bibnamefont {Das}}, \bibinfo {author} {\bibfnamefont {S.}~\bibnamefont {Rao}},\ and\ \bibinfo {author} {\bibfnamefont {D.}~\bibnamefont {Sen}},\ }\bibfield  {title} {\bibinfo {title} {Renormalization group study of the conductances of interacting quantum wire systems with different geometries},\ }\href {https://doi.org/10.1103/PhysRevB.70.085318} {\bibfield  {journal} {\bibinfo  {journal} {Phys. Rev. B}\ }\textbf {\bibinfo {volume} {70}},\ \bibinfo {pages} {085318} (\bibinfo {year} {2004})}\BibitemShut {NoStop}%
\bibitem [{\citenamefont {Nazarov}\ and\ \citenamefont {Glazman}(2003)}]{NazarovGlazman2003}%
  \BibitemOpen
  \bibfield  {author} {\bibinfo {author} {\bibfnamefont {Y.~V.}\ \bibnamefont {Nazarov}}\ and\ \bibinfo {author} {\bibfnamefont {L.~I.}\ \bibnamefont {Glazman}},\ }\bibfield  {title} {\bibinfo {title} {Resonant tunneling of interacting electrons in a one-dimensional wire},\ }\href {https://doi.org/10.1103/PhysRevLett.91.126804} {\bibfield  {journal} {\bibinfo  {journal} {Phys. Rev. Lett.}\ }\textbf {\bibinfo {volume} {91}},\ \bibinfo {pages} {126804} (\bibinfo {year} {2003})}\BibitemShut {NoStop}%
\bibitem [{\citenamefont {Polyakov}\ and\ \citenamefont {Gornyi}(2003)}]{PolyakovGornyi2003}%
  \BibitemOpen
  \bibfield  {author} {\bibinfo {author} {\bibfnamefont {D.~G.}\ \bibnamefont {Polyakov}}\ and\ \bibinfo {author} {\bibfnamefont {I.~V.}\ \bibnamefont {Gornyi}},\ }\bibfield  {title} {\bibinfo {title} {Transport of interacting electrons through a double barrier in quantum wires},\ }\href {https://doi.org/10.1103/PhysRevB.68.035421} {\bibfield  {journal} {\bibinfo  {journal} {Phys. Rev. B}\ }\textbf {\bibinfo {volume} {68}},\ \bibinfo {pages} {035421} (\bibinfo {year} {2003})}\BibitemShut {NoStop}%
\bibitem [{\citenamefont {Shi}\ and\ \citenamefont {Affleck}(2016)}]{ShiAffleck2016}%
  \BibitemOpen
  \bibfield  {author} {\bibinfo {author} {\bibfnamefont {Z.}~\bibnamefont {Shi}}\ and\ \bibinfo {author} {\bibfnamefont {I.}~\bibnamefont {Affleck}},\ }\bibfield  {title} {\bibinfo {title} {Fermionic approach to junctions of multiple quantum wires attached to tomonaga-luttinger liquid leads},\ }\href {https://doi.org/10.1103/PhysRevB.94.035106} {\bibfield  {journal} {\bibinfo  {journal} {Phys. Rev. B}\ }\textbf {\bibinfo {volume} {94}},\ \bibinfo {pages} {035106} (\bibinfo {year} {2016})}\BibitemShut {NoStop}%
\bibitem [{\citenamefont {Shi}\ and\ \citenamefont {Komijani}(2017)}]{ShiKomijani2017}%
  \BibitemOpen
  \bibfield  {author} {\bibinfo {author} {\bibfnamefont {Z.}~\bibnamefont {Shi}}\ and\ \bibinfo {author} {\bibfnamefont {Y.}~\bibnamefont {Komijani}},\ }\bibfield  {title} {\bibinfo {title} {Conductance of closed and open long aharonov-bohm-kondo rings},\ }\href {https://doi.org/10.1103/PhysRevB.95.075147} {\bibfield  {journal} {\bibinfo  {journal} {Phys. Rev. B}\ }\textbf {\bibinfo {volume} {95}},\ \bibinfo {pages} {075147} (\bibinfo {year} {2017})}\BibitemShut {NoStop}%
\bibitem [{\citenamefont {Parra-Rodriguez}(2021)}]{parrarodriguez_PhD_thesis}%
  \BibitemOpen
  \bibfield  {author} {\bibinfo {author} {\bibfnamefont {A.}~\bibnamefont {Parra-Rodriguez}},\ }\href {https://arxiv.org/abs/2104.09410} {\bibinfo {title} {Canonical quantization of superconducting circuits}} (\bibinfo {year} {2021}),\ \Eprint {https://arxiv.org/abs/2104.09410} {arXiv:2104.09410 [quant-ph]} \BibitemShut {NoStop}%
\bibitem [{\citenamefont {Parra-Rodriguez}\ \emph {et~al.}(2019)\citenamefont {Parra-Rodriguez}, \citenamefont {Egusquiza}, \citenamefont {DiVincenzo},\ and\ \citenamefont {Solano}}]{ParraRodriguez2019canonical}%
  \BibitemOpen
  \bibfield  {author} {\bibinfo {author} {\bibfnamefont {A.}~\bibnamefont {Parra-Rodriguez}}, \bibinfo {author} {\bibfnamefont {I.~L.}\ \bibnamefont {Egusquiza}}, \bibinfo {author} {\bibfnamefont {D.~P.}\ \bibnamefont {DiVincenzo}},\ and\ \bibinfo {author} {\bibfnamefont {E.}~\bibnamefont {Solano}},\ }\bibfield  {title} {\bibinfo {title} {Canonical circuit quantization with linear nonreciprocal devices},\ }\href {https://doi.org/10.1103/PhysRevB.99.014514} {\bibfield  {journal} {\bibinfo  {journal} {Phys. Rev. B}\ }\textbf {\bibinfo {volume} {99}},\ \bibinfo {pages} {014514} (\bibinfo {year} {2019})}\BibitemShut {NoStop}%
\bibitem [{\citenamefont {Parra-Rodriguez}\ and\ \citenamefont {Egusquiza}(2022)}]{ParraRodriguez2022canonical}%
  \BibitemOpen
  \bibfield  {author} {\bibinfo {author} {\bibfnamefont {A.}~\bibnamefont {Parra-Rodriguez}}\ and\ \bibinfo {author} {\bibfnamefont {I.~L.}\ \bibnamefont {Egusquiza}},\ }\bibfield  {title} {\bibinfo {title} {Canonical quantisation of telegrapher's equations coupled by ideal nonreciprocal elements},\ }\href {https://doi.org/10.22331/q-2022-04-04-681} {\bibfield  {journal} {\bibinfo  {journal} {{Quantum}}\ }\textbf {\bibinfo {volume} {6}},\ \bibinfo {pages} {681} (\bibinfo {year} {2022})}\BibitemShut {NoStop}%
\bibitem [{\citenamefont {Egusquiza}\ and\ \citenamefont {Parra-Rodriguez}(2022)}]{Egusquiza2022algebraic}%
  \BibitemOpen
  \bibfield  {author} {\bibinfo {author} {\bibfnamefont {I.~L.}\ \bibnamefont {Egusquiza}}\ and\ \bibinfo {author} {\bibfnamefont {A.}~\bibnamefont {Parra-Rodriguez}},\ }\bibfield  {title} {\bibinfo {title} {Algebraic canonical quantization of lumped superconducting networks},\ }\href {https://doi.org/10.1103/PhysRevB.106.024510} {\bibfield  {journal} {\bibinfo  {journal} {Phys. Rev. B}\ }\textbf {\bibinfo {volume} {106}},\ \bibinfo {pages} {024510} (\bibinfo {year} {2022})}\BibitemShut {NoStop}%
\bibitem [{\citenamefont {Parra-Rodriguez}\ and\ \citenamefont {Egusquiza}(2025)}]{ParraRodriguez2025exact}%
  \BibitemOpen
  \bibfield  {author} {\bibinfo {author} {\bibfnamefont {A.}~\bibnamefont {Parra-Rodriguez}}\ and\ \bibinfo {author} {\bibfnamefont {I.~L.}\ \bibnamefont {Egusquiza}},\ }\bibfield  {title} {\bibinfo {title} {Exact quantization of nonreciprocal quasilumped electrical networks},\ }\href {https://doi.org/10.1103/PhysRevX.15.011072} {\bibfield  {journal} {\bibinfo  {journal} {Phys. Rev. X}\ }\textbf {\bibinfo {volume} {15}},\ \bibinfo {pages} {011072} (\bibinfo {year} {2025})}\BibitemShut {NoStop}%
\bibitem [{\citenamefont {Ranzani}\ and\ \citenamefont {Aumentado}(2015)}]{AumentadoETAL2015}%
  \BibitemOpen
  \bibfield  {author} {\bibinfo {author} {\bibfnamefont {L.}~\bibnamefont {Ranzani}}\ and\ \bibinfo {author} {\bibfnamefont {J.}~\bibnamefont {Aumentado}},\ }\bibfield  {title} {\bibinfo {title} {Graph-based analysis of nonreciprocity in coupled-mode systems},\ }\href {https://doi.org/10.1088/1367-2630/17/2/023024} {\bibfield  {journal} {\bibinfo  {journal} {New J. Phys.}\ }\textbf {\bibinfo {volume} {17}},\ \bibinfo {pages} {023024} (\bibinfo {year} {2015})}\BibitemShut {NoStop}%
\bibitem [{\citenamefont {Yan}\ \emph {et~al.}(2023)\citenamefont {Yan}, \citenamefont {Wu}, \citenamefont {Lingenfelter}, \citenamefont {Joshi}, \citenamefont {Andersson}, \citenamefont {Conner}, \citenamefont {Chou}, \citenamefont {Grebel}, \citenamefont {Miller}, \citenamefont {Povey}, \citenamefont {Qiao}, \citenamefont {Clerk},\ and\ \citenamefont {Cleland}}]{coupled_modes_Purcell_filter}%
  \BibitemOpen
  \bibfield  {author} {\bibinfo {author} {\bibfnamefont {H.}~\bibnamefont {Yan}}, \bibinfo {author} {\bibfnamefont {X.}~\bibnamefont {Wu}}, \bibinfo {author} {\bibfnamefont {A.}~\bibnamefont {Lingenfelter}}, \bibinfo {author} {\bibfnamefont {Y.~J.}\ \bibnamefont {Joshi}}, \bibinfo {author} {\bibfnamefont {G.}~\bibnamefont {Andersson}}, \bibinfo {author} {\bibfnamefont {C.~R.}\ \bibnamefont {Conner}}, \bibinfo {author} {\bibfnamefont {M.-H.}\ \bibnamefont {Chou}}, \bibinfo {author} {\bibfnamefont {J.}~\bibnamefont {Grebel}}, \bibinfo {author} {\bibfnamefont {J.~M.}\ \bibnamefont {Miller}}, \bibinfo {author} {\bibfnamefont {R.~G.}\ \bibnamefont {Povey}}, \bibinfo {author} {\bibfnamefont {H.}~\bibnamefont {Qiao}}, \bibinfo {author} {\bibfnamefont {A.~A.}\ \bibnamefont {Clerk}},\ and\ \bibinfo {author} {\bibfnamefont {A.~N.}\ \bibnamefont {Cleland}},\ }\bibfield  {title} {\bibinfo {title} {Broadband bandpass purcell filter for circuit quantum electrodynamics},\ }\href {https://doi.org/10.1063/5.0161893}
  {\bibfield  {journal} {\bibinfo  {journal} {Appl. Phys. Lett.}\ }\textbf {\bibinfo {volume} {123}},\ \bibinfo {pages} {134001} (\bibinfo {year} {2023})}\BibitemShut {NoStop}%
\bibitem [{\citenamefont {Lin}\ \emph {et~al.}(2013)\citenamefont {Lin}, \citenamefont {Inomata}, \citenamefont {Oliver}, \citenamefont {Koshino}, \citenamefont {Nakamura}, \citenamefont {Tsai},\ and\ \citenamefont {Yamamoto}}]{LinETAL2013_JPA}%
  \BibitemOpen
  \bibfield  {author} {\bibinfo {author} {\bibfnamefont {Z.~R.}\ \bibnamefont {Lin}}, \bibinfo {author} {\bibfnamefont {K.}~\bibnamefont {Inomata}}, \bibinfo {author} {\bibfnamefont {W.~D.}\ \bibnamefont {Oliver}}, \bibinfo {author} {\bibfnamefont {K.}~\bibnamefont {Koshino}}, \bibinfo {author} {\bibfnamefont {Y.}~\bibnamefont {Nakamura}}, \bibinfo {author} {\bibfnamefont {J.~S.}\ \bibnamefont {Tsai}},\ and\ \bibinfo {author} {\bibfnamefont {T.}~\bibnamefont {Yamamoto}},\ }\bibfield  {title} {\bibinfo {title} {Single-shot readout of a superconducting flux qubit with a flux-driven josephson parametric amplifier},\ }\href {https://doi.org/10.1063/1.4821822} {\bibfield  {journal} {\bibinfo  {journal} {Appl. Phys. Lett.}\ }\textbf {\bibinfo {volume} {103}},\ \bibinfo {pages} {132602} (\bibinfo {year} {2013})}\BibitemShut {NoStop}%
\bibitem [{\citenamefont {Vijay}\ \emph {et~al.}(2011)\citenamefont {Vijay}, \citenamefont {Slichter},\ and\ \citenamefont {Siddiqi}}]{VijayETAL2011}%
  \BibitemOpen
  \bibfield  {author} {\bibinfo {author} {\bibfnamefont {R.}~\bibnamefont {Vijay}}, \bibinfo {author} {\bibfnamefont {D.~H.}\ \bibnamefont {Slichter}},\ and\ \bibinfo {author} {\bibfnamefont {I.}~\bibnamefont {Siddiqi}},\ }\bibfield  {title} {\bibinfo {title} {Observation of quantum jumps in a superconducting artificial atom},\ }\href {https://doi.org/10.1103/PhysRevLett.106.110502} {\bibfield  {journal} {\bibinfo  {journal} {Phys. Rev. Lett.}\ }\textbf {\bibinfo {volume} {106}},\ \bibinfo {pages} {110502} (\bibinfo {year} {2011})}\BibitemShut {NoStop}%
\bibitem [{\citenamefont {Simoen}\ \emph {et~al.}(2015)\citenamefont {Simoen}, \citenamefont {Chang}, \citenamefont {Krantz}, \citenamefont {Bylander}, \citenamefont {Wustmann}, \citenamefont {Shumeiko}, \citenamefont {Delsing},\ and\ \citenamefont {Wilson}}]{SimoenETAL2015}%
  \BibitemOpen
  \bibfield  {author} {\bibinfo {author} {\bibfnamefont {M.}~\bibnamefont {Simoen}}, \bibinfo {author} {\bibfnamefont {C.~W.~S.}\ \bibnamefont {Chang}}, \bibinfo {author} {\bibfnamefont {P.}~\bibnamefont {Krantz}}, \bibinfo {author} {\bibfnamefont {J.}~\bibnamefont {Bylander}}, \bibinfo {author} {\bibfnamefont {W.}~\bibnamefont {Wustmann}}, \bibinfo {author} {\bibfnamefont {V.}~\bibnamefont {Shumeiko}}, \bibinfo {author} {\bibfnamefont {P.}~\bibnamefont {Delsing}},\ and\ \bibinfo {author} {\bibfnamefont {C.~M.}\ \bibnamefont {Wilson}},\ }\bibfield  {title} {\bibinfo {title} {Characterization of a multimode coplanar waveguide parametric amplifier},\ }\href {https://doi.org/10.1063/1.4933265} {\bibfield  {journal} {\bibinfo  {journal} {Journal of Applied Physics}\ }\textbf {\bibinfo {volume} {118}},\ \bibinfo {pages} {154501} (\bibinfo {year} {2015})}\BibitemShut {NoStop}%
\bibitem [{\citenamefont {Bengtsson}\ \emph {et~al.}(2018)\citenamefont {Bengtsson}, \citenamefont {Krantz}, \citenamefont {Simoen}, \citenamefont {Svensson}, \citenamefont {Schneider}, \citenamefont {Shumeiko}, \citenamefont {Delsing},\ and\ \citenamefont {Bylander}}]{BengtssonETAL2018}%
  \BibitemOpen
  \bibfield  {author} {\bibinfo {author} {\bibfnamefont {A.}~\bibnamefont {Bengtsson}}, \bibinfo {author} {\bibfnamefont {P.}~\bibnamefont {Krantz}}, \bibinfo {author} {\bibfnamefont {M.}~\bibnamefont {Simoen}}, \bibinfo {author} {\bibfnamefont {I.-M.}\ \bibnamefont {Svensson}}, \bibinfo {author} {\bibfnamefont {B.}~\bibnamefont {Schneider}}, \bibinfo {author} {\bibfnamefont {V.}~\bibnamefont {Shumeiko}}, \bibinfo {author} {\bibfnamefont {P.}~\bibnamefont {Delsing}},\ and\ \bibinfo {author} {\bibfnamefont {J.}~\bibnamefont {Bylander}},\ }\bibfield  {title} {\bibinfo {title} {Nondegenerate parametric oscillations in a tunable superconducting resonator},\ }\href {https://doi.org/10.1103/PhysRevB.97.144502} {\bibfield  {journal} {\bibinfo  {journal} {Phys. Rev. B}\ }\textbf {\bibinfo {volume} {97}},\ \bibinfo {pages} {144502} (\bibinfo {year} {2018})}\BibitemShut {NoStop}%
\bibitem [{\citenamefont {Walter}(1973)}]{walter1973regular}%
  \BibitemOpen
  \bibfield  {author} {\bibinfo {author} {\bibfnamefont {J.}~\bibnamefont {Walter}},\ }\bibfield  {title} {\bibinfo {title} {Regular eigenvalue problems with eigenvalue parameter in the boundary condition},\ }\href {https://doi.org/10.1007/BF01177870} {\bibfield  {journal} {\bibinfo  {journal} {Mathematische Zeitschrift}\ }\textbf {\bibinfo {volume} {133}},\ \bibinfo {pages} {301} (\bibinfo {year} {1973})}\BibitemShut {NoStop}%
\bibitem [{\citenamefont {Yurke}(2004)}]{Yurke_DruFic2004}%
  \BibitemOpen
  \bibfield  {author} {\bibinfo {author} {\bibfnamefont {B.}~\bibnamefont {Yurke}},\ }\bibfield  {title} {\bibinfo {title} {Input-output theory},\ }in\ \href@noop {} {\emph {\bibinfo {booktitle} {Quantum Squeezing}}},\ \bibinfo {editor} {edited by\ \bibinfo {editor} {\bibfnamefont {P.}~\bibnamefont {Drummond}}\ and\ \bibinfo {editor} {\bibfnamefont {Z.}~\bibnamefont {Fizek}}}\ (\bibinfo  {publisher} {Springer},\ \bibinfo {year} {2004})\ pp.\ \bibinfo {pages} {53--96}\BibitemShut {NoStop}%
\bibitem [{Note1()}]{Note1}%
  \BibitemOpen
  \bibinfo {note} {Because of the oscillating factor, the contour integral along either the upper or lower semicircle with radius $\to \infty $ diverges. Although one can calculate the integral $\DOTSI \intop \ilimits@ _{-\infty }^{\infty }$ using the non-diverging semicircle, this is in general of little help to determine $\DOTSI \intop \ilimits@ _{0}^{\infty }$. While a branch cut trick would in principle allow to calculate $\DOTSI \intop \ilimits@ _{0}^{\infty }$, it also leads to divergence because it uses the full circle contour integral.}\BibitemShut {Stop}%
\bibitem [{\citenamefont {Nayfeh}\ and\ \citenamefont {Mook}(2008)}]{NayfehMook_book}%
  \BibitemOpen
  \bibfield  {author} {\bibinfo {author} {\bibfnamefont {A.}~\bibnamefont {Nayfeh}}\ and\ \bibinfo {author} {\bibfnamefont {D.}~\bibnamefont {Mook}},\ }\href@noop {} {\emph {\bibinfo {title} {Nonlinear Oscillations}}},\ Wiley Classics Library\ (\bibinfo  {publisher} {Wiley},\ \bibinfo {year} {2008})\BibitemShut {NoStop}%
\bibitem [{Note2()}]{Note2}%
  \BibitemOpen
  \bibinfo {note} {If the combined Hamiltonians are instead approximated in the usual weak-coupling approximation and with coupling-independent TL modes, then the coupling induced frequency shifts may take other forms, e.g.~with $C_\protect \text {eff}= C_s + C_c$ for the purely capacitive coupling.}\BibitemShut {Stop}%
\bibitem [{Note3()}]{Note3}%
  \BibitemOpen
  \bibinfo {note} {Terminating the TL with a capacitor of size $C_0 a$ will apparently remove the singularity. However, the resulting Hamiltonian becomes pathological in the continuum limit $a\to 0$, unless $\Phi _a$ is integrated into the TL part of the Hamiltonian which is equivalent to our treatment in this appendix.}\BibitemShut {Stop}%
\bibitem [{\citenamefont {Jeffreys}\ and\ \citenamefont {Jeffreys}(1999)}]{complexanalysis_book}%
  \BibitemOpen
  \bibfield  {author} {\bibinfo {author} {\bibfnamefont {H.}~\bibnamefont {Jeffreys}}\ and\ \bibinfo {author} {\bibfnamefont {B.}~\bibnamefont {Jeffreys}},\ }\href@noop {} {\emph {\bibinfo {title} {Methods of Mathematical Physics}}},\ Cambridge Mathematical Library\ (\bibinfo  {publisher} {Cambridge University Press},\ \bibinfo {year} {1999})\BibitemShut {NoStop}%
\end{thebibliography}

\begin{thebibliography}{}

\bibitem{JPA_StateoftheArt}
J.~Aumentado,
{\em Superconducting parametric amplifiers: The state of the art in Josephson parametric amplifiers},
IEEE Microw.~Mag.~{\bf 21}, 45 (2020).

\bibitem{ClerkETAL_review2010}
A.A.~Clerk, M.H.~Devoret, S.M.~Girvin, F.~Marquardt, and R.J.~Schoelkopf,
Rev.~Mod.~Phys.~{\bf 82}, 1155 (2010).

\bibitem{KrantzETAL2019_review}
P.~Krantz, M.~Kjaergaard, F.~Yan, T.P.~Orlando, S.~Gustavsson, and W.D.~Oliver,
{\em A quantum engineer's guide to superconducting qubits.},
Appl.~Phys.~Rev.~{\bf 6}, 021318 (2019).

\bibitem{LehnertETAL2007}
M.A.~Castellanos-Beltran and K.W.~Lehnert,
{\em Widely tunable parametric amplifier based on a superconducting quantum interference device array resonator},
Appl.~Phys.~Lett.~{\bf 91}, 083509 (2007).

\bibitem{LehnertETAL2008}
M.A.~Castellanos-Beltran, K.D.~Irwin, G.C.~Hilton, L.R.~Vale, and K.W.~Lehnert,
{\em Amplification and squeezing of quantum noise with a tunable Josephson metamaterial},
Nat.~Phys.~{\bf  4}, 929 (2008).

\bibitem{YamamotoETAL2008}
T.~Yamamoto K.~Inomata, M.~Watanabe, K.~Matsuba, T.~Miyazaki, W.D.~Oliver, Y.~Nakamura, J.S.~ Tsai,
{\em Flux-driven Josephson parametric amplifier},
Appl.~Phys.~Lett.~{\bf 93}, 042510 (2008).

\bibitem{WalterETAL2017}
T.~Walter, P.~Kurpiers, S.~Gasparinetti, P.~Magnard, A. Poto\u{c}nik, Y.~Salath\'{e}, M.~Pechal, M.~Mondal, M.~Oppliger, C.~Eichler, and A.~Wallraff,
{\em Rapid high-fidelity single-shot dispersive readout of superconducting qubits},
Phys.~Rev.~Appl.~{\bf 7}, 054020 (2017).


\bibitem{SliwaETAL2015}
K.M.~Sliwa, M.~Hatridge, A.~Narla, S.~Shankar, L.~Frunzio, R.J.~Schoelkopf, and M.H.~Devoret,
{\em Reconfigurable Josephson circulator/directional amplifier},
Phys.~Rev.~X  {\bf 5}, 041020 (2015).

\bibitem{ChapmanETAL2017b}
B.J.~Chapman, E.I.~Rosenthal,  J.~Kerckhoff, B.A.~Moores, L.R.~Vale, J.~Mates, G.C.~Hilton, K.~Lalumiere, A.~Blais, and K.~Lehnert,
{\em Widely Tunable On-Chip Microwave Circulator for Superconducting Quantum Circuits}, Phys.~Rev.~X {\bf 7}, 041043 (2017).

\bibitem{LecocqETAL2021}
F.~Lecocq, L.~Ranzani, G.~Peterson, K.~Cicak, X.~Jin, R.~Simmonds, J.~Teufel, and J.~Aumentado,
{\em Efficient Qubit Measurement with a Nonreciprocal Microwave Amplifier},
Phys.~Rev.~Lett.~{\bf 126}, 020502 (2021).

\bibitem{LecocqETAL2017}
F.~Lecocq, L.~Ranzani, G.A.~Peterson, K.~Cicak, R.W.~Simmonds, J.D.~Teufel, and J.~Aumentado,
{\em Nonreciprocal microwave signal processing with a field-programmable Josephson amplifier},
Phys.~Rev.~Appl.~{\bf 7}, 024028 (2017).

\bibitem{AbdoETAL2021}
B.~Abdo, O.~Jinka , N.T.~Bronn , S.~Olivadese, and M.~Brink,
{\em High-Fidelity Qubit Readout Using Interferometric Directional Josephson
Devices}
PRX Quantum {\bf 2}, 040360 (2021).


\bibitem{YurkeETAL1988}
B.~Yurke, P.G.~Kaminsky, R.E.~Miller, E.A.~Whittaker, A.D.~Smith, A.H.~Silver, and R.W.~Simon,
{\em Observation of 4.2-K equilibrium-noise squeezing via a Josephson-parametric amplifier}
Phys.~Rev.~Lett.~{\bf 60}, 764 (1988).

\bibitem{EichlerETAL2011}
C.~Eichler, D.~Bozyigit, C.~Lang, M.~Baur, L.~Steffen, J.M.~Fink, S.~Filipp, and A.~Wallraff,
{\em Observation of two-mode squeezing in the microwave frequency domain},
Phys.~Rev.~Lett.~{\bf 107}, 113601 (2011).

\bibitem{OBrienETAL2023}
J.Y.~Qiu, A.~Grimsmo, K.~Peng, B.~Kannan, B.~Lienhard, Y.~Sung, P.~Krantz, V.~Bolkhovsky,
G.~Calusine, D.~Kim, A.~Melville, B.M.~Niedzielski, J.~Yoder, M.E.~Schwartz, T.P.~Orlando, I.~Siddiqi, S.~Gustavsson, K.P.~O'Brien, and W.D.~Oliver,
{\em Broadband squeezed microwaves and amplification with a Josephson travelling-wave parametric amplifier},
Nat.~Phys.~{\bf 19}, 706 (2023).


\bibitem{BergealETAL2010b}
N.~Bergeal, F.~Schackert, M.~Metcalfe, R.~Vijay, V.E.~Manucharyan, L.~Frunzio, D.E.~Prober, R.J.~Schoelkopf, S.M.~Girvin, and M.H.~Devoret,
{\em Phase-preserving amplification near the quantum limit with a Josephson ring modulator},
Nature {\bf 465}, 64 (2010).

\bibitem{NaamanAumentado2022}
O.~Naaman and J.~Aumentado,
{\em Synthesis of Parametrically Coupled Networks},
PRX Quantum {\bf 3}, 020201 (2022).


\bibitem{BraunsteinETAL2005}
S.L.~Braunstein, P.~van Loock, 
Rev.~Mod.~Phys. 2005, 77, 513

\bibitem{FlurinETAL2012}
E.~Flurin, N.~Roch, F.~Mallet, M.H.~Devoret, and B.~Huard,
{\em Generating Entangled Microwave Radiation Over Two Transmission Lines}
Phys.~Rev.~Lett.~{\bf 109}, 183901 (2012).

\bibitem{PetrovninETAL2022}
K.V.~Petrovnin, M.R.~Perelshtein, T.~Korkalainen, V.~Vesterinen, I.~Lilja, G.S.~Paraoanu, and P.J.~Hakonen,
{\em Generation and Structuring of Multipartite Entanglement in a Josephson Parametric System},
Adv.~Quantum.~Technol.~{\bf 6}, 2200031 (2022).


\bibitem{WusShu2013}
W.~Wustmann and V.~Shumeiko,
{\em Parametric resonance in tunable superconducting cavities},
Phys.~Rev.~B {\bf 87}, 184501 (2013).

\bibitem{WusShu2017}
W.~Wustmann and V.~Shumeiko,
{\em Nondegenerate Parametric Resonance in a Tunable Superconducting Cavity},
Phys.~Rev.~Applied {\bf 8}, 024018 (2017).

\bibitem{SvenssonETAL2017}
I.-M.~Svensson, A.~Bengtsson, P.~Krantz, J.~Bylander, V.~Shumeiko, and P.~Delsing, 
{\em Period-tripling subharmonic oscillations in a driven superconducting resonator},
Phys.~Rev.~B {\bf 96}, 174503 (2017).

\bibitem{SvenssonETAL2018}
I.-M.~Svensson, A.~Bengtsson, J.~Bylander, V.~Shumeiko, and P.~Delsing,
{\em Period multiplication in a parametrically driven superconducting resonator}, 
Appl.~Phys.~Lett.~{\bf 113}, 022602 (2018).


\bibitem{YamamotoETAL2014}
Z.R.~Lin, K.~Inomata, K.~Koshino, W.D.~Oliver, Y.~Nakamura, J.S.~Tsai, and T.~Yamamoto,
{\em Josephson parametric phase-locked oscillator and its application to dispersive readout of superconducting qubits},
Nat.~Commun.~{\bf 5}, 4480 (2014).

\bibitem{RosenthalETAL2021}
E.~Rosenthal, C.M.~Schneider, M.~Malnou, Z.~Zhao, F.~Leditzky, B.~Chapman, W.~Wustmann, X.~Ma, D.A.~Palken, L.R.~Vale, G.C.~Hilton, J.~Gao, G.~Smith, G.~Kirchmair, K.~Lehnert, 
{\em Efficient and Low-Backaction Quantum Measurement Using a Chip-Scale Detector},
Phys.~Rev.~Lett.~{\bf 126}, 090503 (2021).



\bibitem{JBA}
I.~Siddiqi, R.~Vijay, F.~Pierre, C.M.~Wilson, M.~Metcalfe, C.~Rigetti, L.~Frunzio, and M.H.~Devoret,
{\em RF-Driven Josephson Bifurcation Amplifier for Quantum Measurement},
Phys.~Rev.~Lett.~{\bf 93}, 207002 (2004).

\bibitem{JBA2}
F.~Mallet, F.~Ong, A.~Palacios-Laloy, F.~Nguyen, P.~Bertet, D.~Vion, and D.~Esteve,
{\em Single-shot qubit readout in circuit quantum electrodynamics},
Nat.~Phys.~{\bf 5}, 791 (2009).

\bibitem{KrantzETAL2016_NatComm}
P.~Krantz, A.~Bengtsson, M.~Simoen, S.~Gustavsson, V.~Shumeiko, W.D.~Oliver,  C.M.~Wilson, P.~Delsing, and J.~Bylander,
{\em Single-shot readout of a superconducting qubit using a Josephson parametric oscillator},
Nat.~Commun.~{\bf 7}, 11417 (2016).


\bibitem{Kerrcatqubit}
A.~Grimm, N.E.~Frattini, S.~Puri, S.O.~Mundhada, S.~Touzard, M.~Mirrahimi, S.M.~Girvin, S.~Shankar, and M.H.~Devoret.
{\em Stabilization and operation of a Kerr-cat qubit},
Nature {\bf 584}, 205 (2020).

\bibitem{Kerrcatqubit2}
X.L.~He, Y.~Lu, D.Q.~Bao, H.~Xue, W.B.~Jiang, Z.~Wang, A.F.~Roudsari, P.~Delsing, J.S.~Tsai, and Z.R.~Lin,
{\em Fast generation of Schr{\"o}dinger cat states using a Kerr-tunable superconducting resonator},
Nat.~Commun.~{\bf 14}, 6358 (2023).


\bibitem{PuriETAL2017_JPO}
S.~Puri, C.K.~Andersen, A.L.~Grimsmo, and A.~Blais,
{\em Quantum annealing with all-to-all connected nonlinear oscillators},
Nat.~Commun.~{\bf 8}, 15785 (2017).


\bibitem{MacklinETAL2015}
C.~Macklin, K.~O’Brien, D.~Hover, M.E.~Schwartz, V.~Bolkhovsky, X.~Zhang, W.D.~Oliver, and I.~Siddiqi, 
{\em A near–quantum-limited Josephson traveling-wave parametric amplifier},
Science {\bf 350}, 307 (2015).

\bibitem{WhiteETAL2015}
T.C.~White, J.Y.~Mutus, I.-C.~Hoi, R.~Barends, B.~Campbell, Y.~Chen, Z.~Chen, B.~Chiaro, A.~Dunsworth, E.~Jeffrey, J.~Kelly, A.~Megrant, C.~Neill, P.J.J.~O'Malley, P.~Roushan, D.~Sank, A.~Vainsencher, J.~Wenner, S.~Chaudhuri, J.~Gao, J.M.~Martinis, 
{\em Traveling wave parametric amplifier with Josephson junctions using minimal resonator phase matching}, 
Appl.~Phys.~Lett.~{\bf 106}, 242601 (2015).



\bibitem{ChenETAL2012}
Y.~Chen, D.~Sank, P.~O'Malley, T.~White, R.~Barends, B.~Chiaro,  J.~Kelly, E.~Lucero, M.~Mariantoni, A.~Megrant, C.~Neill, A.~Vainsencher, J.~Wenner, Y.~Yin, A.N.~Cleland, and J.M.~Martinis,
{\em Multiplexed dispersive readout of superconducting phase qubits},
Appl.~Phys.~Lett.~{\bf 101}, 182601 (2012).

\bibitem{JergerETAL2012}
M.~Jerger, S.~Poletto, P.~Macha,  U.~H{\"u}bner, E.~Il'ichev, and A.V.~Ustinov,
{\em Frequency division multiplexing readout and simultaneous manipulation of an array of flux qubits},
Appl.~Phys.~Lett.~{\bf 101}, 042604 (2012).

\bibitem{ChapmanETAL2017}
B.J.~Chapman, E.I.~Rosenthal,  J.~Kerckhoff, L.R.~Vale, G.C.~Hilton, and K.W.~Lehnert,
{\em Single-sideband modulator for frequency domain multiplexing of superconducting qubit readout},
Appl.~Phys.~Lett.~{\bf 110}, 162601 (2017).



\bibitem{RoyETAL2015}
T.~Roy, S.~Kundu, M.~Chand, A.M.~Vadiraj, A.~Ranadive, N.~Nehra, M.P.~Patankar,
J.~Aumentado, A.A.~Clerk, and R.~Vijay,
{\em Broadband parametric amplification with impedance engineering: Beyond the gain-bandwidth product},
Appl.~Phys.~Lett.~{\bf 107}, 262601 (2015).

\bibitem{MutusETAL2014}
J.Y.~Mutus, T.C.~White, R.~Barends Y.~Chen, Z.~Chen, B.~Chiaro, A.~Dunsworth, E.~Jeffrey, J.~Kelly, A.~Megrant, C.~Neill, P.J.J.~O'Malley, P.~Roushan, D.~Sank, A.~Vainsencher, J.~Wenner, K.M.~Sundqvist, A.N.~Cleland, J.M.~Martinis,
{\em Strong environmental coupling in a Josephson parametric amplifier},
Appl.~Phys.~Lett.~{\bf 104}, 263513 (2014).

\bibitem{GrebelETAL2021}
J.~Grebel, A.~Bienfait, \'{E}.~Dumur, H.-S.~Chang, R.G.~Povey, Y.P.~Zhong, and A.N.~Cleland,
{\em Flux-pumped impedance-engineered broadband Josephson parametric amplifier},
Appl.~Phys.~Lett.~{\bf 118}, 142601 (2021).

\bibitem{DuanETAL2021}
P.~Duan, Z.~Jia, C.~Zhang, L.~Du, H.~Tao, X.~Yang, L.~Guo, Y.~Chen, H.~Zhang, Z.~Peng, W.~Kong, H.~Li, G.~Cao, and G.~Guo,
{\em Broadband flux-pumped Josephson parametric amplifier with an on-chip coplanar
waveguide impedance transformer},
Appl.~Phys.~Express {\bf 14}, 042011 (2021).

\bibitem{WhiteETAL2023}
T.C.~White et al.,
{\em Readout of a quantum processor with high dynamic range Josephson parametric amplifiers},
Appl.~Phys.~Lett.~{\bf 122}, 014001 (2023).


\bibitem{BruusFlensberg_book}
H.~Bruus and K.~Flensberg,
{\em Many-Body Quantum Theory in Condensed Matter Physics},
(Oxford University Press, Oxford, 2013).

\bibitem{GarZol_book}
C.W.~Gardiner, P.~Zoller,
{\em Quantum noise} (Springer, Berlin, 2004).


\bibitem{ParraRodriguezETAL2018}
A.~Parra-Rodriguez, E.~Rico, E.~Solano, and I.L.~Egusquiza:
{\em Quantum networks in divergence-free circuit QED},
Quantum Sci.~Technol.~{\bf 3}, 024012 (2018).

\bibitem{AumentadoETAL2015}
L.~Ranzani and J.~Aumentado, 
{\em Graph-based analysis of nonreciprocity in coupled-mode systems}, 
New J.~Phys.~{\bf 17}, 023024 (2015).

\bibitem{coupled_modes_Purcell_filter}
H.~Yan, X.~Wu, A.~Lingenfelter, Y.J.~Joshi, G.~Andersson,
C.R.~Conner, M.~Chou, J.~Grebel, J.M.~Miller,
R.G.~Povey, H.~Qiao, A.A.~Clerk and A.N.~Cleland.
{\em Broadband bandpass Purcell filter for circuit quantum electrodynamics},
Appl.~Phys.~Lett.~{\bf 123}, 134001 (2023).


\bibitem{LinETAL2013_JPA}
Z.R.~Lin, K.~Inomata, W.D.~Oliver, K.~Koshino, Y.~Nakamura, J.S.~Tsai, T.~Yamamoto, 
{\em Single-shot readout of a superconducting flux qubit with a flux-driven Josephson parametric amplifier},
Appl.~Phys.~Lett.~{\bf 103}, 132602 (2013).


\bibitem{VijayETAL2011}
R.~Vijay, D.H.~Slichter, and I.~Siddiqi,
{\em Observation of quantum jumps in a superconducting artificial atom},
Phys.~Rev.~Lett.~{\bf 106}, 110502 (2011).


\bibitem{SimoenETAL2015}
M.~Simoen, C.W.S.~Chang, P.~Krantz, J.~Bylander, W.~Wustmann, V.~Shumeiko, P.~Delsing, C.M.~Wilson,
{\em Characterization of a multimode coplanar waveguide parametric amplifier},
J.~Appl.~Phys.~{\bf 118}, 154501 (2015).

\bibitem{BengtssonETAL2018}
A.~Bengtsson, P.~Krantz, M.~Simoen, I.-M.~Svensson, B.~Schneider,
V.~Shumeiko, P.~Delsing, and J.~Bylander,
{\em Nondegenerate parametric oscillations in a tunable superconducting resonator},
Phys.~Rev.~B {\bf 97}, 144502 (2018).


\bibitem{Yurke_DruFic2004}
B.~Yurke 
in {\em Quantum Squeezing},
edited by P.D.~Drummond and Z.~Fizek (Springer, Berlin, 2004).


\bibitem{NayfehMook_book}
A.H.~Nayfeh and D.T.~Mook,
{\em Nonlinear Oscillations},
(Wiley, New York, 1995).



\bibitem{complexanalysis_book}
H.~Jeffreys and B.~Swirles,
{\em Methods of mathematical physics}
(Cambridge University Press, Cambridge, 1999).






\end{thebibliography}

%

\end{document}